\documentclass[
  aps,
  prb,
  reprint,
  amsmath,amssymb,
]{revtex4-2}

\usepackage{graphicx}% Include figure files
\usepackage{cleveref}
\usepackage{dcolumn}% Align table columns on decimal point
\usepackage{bm}% bold math
\usepackage{pgfplots}
\usepackage{filecontents}
\usepackage{geometry}
\usepackage{dsfont}
\usepackage{calc}
\usepackage{adjustbox}
\usepackage{physics}
\usepackage{pgfplots}
\usepackage{tikz-3dplot}
\usepackage{bbold}

\pgfplotsset{major grid style={dashed,gray}}

\begin{document}

%\preprint{APS/123-QED}

\title{Finite Projected Entangled Pair States for the Hubbard model}
\author{M. Scheb}
\author{R. M. Noack}%
\affiliation{Fachbereich Physik, Philipps-Universit\"at Marburg, 
	35032 Marburg, Germany}

\date{\today}
\begin{abstract}
We adapt and optimize the projected-pair-entangled-state (PEPS) algorithm on
finite lattices (fPEPS) for two-dimensional Hubbard models and apply
the algorithm to the Hubbard model with
nearest-neighbor hopping on a square lattice.
In particular, we formulate the PEPS algorithm using projected entangled
pair operators, incorporate SU(2) symmetry in all tensor
indices, and optimize the PEPS using both
iterative-diagonalization-based local bond optimization and gradient-based
optimization of the PEPS.
We discuss the performance and convergence of the algorithm for the
Hubbard model on lattice
sizes of up to $8\times 8$ for PEPS states with U(1) symmetric bond dimensions
of up to $D=8$ and SU(2) symmetric bond
dimensions of up to $D=6$.
Finally, we comment on the relative and overall efficiency of schemes
for optimizing fPEPS.
\end{abstract}

\maketitle

\section{Introduction}
One of the most important models used to study strongly correlated
electron systems is the Hubbard model \cite{hubbard}.
Originally, it was conceived to model
narrow energy bands within many-body
quantum systems in order to study the behavior of observables at low
temperatures and, in particular, to study the Mott transition 
\cite{Mott1948,Mott1968Oct}.
With the discovery of high-temperature 
superconductors in the cuprates in 1986 \cite{Bednorz1986Jun},
the focus of interest shifted
to using the
two-dimensional Hubbard model
to model
the electronic transport within
copper-dioxide planes common to all the high-$T_c$ cuprates
\cite{Scalapino1986Dec,Scalapino1987May,Scalapino1995Jan,Monthoux1991Dec}.
A tendency towards superconducting
pairing with d-wave symmetry found in numerical and in
perturbation-theory calculations \cite{White1989Jan} reinforced 
the picture of the two-dimensional Hubbard model as a minimal model
for high-$T_c$ superconductivity in the cuprates.
Experimental studies by Tranquada and coworkers
\cite{Tranquada1995Jun,Tranquada1996Sep,Tranquada1997Jan} further expanded 
knowledge of 
phenomena found in cuprates
in that stripe structures, consisting of oscillating
spins and charge densities, were found to be present in the underdoped
region in a variety of high-$T_c$ cuprates.
These stripe phases generally compete
with the superconducting pairing, so that superconductivity is
suppressed, especially at particular commensurate band fillings.
Subsequent studies of the t-J-model \cite{White1997Jun} and the underdoped 
Hubbard model \cite{Zheng2017Dec} 
were able to qualitatively reproduce aspects of this characteristic
behavior.

Despite these and many other insights gained into the behavior of the
two-dimensional Hubbard 
model within and outside of 
the context of high-temperature
superconductors,
many
aspects of the low-temperature behavior
remain
contentious, such as the competition of different pairing
orders \cite{Zheng2017Dec}, the proper description of topological phases, and the overall behavior
in the thermodynamic limit in two dimensions \cite{Arovas2022Mar,Qin2022Mar}.

Many analytical and numerical methods exist for studying
strongly correlated quantum systems in general and the Hubbard model
in particular.
For the ground-state 
calculation in one dimension, the most efficient numerical method
is the density 
matrix renormalization group (DMRG) \cite{White1992Nov,White1993Oct}, 
in which 
the wave function is
efficiently approximated
as a matrix product state (MPS) 
\cite{Ostlund1995Nov}.
Coupled chains as well as
more general two-dimensional systems of limited width
have been studied extensively using the DMRG 
\cite{Noack1994Aug,White1994Aug,Noack1995Apr,
Noack1996Oct,White1998Feb,Noack1997Sep,
White1998Oct,White1999Jul,Yan2011Jun,Stoudenmire2012Feb,Depenbrock2012Aug,leblanc,
Ehlers2015Dec,Ehlers2017Mar,Zheng2017Dec,Ehlers2018Jan} by mapping
the two-dimensional lattice onto the intrinsically one-dimensional MPS.
However, due to the linear increase of entropy and thus exponential
increase of computational effort
with the width of the lattice in two dimensions, 
the thermodynamic limit will probably remain inaccessible
\cite{Hastings2007Aug,Wolf2008Feb}.
In order to circumvent this exponential computational hurdle,
Verstraete \textit{et al.}\
introduced Projected Entangled Pair
States (PEPS) \cite{Verstraete2004Jul}, which are the natural generalization of 
MPSs to two dimensions.
It has been
shown that PEPSs are the ground
states of local Hamiltonians
\cite{Perez-Garcia2007Jul,Schuch2010Oct,Verstraete2006Jun} and are thus promising
candidates for studying two-dimensional quantum systems. 
Murg \textit{et~al.}\ conducted one of the first numerical simulations and provided
benchmark results for hard-core bosons \cite{Murg2007Mar} and
frustrated spin systems \cite{Murg2009May}.
Further PEPS-based simulations were conducted by Lubasch and coworkers
\cite{Lubasch2014Mar,Lubasch2014Aug}, who focused on the
Heisenberg model.

Development and application of methods to treat fermions
using PEPS has
been led by Corboz and coworkers, who have concentrated primarily on
the infinite PEPS (iPEPS) algorithm
originally introduced by Jordan \textit{et al.}\ \cite{Jordan2008Dec}.
In this method, translational invariance
is emulated by replicating a unit cell infinitely in all
spatial dimensions.
To distinguish the original version of PEPS with open boundary conditions from 
iPEPS,
we designate it finite PEPS (fPEPS) here.
In a series of papers \cite{Corboz2010Apr,Corboz2011Jul,Corboz2010Dec,Corboz2014Jul},
Corboz \textit{et~al.}\ used both fPEPS and iPEPS to examine the t-J-model,
a simplified version of the Hubbard model.
They found the
stripe structure, in which both the charge and the 
spin density of electrons oscillate over multiple sites and compete with d-wave pairing.
In recent years, the most prominent variation of PEPS algorithms has
been 
iPEPS optimized via imaginary time evolution. In addition, gradient-optimizations
of the entire PEPS have proven to be successful 
\cite{Vanderstraeten2016Oct,Liu2017May,Dong2019May}.

In this work, we revisit the original finite PEPS algorithm, which treats a finite
lattice with open boundary conditions.
Working on a specific finite lattice  has the decided advantage that
energies and observables can be compared directly to those obtained
with other finite-lattice numerical methods, in particular, with exact
diagonalization and with the DMRG, 
both of which are strictly variational.
We note that PEPS-based algorithms are not strictly variational in
general because the contraction of a two-dimensional tensor network
must be carried out approximately in order to be efficient
\cite{Verstraete2004Jul,Verstraete2006Jun}.
In addition, optimization of PEPS states has typically been carried
out using imaginary-time evolution within the Trotter approximation,
which introduces an additional systematic Trotter error, further complicating
a variational comparison.
Therefore, our comparison will give us a stringent test bed for
evaluating the accuracy and convergence behavior of the fPEPS as well
as determining
to what extent it obeys the variational principle.

In developing our variant of the fPEPS algorithm, we have made an
effort 
to incorporate as large a scope of modern ideas and methods as
possible. 
In particular, the fact that MPS-based algorithms have profited significantly
from the formulation of matrix-product operators (MPOs) has
motivated us to develop a general scheme for formulating fPEPS
algorithms in terms of the 
generalization of MPOs, projected entangled pair operators (PEPOs)
\footnote{To our knowledge, the PEPO idea was first introduced by Crosswhite in
Ref.~\cite{Crosswhite2008Jul}}.
Since the Hubbard model at general band filling and zero magnetic
field has SU(2) symmetry in the spin sector and U(1) symmetry in the
charge sector, we incorporate these symmetries explicitly into our
construction of PEPSs and PEPOs.
In addition, we build bookkeeping for the fermionic sign into our 
PEPOs, 
which is a different formulation than the original
one of Refs.~\cite{Corboz2009Oct,Corboz2010Apr} in terms of fermionic
swap gates.
For MPS algorithms, variational iterative diagonalization of a local
effective Hamiltonian is a powerful tool for optimizing an MPS and
forms the basis of the DMRG algorithm.
We adapt a scheme for carrying out a local-bond-based
variational optimization of a PEPS \cite{Corboz2016Jul} to fPEPS,
formulating variants that carry out single-site as well as bond
optimization.
In order to reduce the size of the effect Hilbert spaces that must be
treated within local variational optimization, we generalize the
recently formulated controlled bond expansion (CBE) of
Gleis \textit{et~al.}~\cite{Gleis2022Jul}
to reduce the numerical cost of  local bond optimization in fPEPS.
We also apply an alternative optimization scheme for PEPS based on a gradient of the
energy functional that was first applied by Vanderstraeten
\textit{et al.}~\cite{Vanderstraeten2016Oct} to PEPS algorithms
to fPEPS.
In addition, we have reexamined and reworked the scheme for
approximately contracting an fPEPS network first described in
Ref.~\cite{Verstraete2004Jul}, concentrating on improving stability and
efficiency; here we have found a scheme to reduce the numerical effort
needed to optimize the contracted environment using an adaptation of the
CBE mentioned above~\cite{Gleis2022Jul}.

Clearly, our method is tailored for short-ranged two-dimensional
strongly interacting quantum lattice models, especially ones with
fermionic degrees of freedom and SU(2) spin symmetry.
Thus, our method is intended to treat the two-dimensional Hubbard
model, its extensions, as well as other related models (e.g.,
Heisenberg models).
As a benchmark system, we take the usual two-dimensional Hubbard model
with nearest-neighbor hopping on a finite square lattice with open
boundary conditions, i.e., treat the Hamiltonian
\begin{align}
\mathcal{H} &= - \, t\sum\limits_{\left\langle i,j\right \rangle,\sigma} \left( c^{\dagger}_{i,\sigma} \,
   c_{j,\sigma} + \text{h.c.}\right) \nonumber \\
 & \quad + U\sum\limits_{i} n_{i,\uparrow} \, n_{i,\downarrow} \, ,
   \label{eq:usualhubbardham}
\end{align}
where
\begin{equation*}
\quad  n_{i,\sigma} = c^{\dagger}_{i,\sigma} \, c_{i,\sigma} 
\end{equation*}
is the local particle-density operator.
The notation $\left\langle i,j\right\rangle$ indicates
nearest-neighbor sites on an $L\times L$ lattice with open boundary
conditions.
The total number of sites is then $V=L^2$.
We work in the canonical ensemble, so that the numbers of spin-up
electrons  $N_\uparrow$ and spin-down electrons $N_\downarrow$ or,
equivalently, the total particle number
$N \equiv N_\uparrow + N_\downarrow$ and the $z$-component of the total
spin $S_z \equiv (N_\uparrow - N_\downarrow)/2$, are conserved.
Conservation of total particle number corresponds also to the
conservation of the deviation from half filling
$C_z \equiv (N_\uparrow + N_\downarrow - V)/2$, which is the charge-sector
analog of $S_z$.
Here we will be interested in non-magnetized ground states, so we will
take $S_z=0$ in this work.
Ground states in the $S_z=0$ sector generically have total spin $S=0$,
so that we will concentrate on this case for calculations that
explicitly take the SU(2) spin symmetry into account.
We additionally specify the overall band filling using the average particle number
$\langle n \rangle = N/V$
and take $t=1$ to be the scale of the Hamiltonian so
that there are two independent physical parameters, the on-site
interaction $U$ and the average particle number $\langle n \rangle$.

This paper is organized as follows:
Sec.~\ref{sec:peps} gives a concise definition of PEPSs 
and their entanglement structure.
In Sec.~\ref{sec:pepo}, we construct 
the corresponding Projected Entangled Pair Operator (PEPO), which is the Hamiltonian
adapted to the topology of the PEPS.
Sec.~\ref{sec:su2} introduces a generic
scheme to incorporate SU(2)-symmetries into the PEPS-PEPO framework.
In Sec.~\ref{sec:pepsexp}, we explain in
detail how to calculate expectation values approximately within the
PEPS-PEPO scheme.
Sec.~\ref{sec:varoptpeps} presents
two different 
optimization procedures; the first is
comprised of local updates implemented using a sequence of iterative diagonalizations
of local effective Hamiltonians, similar to that used in the DMRG.
The second procedure is
based on direct gradient optimization
of the PEPS tensors.
We find that a combination of the two procedures provides the best
results.
In Sec.~\ref{sec:results}, we present
benchmark results for the two-dimensional Hubbard model, discussing
accuracy and the convergence and its dependence on overall bond
dimension and on environment dimension for lattices sizes ranging from
$3 \times 3$ to $8 \times 8$.
Finally, we discuss the convergence issues and
the state of development of 
the fPEPS-PEPO scheme in Sec.~\ref{sec:discussion}.

\section{Projected entangled pair state scheme}

In this section, we will define the states and operators used in
PEPS-based methods, concentrating on the finite-lattice PEPS
algorithm, as first proposed in Ref.~\cite{Verstraete2004Jul}.
We use a tensor-network formulation to define operators and apply them
to a PEPS, i.e., we work in terms of
PEPOs, 
analogous to Matrix Product Operators (MPOs).
The extension of an MPO to a more general tensor network can be termed
a Tensor Network Operator (TNO), of which a PEPO is a special case.

We will first give a general definition of an fPEPS,
followed by a definition of a PEPO and its construction.

\subsection{Finite projected entangled pair state}
 \label{sec:peps}

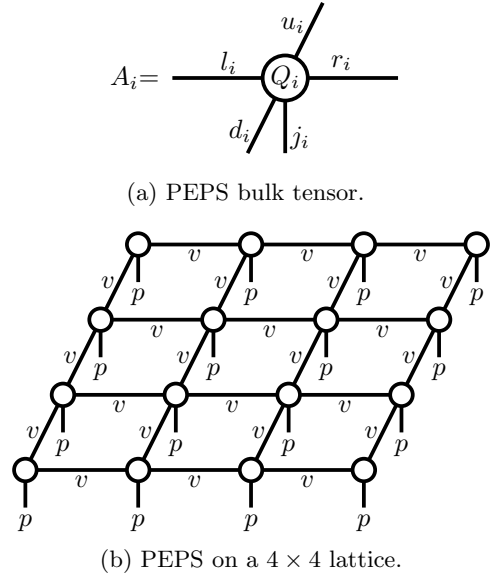
\begin{figure}[!htb]
\begin{minipage}{0.90\columnwidth}
\begin{center}
\begin{tikzpicture}[font=\tiny]
\node[align=center] at (-2,0) {\normalsize $A_i$\color{black}{$=$}};

\draw[line width=0.5mm] (0,0) -- (1.5,0);
\node[align=center] at (0.75,0.2) {\normalsize $r_i$};
\draw[line width=0.5mm] (0,0) -- (-1.5,0);
\node[align=center] at (-0.75,0.2) {\normalsize $l_i$};
\draw[line width=0.5mm] (0,0) -- (0.5,1);
\node[align=center] at (0.1,0.7) {\normalsize $u_i$};
\draw[line width=0.5mm] (0,0) -- (-0.5,-1);
\node[align=center] at (-0.6,-0.7) {\normalsize $d_i$};

\draw[line width=0.5mm] (0,0) -- (0,-1);
\draw[fill=white, line width=0.5mm] (0,0) circle [radius=0.3];
\node[align=center] at (0,0) {\normalsize $Q_i$};
\node[align=center] at (0.2,-0.8) {\normalsize $j_i$};
\end{tikzpicture}

\vspace*{0.2cm}
{\small (a) PEPS bulk tensor.}
\vspace*{0.3cm}
\end{center}
 \end{minipage}
\hfill
\begin{minipage}{0.9\columnwidth}
\begin{tikzpicture}[font=\tiny]

\foreach \x in {0,1.5,3}{
   \draw[line width=0.5mm] (\x,0) -- (\x+1.5,0);
   \node[align=center] at (\x+0.75,-0.15) {\normalsize $v$};
}
\foreach \x in {0.5,2,3.5}{
   \draw[line width=0.5mm] (\x,1) -- (\x+1.5,1);
   \node[align=center] at (\x+0.75,0.85) {\normalsize $v$};
}
\foreach \x in {1,2.5,4}{
   \draw[line width=0.5mm] (\x,2) -- (\x+1.5,2);
   \node[align=center] at (\x+0.75,1.85) {\normalsize $v$};
}
\foreach \x in {1.5,3,4.5}{
   \draw[line width=0.5mm] (\x,3) -- (\x+1.5,3);
   \node[align=center] at (\x+0.75,2.85) {\normalsize $v$};
}

\foreach \x in {0,1.5,3,4.5}{
   \draw[line width=0.5mm] (\x,0) -- (\x,-0.5);
   \draw[line width=0.5mm] (\x,0) -- (\x+0.5,1);
   \draw[fill=white, line width=0.5mm] (\x,0) circle [radius=0.15];
   \node[align=center] at (\x,-0.7) {\normalsize $p$};
   \node[align=center] at (\x+0.1,0.5) {\normalsize $v$};
}
\foreach \x in {0.5,2,3.5,5}{
   \draw[line width=0.5mm] (\x,1) -- (\x,0.5);
   \draw[line width=0.5mm] (\x,1) -- (\x+0.5,2);
   \draw[fill=white, line width=0.5mm] (\x,1) circle [radius=0.15];
   \node[align=center] at (\x,0.3) {\normalsize $p$};
   \node[align=center] at (\x+0.1,1.5) {\normalsize $v$};
}
\foreach \x in {1,2.5,4,5.5}{
   \draw[line width=0.5mm] (\x,2) -- (\x,1.5);
   \draw[line width=0.5mm] (\x,2) -- (\x+0.5,3);
   \draw[fill=white, line width=0.5mm] (\x,2) circle [radius=0.15];
   \node[align=center] at (\x,1.3) {\normalsize $p$};
   \node[align=center] at (\x+0.1,2.5) {\normalsize $v$};
}
\foreach \x in {1.5,3,4.5,6}{
   \draw[line width=0.5mm] (\x,3) -- (\x,2.5);
   \draw[fill=white, line width=0.5mm] (\x,3) circle [radius=0.15];
   \node[align=center] at (\x,2.3) {\normalsize $p$};
}

\end{tikzpicture}
\vspace*{0.2cm}
\centerline{\small (b) PEPS on a $4\times 4$ lattice.}
\end{minipage}

\caption{Projected Entangled Pair State.}
\label{fig:pepsdef}
\end{figure}

Consider a generic many-body wave function on N sites
\begin{align}
\left| \psi \right> = \sum\limits_{j_1,...,j_N} \psi_{j_1,...,j_N}
	\left| j_1,...,j_N \right> \, . \label{eq:state}
\end{align}
Here $\left| j_1,...,j_N \right> = 
	\left| j_1 \right> \otimes ... \otimes \left| j_N \right>$
are many-body basis states with $j_i=1,...,d$, where $d$ is the dimension of 
a local Hilbert space.
A convenient way to store this wave function on a 
two-dimensional lattice is as a
PEPS 
as 
introduced by Verstraete 
and Cirac in 2004 \cite{Verstraete2004Jul}. 
In a PEPS, 
each bond between adjacent sites is associated with a 
maximally entangled state
\begin{align}
\left| \phi \right> = \sum\limits_{k=1}^D \left| k, k \right> \label{eq:maxstate}
\end{align}
in a virtual Hilbert space, where $D$ is the maximum bond dimension.
Conversely, this means that a bulk site $i$ is connected to four virtual 
bonds $u_i$, $d_i$, $l_i$, and $r_i$.
If we define a 
projector $Q_i$ that 
maps virtual bonds at site $i$ to the physical bond $j_i$, 
we can define the 
rank-5 PEPS tensor
\begin{align}
A_i = A^{j_i}_{u_i,d_i,l_i,r_i} = \left< j_i \right| Q_i \left| u_i,d_i,l_i,r_i \right> \, , \label{eq:pepstensor}
\end{align}
which is depicted graphically \cite{Penrose1971}
in Fig.~\ref{fig:pepsdef}(a).  
Tensors at the edges or the corners of the lattice are rank-4 and rank-3, 
accordingly. There is no straightforward way of writing the entire PEPS as an 
explicit equation due to the two-dimensional arrangement of tensors, 
which is why we define it simply as
\begin{align}
\left| \psi \right> = \sum\limits_{j_1,...,j_N} \mathcal{F}
\left( A_1 \, A_2 \, ... \, A_N \right) \left| j_1, ..., j_N \right> , \nonumber
\\ \quad 
\mathcal{F}\left( A_i  \, A_2 \, ... \, A_N \right) = \psi_{j_1,...,j_N} 
\end{align}
for a lattice of $N$ sites, where $\mathcal{F}$ is a function that contracts 
common indices in PEPSs. Fig.~\ref{fig:pepsdef}(b) 
illustrates this arrangement 
on a $4\times 4$ lattice with virtual indices $v$ and physical indices
$p$.
Note that the construction of $A_i$ via $Q_i$ is mainly academic; 
in practical simulations, a PEPS will be initialized as a product state with 
virtual indices of dimension one or via random tensors.
As the optimization proceeds, the dimensions 
then may change and lead to states which are, in general, not maximally entangled.

PEPSs are promising candidates for describing ground states in 
two-dimensional quantum systems 
\cite{Schuch2010Oct,Schuch2011Oct,Verstraete2006Jun,Lubasch2014Aug,
Corboz2016Jul,Cirac2021Dec}. 
Their introduction, however, leads to two complications: 
First, the exact calculation of 
expectation values $\left< \psi \right| O \left| \psi \right>$ scales
exponentially with system size, 
which is why it has to be
carried out 
approximately, and, second, due to the high 
rank of bulk tensors, PEPS-based algorithms scale with a high power of the 
number of virtual states $D$.
We will discuss these complications as well as ways to deal with them
in more detail in the following.

\subsection{Projected entangled pair operators} \label{sec:pepo}
Consider an electronic many-body system on a two-dimensional square
lattice that 
is described by the Hubbard Hamiltonian \eqref{eq:usualhubbardham},
which we rewrite slightly as
\begin{align}
\mathcal{H} = -t\sum\limits_{\left\langle i,j\right\rangle,\sigma} \left(
c^\dagger_{i,\sigma} \, c_{j,\sigma} + \text{h.c.} \right) \nonumber \\
+ \, U\sum\limits_i c^\dagger_{i,\uparrow} c^\dagger_{i,\downarrow} c_{i,\downarrow} c_{i,\uparrow} \, .
\label{eq:hubbardham}
\end{align}
The fermionic creation and annihilation operators can be written in
terms of their action on the four basis states 
$\left\{\left|0\right>,\left|\uparrow\right>,\left|\downarrow\right>,\left|\uparrow\downarrow\right>\right\}$
spanning the local Hilbert space:
\begin{align*}
c^\dagger_{\uparrow} &= \left|\uparrow\right>\left<0\right| + \left|\uparrow\downarrow\right>\left<\downarrow\right| \ , \\
c_{\uparrow} &= \left|0\right>\left<\uparrow\right| + \left|\downarrow\right>\left<\uparrow\downarrow\right| \, ,\\
c^\dagger_{\downarrow} &= \left|\downarrow\right>\left<0\right| - \left|\uparrow\downarrow\right>\left<\uparrow\right| \, ,\\
c_{\downarrow} &= \left|0\right>\left<\downarrow\right| -
\left|\uparrow\right>\left<\uparrow\downarrow\right| \, .
\end{align*}

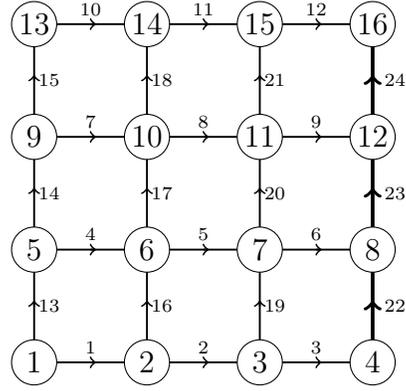
\begin{figure}[!hb]
\centering
\begin{tikzpicture}[font=\large]
  \def \fac {1.5}

	\newcounter{c}
	\stepcounter{c}
	\foreach \y in {0,1,2,3}{
		\foreach \x in {0,1,2}{
			\draw [->,line width=0.25mm] (\fac*\x,\fac*\y) -- (\fac*\x+0.55*\fac,\fac*\y);
			\draw [->,line width=0.25mm] (\fac*\x+0.5*\fac,\fac*\y) -- (\fac*\x+\fac,\fac*\y);
			\node[align=center] at (\fac*\x+\fac*0.5,\fac*\y+0.2) {\scriptsize \arabic{c}};
			\stepcounter{c}
		}
	}

	\foreach \x in {0,1,2}{
		\foreach \y in {0,1,2}{
			\draw [->,line width=0.25mm] (\fac*\x,\fac*\y) -- (\fac*\x,\fac*\y+0.55*\fac);
			\draw [->,line width=0.25mm] (\fac*\x,\fac*\y+0.5*\fac) -- (\fac*\x,\fac*\y+\fac);
			\node[align=center] at (\fac*\x+0.2,\fac*\y+\fac*0.5) {\scriptsize \arabic{c}};
			\stepcounter{c}
		}
	}

	\foreach \y in {0,1,2}{
		\draw [->,line width=0.5mm] (\fac*3,\fac*\y) -- (\fac*3,\fac*\y+0.55*\fac);
		\draw [->,line width=0.5mm] (\fac*3,\fac*\y+0.5*\fac) -- (\fac*3,\fac*\y+\fac);
		\node[align=center] at (\fac*3+0.3,\fac*\y+\fac*0.5) {\scriptsize \arabic{c}};
		\stepcounter{c}
	}

	\newcounter{counter}
	\stepcounter{counter}
	\foreach \y in {0,1,2,3}{
		\foreach \x in {0,1,2,3}{
			\draw[fill=white] (\fac*\x,\fac*\y) circle [radius=0.3];
			\node[align=center] at (\fac*\x,\fac*\y) {\arabic{counter}};
			\stepcounter{counter}
		}
	}
\end{tikzpicture}
\caption{\label{fig:pepo} Sites and signaling channels in between for a $4 \times 4$ lattice}
\end{figure}

We wish to store the
expression corresponding to Eq.~\eqref{eq:hubbardham}
as a
PEPO,
which has the same topology as the PEPS to which it is supposed to be 
applied. 
We start by expanding the full Hamiltonian for a lattice of a given width 
and height. For example, for a $4 \times 4$ lattice with open boundary conditions, the full operator reads, schematically,
\begin{align*}
\mathcal{H} &= \left(U c^\dagger_{\uparrow} c^\dagger_{\downarrow} c_{\downarrow} c_{\uparrow} \right)_1
	\otimes \mathds{1}_2 \otimes \mathds{1}_3 \otimes \, \text{...} \, \otimes \mathds{1}_{16} \\
 &+ \mathds{1}_1 \otimes \left(U c^\dagger_{\uparrow} c^\dagger_{\downarrow} c_{\downarrow} c_{\uparrow} \right)_2
	\otimes \mathds{1}_3 \otimes \, \text{...} \, \otimes \mathds{1}_{16} \\
 & ... \\
 &+ \mathds{1}_1 \otimes \mathds{1}_2 \otimes \, \text{...} \, \otimes \mathds{1}_{15} \otimes
	\left(U c^\dagger_{\uparrow} c^\dagger_{\downarrow} c_{\downarrow} c_{\uparrow} \right)_{16} \\
& + \left(-t \mathcal{P} c^\dagger_\uparrow\right)_1 \otimes \left(c_\uparrow\right)_2 \otimes \mathds{1}_3 \otimes \, ... \, 
	\otimes \mathds{1}_{16} \\
& + \mathds{1}_1 \otimes \left(-t \mathcal{P} c^\dagger_\uparrow\right)_2 \otimes \left(c_\uparrow\right)_3 \otimes \mathds{1}_4 \otimes \, ... \, 
	\otimes \mathds{1}_{16} \\
& ... \\
& + \left(-t \mathcal{P} c^\dagger_\uparrow\right)_1 \otimes \mathcal{P}_2 \otimes \mathcal{P}_3 \otimes \mathcal{P}_4
		\otimes \left(c_\uparrow\right)_5 \\
& \quad \otimes \mathds{1}_6 \otimes \, ... \, \otimes \mathds{1}_{16} \\
& ... 
\end{align*}
where the parity operator
\begin{align*}
\mathcal{P} = \left|0\right>\left<0\right| - \left|\uparrow\right>\left<\uparrow\right|
- \left|\downarrow\right>\left<\downarrow\right| + \left|\uparrow \downarrow\right>\left<\uparrow \downarrow\right|
\end{align*}
encodes the effect of the fermionic sign. (Alternatively, one could incorporate the sign 
into the PEPS through fermionic swap gates \cite{Corboz2009Oct,Corboz2010Apr}.)
The subscripts denote the physical sites on which the operators act, 
whereas the enumeration is chosen according to the scheme depicted 
in Fig.~\ref{fig:pepo}.\\

Inspired by the construction of matrix product operators via
finite-state machines 
(FSMs), which was introduced by Crosswhite
\textit{et al.}~\cite{Crosswhite2008Jul,Crosswhite2008Jul2}
and refined by Paeckel \textit{et al.}~\cite{Paeckel2017Nov}, we now conceptualize
$\mathcal{H}$ as a regular expression. In particular, for each unique sequence of operators 
on one site, we substitute one symbol
($U c^\dagger_{\uparrow} c^\dagger_{\downarrow} c_{\downarrow}
c_{\uparrow} \rightarrow \text{U}, \ 
-t \mathcal{P} c^\dagger_\uparrow \rightarrow \text{A}, \  
c_\uparrow \rightarrow \text{B}, \ 
-t \mathcal{P} c^\dagger_\downarrow \rightarrow \text{C}, \ 
c_\downarrow \rightarrow \text{D}, \ 
\mathds{1} \rightarrow \text{I}, \ 
\mathcal{P} \rightarrow \text{P}$). Tensor products connecting different local Hilbert spaces are then interpreted as concatenations,
whereas the sums become unions. In this way, all symbols form an alphabet, each summand of the many-body Hamiltonian
becomes a word, and the set of all words forms a regular language:
\begin{center}
H = \{ UIIIIIIIIIIIIIII, IUIIIIIIIIIIIIII, ... , IIIIIIIIIIIIIIIU, ABIIIIIIIIIIIIII, IABIIIIIIIIIIIII, ... , APPPBIIIIIIIIIII, ...\} \, .
\end{center}
Here we drop the subscripts---they are determined by the position 
of a symbol in the sequence.

The PEPO we want to form will be a set of interconnected FSMs, which generate 
all of the words of the language in a two-dimensional fashion. 
In order to do this, we connect adjacent 
sites through directed signaling channels, as depicted in Fig.~\ref{fig:pepo}.
The flow of information is defined as going upwards and to the right. Information at the
upper boundary is dropped, while information at the right boundary is 
redirected to the top,
which makes the rightmost vertical channels of the lattice the trunk. 
The upper-right corner (site $16$ in Fig.~\ref{fig:pepo}) is thus the
sink, i.e., the location at which all the
information passed by the FSMs is gathered.
Here it is determined whether a string of symbols
is actually a part of the language and, hence, is a meaningful term of
the Hamiltonian.
For example, channel $5$ reports what symbols have occurred on
sites $5$ and $6$, and channel $21$ what symbols have occurred on
sites $3$, $7$, and $11$. 
Channel $23$, as a trunk channel, 
carries information from the entire block below, i.e., from all sites from $1$ to $8$. 
In this way, each FSM can be seen as an associative tensor or table in which a set of 
incoming and outgoing states forms a key, the associated symbol represents 
their value, and 
the pair of both is an element.
The rank is equal to the coordination number, which is two at the corners, 
three at the edges, and four otherwise. 
As an example, the FSM on site $6$ is schematically described in Table
\ref{tab:fsm6}.
\begin{table}[!hb]
\caption{\label{tab:fsm6}
  Finite-state machine for site $6$ implemented as an associative
  tensor.
  For the incoming states $s_{4,i}$ from channel $4$ and $s_{16,i}$ from
  channel $16$, the symbol $\sigma_i$ is inserted into the current word,
  and $s_{5,i}$ and $s_{17,i}$ are emitted via channels $5$ and $17$.}
\begin{tabular}{cccc|c}
C 4& C 16& C 5 & C 17 & S 6 \\
\hline
$s_{4,1}$&$s_{16,1}$&$s_{5,1}$&$s_{17,1}$& $\sigma_1$\\
$s_{4,2}$&$s_{16,2}$&$s_{5,2}$&$s_{17,2}$& $\sigma_2$\\
$s_{4,3}$&$s_{16,3}$&$s_{5,3}$&$s_{17,3}$& $\sigma_3$\\
$\ldots$
\end{tabular}

\end{table}

We define three distinct
states, $s_i$, $s_f$, and $s_P$,
which designate the initial state, 
the final state, and the parity state, respectively.
The symbol $s_i$ indicates that,
up to this point,
only identities (designated as ``I'') have appeared,
and the first occurrence of a nontrivial symbol is pending.
The symbol $s_f$ designates the complement,
namely that a valid combination of nontrivial symbols has already appeared 
and, after the current point, 
only identity symbols 
are allowed to be attached to the word. 
Finally, the symbol $s_P$ takes the parity operators (designated as ``P'')
from the Jordan-Wigner transformation into account and connects vertical
hopping terms in a way that will be elucidated via an example
later on.
These states, together with the two trivial symbols ``I'' and ``P'', 
are used to initialize the PEPO as depicted in Tables 
\ref{tab:centerinit} and \ref{tab:steminit}.
Note how there are no elements with two incoming final states $s_f$, 
which prevents different words from mixing with each other.
\begin{table}[!hb]
\centering
\caption{\label{tab:centerinit}Initial elements of a bulk tensor/FSM.
  Here C $I_l$ and C $I_b$ are the two incoming channels from left and
  bottom, and C $O_r$ and C $O_t$ are the outgoing channels to the
  right and top, respectively.}
\begin{tabular}{cccc|c}
C $I_l$& C $I_b$& C $O_r$ & C $O_t$ & S \\
\hline
$s_{i}$&$s_{i}$&$s_{i}$&$s_{i}$& I \\
$s_{i}$&$s_{f}$&$s_{i}$&$s_{f}$& I \\
$s_{f}$&$s_{i}$&$s_{f}$&$s_{i}$& I \\
$s_{i}$&$s_{P}$&$s_{i}$&$s_{P}$& I \\
$s_{f}$&$s_{P}$&$s_{f}$&$s_{P}$& I \\
$s_{P}$&$s_{i}$&$s_{P}$&$s_{P}$& P
\end{tabular}
\end{table}
\begin{table}[!hb]
\centering
\caption{\label{tab:steminit}Initial elements of a trunk tensor/FSM.
  Here C $I_l$ and C $I_b$ are the two incoming channels from left and
  bottom, and $O_t$ is the outgoing channel to the top.}
\begin{tabular}{ccc|c}
C $I_l$& C $I_b$& C $O_t$ & S \\
\hline
$s_{i}$&$s_{i}$&$s_{i}$& I \\
$s_{f}$&$s_{i}$&$s_{f}$& I \\
$s_{i}$&$s_{f}$&$s_{f}$& I \\
$s_{f}$&$s_{P}$&$s_{f}$& I \\
$s_{P}$&$s_{i}$&$s_{P}$& P
\end{tabular}
\end{table}

After all of these preparations, 
the Hamiltonian can now be converted into a PEPO by recasting every
word into a set of nontrivial pairs, with the first element being a 
nontrivial symbol and the second the index of the site it acts on. 
Each pair is then inserted into the PEPO by adding an element to 
the tensor of the respective 
site.
In this way, the entire PEPO can be systematically constructed word for word.

For example, the word ``IIIIIUIIIIIIIIII'' translates
as $\left\{\left(U,6\right)\right\}$ and is incorporated into the PEPO by adding the element
$\left\{\left(s_i,s_i,s_f,s_f\right),U\right\}$ to the tensor at site $6$ in accordance with 
Table \ref{tab:fsm6}.
The remaining $15$ identity symbols
not explicitly contained
in the set are taken care of by the previous initialization.

The next type of words are those which represent horizontal hopping terms, such as
``IIIIIABIIIIIIIII'', as
is depicted in Fig.~\ref{fig:AB}.
The set of pairs for this case reads 
$\left\{\left(A,6\right),\left(B,7\right)\right\}$ and the elements which need to be 
attached to the tensors of site $6$ and $7$ are $\left\{\left(s_i,s_i,\left\{\left(A,6\right)\right\},s_f\right),A\right\}$
and
$\left\{\left(\left\{\left(A,6\right)\right\},s_i,s_f,s_f\right),B\right\}$,
respectively.
\begin{figure}[!hb]
\centering
\begin{tikzpicture}[font=\large]
	\def \fac {1.5}

	\foreach \y in {0,1,2,3}{
		\foreach \x in {0,1,2}{
			\draw [->,line width=0.25mm] (\fac*\x,\fac*\y) -- (\fac*\x+0.55*\fac,\fac*\y);
			\draw [->,line width=0.25mm] (\fac*\x+0.5*\fac,\fac*\y) -- (\fac*\x+\fac,\fac*\y);
		}
	}
	\node[align=center] at (\fac*0+\fac*0.5,\fac*0+0.2) {\scriptsize $s_i$};
	\node[align=center] at (\fac*1+\fac*0.5,\fac*0+0.2) {\scriptsize $s_i$};
	\node[align=center] at (\fac*2+\fac*0.5,\fac*0+0.2) {\scriptsize $s_i$};

	\node[align=center] at (\fac*0+\fac*0.5,\fac*1+0.2) {\scriptsize $s_i$};
	\node[align=center] at (\fac*1+\fac*0.5,\fac*1+0.2) {\scriptsize $\left\{\left(A,6\right)\right\}$};
	\node[align=center] at (\fac*2+\fac*0.5,\fac*1+0.2) {\scriptsize $s_f$};

	\node[align=center] at (\fac*0+\fac*0.5,\fac*2+0.2) {\scriptsize $s_i$};
	\node[align=center] at (\fac*1+\fac*0.5,\fac*2+0.2) {\scriptsize $s_i$};
	\node[align=center] at (\fac*2+\fac*0.5,\fac*2+0.2) {\scriptsize $s_i$};

	\node[align=center] at (\fac*0+\fac*0.5,\fac*3+0.2) {\scriptsize $s_i$};
	\node[align=center] at (\fac*1+\fac*0.5,\fac*3+0.2) {\scriptsize $s_i$};
	\node[align=center] at (\fac*2+\fac*0.5,\fac*3+0.2) {\scriptsize $s_i$};

	\foreach \x in {0,1,2}{
		\foreach \y in {0,1,2}{
			\draw [->,line width=0.25mm] (\fac*\x,\fac*\y) -- (\fac*\x,\fac*\y+0.55*\fac);
			\draw [->,line width=0.25mm] (\fac*\x,\fac*\y+0.5*\fac) -- (\fac*\x,\fac*\y+\fac);
		}
	}
	\node[align=center] at (\fac*0+0.2,\fac*0+\fac*0.5) {\scriptsize $s_i$};
	\node[align=center] at (\fac*0+0.2,\fac*1+\fac*0.5) {\scriptsize $s_i$};
	\node[align=center] at (\fac*0+0.2,\fac*2+\fac*0.5) {\scriptsize $s_i$};

	\node[align=center] at (\fac*1+0.2,\fac*0+\fac*0.5) {\scriptsize $s_i$};
	\node[align=center] at (\fac*1+0.2,\fac*1+\fac*0.5) {\scriptsize $s_f$};
	\node[align=center] at (\fac*1+0.2,\fac*2+\fac*0.5) {\scriptsize $s_f$};

	\node[align=center] at (\fac*2+0.2,\fac*0+\fac*0.5) {\scriptsize $s_i$};
	\node[align=center] at (\fac*2+0.2,\fac*1+\fac*0.5) {\scriptsize $s_f$};
	\node[align=center] at (\fac*2+0.2,\fac*2+\fac*0.5) {\scriptsize $s_f$};

	\foreach \y in {0,1,2}{
		\draw [->,line width=0.5mm] (\fac*3,\fac*\y) -- (\fac*3,\fac*\y+0.55*\fac);
		\draw [->,line width=0.5mm] (\fac*3,\fac*\y+0.5*\fac) -- (\fac*3,\fac*\y+\fac);
	}
	\node[align=center] at (\fac*3+0.3,\fac*0+\fac*0.5) {\scriptsize $s_i$};
	\node[align=center] at (\fac*3+0.3,\fac*1+\fac*0.5) {\scriptsize $s_f$};
	\node[align=center] at (\fac*3+0.3,\fac*2+\fac*0.5) {\scriptsize $s_f$};

	\foreach \y in {0,1,2,3}{
		\foreach \x in {0,1,2,3}{
			\draw[fill=white] (\fac*\x,\fac*\y) circle [radius=0.3];
		}
	}

	\foreach \y in {0,2,3}{
		\foreach \x in {0,1,2,3}{
			\node[align=center] at (\fac*\x,\fac*\y) {I};
		}
	}
	\node[align=center] at (\fac*0,\fac*1) {I};
	\node[align=center] at (\fac*1,\fac*1) {A};
	\node[align=center] at (\fac*2,\fac*1) {B};
	\node[align=center] at (\fac*3,\fac*1) {I};
\end{tikzpicture}
\caption{\label{fig:AB} Graphical depiction of the word
  ``IIIIIABIIIIIIIII'' embedded in
  the PEPO.} 
\end{figure}
This means that at site $6$, a word is initialized by inserting the
symbol ``A''.
A final state is emitted upwards,
meaning that no other
nontrivial symbol is expected above, while the intermediate state $\left\{\left(A,6\right)\right\}$
is sent to the right.
At site $7$, the latter
state is received from the left, the second symbol $B$
is attached to the word, and a final state is emitted
both upwards and to the right.

Finally, we consider terms that span multiple rows,
taking
``IAPPPBIIIIIIIIII'',  which describes vertical hopping between sites
$2$ and $6$, as a concrete example;
this case is depicted
in Fig.~\ref{fig:APB}.
\begin{figure}[!htb]
\centering
\begin{tikzpicture}[font=\large]
	\def \fac {1.5}

	\foreach \y in {0,1,2,3}{
		\foreach \x in {0,1,2}{
			\draw [->,line width=0.25mm] (\fac*\x,\fac*\y) -- (\fac*\x+0.55*\fac,\fac*\y);
			\draw [->,line width=0.25mm] (\fac*\x+0.5*\fac,\fac*\y) -- (\fac*\x+\fac,\fac*\y);
		}
	}
	\node[align=center] at (\fac*0+\fac*0.5,\fac*0+0.2) {\scriptsize $s_i$};
	\node[align=center] at (\fac*1+\fac*0.5,\fac*0+0.2) {\scriptsize $s_P$};
	\node[align=center] at (\fac*2+\fac*0.5,\fac*0+0.2) {\scriptsize $s_P$};

	\node[align=center] at (\fac*0+\fac*0.5,\fac*1+0.2) {\scriptsize $s_P$};
	\node[align=center] at (\fac*1+\fac*0.5,\fac*1+0.2) {\scriptsize $s_f$};
	\node[align=center] at (\fac*2+\fac*0.5,\fac*1+0.2) {\scriptsize $s_f$};

	\node[align=center] at (\fac*0+\fac*0.5,\fac*2+0.2) {\scriptsize $s_i$};
	\node[align=center] at (\fac*1+\fac*0.5,\fac*2+0.2) {\scriptsize $s_i$};
	\node[align=center] at (\fac*2+\fac*0.5,\fac*2+0.2) {\scriptsize $s_i$};

	\node[align=center] at (\fac*0+\fac*0.5,\fac*3+0.2) {\scriptsize $s_i$};
	\node[align=center] at (\fac*1+\fac*0.5,\fac*3+0.2) {\scriptsize $s_i$};
	\node[align=center] at (\fac*2+\fac*0.5,\fac*3+0.2) {\scriptsize $s_i$};

	\foreach \x in {0,1,2}{
		\foreach \y in {0,1,2}{
			\draw [->,line width=0.25mm] (\fac*\x,\fac*\y) -- (\fac*\x,\fac*\y+0.55*\fac);
			\draw [->,line width=0.25mm] (\fac*\x,\fac*\y+0.5*\fac) -- (\fac*\x,\fac*\y+\fac);
		}
	}
	\node[align=center] at (\fac*0+0.3,\fac*0+\fac*0.5) {\scriptsize $s_i$};
	\node[align=center] at (\fac*0+0.3,\fac*1+\fac*0.5) {\scriptsize $s_P$};
	\node[align=center] at (\fac*0+0.3,\fac*2+\fac*0.5) {\scriptsize $s_P$};

	\node[align=center] at (\fac*1+0.6,\fac*0+\fac*0.5) {\tiny $\left\{\left(A,2\right)\right\}$};
	\node[align=center] at (\fac*1+0.3,\fac*1+\fac*0.5) {\scriptsize $s_f$};
	\node[align=center] at (\fac*1+0.3,\fac*2+\fac*0.5) {\scriptsize $s_f$};

	\node[align=center] at (\fac*2+0.5,\fac*0+\fac*0.5) {\scriptsize $s_P$};
	\node[align=center] at (\fac*2+0.3,\fac*1+\fac*0.5) {\scriptsize $s_P$};
	\node[align=center] at (\fac*2+0.3,\fac*2+\fac*0.5) {\scriptsize $s_P$};

	\foreach \y in {0,1,2}{
		\draw [->,line width=0.5mm] (\fac*3,\fac*\y) -- (\fac*3,\fac*\y+0.55*\fac);
		\draw [->,line width=0.5mm] (\fac*3,\fac*\y+0.5*\fac) -- (\fac*3,\fac*\y+\fac);
	}
	\node[align=center] at (\fac*3+0.5,\fac*0.5) {\scriptsize $s_P$};

	\node[align=center] at (\fac*3+0.3,\fac*1+\fac*0.5) {\scriptsize $s_f$};
	\node[align=center] at (\fac*3+0.3,\fac*2+\fac*0.5) {\scriptsize $s_f$};

	\foreach \y in {0,1,2,3}{
		\foreach \x in {0,1,2,3}{
			\draw[fill=white] (\fac*\x,\fac*\y) circle [radius=0.3];
		}
	}

	\foreach \y in {2,3}{
		\foreach \x in {0,1,2,3}{
			\node[align=center] at (\fac*\x,\fac*\y) {I};
		}
	}
	\node[align=center] at (\fac*0,\fac*0) {I};
	\node[align=center] at (\fac*1,\fac*0) {A};
	\node[align=center] at (\fac*2,\fac*0) {P};
	\node[align=center] at (\fac*3,\fac*0) {P};
	\node[align=center] at (\fac*0,\fac*1) {P};
	\node[align=center] at (\fac*1,\fac*1) {B};
	\node[align=center] at (\fac*2,\fac*1) {I};
	\node[align=center] at (\fac*3,\fac*1) {I};

\end{tikzpicture}
\caption{\label{fig:APB} Graphical depiction of the word
  ``IAPPPBIIIIIIIIII'' embedded in
  the PEPO}
\end{figure}
Since the parity operators have been
already taken into account through
proper initialization, 
the term is effectively local and can be
represented by inserting
just two pairs
$\left\{\left(A,2\right),\left(B,6\right)\right\}$ and the according tensor elements
$\left\{\left(s_i,s_P,\left\{\left(A,2\right)\right\}\right)\right.$$\left.,A\right\}$ and 
$\left\{\left(s_P,\left\{\left(A,2\right)\right\},s_f,s_f\right),B\right\}$ at 
sites $2$ and $6$, respectively.

When a Hamiltonian contains non-local terms (a case that we
will not further cover explicitly in this paper), elements may have
been generated by a previous update of another (non-local) term.
In this case, a single element can be used as a component of two or
more words. 
Finally, it is crucial to note that 
intermediate states must be defined
as unordered
sets, meaning that  
$\left\{\left(A,i\right),\left(B,j\right)\right\} =
\left\{\left(B,j\right),\left(A,i\right)\right\}$.

The three examples considered above should
cover all possible variants 
of words needed for
local Hamiltonians
and describe how
to insert them into the appropriate
tensors.
After the PEPO has been
fully assigned,
all of the intermediate states within keys, which are symbolic, can be
encoded as 
unique numbers so that the PEPO can be stored efficiently.
(For the case of the Hubbard model,
the number of states required to represent the Hamiltonian
is exactly seven, consisting of
$s_i$, $s_f$, $s_p$,
and the four nontrivial operators.)
This numbering can then be 
translated into the corresponding quantum numbers,
which for the Hubbard model treated here are
the spin $S$, its
$z$-component $S_z$, and the deviation of the particle number
from half filling $C_z$.
Once these steps have been completed, 
the symbols within the tensors are replaced
by the full 
quantum-mechanical operators that they represent.
To take full advantage of the SU(2)-spin symmetry, the PEPO tensors
must 
be compactified using
Eq. \eqref{eq:su2compact}.
The PEPO is now ready to be used to implement
the tensor-network algorithm of one's choice, such as variational
optimization or imaginary- or real-time evolution.

\section{SU(2)-symmetric tensors} \label{sec:su2}

In order to optimize PEPSs and calculate their expectation values,
we need to work efficiently with the tensors of which they and their 
corresponding PEPOs are composed.
Especially 
useful for this purpose is the 
exploitation of continuous symmetries, which are present in most local
Hamiltonians. The Hubbard model, in particular, exhibits a U(1) symmetry 
for the charge degrees of freedom and a SU(2) symmetry for the spin degrees
of freedom in the absence of an external magnetic field. The 
implementation of U(1) symmetries in tensor networks has already been covered
thoroughly, for instance in Refs.~\citep{Bauer2011Mar,Singh2011Mar}, and will 
not be discussed in this paper. Instead, we give a generic and concise 
documentation of how to incorporate 
SU(2) symmetries into a networks of PEPS tensors.
Some of our concepts are similar to those presented in previous work
on SU(2) \citep{Singh2012Nov,Weichselbaum2012Dec,Weichselbaum2020Jun}, but 
several techniques differ in ways that 
may impact both implementation and performance.

In the following, we first define SU(2)-symmetrized tensors in
Sec.~\ref{sec:tensor_definition}, then define additional manipulations
of the tensors and their indexes that we will need to evaluate and
optimize PEPS/PEPO tensor networks:
permutation (Sec.~\ref{sec:permutation}), contraction
(Sec.~\ref{sec:paircont}), charge fusion
(Sec.~\ref{sec:fusesplit}), and index reversal
(Sec.~\ref{sec:indreversal}).

\subsection{Definition}
\label{sec:tensor_definition}

\begin{figure*}[!htb] 
\centering
\scalebox{0.8}{
\begin{tikzpicture}[font=\tiny]

\draw[line width=0.5mm] (-1.5,0) -- (0,0);
\node[align=center] at (-1.2,0.3) {\footnotesize $\left(j_1\,m_1\,t_1\right)$};
\draw[line width=0.5mm] (1.5,0) -- (0,0);
\node[align=center] at (1.2,0.3) {\footnotesize $\left(j_4\,m_4\,t_4\right)$};
\draw[line width=0.5mm] (-1,-1) -- (0,0);
\node[align=center] at (-1,-1.2) {\footnotesize $\left(j_2\,m_2\,t_2\right)$};
\draw[line width=0.5mm] (1,-1) -- (0,0);
\node[align=center] at (1,-1.2) {\footnotesize $\left(j_3\,m_3\,t_3\right)$};
\draw[fill=white, line width=0.5mm] (0,0) circle [radius=0.35];
\node[align=center] at (0,0) {\large $T$};

\draw[line width=0.5mm,->] (-0.9,0) -- (-0.8,0);
\draw[line width=0.5mm,->] (-0.8,0) -- (-0.9,0);

\draw[line width=0.5mm,->] (0.9,0) -- (0.8,0);
\draw[line width=0.5mm,->] (0.8,0) -- (0.9,0);

\draw[line width=0.5mm,->] (-0.6,-0.6) -- (-0.7,-0.7);
\draw[line width=0.5mm,->] (-0.7,-0.7) -- (-0.6,-0.6);

\draw[line width=0.5mm,->] (0.6,-0.6) -- (0.7,-0.7);
\draw[line width=0.5mm,->] (0.7,-0.7) -- (0.6,-0.6);

\node[align=center] at (3,-0.4) {\footnotesize decompose};
\draw[line width=0.5mm,->] (2.5,-0.7) -- (3.5,-0.7);

\draw[line width=0.5mm] (4,0) -- (13,0);
\draw[line width=0.5mm,->] (4,0) -- (4.7,0);
\draw[line width=0.5mm,->] (10,0) -- (12.5,0);
\draw[line width=0.5mm,->] (5,0) -- (6.6,0);
\draw[line width=0.5mm,->] (7,0) -- (8.6,0);
\draw[line width=0.5mm,->] (9,0) -- (10.6,0);
\node[align=center] at (4.5,0.3) {\footnotesize $\left(0\,0\,0\right)$};
\node[align=center] at (12.5,0.3) {\footnotesize $\left(0\,0\,0\right)$};
\node[align=center] at (6.5,0.3) {\footnotesize $\left(j_1\,m_1\,t_1\right)$};
\node[align=center] at (8.5,0.3) {\footnotesize $\left(j_a\,m_a\,t_a\right)$};
\node[align=center] at (10.5,0.3) {\footnotesize $\left(j_4\,m_4\,t_4\right)$};

\draw[line width=0.5mm] (5.5,-1.5) -- (5.5,0);
\node[align=center] at (5.5,-1.7) {\footnotesize $\left(j_1\,m_1\,t_1\right)$};
\draw[fill=white, line width=0.5mm] (5.5,0) circle [radius=0.35];
\node[align=center] at (5.5,0) {\large $T$};

\draw[line width=0.5mm] (7.5,-1.5) -- (7.5,0);
\node[align=center] at (7.5,-1.7) {\footnotesize $\left(j_2\,m_2\,t_2\right)$};
\draw[fill=white, line width=0.5mm] (7.5,0) circle [radius=0.35];
\node[align=center] at (7.5,0) {\large $T$};

\draw[line width=0.5mm] (9.5,-1.5) -- (9.5,0);
\node[align=center] at (9.5,-1.7) {\footnotesize $\left(j_3\,m_3\,t_3\right)$};
\draw[fill=white, line width=0.5mm] (9.5,0) circle [radius=0.35];
\node[align=center] at (9.5,0) {\large $T$};

\draw[line width=0.5mm] (11.5,-1.5) -- (11.5,0);
\node[align=center] at (11.5,-1.7) {\footnotesize $\left(j_4\,m_4\,t_4\right)$};

\draw[fill=white, line width=0.5mm] (11.5,0) circle [radius=0.35];
\node[align=center] at (11.5,0) {\large $T$};

\draw[line width=0.5mm,->] (5.5,-0.9) -- (5.5,-0.8);
\draw[line width=0.5mm,->] (5.5,-0.8) -- (5.5,-0.9);

\draw[line width=0.5mm,->] (7.5,-0.9) -- (7.5,-0.8);
\draw[line width=0.5mm,->] (7.5,-0.8) -- (7.5,-0.9);

\draw[line width=0.5mm,->] (9.5,-0.9) -- (9.5,-0.8);
\draw[line width=0.5mm,->] (9.5,-0.8) -- (9.5,-0.9);

\draw[line width=0.5mm,->] (11.5,-0.9) -- (11.5,-0.8);
\draw[line width=0.5mm,->] (11.5,-0.8) -- (11.5,-0.9);

%%%%%%%% NEXT ROW

\node[align=center] at (-2,-3) {\footnotesize Wigner-Eckart \\ \footnotesize theorem};
\draw[line width=0.5mm,->] (-2.5,-3.5) -- (-1.5,-3.5);

\draw[line width=0.5mm] (-0.5,-3) -- (6,-3);
\draw[line width=0.5mm,->] (0.5,-3) -- (1.35,-3);
\draw[line width=0.5mm,->] (2,-3) -- (2.85,-3);
\draw[line width=0.5mm,->] (-0.5,-3) -- (0,-3);
\draw[line width=0.5mm,->] (3,-3) -- (4.35,-3);
\draw[line width=0.5mm,->] (5,-3) -- (5.7,-3);
\node[align=center] at (-0.2,-2.7) {\footnotesize $\left(0\,0\right)$};
\node[align=center] at (5.6,-2.7) {\footnotesize $\left(0\,0\right)$};
\node[align=center] at (1.25,-2.7) {\footnotesize $\left(j_1\,t_1\right)$};
\node[align=center] at (2.75,-2.7) {\footnotesize $\left(j_a\,t_a\right)$};
\node[align=center] at (4.3,-2.7) {\footnotesize $\left(j_4\,t_4\right)$};

\draw[line width=0.5mm] (0.5,-4) -- (0.5,-3);
\draw[line width=0.5mm,->] (0.5,-3.6) -- (0.5,-3.65);
\draw[line width=0.5mm,->] (0.5,-3.65) -- (0.5,-3.6);
\node[align=center] at (0.5,-4.2) {\footnotesize $\left(j_1\,t_1\right)$};
\draw[fill=white, line width=0.5mm] (0.5,-3) circle [radius=0.3];
\node[align=center] at (0.5,-3) {\large $P$};

\draw[line width=0.5mm] (2,-4) -- (2,-3);
\draw[line width=0.5mm,->] (2,-3.6) -- (2,-3.65);
\draw[line width=0.5mm,->] (2,-3.65) -- (2,-3.6);
\node[align=center] at (2,-4.2) {\footnotesize $\left(j_2\,t_2\right)$};
\draw[fill=white, line width=0.5mm] (2,-3) circle [radius=0.3];
\node[align=center] at (2,-3) {\large $P$};

\draw[line width=0.5mm] (3.5,-4) -- (3.5,-3);
\draw[line width=0.5mm,->] (3.5,-3.6) -- (3.5,-3.65);
\draw[line width=0.5mm,->] (3.5,-3.65) -- (3.5,-3.6);
\node[align=center] at (3.5,-4.2) {\footnotesize $\left(j_3\,t_3\right)$};
\draw[fill=white, line width=0.5mm] (3.5,-3) circle [radius=0.3];
\draw[line width=0.5mm,->] (5,-3.6) -- (5,-3.65);
\draw[line width=0.5mm,->] (5,-3.65) -- (5,-3.6);
\node[align=center] at (3.5,-3) {\large $P$};

\draw[line width=0.5mm] (5,-4) -- (5,-3);
\node[align=center] at (5,-4.2) {\footnotesize $\left(j_4\,t_4\right)$};
\draw[fill=white, line width=0.5mm] (5,-3) circle [radius=0.3];
\node[align=center] at (5,-3) {\large $P$};

\node[align=center] at (6.5,-3.5) {\large $\times$};

\draw[line width=0.5mm] (7,-3) -- (13.5,-3);
\draw[line width=0.5mm,->] (8,-3) -- (8.85,-3);
\draw[line width=0.5mm,->] (9.5,-3) -- (10.35,-3);
\draw[line width=0.5mm,->] (7,-3) -- (7.5,-3);
\draw[line width=0.5mm,->] (10.5,-3) -- (11.85,-3);
\draw[line width=0.5mm,->] (12.5,-3) -- (13.2,-3);
\node[align=center] at (7.3,-2.7) {\footnotesize $\left(0\,0\right)$};
\node[align=center] at (13.1,-2.7) {\footnotesize $\left(0\,0\right)$};
\node[align=center] at (8.75,-2.7) {\footnotesize $\left(j_1\,m_1\right)$};
\node[align=center] at (10.25,-2.7) {\footnotesize $\left(j_a\,m_a\right)$};
\node[align=center] at (11.8,-2.7) {\footnotesize $\left(j_4\,m_4\right)$};

\draw[line width=0.5mm] (8,-4) -- (8,-3);
\draw[line width=0.5mm,->] (8,-3.6) -- (8,-3.65);
\draw[line width=0.5mm,->] (8,-3.65) -- (8,-3.6);
\node[align=center] at (8,-4.2) {\footnotesize $\left(j_1\,m_1\right)$};
\draw[fill=white, line width=0.5mm] (8,-3) circle [radius=0.3];
\node[align=center] at (8,-3) {\large $C$};

\draw[line width=0.5mm] (9.5,-4) -- (9.5,-3);
\draw[line width=0.5mm,->] (9.5,-3.6) -- (9.5,-3.65);
\draw[line width=0.5mm,->] (9.5,-3.65) -- (9.5,-3.6);
\node[align=center] at (9.5,-4.2) {\footnotesize $\left(j_2\,m_2\right)$};
\draw[fill=white, line width=0.5mm] (9.5,-3) circle [radius=0.3];
\node[align=center] at (9.5,-3) {\large $C$};

\draw[line width=0.5mm] (11,-4) -- (11,-3);
\draw[line width=0.5mm,->] (11,-3.6) -- (11,-3.65);
\draw[line width=0.5mm,->] (11,-3.65) -- (11,-3.6);
\node[align=center] at (11,-4.2) {\footnotesize $\left(j_3\,m_3\right)$};
\draw[fill=white, line width=0.5mm] (11,-3) circle [radius=0.3];
\draw[line width=0.5mm,->] (12.5,-3.6) -- (12.5,-3.65);
\draw[line width=0.5mm,->] (12.5,-3.65) -- (12.5,-3.6);
\node[align=center] at (11,-3) {\large $C$};

\draw[line width=0.5mm] (12.5,-4) -- (12.5,-3);
\node[align=center] at (12.5,-4.2) {\footnotesize $\left(j_4\,m_4\right)$};
\draw[fill=white, line width=0.5mm] (12.5,-3) circle [radius=0.3];
\node[align=center] at (12.5,-3) {\large $C$};

%%%%%%%% NEXT ROW

\node[align=center] at (-2,-6.2) {\footnotesize contract};
\draw[line width=0.5mm,->] (-2.5,-6.5) -- (-1.5,-6.5);

\draw[line width=0.5mm] (0.5,-6) -- (2,-6);
\draw[line width=0.5mm,->] (1.2,-6) -- (1.1,-6);
\draw[line width=0.5mm,->] (1.1,-6) -- (1.2,-6);
\node[align=center] at (0.7,-5.7) {\footnotesize $\left(j_1\,t_1\right)$};

\draw[line width=0.5mm] (3.5,-6) -- (2,-6);
\draw[line width=0.5mm,->] (2.8,-6) -- (2.9,-6);
\draw[line width=0.5mm,->] (2.9,-6) -- (2.8,-6);
\node[align=center] at (3.3,-5.7) {\footnotesize $\left(j_4\,t_4\right)$};

\draw[line width=0.5mm] (1,-7) -- (2,-6);
\draw[line width=0.5mm,->] (1.3,-6.7) -- (1.4,-6.6);
\draw[line width=0.5mm,->] (1.4,-6.6) -- (1.3,-6.7);
\node[align=center] at (0.7,-6.7) {\footnotesize $\left(j_2\,t_2\right)$};

\draw[line width=0.5mm] (3,-7) -- (2,-6);
\draw[line width=0.5mm,->] (2.7,-6.7) -- (2.6,-6.6);
\draw[line width=0.5mm,->] (2.6,-6.6) -- (2.7,-6.7);
\node[align=center] at (3.3,-6.7) {\footnotesize $\left(j_3\,t_3\right)$};

\draw[line width=0.5mm] (2,-5) -- (2,-6);
\node[align=center] at (2.3,-5.4) {\footnotesize $j_a$};

\draw[fill=white, line width=0.5mm] (2,-6) circle [radius=0.3];
\node[align=center] at (2,-6) {\large $P$};

\node[align=center] at (5,-6.5) {\large $\times$};

\draw[line width=0.5mm] (6.5,-6) -- (8,-6);
\draw[line width=0.5mm,->] (7.2,-6) -- (7.1,-6);
\draw[line width=0.5mm,->] (7.1,-6) -- (7.2,-6);
\node[align=center] at (6.7,-5.7) {\footnotesize $\left(j_1\,m_1\right)$};

\draw[line width=0.5mm] (9.5,-6) -- (8,-6);
\draw[line width=0.5mm,->] (8.8,-6) -- (8.9,-6);
\draw[line width=0.5mm,->] (8.9,-6) -- (8.8,-6);
\node[align=center] at (9.3,-5.7) {\footnotesize $\left(j_4\,m_4\right)$};

\draw[line width=0.5mm] (7,-7) -- (8,-6);
\draw[line width=0.5mm,->] (7.3,-6.7) -- (7.4,-6.6);
\draw[line width=0.5mm,->] (7.4,-6.6) -- (7.3,-6.7);
\node[align=center] at (6.6,-6.7) {\footnotesize $\left(j_2\,m_2\right)$};

\draw[line width=0.5mm] (9,-7) -- (8,-6);
\draw[line width=0.5mm,->] (8.7,-6.7) -- (8.6,-6.6);
\draw[line width=0.5mm,->] (8.6,-6.6) -- (8.7,-6.7);
\node[align=center] at (9.4,-6.7) {\footnotesize $\left(j_3\,m_3\right)$};

\draw[line width=0.5mm] (8,-5) -- (8,-6);
\node[align=center] at (8.3,-5.4) {\footnotesize $j_a$};

\draw[fill=white, line width=0.5mm] (8,-6) circle [radius=0.3];
\node[align=center] at (8,-6) {\large $C$};

\end{tikzpicture}
}
\caption{Compactification of an SU(2)-invariant rank-4 tensor.}
\label{fig:su2compact}
\end{figure*}
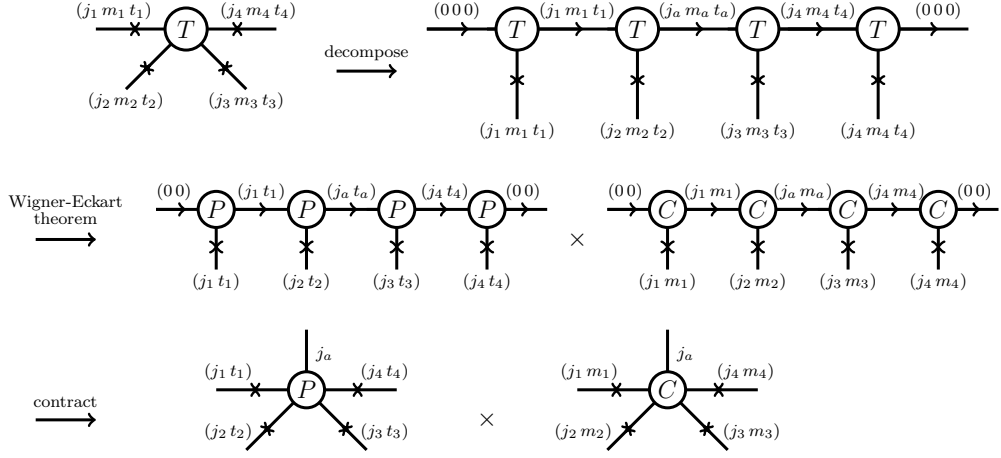

Consider the tensor operator $T^k_q$, with total angular momentum $k$ and
its $z$-component $q \in \left\{-k,-k+1,...,k-1,k\right\}$.
Consider in particular its transformation properties under the 
rotation
\begin{align}
  R = R\left(\boldsymbol\theta\right) = \exp{\left(-i \;
    {\boldsymbol\theta} \cdot
  {\boldsymbol j}\right)} 
\, ,
\end{align}
where $\boldsymbol\theta$
is a vectorial angle, and the components of $\boldsymbol j$
satisfy the angular-momentum algebra
\begin{align}
\left[ j_k, j_l \right] = \text{i} \, \epsilon_{klm} \, j_m \, .
\end{align}
If
\begin{align}
R \, T^k_q \, R^{-1} & = \sum\limits_{q'} T^k_{q'} \, R^k_{q',q} \, ,
\end{align}
with
\begin{align}
\quad R^k_{q',q} & = \left< k\, q' \right| R \left| k \, q \right> \, ,
\nonumber
\end{align}
we can apply the Wigner-Eckart theorem
\begin{align}
& \left< t_1 \, j_1 \, m_1 \right| T^k_{q} \left| t_2 \, j_2 \, m_2
\right> \nonumber \\
   &= \left< t_1 \, j_1 \| T^k \| t_2 \, j_2 \right>
      \bra{ j_2 \, k \, m_2 \, q} \ket{ j_1 \, m_1} \, , \label{eq:wigner}
\end{align}
where $j_i$ and $m_i$ again parameterize angular momentum and its $z$-component,
respectively, and $t_i$ denotes some additional degeneracy.
The right-hand 
side consists of the reduced matrix element,
$\left< t_1 \, j_1 \| T^k \| t_2 \, j_2 \right>$,
which has no  $m$-dependence,
and the Clebsch-Gordan coefficent (CGC),
$\bra{ j_2 \, k \, m_2 \, q} \ket{ j_1 \, m_1}$.
If we substitute
 $\left(j_3,m_3,t_3\right)$ for $\left(k,q\right)$,
$P$ for $T^k$, and $C$ for the CGC,
we obtain 
the representation of an SU(2)-invariant rank-3 tensor:
\begin{align}
T_{i_1 i_2 i_3} & = 
T_{\left(j_1 m_1 t_1\right),\left(j_2 m_2 t_2\right),\left(j_3 m_3
  t_3\right)} \nonumber \\
 & = P_{\left(j_1 t_1\right),\left(j_2 t_2\right),\left(j_3 t_3\right)}
  \, C_{\left(j_1 m_1\right),\left(j_2 m_2\right),\left(j_3 m_3\right)} \, . \label{eq:rank3}
\end{align}
\quad The entire machinery of SU(2)-symmetric tensors is based on 
Eq.~\eqref{eq:rank3}, as a tensor of arbitrary rank can now be constructed by
multiplying rank-3 tensors and summing over common
indices. Fig.~\ref{fig:su2compact} illustrates the resulting
compactification for a higher-dimensional tensor.
In the upper-left corner, we start with a rank-4 tensor with unspecified
directions of bonds, depicted as bidirectional arrows. The first step is
to think of this object as a sequence of four rank-three tensors, leading to 
external indices, $(1,2,3,4)$, given by the original tensor and internal 
indices, $(0,1,a,4,0)$.
The directions of the latter are arbitrary;
we set them here 
as going from left to right.
We now apply the Wigner-Eckart theorem in the form of
Eq.~\eqref{eq:rank3} 
to each of the four tensors, leading to the factorization in 
the second line of Fig.~\ref{fig:su2compact}. 
The final step is to sum over common indices
$j_i$, $t_i$, and $m_i$
that do not connect $P$ and $C$, which leads to the last line 
and is equivalent to 
the following formula for a rank-4 tensor:
\begin{align}
& T_{\left(j_1 m_1 t_1\right),\left(j_2 m_2 t_2\right),\left(j_3 m_3 t_3\right),
	\left(j_4 m_4 t_4\right)} \nonumber \\
	& \quad = \sum\limits_{j_a} 
		P^{j_a}_{\left(j_1 t_1\right),\left(j_2 t_2\right),\left(j_3 t_3\right),
		\left(j_4 t_4\right)} \nonumber \\
	& \qquad \quad \cdot C^{j_a}_{\left(j_1 m_1\right),\left(j_2 m_2\right),\left(j_3 m_3\right),
		\left(j_4 m_4\right)} \, .
	\label{eq:su2def4}
\end{align}
\quad The scheme depicted in Fig.~\ref{fig:su2compact} can easily be generalized
to tensors of arbitrary rank $r$, leading to
\begin{align}
& T_{\left(j_1 m_1 t_1\right),...,\left(j_r m_r t_r\right)} = \nonumber \\
	& \sum\limits_{j_{a_1},...,j_{a_{r-3}}} 
	P^{j_{a_1},...,j_{a_{r-3}}}_{\left(j_1 t_1\right),...,\left(j_r t_r\right)} \,
	C^{j_{a_1},...,j_{a_{r-3}}}_{\left(j_1 m_1\right),...,\left(j_r m_r\right)} \, . \label{eq:su2def}
\end{align}
In addition to $r$ external indices, we thus obtain 
$r-3$ independent, internal, indices 
connecting the reduced tensor elements with the CGCs.
This degeneracy is a 
corollary of the triangle inequality of angular momenta,
$j_1-j_2 \leq j \leq j_1+j_2$, and stands in contrast to a plain
$\text{U}\left(1\right)$-symmetry, where Abelian quantum numbers just add up
and lead to a Kronecker delta.

Adopting the terminology of 
Singh and Vidal
\citep{Singh2012Nov}, who performed the same decomposition using 
fusion-splitting trees, we term 
$C^{j_{a_1},...,j_{a_{r-3}}}_{\left(j_1 m_1\right),...,\left(j_r m_r\right)}$
an \textit{intertwiner}.
Intertwiners 
inherit the orthogonality relations of CGCs in 
the form 
\begin{align}
\sum\limits_{m_1,...,m_r}
	& C^{j_{a_1},...,j_{a_{r-3}}}_{\left(j_1 m_1\right),...,\left(j_r m_r\right)} \,
	C^{j_{a'_1},...,j_{a'_{r-3}}}_{\left(j_1
          m_1\right),...,\left(j_r m_r\right)} \nonumber \\
 & = N^{j_{a_1},...,j_{a_{r-3}}}_{j_1,...,j_r} \, 
		\delta_{j_{a_1},j_{a'_1}},...,\delta_{j_{a_{r-3}},j_{a'_{r-3}}}
                \, ,
\end{align}
where
\begin{align}
  & N^{j_{a_1},...,j_{a_{r-3}}}_{j_1,...,j_r} \nonumber \\
  & = \sum\limits_{m_1,...,m_r}
	C^{j_{a_1},...,j_{a_{r-3}}}_{\left(j_1 m_1\right),...,\left(j_r m_r\right)} \,
	C^{j_{a_1},...,j_{a_{r-3}}}_{\left(j_1 m_1\right),...,\left(j_r m_r\right)} \, ,
\end{align}
which can be used to flesh out the reduced tensors $P$
for a given full tensor $T$:
\begin{align}
& P^{j_{a_1},...,j_{a_{r-3}}}_{\left(j_1 t_1\right),...,\left(j_r t_r\right)} 
= 
 \left(N^{j_{a_1},...,j_{a_{r-3}}}_{j_1,...,j_r} \right)^{-1} \, \nonumber \\
& \times \sum\limits_{m_1,...,m_r} \,
C^{j_{a_1},...,j_{a_{r-3}}}_{\left(j_1 m_1\right),...,\left(j_r m_r\right)} \,
	T_{\left(j_1 m_1 t_1\right),...,\left(j_r m_r t_r\right)} \, 
	\label{eq:su2compact}.
\end{align}
To verify SU(2) symmetry, one can then reinsert $P$ into
Eq.~\eqref{eq:su2def} and
check if the initial and final $T$ are equal.

The purpose of this scheme is thus to be able to operate on a 
compressed object $P$ instead of the full tensor $T$, given that the
model under investigation is SU(2)-invariant.
Except for the initialization of the algorithm, the intertwiners $C$ are not 
actually calculated, but serve as placeholding aids 
to derive the elementary tensor operations discussed in the following.

\subsection{Permutation} \label{sec:permutation}
\begin{figure}[!htb]
\begin{minipage}{0.5\textwidth}
\scalebox{0.6}{
\begin{tikzpicture}[font=\Large]
\draw[line width=0.5mm] (0,0) -- (0,2);
\node[align=center] at (0.3,1) {\Large $j_1$};
\draw[line width=0.5mm] (2,0) -- (2,2);
\node[align=center] at (2.3,1) {\Large $j_3$};
\draw[line width=0.5mm] (4,0) -- (4,2);
\node[align=center] at (4.3,1) {\Large $j_2$};
\draw[line width=0.5mm] (6,0) -- (6,2);
\node[align=center] at (6.3,1) {\Large $j_4$};
\draw[line width=0.5mm] (0,0) -- (6,0);
\draw[line width=0.5mm] (0,2) -- (6,2);
\node[align=center] at (1,2.35) {\Large $j_1$};
\node[align=center] at (1,0.35) {\Large $j_1$};
\node[align=center] at (3,2.35) {\Large $j'_a$};
\node[align=center] at (3,0.35) {\Large $j'_a$};
\node[align=center] at (5,2.35) {\Large $j_4$};
\node[align=center] at (5,0.35) {\Large $j_4$};
\node[align=center] at (-1.7,1) {\Large $j=0$};
\node[align=center] at (7.7,1) {\Large $j=0$};
\draw[line width=0.5mm] (0,0) to[out=180,in=270] (-1,1);
\draw[line width=0.5mm] (-1,1) to[out=90,in=180] (0,2);
\draw[line width=0.5mm] (6,0) to[out=0,in=270] (7,1);
\draw[line width=0.5mm] (7,1) to[out=90,in=0] (6,2);

\draw[line width=0.5mm,->] (-1,0.95) -- (-1,1.05);
\draw[line width=0.5mm,->] (7,1.05) -- (7,0.95);

\draw[line width=0.5mm,->] (0.95,2) -- (1.05,2);
\draw[line width=0.5mm,->] (2.95,2) -- (3.05,2);
\draw[line width=0.5mm,->] (4.95,2) -- (5.05,2);

\draw[line width=0.5mm,->] (1.05,0) -- (0.95,0);
\draw[line width=0.5mm,->] (3.05,0) -- (2.95,0);
\draw[line width=0.5mm,->] (5.05,0) -- (4.95,0);

\draw[line width=0.5mm,->] (0,0.95) -- (0,1.05);
\draw[line width=0.5mm,->] (0,1.05) -- (0,0.95);

\draw[line width=0.5mm,->] (2,0.95) -- (2,1.05);
\draw[line width=0.5mm,->] (2,1.05) -- (2,0.95);

\draw[line width=0.5mm,->] (4,0.95) -- (4,1.05);
\draw[line width=0.5mm,->] (4,1.05) -- (4,0.95);

\draw[line width=0.5mm,->] (6,0.95) -- (6,1.05);
\draw[line width=0.5mm,->] (6,1.05) -- (6,0.95);

\draw[fill=white, line width=0.5mm] (0,2) circle [radius=0.35];
\draw[fill=white, line width=0.5mm] (2,2) circle [radius=0.35];
\draw[fill=white, line width=0.5mm] (4,2) circle [radius=0.35];
\draw[fill=white, line width=0.5mm] (6,2) circle [radius=0.35];
\draw[fill=white, line width=0.5mm] (0,0) circle [radius=0.35];
\draw[fill=white, line width=0.5mm] (2,0) circle [radius=0.35];
\draw[fill=white, line width=0.5mm] (4,0) circle [radius=0.35];
\draw[fill=white, line width=0.5mm] (6,0) circle [radius=0.35];

\end{tikzpicture}
}

\vspace*{0.2cm}
\centerline{\small (a) $N^{j'_a}_{j_1,j_3,j_2,j_4}$}
\vspace*{0.1cm}
\end{minipage}
\\ \mbox{} \\
\begin{minipage}{0.5\textwidth}
\scalebox{0.6}{
\begin{tikzpicture}[font=\Large]

\draw[fill=white, color=white] (0,3) circle [radius=0.35];

\draw[line width=0.5mm] (0,0) -- (0,2);
\node[align=center] at (0.3,1) {\Large $j_1$};

\draw[line width=0.5mm] (2,0) -- (4,2);
\node[align=center] at (2.3,1.3) {\Large $j_3$};
\draw[line width=0.5mm] (4,0) -- (3.1,0.9);
\draw[line width=0.5mm] (2.9,1.1) -- (2,2);
\node[align=center] at (3.8,1.3) {\Large $j_2$};

\draw[line width=0.5mm] (6,0) -- (6,2);
\node[align=center] at (6.3,1) {\Large $j_4$};
\draw[line width=0.5mm] (0,0) -- (6,0);
\draw[line width=0.5mm] (0,2) -- (6,2);
\node[align=center] at (1,2.35) {\Large $j_1$};
\node[align=center] at (1,0.35) {\Large $j_1$};
\node[align=center] at (3,2.35) {\Large $j'_a$};
\node[align=center] at (3,0.35) {\Large $j_a$};
\node[align=center] at (5,2.35) {\Large $j_4$};
\node[align=center] at (5,0.35) {\Large $j_4$};
\node[align=center] at (-1.7,1) {\Large $j=0$};
\node[align=center] at (7.7,1) {\Large $j=0$};
\draw[line width=0.5mm] (0,0) to[out=180,in=270] (-1,1);
\draw[line width=0.5mm] (-1,1) to[out=90,in=180] (0,2);
\draw[line width=0.5mm] (6,0) to[out=0,in=270] (7,1);
\draw[line width=0.5mm] (7,1) to[out=90,in=0] (6,2);

\draw[line width=0.5mm,->] (-1,0.95) -- (-1,1.05);
\draw[line width=0.5mm,->] (7,1.05) -- (7,0.95);

\draw[line width=0.5mm,->] (0.95,2) -- (1.05,2);
\draw[line width=0.5mm,->] (2.95,2) -- (3.05,2);
\draw[line width=0.5mm,->] (4.95,2) -- (5.05,2);

\draw[line width=0.5mm,->] (1.05,0) -- (0.95,0);
\draw[line width=0.5mm,->] (3.05,0) -- (2.95,0);
\draw[line width=0.5mm,->] (5.05,0) -- (4.95,0);

\draw[line width=0.5mm,->] (0,0.95) -- (0,1.05);
\draw[line width=0.5mm,->] (0,1.05) -- (0,0.95);

\draw[line width=0.5mm,->] (2,2) -- (2.5,1.5);
\draw[line width=0.5mm,->] (2.7,1.3) -- (2.5,1.5);

\draw[line width=0.5mm,->] (2,0) -- (3.5,1.5);
\draw[line width=0.5mm,->] (4,2) -- (3.5,1.5);

\draw[line width=0.5mm,->] (6,0.95) -- (6,1.05);
\draw[line width=0.5mm,->] (6,1.05) -- (6,0.95);

\draw[fill=white, line width=0.5mm] (0,2) circle [radius=0.35];
\draw[fill=white, line width=0.5mm] (2,2) circle [radius=0.35];
\draw[fill=white, line width=0.5mm] (4,2) circle [radius=0.35];
\draw[fill=white, line width=0.5mm] (6,2) circle [radius=0.35];
\draw[fill=white, line width=0.5mm] (0,0) circle [radius=0.35];
\draw[fill=white, line width=0.5mm] (2,0) circle [radius=0.35];
\draw[fill=white, line width=0.5mm] (4,0) circle [radius=0.35];
\draw[fill=white, line width=0.5mm] (6,0) circle [radius=0.35];

\end{tikzpicture}
}

\vspace*{0.2cm}
\centerline{\small (b) $X^{j_a,j'_a}_{\left(j_1,j_2,j_3,j_4\right),\left(j_1,j_3,j_2,j_4\right)}$}
\vspace*{0.1cm}
\end{minipage}

\begin{minipage}{0.5\textwidth}
\scalebox{0.6}{
\begin{tikzpicture}[font=\tiny]

\draw[fill=white, color=white] (0,3) circle [radius=0.35];
%% NORM
\draw[line width=0.5mm,rounded corners] (-5,-0.5) -- (-5.6,-0.5) -- (-5.6,2.5) -- (-5,2.5);

\draw[line width=0.5mm,rounded corners] (-1,-0.5) -- (-0.4,-0.5) -- (-0.4,2.5) -- (-1,2.5);

\node[align=center] at (-0.2,2.6) {\Large $-1$};

\node[align=center] at (0.1,1) {\Large $\cdot$};

\draw[line width=0.5mm] (-4,0) -- (-4,2);
\node[align=center] at (-3.7,1) {\Large $j_3$};
\draw[line width=0.5mm] (-2,0) -- (-2,2);
\node[align=center] at (-2.3,1) {\Large $j_2$};

\draw[line width=0.5mm] (-4,0) -- (-2,0);
\draw[line width=0.5mm] (-4,2) -- (-2,2);
\node[align=center] at (-3,2.35) {\Large $j'_a$};
\node[align=center] at (-3,0.35) {\Large $j'_a$};
\node[align=center] at (-5.3,1) {\Large $j_1$};
\node[align=center] at (-0.7,1) {\Large $j_4$};
\draw[line width=0.5mm] (-4,0) to[out=180,in=270] (-5,1);
\draw[line width=0.5mm] (-5,1) to[out=90,in=180] (-4,2);
\draw[line width=0.5mm] (-2,0) to[out=0,in=270] (-1,1);
\draw[line width=0.5mm] (-1,1) to[out=90,in=0] (-2,2);

\draw[line width=0.5mm,->] (-5,0.95) -- (-5,1.05);
\draw[line width=0.5mm,->] (-1,1.05) -- (-1,0.95);

\draw[line width=0.5mm,->] (-3.05,2) -- (-2.95,2);

\draw[line width=0.5mm,->] (-2.95,0) -- (-3.05,0);

\draw[line width=0.5mm,->] (-4,0.95) -- (-4,1.05);
\draw[line width=0.5mm,->] (-4,1.05) -- (-4,0.95);

\draw[line width=0.5mm,->] (-2,0.95) -- (-2,1.05);
\draw[line width=0.5mm,->] (-2,1.05) -- (-2,0.95);

\draw[fill=white, line width=0.5mm] (-4,2) circle [radius=0.35];
\draw[fill=white, line width=0.5mm] (-2,2) circle [radius=0.35];
\draw[fill=white, line width=0.5mm] (-4,0) circle [radius=0.35];
\draw[fill=white, line width=0.5mm] (-2,0) circle [radius=0.35];

%% EXCHANGE
\draw[line width=0.5mm] (2,0) -- (4,2);
\node[align=center] at (2.3,1.3) {\Large $j_3$};
\draw[line width=0.5mm] (4,0) -- (3.1,0.9);
\draw[line width=0.5mm] (2.9,1.1) -- (2,2);
\node[align=center] at (3.8,1.3) {\Large $j_2$};

\draw[line width=0.5mm] (2,0) -- (4,0);
\draw[line width=0.5mm] (2,2) -- (4,2);
\node[align=center] at (3,2.35) {\Large $j'_a$};
\node[align=center] at (3,0.35) {\Large $j_a$};
\node[align=center] at (0.7,1) {\Large $j_1$};
\node[align=center] at (5.3,1) {\Large $j_4$};
\draw[line width=0.5mm] (2,0) to[out=180,in=270] (1,1);
\draw[line width=0.5mm] (1,1) to[out=90,in=180] (2,2);
\draw[line width=0.5mm] (4,0) to[out=0,in=270] (5,1);
\draw[line width=0.5mm] (5,1) to[out=90,in=0] (4,2);

\draw[line width=0.5mm,->] (1,0.95) -- (1,1.05);
\draw[line width=0.5mm,->] (5,1.05) -- (5,0.95);

\draw[line width=0.5mm,->] (2.95,2) -- (3.05,2);

\draw[line width=0.5mm,->] (3.05,0) -- (2.95,0);

\draw[line width=0.5mm,->] (2,2) -- (2.5,1.5);
\draw[line width=0.5mm,->] (2.7,1.3) -- (2.5,1.5);

\draw[line width=0.5mm,->] (2,0) -- (3.5,1.5);
\draw[line width=0.5mm,->] (4,2) -- (3.5,1.5);

\draw[fill=white, line width=0.5mm] (2,2) circle [radius=0.35];
\draw[fill=white, line width=0.5mm] (4,2) circle [radius=0.35];
\draw[fill=white, line width=0.5mm] (2,0) circle [radius=0.35];
\draw[fill=white, line width=0.5mm] (4,0) circle [radius=0.35];

\end{tikzpicture}
}

\vspace*{0.2cm}
\centerline{\small (c) $\left( N^{j'_a}_{j_1,j_3,j_2,j_4} \right)^{-1} \cdot
X^{j_a,j'_a}_{\left(j_1,j_2,j_3,j_4\right),\left(j_1,j_3,j_2,j_4\right)}$}
\vspace*{0.1cm}
\end{minipage}

\caption{Permutation of an SU(2)-invariant rank-4 tensor}
\label{fig:permcluster}
\end{figure}
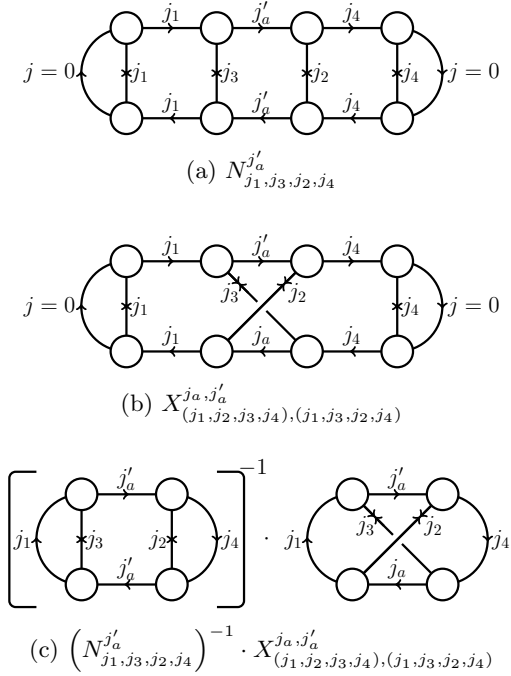

Consider the permutation of two adjacent indices in a rank-4 tensor. 
Through Eq.~\eqref{eq:su2def},
two 
orders are defined as follows:
\begin{align}
& T_{\left(j_1 m_1 t_1\right),\left(j_2 m_2 t_2\right),
			\left(j_3 m_3 t_3\right),\left(j_4 m_4 t_4\right)} \nonumber \\
& \quad = \sum\limits_{j_a} P^{j_a}_{\left(j_1 t_1\right),\left(j_2 t_2\right),
			\left(j_3 t_3\right),\left(j_4 t_4\right)} \, \nonumber \\
& \qquad \quad \cdot C^{j_a}_{\left(j_1 m_1\right),\left(j_2 m_2\right),
			\left(j_3 m_3\right),\left(j_4 m_4\right)} \, ,
\end{align}
\begin{align}
&T_{\left(j_1 m_1 t_1\right),\left(j_3 m_3 t_3\right),
			\left(j_2 m_2 t_2\right),\left(j_4 m_4 t_4\right)} \nonumber \\
& \quad = \sum\limits_{j'_a} P^{j'_a}_{\left(j_1 t_1\right),\left(j_3 t_3\right),
			\left(j_2 t_2\right),\left(j_4 t_4\right)} \nonumber \\
& \qquad \quad \cdot C^{j'_a}_{\left(j_1 m_1\right),\left(j_3 m_3\right),
			\left(j_2 m_2\right),\left(j_4 m_4\right)} \, .
\end{align}
Note how the first internal index, $j_a$, differs from the second, $j'_a$.
Since we want all orders of indices to be consistent with respect to the
decomposition in Fig.~\ref{fig:su2compact}, we think of a given
$P^{j_a}_{\left(j_1 t_1\right),\left(j_2 t_2\right), 
\left(j_3 t_3\right),\left(j_4 t_4\right)}$ as being multiplied by 
the corresponding intertwiner, then permuted as usual,
and finally compactified through Eq.~\eqref{eq:su2compact}, leading to the 
new reduced tensor $P^{j'_a}_{\left(j_1 t_1\right),\left(j_3 t_3\right), 
\left(j_2 t_2\right),\left(j_4 t_4\right)}$:
\begin{align}
& P^{j'_a}_{\left(j_1 t_1\right),\left(j_3 t_3\right),
			\left(j_2 t_2\right),\left(j_4 t_4\right)} \nonumber \\
& \quad = \left( N^{j'_a}_{j_1,j_3,j_2,j_4} \right)^{-1}
 \sum\limits_{j_a} P^{j_a}_{\left(j_1 t_1\right),\left(j_2 t_2\right),
	\left(j_3 t_3\right),\left(j_4 t_4\right)} \nonumber \\
& \qquad \quad \cdot X^{j_a,j'_a}_{\left(j_1,j_2,j_3,j_4\right),\left(j_1,j_3,j_2,j_4\right)} \, , \label{eq:permsu2}
\end{align}
where
\begin{align}
& X^{j_a,j'_a}_{\left(j_1,j_2,j_3,j_4\right),\left(j_1,j_3,j_2,j_4\right)} \nonumber \\
& \quad =  \sum\limits_{m_1,m_2,m_3,m_4}
			C^{j_a}_{\left(j_1 m_1\right),\left(j_2 m_2\right),
				\left(j_3 m_3\right),\left(j_4 m_4\right)} \nonumber \\
& \qquad \quad \cdot C^{j'_a}_{\left(j_1 m_1\right),\left(j_3 m_3\right),
				\left(j_2 m_2\right),\left(j_4 m_4\right)} \, .
\end{align}
Figs.~\ref{fig:permcluster}(a) 
and (b) 
depict 
the clusters of CGCs that make up $N$ and $X$, respectively, where each circle
represents a CGC and each line a common index. 
Since Eq.~\eqref{eq:permsu2} requires only their ratio, the CGCs at
the edges cancel 
out and lead to Fig.~\ref{fig:permcluster}(c). 
The left cluster can easily be
calculated using the orthogonality relations of CGCs.
The right cluster is proportional 
to a Wigner 6-j symbol and can be calculated analytically using the
Racah-formula or numerically by generating the CGCs and summing over indices
according to the sketch.
Note that for 
every single $j_a$, there are potentially
multiple $j'_a$.

The best way to implement an arbitrary permutation is
through successive permutations of two adjacent indices.
These computations are the only expensive ones within the SU(2)
bookkeeping, which is why  
the resulting prefactors $N^{-1} \cdot X$ are best calculated on demand and then
cached for future retrieval. 

\subsection{Contraction} \label{sec:paircont}
\begin{figure*}[!htb]
\begin{minipage}{\textwidth}
\scalebox{0.8}{
\begin{tikzpicture}[font=\tiny]
\draw[line width=0.5mm] (0,0) -- (0,2);
\node[align=center] at (0.3,1) {\small $j_1$};
\draw[line width=0.5mm] (2,0) -- (2,2);
\node[align=center] at (2.3,1) {\small $j_2$};
\draw[line width=0.5mm] (4,0) -- (4,2);
\node[align=center] at (4.3,1) {\small $j_5$};
\draw[line width=0.5mm] (6,0) -- (6,2);
\node[align=center] at (6.3,1) {\small $j_6$};
\draw[line width=0.5mm] (0,0) -- (6,0);
\draw[line width=0.5mm] (0,2) -- (6,2);
\node[align=center] at (1,2.35) {\small $j_1$};
\node[align=center] at (1,0.35) {\small $j_1$};
\node[align=center] at (3,2.35) {\small $j_a$};
\node[align=center] at (3,0.35) {\small $j_a$};
\node[align=center] at (5,2.35) {\small $j_6$};
\node[align=center] at (5,0.35) {\small $j_6$};
\node[align=center] at (-1.6,1) {\small $j=0$};
\node[align=center] at (7.6,1) {\small $j=0$};
\draw[line width=0.5mm] (0,0) to[out=180,in=270] (-1,1);
\draw[line width=0.5mm] (-1,1) to[out=90,in=180] (0,2);
\draw[line width=0.5mm] (6,0) to[out=0,in=270] (7,1);
\draw[line width=0.5mm] (7,1) to[out=90,in=0] (6,2);

\draw[line width=0.5mm,->] (-1,0.95) -- (-1,1.05);
\draw[line width=0.5mm,->] (7,1.05) -- (7,0.95);

\draw[line width=0.5mm,->] (0.95,2) -- (1.05,2);
\draw[line width=0.5mm,->] (2.95,2) -- (3.05,2);
\draw[line width=0.5mm,->] (4.95,2) -- (5.05,2);

\draw[line width=0.5mm,->] (1.05,0) -- (0.95,0);
\draw[line width=0.5mm,->] (3.05,0) -- (2.95,0);
\draw[line width=0.5mm,->] (5.05,0) -- (4.95,0);

\draw[line width=0.5mm,->] (0,0.95) -- (0,1.05);
\draw[line width=0.5mm,->] (0,1.05) -- (0,0.95);

\draw[line width=0.5mm,->] (2,0.95) -- (2,1.05);
\draw[line width=0.5mm,->] (2,1.05) -- (2,0.95);

\draw[line width=0.5mm,->] (4,0.95) -- (4,1.05);
\draw[line width=0.5mm,->] (4,1.05) -- (4,0.95);

\draw[line width=0.5mm,->] (6,0.95) -- (6,1.05);
\draw[line width=0.5mm,->] (6,1.05) -- (6,0.95);

\draw[fill=white, line width=0.5mm] (0,2) circle [radius=0.35];
\draw[fill=white, line width=0.5mm] (2,2) circle [radius=0.35];
\draw[fill=white, line width=0.5mm] (4,2) circle [radius=0.35];
\draw[fill=white, line width=0.5mm] (6,2) circle [radius=0.35];
\draw[fill=white, line width=0.5mm] (0,0) circle [radius=0.35];
\draw[fill=white, line width=0.5mm] (2,0) circle [radius=0.35];
\draw[fill=white, line width=0.5mm] (4,0) circle [radius=0.35];
\draw[fill=white, line width=0.5mm] (6,0) circle [radius=0.35];

\end{tikzpicture}
}

\vspace*{0.2cm}
\centerline{(a) \small $N^{j_a}_{j_1,j_2,j_5,j_6}$}
\vspace*{0.1cm}
\end{minipage}
\\ \mbox{} \\
\begin{minipage}{\textwidth}
\scalebox{0.8}{
\begin{tikzpicture}[font=\tiny]

\draw[line width=0.5mm] (0,0) -- (0,2);
\node[align=center] at (0.3,1) {\small $j_1$};
\draw[line width=0.5mm] (2,0) -- (2,2);
\node[align=center] at (2.3,1) {\small $j_2$};

\draw[line width=0.5mm] (12,0) -- (12,2);
\node[align=center] at (12.3,1) {\small $j_5$};
\draw[line width=0.5mm] (14,0) -- (14,2);
\node[align=center] at (14.3,1) {\small $j_6$};

\draw[line width=0.5mm] (0,0) -- (14,0);
\draw[line width=0.5mm] (0,2) -- (14,2);

\node[align=center] at (1,2.35) {\small $j_1$};
\node[align=center] at (3,2.35) {\small $j_a$};
\node[align=center] at (5,2.35) {\small $j_4$};
\node[align=center] at (7,2.35) {\small $j=0$};
\node[align=center] at (9,2.35) {\small $j_4$};
\node[align=center] at (11,2.35) {\small $j_a$};
\node[align=center] at (13,2.35) {\small $j_6$};

\node[align=center] at (1,-0.35) {\small $j_1$};
\node[align=center] at (7,-0.35) {\small $j_a$};
\node[align=center] at (13,-0.35) {\small $j_6$};

\node[align=center] at (7,1.2) {\small $j_4$};
\node[align=center] at (7,0.4) {\small $j_3$};

\node[align=center] at (-1.6,1) {\small $j=0$};
\node[align=center] at (15.6,1) {\small $j=0$};

\draw[line width=0.5mm] (0,0) to[out=180,in=270] (-1,1);
\draw[line width=0.5mm] (-1,1) to[out=90,in=180] (0,2);
\draw[line width=0.5mm] (14,0) to[out=0,in=270] (15,1);
\draw[line width=0.5mm] (15,1) to[out=90,in=0] (14,2);

\draw[line width=0.5mm] (6,2) to[out=315,in=180] (7,1.5);
\draw[line width=0.5mm] (7,1.5) to[out=0,in=225] (8,2);

\draw[line width=0.5mm] (4,2) to[out=315,in=180] (7,0.7);
\draw[line width=0.5mm] (7,0.7) to[out=0,in=225] (10,2);

\draw[line width=0.5mm,->] (-1,0.95) -- (-1,1.05);
\draw[line width=0.5mm,->] (15,1.05) -- (15,0.95);

\draw[line width=0.5mm,->] (0.95,2) -- (1.05,2);
\draw[line width=0.5mm,->] (2.95,2) -- (3.05,2);
\draw[line width=0.5mm,->] (4.95,2) -- (5.05,2);
\draw[line width=0.5mm,->] (6.95,2) -- (7.05,2);
\draw[line width=0.5mm,->] (8.95,2) -- (9.05,2);
\draw[line width=0.5mm,->] (10.95,2) -- (11.05,2);
\draw[line width=0.5mm,->] (12.95,2) -- (13.05,2);

\draw[line width=0.5mm,->] (1.05,0) -- (0.95,0);
\draw[line width=0.5mm,->] (7.05,0) -- (6.95,0);
\draw[line width=0.5mm,->] (13.05,0) -- (12.95,0);

\draw[line width=0.5mm,->] (0,0.95) -- (0,1.05);
\draw[line width=0.5mm,->] (0,1.05) -- (0,0.95);

\draw[line width=0.5mm,->] (2,0.95) -- (2,1.05);
\draw[line width=0.5mm,->] (2,1.05) -- (2,0.95);

\draw[line width=0.5mm,->] (12,0.95) -- (12,1.05);
\draw[line width=0.5mm,->] (12,1.05) -- (12,0.95);

\draw[line width=0.5mm,->] (14,0.95) -- (14,1.05);
\draw[line width=0.5mm,->] (14,1.05) -- (14,0.95);

\draw[line width=0.5mm,->] (6.95,1.5) -- (7.05,1.5);
\draw[line width=0.5mm,->] (7.05,1.5) -- (6.95,1.5);

\draw[line width=0.5mm,->] (6.95,0.7) -- (7.05,0.7);
\draw[line width=0.5mm,->] (7.05,0.7) -- (6.95,0.7);

\draw[fill=white, line width=0.5mm] (0,2) circle [radius=0.35];
\draw[fill=white, line width=0.5mm] (2,2) circle [radius=0.35];
\draw[fill=white, line width=0.5mm] (4,2) circle [radius=0.35];
\draw[fill=white, line width=0.5mm] (6,2) circle [radius=0.35];
\draw[fill=white, line width=0.5mm] (8,2) circle [radius=0.35];
\draw[fill=white, line width=0.5mm] (10,2) circle [radius=0.35];
\draw[fill=white, line width=0.5mm] (12,2) circle [radius=0.35];
\draw[fill=white, line width=0.5mm] (14,2) circle [radius=0.35];
\draw[fill=white, line width=0.5mm] (0,0) circle [radius=0.35];
\draw[fill=white, line width=0.5mm] (2,0) circle [radius=0.35];
\draw[fill=white, line width=0.5mm] (12,0) circle [radius=0.35];
\draw[fill=white, line width=0.5mm] (14,0) circle [radius=0.35];

\end{tikzpicture}
}

\vspace*{0.2cm}
\centerline{\small (b) $Y^{j_a}_{\left(j_1,j_2,j_3,j_4\right),
	\left(j_4,j_3,j_5,j_6\right),\left(j_1,j_2,j_5,j_6\right)}$}
\vspace*{0.1cm}
\end{minipage}
\\ \mbox{} \\
\begin{minipage}{\textwidth}
\begin{tikzpicture}[font=\tiny]

\draw[line width=0.5mm] (4,0) -- (10,0);
\draw[line width=0.5mm] (4,2) -- (10,2);

\node[align=center] at (5,2.35) {\small $j_4$};
\node[align=center] at (7,2.35) {\small $j=0$};
\node[align=center] at (9,2.35) {\small $j_4$};

\node[align=center] at (7,-0.35) {\small $j_a$};
\node[align=center] at (7,1.2) {\small $j_4$};
\node[align=center] at (7,0.4) {\small $j_3$};

\draw[line width=0.5mm] (4,0) to[out=180,in=270] (3,1);
\draw[line width=0.5mm] (3,1) to[out=90,in=180] (4,2);
\draw[line width=0.5mm] (10,0) to[out=0,in=270] (11,1);
\draw[line width=0.5mm] (11,1) to[out=90,in=0] (10,2);

\draw[line width=0.5mm] (6,2) to[out=315,in=180] (7,1.5);
\draw[line width=0.5mm] (7,1.5) to[out=0,in=225] (8,2);

\draw[line width=0.5mm] (4,2) to[out=315,in=180] (7,0.7);
\draw[line width=0.5mm] (7,0.7) to[out=0,in=225] (10,2);

\draw[line width=0.5mm,->] (3,0.95) -- (3,1.05);
\draw[line width=0.5mm,->] (11,1.05) -- (11,0.95);

\draw[line width=0.5mm,->] (4.95,2) -- (5.05,2);
\draw[line width=0.5mm,->] (6.95,2) -- (7.05,2);
\draw[line width=0.5mm,->] (7.05,0) -- (6.95,0);
\draw[line width=0.5mm,->] (8.95,2) -- (9.05,2);

\draw[line width=0.5mm,->] (6.95,1.5) -- (7.05,1.5);
\draw[line width=0.5mm,->] (7.05,1.5) -- (6.95,1.5);

\draw[line width=0.5mm,->] (6.95,0.7) -- (7.05,0.7);
\draw[line width=0.5mm,->] (7.05,0.7) -- (6.95,0.7);

\draw[fill=white, line width=0.5mm] (4,2) circle [radius=0.35];
\draw[fill=white, line width=0.5mm] (6,2) circle [radius=0.35];
\draw[fill=white, line width=0.5mm] (8,2) circle [radius=0.35];
\draw[fill=white, line width=0.5mm] (10,2) circle [radius=0.35];

\end{tikzpicture}

\vspace*{0.2cm}
\centerline{(c) \small $\left(N^{j_a}_{j_1,j_3,j_2,j_4}\right)^{-1}
	\cdot Y^{j_a}_{\left(j_1,j_2,j_3,j_4\right),
	\left(j_4,j_3,j_5,j_6\right),\left(j_1,j_2,j_5,j_6\right)}$}
\vspace*{0.1cm}
\end{minipage}
\caption{Contraction of two SU(2)-invariant rank-4 tensors.}
\label{fig:su2cont}
\end{figure*}
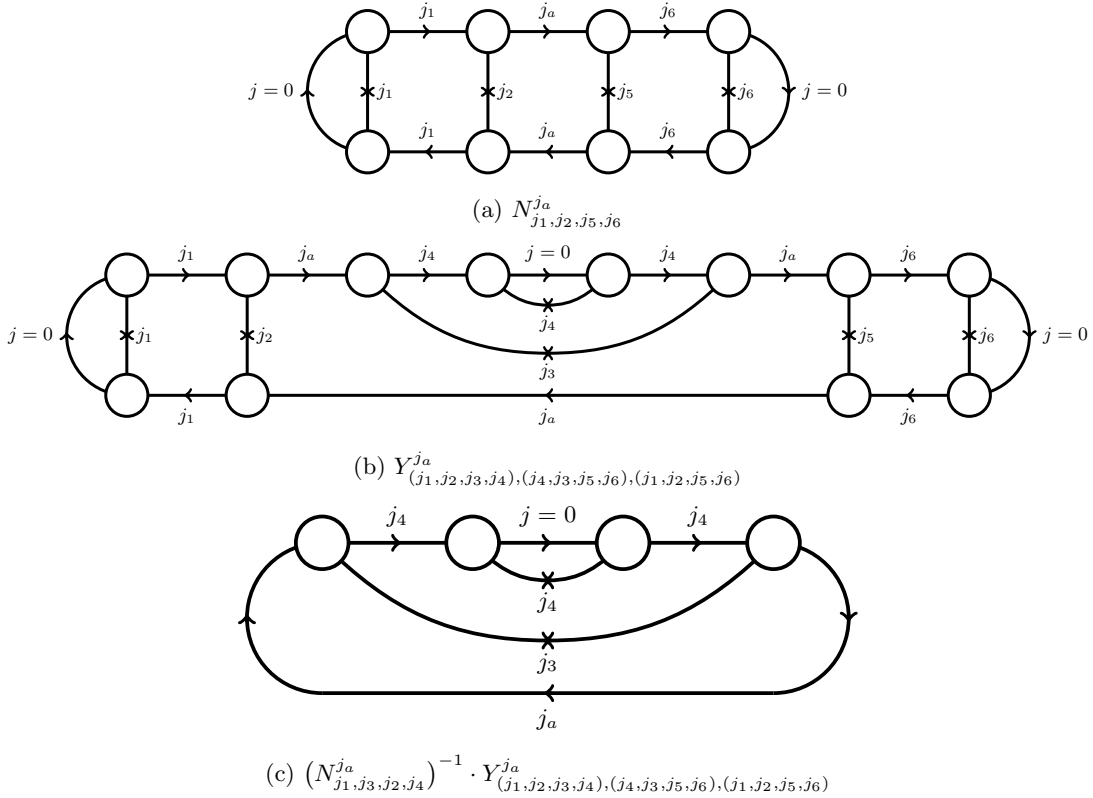

As a minimal, but sufficiently general example for the contraction of
tensors, we consider two rank-4 tensors, 
$T^1$ and $T^2$, summed over two common indices.
The first step is to permute both orders of indices
using the scheme outlined in the 
previous section to 
bring them into the following form:
\begin{align}
& T^1_{\left(j_1 m_1 t_1\right),\left(j_2 m_2 t_2\right),
			\left(j_3 m_3 t_3\right),\left(j_4 m_4 t_4\right)} \nonumber \\
&\quad = \sum\limits_{j_a} P^{1,j_a}_{\left(j_1 t_1\right),\left(j_2 t_2\right),
			\left(j_3 t_3\right),\left(j_4 t_4\right)} \nonumber \\
&\qquad \quad \cdot C^{j_a}_{\left(j_1 m_1\right),\left(j_2 m_2\right),
			\left(j_3 m_3\right),\left(j_4 m_4\right)} \, ,
\end{align}
\begin{align}
& T^2_{\left(j_3 m_3 t_3\right),\left(j_4 m_4 t_4\right),
			\left(j_5 m_5 t_5\right),\left(j_6 m_6 t_6\right)} \nonumber \\
& \quad = \sum\limits_{j_a} P^{2,j_a}_{\left(j_3 t_3\right),\left(j_4 t_4\right),
			\left(j_5 t_5\right),\left(j_6 t_6\right)} \nonumber \\
& \qquad \quad \cdot C^{j_a}_{\left(j_4 m_4\right),\left(j_3 m_3\right),
			\left(j_5 m_5\right),\left(j_6 m_6\right)} \, . \label{eq:su2contorder}
\end{align}
Note the different orders of indices in Eq.~\eqref{eq:su2contorder}. The
indexing $(3,4,5,6)$ on the degenerate level allows for the contraction to be
implemented as a matrix-matrix-multiplication, whereas the order 
$(4,3,5,6)$ on the structural level leads to a trivial cluster of CGCs. 
Analogously to Sec.~\ref{sec:permutation}, we think of the reduced tensors 
$P^1$ and $P^2$ as being multiplied by their intertwiners, contracted, and then
compactified via 
Eq.~\eqref{eq:su2compact}. If the result is defined by
\begin{align}
& T^3_{\left(j_1 m_1 t_1\right),\left(j_2 m_2 t_2\right),
		\left(j_5 m_5 t_5\right),\left(j_6 m_6 t_6\right)} \nonumber \\
& \quad = \sum\limits_{j_a} P^{3,j_a}_{\left(j_1 t_1\right),\left(j_2 t_2\right),
			\left(j_5 t_5\right),\left(j_6 t_6\right)} \nonumber \\
& \qquad \quad \cdot C^{j_a}_{\left(j_1 m_1\right),\left(j_2 m_2\right),
			\left(j_5 m_5\right),\left(j_6 m_6\right)} \, ,
\end{align}
the reduced tensor $P^3$ can be written as
\begin{align}
& P^{3,j_a}_{\left(j_1 t_1\right),\left(j_2 t_2\right),
	\left(j_5 t_5\right),\left(j_6 t_6\right)} \nonumber \\
& \quad = \left( N^{j_a}_{j_1,j_2,j_5,j_6} \right)^{-1} \nonumber \\
& \qquad \sum\limits_{j_3,t_3,j_4,t_4} P^{1,j_a}_{\left(j_1 t_1\right),\left(j_2 t_2\right),
		\left(j_3 t_3\right),\left(j_4 t_4\right)} \nonumber \\
& \qquad \qquad \quad \cdot P^{2,j_a}_{\left(j_3 t_3\right),\left(j_4 t_4\right),
			\left(j_5 t_5\right),\left(j_6 t_6\right)} \, \nonumber \\
& \qquad \qquad \quad \cdot Y^{j_a}_{\left(j_1,j_2,j_3,j_4\right), 
			\left(j_4,j_3,j_5,j_6\right),\left(j_1,j_2,j_5,j_6\right)} 
			\, ,\label{eq:su2contract}
\end{align}
where
\begin{align}
& Y^{j_a}_{\left(j_1,j_2,j_3,j_4\right), 
			\left(j_4,j_3,j_5,j_6\right),\left(j_1,j_2,j_5,j_6\right)}
		\nonumber \\
& \quad = \sum\limits_{m_1,...,m_6} 
	C^{j_a}_{\left(j_1 m_1\right),\left(j_2 m_2\right),
		\left(j_3 m_3\right),\left(j_4 m_4\right)} \nonumber \\
& \qquad \qquad \quad \cdot C^{j_a}_{\left(j_4 m_4\right),\left(j_3 m_3\right),
		\left(j_5 m_5\right),\left(j_6 m_6\right)} \, \nonumber \\
& \qquad \qquad \quad  
	\cdot C^{j_a}_{\left(j_1 m_1\right),\left(j_2 m_2\right),
			\left(j_5 m_5\right),\left(j_6 m_6\right)} \, .\label{eq:itwy}
\end{align}
The clusters $N$ and $Y$ are depicted graphically in
Figs.~\ref{fig:su2cont}(a) 
and (b), 
respectively.
Due to the advantageous order of indices, the cluster $Y$
and thus the ratio given in Fig.~\ref{fig:su2cont}(c)
is now a trivial sum
over CGCs that can be processed using their orthogonality relations and the 
identity
\begin{align}
&\bra{j_1\, m_1\, j_2\, m_2}\ket{j\, m} \nonumber \\
& \qquad = \left(-1\right)^{j_1+j_2-j} \bra{j_2\, m_2\, j_1\, m_1}\ket{j\, m} \, .
\end{align}
The generalization to arbitrary contractions
is straightforward. 

The approach presented above can be shown to be 
equivalent to the use of
$X$-symbols introduced by Weichselbaum \citep{Weichselbaum2020Jun}.

\subsection{Charge fusion} \label{sec:fusesplit}
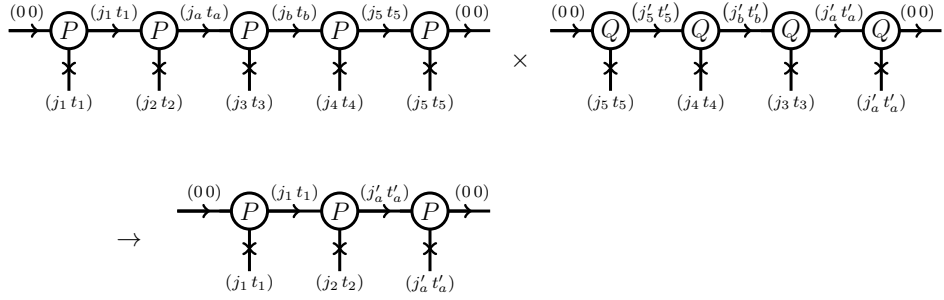
\begin{figure*}[!htb]
\centering
\scalebox{0.8}{
\begin{tikzpicture}[font=\tiny]

\draw[line width=0.5mm] (-0.5,-3) -- (7.5,-3);
\draw[line width=0.5mm,->] (0.5,-3) -- (1.35,-3);
\draw[line width=0.5mm,->] (2,-3) -- (2.85,-3);
\draw[line width=0.5mm,->] (-0.5,-3) -- (0,-3);
\draw[line width=0.5mm,->] (3,-3) -- (4.35,-3);
\draw[line width=0.5mm,->] (5,-3) -- (5.85,-3);
\draw[line width=0.5mm,->] (6,-3) -- (7.2,-3);

\node[align=center] at (-0.2,-2.7) {\footnotesize $\left(0\,0\right)$};
\node[align=center] at (7.2,-2.7) {\footnotesize $\left(0\,0\right)$};
\node[align=center] at (1.25,-2.7) {\footnotesize $\left(j_1\,t_1\right)$};
\node[align=center] at (2.75,-2.7) {\footnotesize $\left(j_a\,t_a\right)$};
\node[align=center] at (4.25,-2.7) {\footnotesize $\left(j_b\,t_b\right)$};
\node[align=center] at (5.75,-2.7) {\footnotesize $\left(j_5\,t_5\right)$};

\draw[line width=0.5mm] (0.5,-4) -- (0.5,-3);
\draw[line width=0.5mm,->] (0.5,-3.6) -- (0.5,-3.65);
\draw[line width=0.5mm,->] (0.5,-3.65) -- (0.5,-3.6);
\node[align=center] at (0.5,-4.2) {\footnotesize $\left(j_1\,t_1\right)$};
\draw[fill=white, line width=0.5mm] (0.5,-3) circle [radius=0.3];
\node[align=center] at (0.5,-3) {\large $P$};

\draw[line width=0.5mm] (2,-4) -- (2,-3);
\draw[line width=0.5mm,->] (2,-3.6) -- (2,-3.65);
\draw[line width=0.5mm,->] (2,-3.65) -- (2,-3.6);
\node[align=center] at (2,-4.2) {\footnotesize $\left(j_2\,t_2\right)$};
\draw[fill=white, line width=0.5mm] (2,-3) circle [radius=0.3];
\node[align=center] at (2,-3) {\large $P$};

\draw[line width=0.5mm] (3.5,-4) -- (3.5,-3);
\draw[line width=0.5mm,->] (3.5,-3.6) -- (3.5,-3.65);
\draw[line width=0.5mm,->] (3.5,-3.65) -- (3.5,-3.6);
\node[align=center] at (3.5,-4.2) {\footnotesize $\left(j_3\,t_3\right)$};
\draw[fill=white, line width=0.5mm] (3.5,-3) circle [radius=0.3];
\draw[line width=0.5mm,->] (5,-3.6) -- (5,-3.65);
\draw[line width=0.5mm,->] (5,-3.65) -- (5,-3.6);
\node[align=center] at (3.5,-3) {\large $P$};

\draw[line width=0.5mm] (5,-4) -- (5,-3);
\node[align=center] at (5,-4.2) {\footnotesize $\left(j_4\,t_4\right)$};
\draw[fill=white, line width=0.5mm] (5,-3) circle [radius=0.3];
\node[align=center] at (5,-3) {\large $P$};

\draw[line width=0.5mm] (6.5,-4) -- (6.5,-3);
\draw[line width=0.5mm,->] (6.5,-3.6) -- (6.5,-3.65);
\draw[line width=0.5mm,->] (6.5,-3.65) -- (6.5,-3.6);
\node[align=center] at (6.5,-4.2) {\footnotesize $\left(j_5\,t_5\right)$};
\draw[fill=white, line width=0.5mm] (6.5,-3) circle [radius=0.3];
\node[align=center] at (6.5,-3) {\large $P$};

\node[align=center] at (8,-3.5) {\large $\times$};

\draw[line width=0.5mm] (8.5,-3) -- (15,-3);
\draw[line width=0.5mm,->] (9.5,-3) -- (10.35,-3);
\draw[line width=0.5mm,->] (11,-3) -- (11.85,-3);
\draw[line width=0.5mm,->] (8.5,-3) -- (9,-3);
\draw[line width=0.5mm,->] (12,-3) -- (13.35,-3);
\draw[line width=0.5mm,->] (14,-3) -- (14.7,-3);
\node[align=center] at (8.8,-2.7) {\footnotesize $\left(0\,0\right)$};
\node[align=center] at (14.6,-2.7) {\footnotesize $\left(0\,0\right)$};
\node[align=center] at (10.25,-2.7) {\footnotesize $\left(j'_5\,t'_5\right)$};
\node[align=center] at (11.75,-2.7) {\footnotesize $\left(j'_b\,t'_b\right)$};
\node[align=center] at (13.3,-2.7) {\footnotesize $\left(j'_a\,t'_a\right)$};

\draw[line width=0.5mm] (9.5,-4) -- (9.5,-3);
\draw[line width=0.5mm,->] (9.5,-3.6) -- (9.5,-3.65);
\draw[line width=0.5mm,->] (9.5,-3.65) -- (9.5,-3.6);
\node[align=center] at (9.5,-4.2) {\footnotesize $\left(j_5\,t_5\right)$};
\draw[fill=white, line width=0.5mm] (9.5,-3) circle [radius=0.3];
\node[align=center] at (9.5,-3) {\large $Q$};

\draw[line width=0.5mm] (11,-4) -- (11,-3);
\draw[line width=0.5mm,->] (11,-3.6) -- (11,-3.65);
\draw[line width=0.5mm,->] (11,-3.65) -- (11,-3.6);
\node[align=center] at (11,-4.2) {\footnotesize $\left(j_4\,t_4\right)$};
\draw[fill=white, line width=0.5mm] (11,-3) circle [radius=0.3];
\node[align=center] at (11,-3) {\large $Q$};

\draw[line width=0.5mm] (12.5,-4) -- (12.5,-3);
\draw[line width=0.5mm,->] (12.5,-3.6) -- (12.5,-3.65);
\draw[line width=0.5mm,->] (12.5,-3.65) -- (12.5,-3.6);
\node[align=center] at (12.5,-4.2) {\footnotesize $\left(j_3\,t_3\right)$};
\draw[fill=white, line width=0.5mm] (12.5,-3) circle [radius=0.3];
\draw[line width=0.5mm,->] (14,-3.6) -- (14,-3.65);
\draw[line width=0.5mm,->] (14,-3.65) -- (14,-3.6);
\node[align=center] at (12.5,-3) {\large $Q$};

\draw[line width=0.5mm] (14,-4) -- (14,-3);
\node[align=center] at (14,-4.2) {\footnotesize $\left(j'_a\,t'_a\right)$};
\draw[fill=white, line width=0.5mm] (14,-3) circle [radius=0.3];
\node[align=center] at (14,-3) {\large $Q$};

%%%%% NEXT ROW

\node[align=center] at (1.5,-6.5) {\large $\rightarrow$};

\draw[line width=0.5mm] (2.3,-6) -- (7.5,-6);
\draw[line width=0.5mm,->] (2.3,-6) -- (2.85,-6);
\draw[line width=0.5mm,->] (3,-6) -- (4.35,-6);
\draw[line width=0.5mm,->] (5,-6) -- (5.85,-6);
\draw[line width=0.5mm,->] (6,-6) -- (7.2,-6);

\node[align=center] at (7.2,-5.7) {\footnotesize $\left(0\,0\right)$};
\node[align=center] at (2.75,-5.7) {\footnotesize $\left(0\,0\right)$};
\node[align=center] at (4.25,-5.7) {\footnotesize $\left(j_1\,t_1\right)$};
\node[align=center] at (5.75,-5.7) {\footnotesize $\left(j'_a\,t'_a\right)$};

\draw[line width=0.5mm] (3.5,-7) -- (3.5,-6);
\draw[line width=0.5mm,->] (3.5,-6.6) -- (3.5,-6.65);
\draw[line width=0.5mm,->] (3.5,-6.65) -- (3.5,-6.6);
\node[align=center] at (3.5,-7.2) {\footnotesize $\left(j_1\,t_1\right)$};
\draw[fill=white, line width=0.5mm] (3.5,-6) circle [radius=0.3];
\draw[line width=0.5mm,->] (5,-6.6) -- (5,-6.65);
\draw[line width=0.5mm,->] (5,-6.65) -- (5,-6.6);
\node[align=center] at (3.5,-6) {\large $P$};

\draw[line width=0.5mm] (5,-7) -- (5,-6);
\node[align=center] at (5,-7.2) {\footnotesize $\left(j_2\,t_2\right)$};
\draw[fill=white, line width=0.5mm] (5,-6) circle [radius=0.3];
\node[align=center] at (5,-6) {\large $P$};

\draw[line width=0.5mm] (6.5,-7) -- (6.5,-6);
\draw[line width=0.5mm,->] (6.5,-6.6) -- (6.5,-6.65);
\draw[line width=0.5mm,->] (6.5,-6.65) -- (6.5,-6.6);
\node[align=center] at (6.5,-7.2) {\footnotesize $\left(j'_a\,t'_a\right)$};
\draw[fill=white, line width=0.5mm] (6.5,-6) circle [radius=0.3];
\node[align=center] at (6.5,-6) {\large $P$};

\end{tikzpicture}
}
\caption{Fusion of three indices in an SU(2)-invariant rank-5 tensor.}
\label{fig:su2fuse}
\end{figure*}

In order to perform singular value decompositions in tensor networks,
indices need to be fused beforehand and split afterwards.
For this purpose, we take 
a rank-5 tensor as an example and conceptualize its reduced 
version as a sequence of rank-three tensors with internal angular momenta 
$j_a$ and $j_b$ and fictitious, internal, dense, indices $t_a$ and $t_b$, 
as depicted in Fig.~\ref{fig:su2fuse}.
Suppose that the indices $3$, $4$, and $5$ have to be fused by an isometry $Q$.
Due to the selection rules of CGCs, 
the external angular momenta, $j_3$, $j_4$, and $j_5$, do not uniquely
identify the outcome of fusion, in contrast to the 
case of U(1) symmetry, for which the outcome would simply be the sum
of Abelian quantum numbers.
Instead, the final index of the 
isometry is dictated by the internal index of the reduced tensor, in this 
case $j_a$. To isolate it, we think of the reduced isometry as a sequence of 
rank-3 tensors as well, shown as four $Q$'s with the reverse order of indices
in Fig.~\ref{fig:su2fuse}. Adjacent $P$'s and $Q$'s are now contracted from the 
inside out. The first two  
yield a result proportional to $\delta_{j_5,j'_5}\, \delta_{t_5,t'_5}$ 
due to SU(2)-invariance, which can be absorbed to either the left or the right. 
The next contraction eliminates $\left(j_b \, t_b\right)$, and the last one 
finally exposes $\left(j_a \, t_a\right)$. The dimension 
$d_a$ of $t_a$ is determined by:
\begin{align}
d_a = \sum\limits_{j_3,j_4,j_5,j_b} \left(d_3 \, d_4 \, d_5 \right)_{
		j_3,j_4,j_5,j_b}\, .
\end{align}
One may again contrast this with U(1)-symmetry, where the sum over $j_b$ would
be missing. The next step is to determine the proper prefactor 
$Q^{j'_b}_{j_5 \, j_4 \, j_3 \, j'_a}$ that yields unitarity.
With this in mind, we note that the reduced isometry
$Q^{j'_b}_{\left(j_5 t_5\right),\left(j_4 t_4\right),\left(j_3 t_3\right),
\left(j'_a t'_a\right)}$ is accompanied by the intertwiner 
$C^{j'_b}_{\left(j_5 m_5\right),\left(j_4 m_4\right),\left(j_3 m_3\right),
\left(j'_a m'_a\right)}$. To ensure $Q \, Q^{\dagger} = \mathds{1}$, 
the following condition must then hold:
\begin{align}
& Q^{j'_b}_{j_5 \, j_4 \, j_3 \, j'_a}
Q^{j'_b}_{j_5 \, j_4 \, j_3 \, j''_a} \nonumber \\
& \quad \cdot \sum\limits_{m_5,m_4,m_3}
 C^{j'_b}_{\left(j_5 m_5\right),\left(j_4 m_4\right),\left(j_3 m_3\right),
\left(j'_a m'_a\right)} \nonumber \\
& \qquad \qquad \qquad C^{j'_b}_{\left(j_5 m_5\right),\left(j_4 m_4\right),\left(j_3 m_3\right),
\left(j''_a m''_a\right)} \nonumber \\
& \qquad \qquad\qquad\qquad= \delta_{j'_a,j''_a} \, \delta_{m'_a,m''_a} \, ,
\end{align}
\begin{align}
\Rightarrow \quad 
	Q^{j'_b}_{j_5 \, j_4 \, j_3 \, j'_a} = 
\sqrt{\frac{2 \, j'_a + 1}{N^{j_b}_{j_5,j_4,j_3,j'_a}}} \, . \label{eq:su2prefactor}
\end{align}
As in the case of permutation and contraction, we assume that this example of 
charge fusion is sufficient to illustrate how to proceed in 
the general case.

\subsection{Index reversal}
\label{sec:indreversal}
\begin{figure}[!htb]
\centering
\begin{minipage}{0.5\textwidth}
\scalebox{0.7}{
\begin{tikzpicture}[font=\large]
\draw[line width=0.5mm] (0,0) -- (0,2);
\node[align=center] at (0.3,1) {\large $j_1$};
\draw[line width=0.5mm] (2,0) -- (2,2);
\node[align=center] at (2.3,1) {\large $j_2$};
\draw[line width=0.5mm] (4,0) -- (4,2);
\node[align=center] at (4.3,1) {\large $j_3$};
\draw[line width=0.5mm] (6,0) -- (6,2);
\node[align=center] at (6.3,1) {\large $j_4$};
\draw[line width=0.5mm] (0,0) -- (6,0);
\draw[line width=0.5mm] (0,2) -- (6,2);
\node[align=center] at (1,2.35) {\large $j_1$};
\node[align=center] at (1,0.35) {\large $j_1$};
\node[align=center] at (3,2.35) {\large $j_a$};
\node[align=center] at (3,0.35) {\large $j_a$};
\node[align=center] at (5,2.35) {\large $j_4$};
\node[align=center] at (5,0.35) {\large $j_4$};
\node[align=center] at (-1.6,1) {\large $j=0$};
\node[align=center] at (7.6,1) {\large $j=0$};
\draw[line width=0.5mm] (0,0) to[out=180,in=270] (-1,1);
\draw[line width=0.5mm] (-1,1) to[out=90,in=180] (0,2);
\draw[line width=0.5mm] (6,0) to[out=0,in=270] (7,1);
\draw[line width=0.5mm] (7,1) to[out=90,in=0] (6,2);

\draw[line width=0.5mm,->] (-1,0.95) -- (-1,1.05);
\draw[line width=0.5mm,->] (7,1.05) -- (7,0.95);

\draw[line width=0.5mm,->] (0.95,2) -- (1.05,2);
\draw[line width=0.5mm,->] (2.95,2) -- (3.05,2);
\draw[line width=0.5mm,->] (4.95,2) -- (5.05,2);

\draw[line width=0.5mm,->] (1.05,0) -- (0.95,0);
\draw[line width=0.5mm,->] (3.05,0) -- (2.95,0);
\draw[line width=0.5mm,->] (5.05,0) -- (4.95,0);

\draw[line width=0.5mm,->] (0,0.95) -- (0,1.05);

\draw[line width=0.5mm,->] (2,1.05) -- (2,0.95);

\draw[line width=0.5mm,->] (4,1.05) -- (4,0.95);

\draw[line width=0.5mm,->] (6,0.95) -- (6,1.05);

\draw[fill=white, line width=0.5mm] (0,2) circle [radius=0.35];
\draw[fill=white, line width=0.5mm] (2,2) circle [radius=0.35];
\draw[fill=white, line width=0.5mm] (4,2) circle [radius=0.35];
\draw[fill=white, line width=0.5mm] (6,2) circle [radius=0.35];
\draw[fill=white, line width=0.5mm] (0,0) circle [radius=0.35];
\draw[fill=white, line width=0.5mm] (2,0) circle [radius=0.35];
\draw[fill=white, line width=0.5mm] (4,0) circle [radius=0.35];
\draw[fill=white, line width=0.5mm] (6,0) circle [radius=0.35];

\end{tikzpicture}
}

\vspace*{0.2cm}
\centerline{\small (a) $\tilde{N}^{j_a}_{j_1,j_2,j_3,j_4}$}
\vspace*{0.2cm}
\end{minipage}
\\ \mbox{} \\
\begin{minipage}{0.5\textwidth}
\scalebox{0.7}{
\begin{tikzpicture}[font=\large]

\draw[line width=0.5mm] (0,0) -- (0,2);
\node[align=center] at (0.3,1) {\large $j_1$};
\draw[line width=0.5mm] (2,0) -- (2,2);
\node[align=center] at (2.3,1) {\large $j_2$};
\draw[line width=0.5mm] (4,0) -- (4,2);
\node[align=center] at (4.3,1) {\large $j_3$};
\draw[line width=0.5mm] (6,0) -- (6,2);
\node[align=center] at (6.3,1) {\large $j_4$};
\draw[line width=0.5mm] (0,0) -- (6,0);
\draw[line width=0.5mm] (0,2) -- (6,2);
\node[align=center] at (1,2.35) {\large $j_1$};
\node[align=center] at (1,0.35) {\large $j_1$};
\node[align=center] at (3,2.35) {\large $j_a$};
\node[align=center] at (3,0.35) {\large $j_a$};
\node[align=center] at (5,2.35) {\large $j_4$};
\node[align=center] at (5,0.35) {\large $j_4$};
\node[align=center] at (-1.6,1) {\large $j=0$};
\node[align=center] at (7.6,1) {\large $j=0$};
\draw[line width=0.5mm] (0,0) to[out=180,in=270] (-1,1);
\draw[line width=0.5mm] (-1,1) to[out=90,in=180] (0,2);
\draw[line width=0.5mm] (6,0) to[out=0,in=270] (7,1);
\draw[line width=0.5mm] (7,1) to[out=90,in=0] (6,2);

\draw[line width=0.5mm,->] (-1,0.95) -- (-1,1.05);
\draw[line width=0.5mm,->] (7,0.95) -- (7,1.05);

\draw[line width=0.5mm,->] (0.95,2) -- (1.05,2);
\draw[line width=0.5mm,->] (2.95,2) -- (3.05,2);
\draw[line width=0.5mm,->] (4.95,2) -- (5.05,2);

\draw[line width=0.5mm,->] (0.95,0) -- (1.05,0);
\draw[line width=0.5mm,->] (2.95,0) -- (3.05,0);
\draw[line width=0.5mm,->] (4.95,0) -- (5.05,0);

\draw[line width=0.5mm,->] (0,1.05) -- (0,0.95);

\draw[line width=0.5mm,->] (2,0.95) -- (2,1.05);

\draw[line width=0.5mm,->] (4,0.95) -- (4,1.05);

\draw[line width=0.5mm,->] (6,1.05) -- (6,0.95);

\draw[fill=white, line width=0.5mm] (0,2) circle [radius=0.35];
\draw[fill=white, line width=0.5mm] (2,2) circle [radius=0.35];
\draw[fill=white, line width=0.5mm] (4,2) circle [radius=0.35];
\draw[fill=white, line width=0.5mm] (6,2) circle [radius=0.35];
\draw[fill=white, line width=0.5mm] (0,0) circle [radius=0.35];
\draw[fill=white, line width=0.5mm] (2,0) circle [radius=0.35];
\draw[fill=white, line width=0.5mm] (4,0) circle [radius=0.35];
\draw[fill=white, line width=0.5mm] (6,0) circle [radius=0.35];

\end{tikzpicture}
}

\vspace*{0.2cm}
\centerline{\small (b) $Z^{j_a,j_a}_{\left(j_1,j_2,j_3,j_4\right),\left(j_1,j_2,j_3,j_4\right)}$}
\vspace*{0.1cm}
\end{minipage}

\caption{Index reversal of an SU(2)-invariant rank-4 tensor.}
\label{fig:reversecluster}
\end{figure}
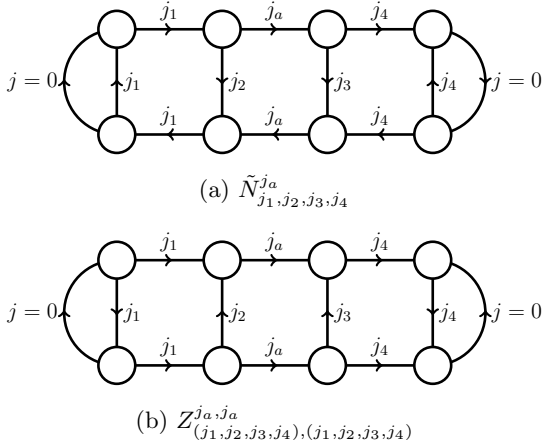

The contraction of two tensors sometimes requires the reversal of
indices, so 
that joint indices to be summed over are always ingoing
for one and outgoing for the other.
If all indices are initialized so that this condition is met in the
beginning of a tensor network algorithm, and their directions are consistently 
reassigned during decompositions,
it is sufficient to 
implement the reversal of all indices of a tensor only.
To this end, we again consider our standard example 
of a rank-4 tensor, this time with specified directions of 
indices, say (out,in,in,out)
\begin{align}
& T_{\left(j_1 m_1 t_1\right),\left(j_2 m_2 t_2\right),
			\left(j_3 m_3 t_3\right),\left(j_4 m_4 t_4\right)} \nonumber \\
& \quad = 
\sum\limits_{j_a} P^{j_a}_{\left(j_1 t_1\right),\left(j_2 t_2\right),
			\left(j_3 t_3\right),\left(j_4 t_4\right)} \nonumber \\
& \qquad \quad \cdot C^{j_a}_{\left(j_1 m_1\right),\left(j_2 m_2\right),
			\left(j_3 m_3\right),\left(j_4 m_4\right)} \, .
\end{align}
We denote the flipped set of directions, (in,out,out,in), by a tilde:
\begin{align}
&\tilde{T}_{\left(j_1 m_1 t_1\right),\left(j_2 m_2 t_2\right),
			\left(j_3 m_3 t_3\right),\left(j_4 m_4 t_4\right)} \nonumber \\
&\quad = \sum\limits_{j_a} \tilde{P}^{j_a}_{\left(j_1 t_1\right),\left(j_2 t_2\right),
			\left(j_3 t_3\right),\left(j_4 t_4\right)} \nonumber \\
&\qquad \quad \cdot \tilde{C}^{j_a}_{\left(j_1 m_1\right),\left(j_2 m_2\right),
			\left(j_3 m_3\right),\left(j_4 m_4\right)} \, .
\end{align}
Since the indices of a full tensor $T$ can be flipped simultaneously without
changing its value, $T=\tilde{T}$.
Following the methods of 
the previous sections, 
we 
multiply a given $P$ with its intertwiner $C$, flip the indices,
and compactify using 
Eq.~\eqref{eq:su2compact}, applying 
the intertwiner $\tilde{C}$ with reversed external indices:
\begin{align}
& \tilde{P}^{j_a}_{\left(j_1 t_1\right),\left(j_2 t_2\right),
			\left(j_3 t_3\right),\left(j_4 t_4\right)} \nonumber \\
& \quad = \left( \tilde{N}^{j_a}_{j_1,j_2,j_3,j_4} \right)^{-1} \nonumber \\
& \qquad \quad \sum\limits_{j_a} P^{j_a}_{\left(j_1 t_1\right),\left(j_2 t_2\right),
		\left(j_3 t_3\right),\left(j_4 t_4\right)} \nonumber \\
& \qquad \quad Z^{j_a,j_a}_{\left(j_1,j_2,j_3,j_4\right),\left(j_1,j_2,j_3,j_4\right)} \, , \label{eq:reversesu2}
\end{align}
\begin{align}
& Z^{j_a,j_a}_{\left(j_1,j_2,j_3,j_4\right),\left(j_1,j_2,j_3,j_4\right)} \nonumber \\
& \quad = \sum\limits_{m_1,m_2,m_3,m_4}
			C^{j_a}_{\left(j_1 m_1\right),\left(j_2 m_2\right),
				\left(j_3 m_3\right),\left(j_4 m_4\right)} \nonumber \\
& \qquad \qquad \qquad \quad \cdot \tilde{C}^{j_a}_{\left(j_1 m_1\right),\left(j_2 m_2\right),
				\left(j_3 m_3\right),\left(j_4 m_4\right)} \, .
\end{align}
The resulting clusters $N$ and $Z$ are depicted in 
Figs.~\ref{fig:reversecluster}(a) 
and (b), respectively, 
and can again be processed by utilizing 
the orthogonality relations of CGCs.

\newcommand{\widthfac}{0.45}
\newcommand{\scalefac}{0.7}
\newcommand{\widthfacorth}{0.45}
\newcommand{\scalefacorth}{0.55}

\section{Evaluating and optimizing a PEPS} \label{sec:pepseval}

We now turn to extracting physical properties from an fPEPS (or, more
precisely, a tensor network composing of fPEPS and PEPOs).
First, we need to describe how to evaluate expectation values of
fPEPS-PEPO tensor networks; 
they
are required to
implement optimization schemes as well as to 
measure
observables.
Subsequently, we describe two different schemes to
variationally optimize fPEPS: local updates based on the iterative
diagonalization of effective Hamiltonians for a single site or a
single bond, and a global optimization of the fPEPS in which an
approximation to the gradient of the energy functional is used as the
basis of a gradient-based
optimization scheme.

\subsection{Expectation values} \label{sec:pepsexp}
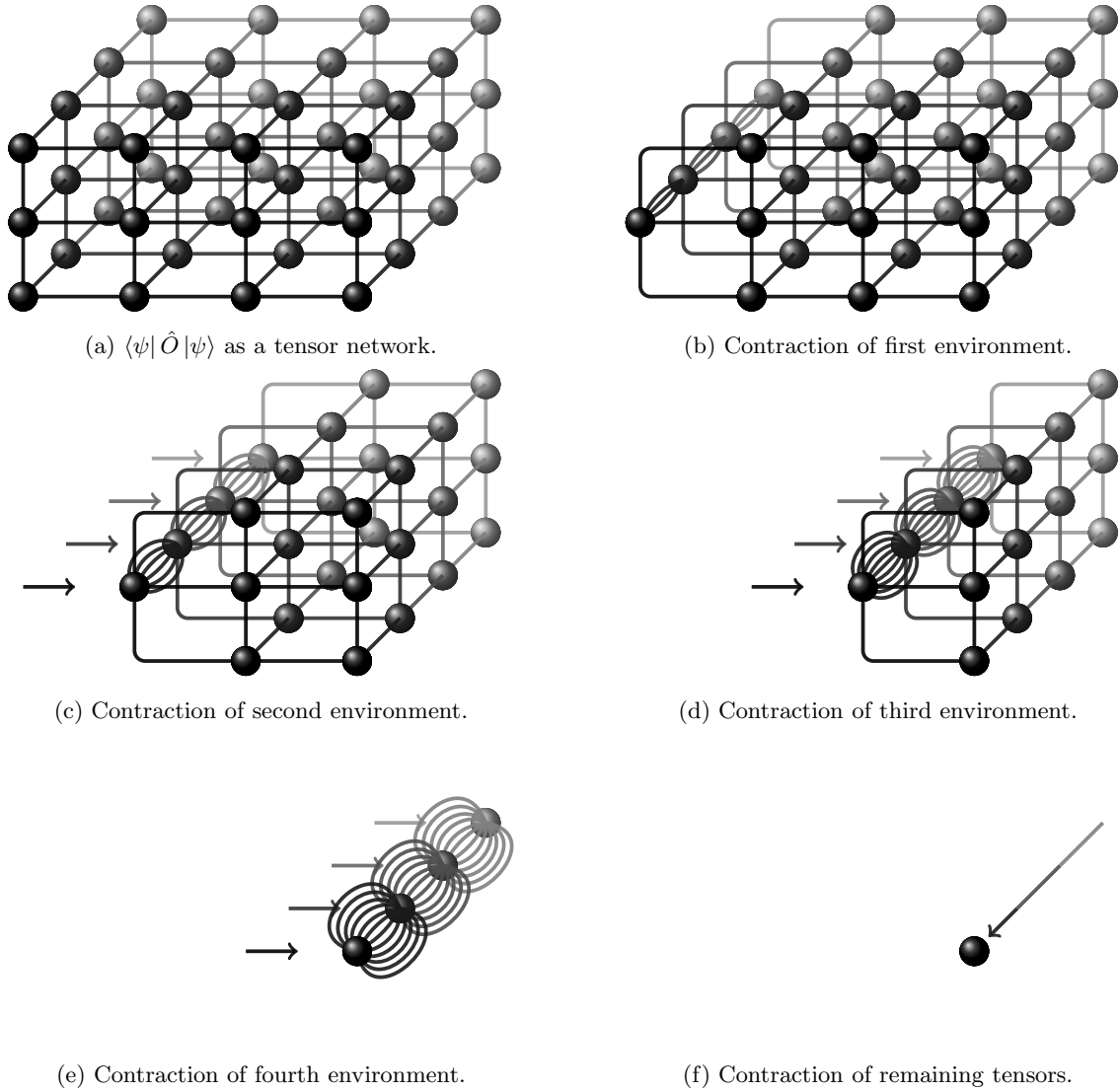
\begin{figure*}[!htb]
\begin{minipage}{\widthfac\textwidth}
\centering
% PIC 1
\begin{tikzpicture}[font=\tiny,opacity=0.9]

% plane 1
\foreach \y in {0,1,2}{
	\draw[line width=0.5mm, black!40, opacity=0.9] (0,\y,0) -- (4.5,\y,0);
}
\foreach \x in {0,1.5,3,4.5}{
	\draw[line width=0.5mm, black!40, opacity=0.9] (\x,0,0) -- (\x,2,0);
}
\foreach \x in {0,1.5,3,4.5}{
	\foreach \y in {0,1,2}{
		\shade[ball color = black!40, opacity = 0.9] (\x,\y,0) circle (0.2cm);
	}
}
\foreach \x in {0,1.5,3,4.5}{
	\foreach \y in {0,1,2}{
		\draw[line width=0.5mm, black!50, opacity=0.9] (\x,\y,0) -- (\x,\y,1.5);
	}
}

% plane 2
\foreach \y in {0,1,2}{
	\draw[line width=0.5mm, black!60, opacity=0.9] (0,\y,1.5) -- (4.5,\y,1.5);
}
\foreach \x in {0,1.5,3,4.5}{
	\draw[line width=0.5mm, black!60, opacity=0.9] (\x,0,1.5) -- (\x,2,1.5);
}
\foreach \x in {0,1.5,3,4.5}{
	\foreach \y in {0,1,2}{
		\shade[ball color = black!60, opacity = 0.9] (\x,\y,1.5) circle (0.2cm);
	}
}
\foreach \x in {0,1.5,3,4.5}{
	\foreach \y in {0,1,2}{
		\draw[line width=0.5mm, black!70, opacity=0.9] (\x,\y,1.5) -- (\x,\y,3);
	}
}

% plane 3
\foreach \y in {0,1,2}{
	\draw[line width=0.5mm, black!80, opacity=0.9] (0,\y,3) -- (4.5,\y,3);
}
\foreach \x in {0,1.5,3,4.5}{
	\draw[line width=0.5mm, black!80, opacity=0.9] (\x,0,3) -- (\x,2,3);
}
\foreach \x in {0,1.5,3,4.5}{
	\foreach \y in {0,1,2}{
		\shade[ball color = black!80, opacity = 0.9] (\x,\y,3) circle (0.2cm);
	}
}
\foreach \x in {0,1.5,3,4.5}{
	\foreach \y in {0,1,2}{
		\draw[line width=0.5mm, black!90, opacity=0.9] (\x,\y,3) -- (\x,\y,4.5);
	}
}

% plane 4
\foreach \y in {0,1,2}{
	\draw[line width=0.5mm, black!100, opacity=0.9] (0,\y,4.5) -- (4.5,\y,4.5);
}
\foreach \x in {0,1.5,3,4.5}{
	\draw[line width=0.5mm, black!100, opacity=0.9] (\x,0,4.5) -- (\x,2,4.5);
}

\foreach \x in {0,1.5,3,4.5}{
	\foreach \y in {0,1,2}{
		\shade[ball color = black!100, opacity = 0.9] (\x,\y,4.5) circle (0.2cm);
	}
}

\shade[ball color = white, opacity = 0] (6,2,0) circle (0.2cm);

\end{tikzpicture}

\vspace*{0.2cm}
\centerline{\small (a) $\left<\psi\right| \hat{O} \left| \psi \right>$ as a
  tensor network.}
\vspace*{0.1cm}
\end{minipage}
\hfill \hspace{0.02\textwidth}
\begin{minipage}{\widthfac\textwidth}
\centering

% PIC 2
\begin{tikzpicture}[font=\tiny,opacity=0.9]

% plane 1
\foreach \y in {0,1,2}{
	\draw[line width=0.5mm, black!40, opacity=0.9] (1.5,\y,0) -- (4.5,\y,0);
}
\foreach \x in {1.5,3,4.5}{
	\draw[line width=0.5mm, black!40, opacity=0.9] (\x,0,0) -- (\x,2,0);
}
\draw[line width=0.5mm, black!40, opacity=0.9,rounded corners] (1.5,2,0) -- (0,2,0) -- (0,0,0) -- (1.5,0,0);
\draw[line width=0.5mm, black!40, opacity=0.9] (0,1,0) -- (1.5,1,0);
\foreach \x in {1.5,3,4.5}{
	\foreach \y in {0,1,2}{
		\shade[ball color = black!40, opacity = 0.9] (\x,\y,0) circle (0.2cm);
	}
}
\shade[ball color = black!40, opacity = 0.9] (0,1,0) circle (0.2cm);
\foreach \x in {1.5,3,4.5}{
	\foreach \y in {0,1,2}{
		\draw[line width=0.5mm, black!50, opacity=0.9] (\x,\y,0) -- (\x,\y,1.5);
	}
}
\draw[line width=0.5mm, black!50, opacity=0.9] (0,1,0) to[out=225,in=45] (0,1,1.5);
\draw[line width=0.5mm, black!50, opacity=0.9] (0,1,0) to[out=200,in=65] (0,1,1.5);
\draw[line width=0.5mm, black!50, opacity=0.9] (0,1,0) to[out=250,in=20] (0,1,1.5);

% plane 2
\foreach \y in {0,1,2}{
	\draw[line width=0.5mm, black!60, opacity=0.9] (1.5,\y,1.5) -- (4.5,\y,1.5);
}
\foreach \x in {1.5,3,4.5}{
	\draw[line width=0.5mm, black!60, opacity=0.9] (\x,0,1.5) -- (\x,2,1.5);
}
\draw[line width=0.5mm, black!60, opacity=0.9,rounded corners] (1.5,2,1.5) -- (0,2,1.5) -- (0,0,1.5) -- (1.5,0,1.5);
\draw[line width=0.5mm, black!60, opacity=0.9] (0,1,1.5) -- (1.5,1,1.5);
\foreach \x in {1.5,3,4.5}{
	\foreach \y in {0,1,2}{
		\shade[ball color = black!60, opacity = 0.9] (\x,\y,1.5) circle (0.2cm);
	}
}
\shade[ball color = black!60, opacity = 0.9] (0,1,1.5) circle (0.2cm);
\foreach \x in {1.5,3,4.5}{
	\foreach \y in {0,1,2}{
		\draw[line width=0.5mm, black!70, opacity=0.9] (\x,\y,1.5) -- (\x,\y,3);
	}
}
\draw[line width=0.5mm, black!70, opacity=0.9] (0,1,1.5) to[out=225,in=45] (0,1,3);
\draw[line width=0.5mm, black!70, opacity=0.9] (0,1,1.5) to[out=200,in=65] (0,1,3);
\draw[line width=0.5mm, black!70, opacity=0.9] (0,1,1.5) to[out=250,in=20] (0,1,3);

% plane 3
\foreach \y in {0,1,2}{
	\draw[line width=0.5mm, black!80, opacity=0.9] (1.5,\y,3) -- (4.5,\y,3);
}
\foreach \x in {1.5,3,4.5}{
	\draw[line width=0.5mm, black!80, opacity=0.9] (\x,0,3) -- (\x,2,3);
}
\draw[line width=0.5mm, black!80, opacity=0.9,rounded corners] (1.5,2,3) -- (0,2,3) -- (0,0,3) -- (1.5,0,3);
\draw[line width=0.5mm, black!80, opacity=0.9] (0,1,3) -- (1.5,1,3);
\foreach \x in {1.5,3,4.5}{
	\foreach \y in {0,1,2}{
		\shade[ball color = black!80, opacity = 0.9] (\x,\y,3) circle (0.2cm);
	}
}
\shade[ball color = black!80, opacity = 0.9] (0,1,3) circle (0.2cm);
\foreach \x in {1.5,3,4.5}{
	\foreach \y in {0,1,2}{
		\draw[line width=0.5mm, black!90, opacity=0.9] (\x,\y,3) -- (\x,\y,4.5);
	}
}
\draw[line width=0.5mm, black!90, opacity=0.9] (0,1,3) to[out=225,in=45] (0,1,4.5);
\draw[line width=0.5mm, black!90, opacity=0.9] (0,1,3) to[out=200,in=65] (0,1,4.5);
\draw[line width=0.5mm, black!90, opacity=0.9] (0,1,3) to[out=250,in=20] (0,1,4.5);

% plane 4
\foreach \y in {0,1,2}{
	\draw[line width=0.5mm, black!100, opacity=0.9] (1.5,\y,4.5) -- (4.5,\y,4.5);
}
\foreach \x in {1.5,3,4.5}{
	\draw[line width=0.5mm, black!100, opacity=0.9] (\x,0,4.5) -- (\x,2,4.5);
}
\draw[line width=0.5mm, black!100, opacity=0.9,rounded corners] (1.5,2,4.5) -- (0,2,4.5) -- (0,0,4.5) -- (1.5,0,4.5);
\draw[line width=0.5mm, black!100, opacity=0.9] (0,1,4.5) -- (1.5,1,4.5);
\foreach \x in {1.5,3,4.5}{
	\foreach \y in {0,1,2}{
		\shade[ball color = black!100, opacity = 0.9] (\x,\y,4.5) circle (0.2cm);
	}
}
\shade[ball color = black!100, opacity = 0.9] (0,1,4.5) circle (0.2cm);

\shade[ball color = white, opacity = 0] (6,2,0) circle (0.2cm);

\end{tikzpicture}

\vspace*{0.2cm}
\centerline{\small (b) Contraction of first environment.}
\vspace{0.1cm}        
\end{minipage}

\begin{minipage}{\widthfac\textwidth}
\centering

% PIC 3

\begin{tikzpicture}[font=\tiny,opacity=0.9]

\draw[line width=0.5mm, black!40, opacity=0.9,->] (0,1,0) -- (0.7,1,0);
\draw[line width=0.5mm, black!60, opacity=0.9,->] (0,1,1.5) -- (0.7,1,1.5);
\draw[line width=0.5mm, black!80, opacity=0.9,->] (0,1,3) -- (0.7,1,3);
\draw[line width=0.5mm, black!100, opacity=0.9,->] (0,1,4.5) -- (0.7,1,4.5);

% plane 1

\foreach \y in {0,1,2}{
	\draw[line width=0.5mm, black!40, opacity=0.9] (3,\y,0) -- (4.5,\y,0);
}
\foreach \x in {3,4.5}{
	\draw[line width=0.5mm, black!40, opacity=0.9] (\x,0,0) -- (\x,2,0);
}
\draw[line width=0.5mm, black!40, opacity=0.9,rounded corners] (3,2,0) -- (1.5,2,0) -- (1.5,0,0) -- (3,0,0);
\draw[line width=0.5mm, black!40, opacity=0.9] (1.5,1,0) -- (3,1,0);
\foreach \x in {3,4.5}{
	\foreach \y in {0,1,2}{
		\shade[ball color = black!40, opacity = 0.9] (\x,\y,0) circle (0.2cm);
	}
}
\shade[ball color = black!40, opacity = 0.9] (1.5,1,0) circle (0.2cm);
\foreach \x in {3,4.5}{
	\foreach \y in {0,1,2}{
		\draw[line width=0.5mm, black!50, opacity=0.9] (\x,\y,0) -- (\x,\y,1.5);
	}
}

\draw[line width=0.5mm, black!50, opacity=0.9] (1.5,1,0) to[out=150,in=130,distance=0.4cm] (1.5,1,1.5);
\draw[line width=0.5mm, black!50, opacity=0.9] (1.5,1,0) to[out=185,in=95] (1.5,1,1.5);
\draw[line width=0.5mm, black!50, opacity=0.9] (1.5,1,0) to[out=210,in=60] (1.5,1,1.5);
\draw[line width=0.5mm, black!50, opacity=0.9] (1.5,1,0) to[out=240,in=30] (1.5,1,1.5);
\draw[line width=0.5mm, black!50, opacity=0.9] (1.5,1,0) to[out=275,in=355] (1.5,1,1.5);
\draw[line width=0.5mm, black!50, opacity=0.9] (1.5,1,0) to[out=310,in=320,distance=0.4cm] (1.5,1,1.5);

\draw[line width=0.5mm, black!50, opacity=0.9] (1.5,1,0) to[out=275,in=355] (1.5,1,1.5);

% plane 2
\foreach \y in {0,1,2}{
	\draw[line width=0.5mm, black!60, opacity=0.9] (3,\y,1.5) -- (4.5,\y,1.5);
}
\foreach \x in {3,4.5}{
	\draw[line width=0.5mm, black!60, opacity=0.9] (\x,0,1.5) -- (\x,2,1.5);
}
\draw[line width=0.5mm, black!60, opacity=0.9,rounded corners] (3,2,1.5) -- (1.5,2,1.5) -- (1.5,0,1.5) -- (3,0,1.5);
\draw[line width=0.5mm, black!60, opacity=0.9] (1.5,1,1.5) -- (3,1,1.5);
\foreach \x in {3,4.5}{
	\foreach \y in {0,1,2}{
		\shade[ball color = black!60, opacity = 0.9] (\x,\y,1.5) circle (0.2cm);
	}
}
\shade[ball color = black!60, opacity = 0.9] (1.5,1,1.5) circle (0.2cm);
\foreach \x in {3,4.5}{
	\foreach \y in {0,1,2}{
		\draw[line width=0.5mm, black!70, opacity=0.9] (\x,\y,1.5) -- (\x,\y,3);
	}
}
\draw[line width=0.5mm, black!70, opacity=0.9] (1.5,1,1.5) to[out=150,in=130,distance=0.4cm] (1.5,1,3);
\draw[line width=0.5mm, black!70, opacity=0.9] (1.5,1,1.5) to[out=185,in=95] (1.5,1,3);
\draw[line width=0.5mm, black!70, opacity=0.9] (1.5,1,1.5) to[out=210,in=60] (1.5,1,3);
\draw[line width=0.5mm, black!70, opacity=0.9] (1.5,1,1.5) to[out=240,in=30] (1.5,1,3);
\draw[line width=0.5mm, black!70, opacity=0.9] (1.5,1,1.5) to[out=275,in=355] (1.5,1,3);
\draw[line width=0.5mm, black!70, opacity=0.9] (1.5,1,1.5) to[out=310,in=320,distance=0.4cm] (1.5,1,3);

% plane 3
\foreach \y in {0,1,2}{
	\draw[line width=0.5mm, black!80, opacity=0.9] (3,\y,3) -- (4.5,\y,3);
}
\foreach \x in {3,4.5}{
	\draw[line width=0.5mm, black!80, opacity=0.9] (\x,0,3) -- (\x,2,3);
}
\draw[line width=0.5mm, black!80, opacity=0.9,rounded corners] (3,2,3) -- (1.5,2,3) -- (1.5,0,3) -- (3,0,3);
\draw[line width=0.5mm, black!80, opacity=0.9] (1.5,1,3) -- (3,1,3);
\foreach \x in {3,4.5}{
	\foreach \y in {0,1,2}{
		\shade[ball color = black!80, opacity = 0.9] (\x,\y,3) circle (0.2cm);
	}
}
\shade[ball color = black!80, opacity = 0.9] (1.5,1,3) circle (0.2cm);
\foreach \x in {3,4.5}{
	\foreach \y in {0,1,2}{
		\draw[line width=0.5mm, black!90, opacity=0.9] (\x,\y,3) -- (\x,\y,4.5);
	}
}
\draw[line width=0.5mm, black!90, opacity=0.9] (1.5,1,3) to[out=150,in=130,distance=0.4cm] (1.5,1,4.5);
\draw[line width=0.5mm, black!90, opacity=0.9] (1.5,1,3) to[out=185,in=95] (1.5,1,4.5);
\draw[line width=0.5mm, black!90, opacity=0.9] (1.5,1,3) to[out=210,in=60] (1.5,1,4.5);
\draw[line width=0.5mm, black!90, opacity=0.9] (1.5,1,3) to[out=240,in=30] (1.5,1,4.5);
\draw[line width=0.5mm, black!90, opacity=0.9] (1.5,1,3) to[out=275,in=355] (1.5,1,4.5);
\draw[line width=0.5mm, black!90, opacity=0.9] (1.5,1,3) to[out=310,in=320,distance=0.4cm] (1.5,1,4.5);

% plane 4
\foreach \y in {0,1,2}{
	\draw[line width=0.5mm, black!100, opacity=0.9] (3,\y,4.5) -- (4.5,\y,4.5);
}
\foreach \x in {3,4.5}{
	\draw[line width=0.5mm, black!100, opacity=0.9] (\x,0,4.5) -- (\x,2,4.5);
}
\draw[line width=0.5mm, black!100, opacity=0.9,rounded corners] (3,2,4.5) -- (1.5,2,4.5) -- (1.5,0,4.5) -- (3,0,4.5);
\draw[line width=0.5mm, black!100, opacity=0.9] (1.5,1,4.5) -- (3,1,4.5);
\foreach \x in {3,4.5}{
	\foreach \y in {0,1,2}{
		\shade[ball color = black!100, opacity = 0.9] (\x,\y,4.5) circle (0.2cm);
	}
}
\shade[ball color = black!100, opacity = 0.9] (1.5,1,4.5) circle (0.2cm);

\shade[ball color = white, opacity = 0] (0,0,4.5) circle (0.2cm);
\shade[ball color = white, opacity = 0] (6,2,0) circle (0.2cm);

\end{tikzpicture}

\vspace*{0.2cm}
\centerline{\small (c) Contraction of second environment.}
\vspace{0.1cm}        
\end{minipage}
\hfill \hspace{0.02\textwidth}
\begin{minipage}{\widthfac\textwidth}
\centering
% PIC 4

\begin{tikzpicture}[font=\tiny,opacity=0.9]

\draw[line width=0.5mm, black!40, opacity=0.9,->] (1.5,1,0) -- (2.2,1,0);
\draw[line width=0.5mm, black!60, opacity=0.9,->] (1.5,1,1.5) -- (2.2,1,1.5);
\draw[line width=0.5mm, black!80, opacity=0.9,->] (1.5,1,3) -- (2.2,1,3);
\draw[line width=0.5mm, black!100, opacity=0.9,->] (1.5,1,4.5) -- (2.2,1,4.5);

% plane 1
\draw[line width=0.5mm, black!40, opacity=0.9] (4.5,0,0) -- (4.5,2,0);

\draw[line width=0.5mm, black!40, opacity=0.9,rounded corners] (4.5,2,0) -- (3,2,0) -- (3,0,0) -- (4.5,0,0);
\draw[line width=0.5mm, black!40, opacity=0.9] (3,1,0) -- (4.5,1,0);
\foreach \y in {0,1,2}{
	\shade[ball color = black!40, opacity = 0.9] (4.5,\y,0) circle (0.2cm);
}
\shade[ball color = black!40, opacity = 0.9] (3,1,0) circle (0.2cm);
\foreach \y in {0,1,2}
	\draw[line width=0.5mm, black!50, opacity=0.9] (4.5,\y,0) -- (4.5,\y,1.5);

\draw[line width=0.5mm, black!50, opacity=0.9] (3,1,0) to[out=120,in=145,distance=0.55cm] (3,1,1.5);
\draw[line width=0.5mm, black!50, opacity=0.9] (3,1,0) to[out=140,in=125,distance=0.4cm] (3,1,1.5);
\draw[line width=0.5mm, black!50, opacity=0.9] (3,1,0) to[out=170,in=95] (3,1,1.5);

\draw[line width=0.5mm, black!50, opacity=0.9] (3,1,0) to[out=200,in=65] (3,1,1.5);
\draw[line width=0.5mm, black!50, opacity=0.9] (3,1,0) to[out=225,in=45] (3,1,1.5);
\draw[line width=0.5mm, black!50, opacity=0.9] (3,1,0) to[out=250,in=20] (3,1,1.5);

\draw[line width=0.5mm, black!50, opacity=0.9] (3,1,0) to[out=280,in=350] (3,1,1.5);
\draw[line width=0.5mm, black!50, opacity=0.9] (3,1,0) to[out=310,in=320,distance=0.4cm] (3,1,1.5);
\draw[line width=0.5mm, black!50, opacity=0.9] (3,1,0) to[out=330,in=300,distance=0.55cm] (3,1,1.5);

% plane 2
\draw[line width=0.5mm, black!60, opacity=0.9] (4.5,0,1.5) -- (4.5,2,1.5);

\draw[line width=0.5mm, black!60, opacity=0.9,rounded corners] (4.5,2,1.5) -- (3,2,1.5) -- (3,0,1.5) -- (4.5,0,1.5);
\draw[line width=0.5mm, black!60, opacity=0.9] (3,1,1.5) -- (4.5,1,1.5);
\foreach \y in {0,1,2}{
	\shade[ball color = black!60, opacity = 0.9] (4.5,\y,1.5) circle (0.2cm);
}
\shade[ball color = black!60, opacity = 0.9] (3,1,1.5) circle (0.2cm);
\foreach \y in {0,1,2}{
	\draw[line width=0.5mm, black!70, opacity=0.9] (4.5,\y,1.5) -- (4.5,\y,3);
}

\draw[line width=0.5mm, black!70, opacity=0.9] (3,1,1.5) to[out=120,in=145,distance=0.55cm] (3,1,3);
\draw[line width=0.5mm, black!70, opacity=0.9] (3,1,1.5) to[out=140,in=125,distance=0.4cm] (3,1,3);
\draw[line width=0.5mm, black!70, opacity=0.9] (3,1,1.5) to[out=170,in=95] (3,1,3);

\draw[line width=0.5mm, black!70, opacity=0.9] (3,1,1.5) to[out=200,in=65] (3,1,3);
\draw[line width=0.5mm, black!70, opacity=0.9] (3,1,1.5) to[out=225,in=45] (3,1,3);
\draw[line width=0.5mm, black!70, opacity=0.9] (3,1,1.5) to[out=250,in=20] (3,1,3);

\draw[line width=0.5mm, black!70, opacity=0.9] (3,1,1.5) to[out=280,in=350] (3,1,3);
\draw[line width=0.5mm, black!70, opacity=0.9] (3,1,1.5) to[out=310,in=320,distance=0.4cm] (3,1,3);
\draw[line width=0.5mm, black!70, opacity=0.9] (3,1,1.5) to[out=330,in=300,distance=0.55cm] (3,1,3);

% plane 3
\draw[line width=0.5mm, black!80, opacity=0.9] (4.5,0,3) -- (4.5,2,3);

\draw[line width=0.5mm, black!80, opacity=0.9,rounded corners] (4.5,2,3) -- (3,2,3) -- (3,0,3) -- (4.5,0,3);
\draw[line width=0.5mm, black!80, opacity=0.9] (3,1,3) -- (4.5,1,3);
\foreach \y in {0,1,2}{
	\shade[ball color = black!80, opacity = 0.9] (4.5,\y,3) circle (0.2cm);
}
\shade[ball color = black!80, opacity = 0.9] (3,1,3) circle (0.2cm);
\foreach \y in {0,1,2}{
	\draw[line width=0.5mm, black!90, opacity=0.9] (4.5,\y,3) -- (4.5,\y,4.5);
}

\draw[line width=0.5mm, black!90, opacity=0.9] (3,1,3) to[out=120,in=145,distance=0.55cm] (3,1,4.5);
\draw[line width=0.5mm, black!90, opacity=0.9] (3,1,3) to[out=140,in=125,distance=0.4cm] (3,1,4.5);
\draw[line width=0.5mm, black!90, opacity=0.9] (3,1,3) to[out=170,in=95] (3,1,4.5);

\draw[line width=0.5mm, black!90, opacity=0.9] (3,1,3) to[out=200,in=65] (3,1,4.5);
\draw[line width=0.5mm, black!90, opacity=0.9] (3,1,3) to[out=225,in=45] (3,1,4.5);
\draw[line width=0.5mm, black!90, opacity=0.9] (3,1,3) to[out=250,in=20] (3,1,4.5);

\draw[line width=0.5mm, black!90, opacity=0.9] (3,1,3) to[out=280,in=350] (3,1,4.5);
\draw[line width=0.5mm, black!90, opacity=0.9] (3,1,3) to[out=310,in=320,distance=0.4cm] (3,1,4.5);
\draw[line width=0.5mm, black!90, opacity=0.9] (3,1,3) to[out=330,in=300,distance=0.55cm] (3,1,4.5);

% plane 4
\draw[line width=0.5mm, black!100, opacity=0.9] (4.5,0,4.5) -- (4.5,2,4.5);

\draw[line width=0.5mm, black!100, opacity=0.9,rounded corners] (4.5,2,4.5) -- (3,2,4.5) -- (3,0,4.5) -- (4.5,0,4.5);
\draw[line width=0.5mm, black!100, opacity=0.9] (3,1,4.5) -- (4.5,1,4.5);
\foreach \y in {0,1,2}{
	\shade[ball color = black!100, opacity = 0.9] (4.5,\y,4.5) circle (0.2cm);
}
\shade[ball color = black!100, opacity = 0.9] (3,1,4.5) circle (0.2cm);

\shade[ball color = white, opacity = 0] (0,0,4.5) circle (0.2cm);
\shade[ball color = white, opacity = 0] (6,2,0) circle (0.2cm);

\end{tikzpicture}

\vspace*{0.2cm}
\centerline{\small (d) Contraction of third environment.}
\vspace{0.1cm}        
\end{minipage}
\begin{minipage}{\widthfac\textwidth}
\centering

% PIC 5

\begin{tikzpicture}[font=\tiny,opacity=0.9]

\draw[line width=0.5mm, black!40, opacity=0.9,->] (3,1,0) -- (3.7,1,0);
\draw[line width=0.5mm, black!60, opacity=0.9,->] (3,1,1.5) -- (3.7,1,1.5);
\draw[line width=0.5mm, black!80, opacity=0.9,->] (3,1,3) -- (3.7,1,3);
\draw[line width=0.5mm, black!100, opacity=0.9,->] (3,1,4.5) -- (3.7,1,4.5);

% plane 1
\shade[ball color = black!40, opacity = 0.9] (4.5,1,0) circle (0.2cm);

\draw[line width=0.5mm, black!50, opacity=0.9] (4.5,1,0) to[out=110,in=170,distance=1cm] (4.5,1,1.5);
\draw[line width=0.5mm, black!50, opacity=0.9] (4.5,1,0) to[out=120,in=160,distance=0.75cm] (4.5,1,1.5);
\draw[line width=0.5mm, black!50, opacity=0.9] (4.5,1,0) to[out=135,in=145,distance=0.55cm] (4.5,1,1.5);
\draw[line width=0.5mm, black!50, opacity=0.9] (4.5,1,0) to[out=150,in=130,distance=0.4cm] (4.5,1,1.5);
\draw[line width=0.5mm, black!50, opacity=0.9] (4.5,1,0) to[out=185,in=95] (4.5,1,1.5);
\draw[line width=0.5mm, black!50, opacity=0.9] (4.5,1,0) to[out=210,in=60] (4.5,1,1.5);
\draw[line width=0.5mm, black!50, opacity=0.9] (4.5,1,0) to[out=240,in=30] (4.5,1,1.5);
\draw[line width=0.5mm, black!50, opacity=0.9] (4.5,1,0) to[out=275,in=355] (4.5,1,1.5);
\draw[line width=0.5mm, black!50, opacity=0.9] (4.5,1,0) to[out=310,in=320,distance=0.4cm] (4.5,1,1.5);
\draw[line width=0.5mm, black!50, opacity=0.9] (4.5,1,0) to[out=325,in=305,distance=0.55cm] (4.5,1,1.5);
\draw[line width=0.5mm, black!50, opacity=0.9] (4.5,1,0) to[out=340,in=290,distance=0.75cm] (4.5,1,1.5);
\draw[line width=0.5mm, black!50, opacity=0.9] (4.5,1,0) to[out=350,in=280,distance=1cm] (4.5,1,1.5);

% plane 2
\shade[ball color = black!60, opacity = 0.9] (4.5,1,1.5) circle (0.2cm);

\draw[line width=0.5mm, black!70, opacity=0.9] (4.5,1,1.5) to[out=110,in=170,distance=1cm] (4.5,1,3);
\draw[line width=0.5mm, black!70, opacity=0.9] (4.5,1,1.5) to[out=120,in=160,distance=0.75cm] (4.5,1,3);
\draw[line width=0.5mm, black!70, opacity=0.9] (4.5,1,1.5) to[out=135,in=145,distance=0.55cm] (4.5,1,3);
\draw[line width=0.5mm, black!70, opacity=0.9] (4.5,1,1.5) to[out=150,in=130,distance=0.4cm] (4.5,1,3);
\draw[line width=0.5mm, black!70, opacity=0.9] (4.5,1,1.5) to[out=185,in=95] (4.5,1,3);
\draw[line width=0.5mm, black!70, opacity=0.9] (4.5,1,1.5) to[out=210,in=60] (4.5,1,3);
\draw[line width=0.5mm, black!70, opacity=0.9] (4.5,1,1.5) to[out=240,in=30] (4.5,1,3);
\draw[line width=0.5mm, black!70, opacity=0.9] (4.5,1,1.5) to[out=275,in=355] (4.5,1,3);
\draw[line width=0.5mm, black!70, opacity=0.9] (4.5,1,1.5) to[out=310,in=320,distance=0.4cm] (4.5,1,3);
\draw[line width=0.5mm, black!70, opacity=0.9] (4.5,1,1.5) to[out=325,in=305,distance=0.55cm] (4.5,1,3);
\draw[line width=0.5mm, black!70, opacity=0.9] (4.5,1,1.5) to[out=340,in=290,distance=0.75cm] (4.5,1,3);
\draw[line width=0.5mm, black!70, opacity=0.9] (4.5,1,1.5) to[out=350,in=280,distance=1cm] (4.5,1,3);

% plane 3
\shade[ball color = black!80, opacity = 0.9] (4.5,1,3) circle (0.2cm);

\draw[line width=0.5mm, black!90, opacity=0.9] (4.5,1,3) to[out=110,in=170,distance=1cm] (4.5,1,4.5);
\draw[line width=0.5mm, black!90, opacity=0.9] (4.5,1,3) to[out=120,in=160,distance=0.75cm] (4.5,1,4.5);
\draw[line width=0.5mm, black!90, opacity=0.9] (4.5,1,3) to[out=135,in=145,distance=0.55cm] (4.5,1,4.5);
\draw[line width=0.5mm, black!90, opacity=0.9] (4.5,1,3) to[out=150,in=130,distance=0.4cm] (4.5,1,4.5);
\draw[line width=0.5mm, black!90, opacity=0.9] (4.5,1,3) to[out=185,in=95] (4.5,1,4.5);
\draw[line width=0.5mm, black!90, opacity=0.9] (4.5,1,3) to[out=210,in=60] (4.5,1,4.5);
\draw[line width=0.5mm, black!90, opacity=0.9] (4.5,1,3) to[out=240,in=30] (4.5,1,4.5);
\draw[line width=0.5mm, black!90, opacity=0.9] (4.5,1,3) to[out=275,in=355] (4.5,1,4.5);
\draw[line width=0.5mm, black!90, opacity=0.9] (4.5,1,3) to[out=310,in=320,distance=0.4cm] (4.5,1,4.5);
\draw[line width=0.5mm, black!90, opacity=0.9] (4.5,1,3) to[out=325,in=305,distance=0.55cm] (4.5,1,4.5);
\draw[line width=0.5mm, black!90, opacity=0.9] (4.5,1,3) to[out=340,in=290,distance=0.75cm] (4.5,1,4.5);
\draw[line width=0.5mm, black!90, opacity=0.9] (4.5,1,3) to[out=350,in=280,distance=1cm] (4.5,1,4.5);

% plane 4
\shade[ball color = black!100, opacity = 0.9] (4.5,1,4.5) circle (0.2cm);

\shade[ball color = white, opacity = 0] (0,0,4.5) circle (0.2cm);
\shade[ball color = white, opacity = 0] (6,2,0) circle (0.2cm);

\end{tikzpicture}

\vspace*{0.2cm}
\centerline{\small (e) Contraction of fourth environment.}
\vspace{0.1cm}        
\end{minipage}
\hfill \hspace{0.02\textwidth}
\begin{minipage}{\widthfac\textwidth}

\centering
% PIC 6

\begin{tikzpicture}[font=\tiny,opacity=0.9]

\draw[line width=0.5mm, black!50, opacity=0.9] (4.5,1,1.5) -- (4.5,1,0);
\draw[line width=0.5mm, black!70, opacity=0.9] (4.5,1,3) -- (4.5,1,1.5);
\draw[line width=0.5mm, black!90, opacity=0.9,->] (4.5,1,3) -- (4.5,1,4);

\shade[ball color = black!100, opacity = 0.9] (4.5,1,4.5) circle (0.2cm);

\shade[ball color = white, opacity = 0] (0,0,4.5) circle (0.2cm);
\shade[ball color = white, opacity = 0] (6,2,0) circle (0.2cm);

\end{tikzpicture}

\vspace*{0.2cm}
\centerline{\small (f) Contraction of remaining tensors.}
\vspace{0.1cm}        
\end{minipage}
\caption{Naive contraction of a PEPS-based tensor network.}
\label{fig:pepstn}
\end{figure*}

Given the representations of PEPSs and PEPOs described in previous
sections, we now consider 
the calculation of expectation values using these representations,
i.e.,
$\left<\psi \right| \hat{O} \left| \psi \right>$.
The straightforward sequential contraction of an
fPEPS-PEPO-fPEPS tensor network one might naively utilize to do this is
illustrated in Fig.~\ref{fig:pepstn} for a $4\times 4$ lattice.
Fig.~\ref{fig:pepstn}(a)
depicts the initial tensor network. The upper, middle, 
and lower planes are the fPEPS, PEPO, and adjoint fPEPS respectively. 
The subsequent contraction steps, depicted in
Figs.~\ref{fig:pepstn}(b) to (e),
successively build environments until the border is reached and 
the remaining tensors are contracted into a single number, yielding 
$\left<\psi \right| \hat{O} \left| \psi \right>$ in
Fig.~\ref{fig:pepstn}(f). 

The difficulty with this
naive approach is evident: 
In every new contraction step, the rank of the intermediate tensors
increases; this explosion of tensor complexity cannot be circumvented by
choosing any other conceivable order of contractions.
The exact contraction of fPEPS-based tensor networks thus scales exponentially 
with respect to system size and is therefore not feasible to carry out
in realistic applications.
However, we can circumvent this problem by constructing a sufficiently
well-controlled
approximation for each environment.

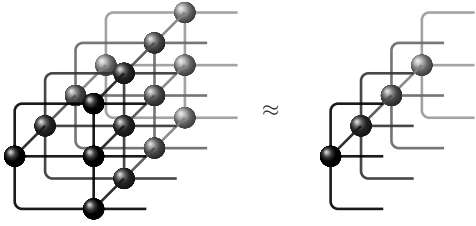
\begin{figure}[!htb]
\centering
\scalebox{0.7}{
\begin{tikzpicture}[font=\tiny,opacity=0.9]

% plane 1
\foreach \y in {0,1,2}{
	\draw[line width=0.5mm, black!40, opacity=0.9] (1.5,\y,0) -- (2.5,\y,0);
}

\draw[line width=0.5mm, black!40, opacity=0.9] (1.5,0,0) -- (1.5,2,0);
\draw[line width=0.5mm, black!40, opacity=0.9,rounded corners] (1.5,2,0) -- (0,2,0) -- (0,0,0) -- (1.5,0,0);
\draw[line width=0.5mm, black!40, opacity=0.9] (0,1,0) -- (1.5,1,0);
\foreach \y in {0,1,2}{
	\shade[ball color = black!40, opacity = 0.9] (1.5,\y,0) circle (0.2cm);
}
\shade[ball color = black!40, opacity = 0.9] (0,1,0) circle (0.2cm);
\foreach \y in {0,1,2}{
	\draw[line width=0.5mm, black!50, opacity=0.9] (1.5,\y,0) -- (1.5,\y,1.5);
}
\draw[line width=0.5mm, black!50, opacity=0.9] (0,1,0) to[out=225,in=45] (0,1,1.5);

% plane 2
\foreach \y in {0,1,2}{
	\draw[line width=0.5mm, black!60, opacity=0.9] (1.5,\y,1.5) -- (2.5,\y,1.5);
}
\draw[line width=0.5mm, black!60, opacity=0.9] (1.5,0,1.5) -- (1.5,2,1.5);
\draw[line width=0.5mm, black!60, opacity=0.9,rounded corners] (1.5,2,1.5) -- (0,2,1.5) -- (0,0,1.5) -- (1.5,0,1.5);
\draw[line width=0.5mm, black!60, opacity=0.9] (0,1,1.5) -- (1.5,1,1.5);
\foreach \y in {0,1,2}{
	\shade[ball color = black!60, opacity = 0.9] (1.5,\y,1.5) circle (0.2cm);
}
\shade[ball color = black!60, opacity = 0.9] (0,1,1.5) circle (0.2cm);
\foreach \y in {0,1,2}{
	\draw[line width=0.5mm, black!70, opacity=0.9] (1.5,\y,1.5) -- (1.5,\y,3);
}
\draw[line width=0.5mm, black!70, opacity=0.9] (0,1,1.5) to[out=225,in=45] (0,1,3);

% plane 3
\foreach \y in {0,1,2}{
	\draw[line width=0.5mm, black!80, opacity=0.9] (1.5,\y,3) -- (2.5,\y,3);
}
\draw[line width=0.5mm, black!80, opacity=0.9] (1.5,0,3) -- (1.5,2,3);
\draw[line width=0.5mm, black!80, opacity=0.9,rounded corners] (1.5,2,3) -- (0,2,3) -- (0,0,3) -- (1.5,0,3);
\draw[line width=0.5mm, black!80, opacity=0.9] (0,1,3) -- (1.5,1,3);
\foreach \y in {0,1,2}{
	\shade[ball color = black!80, opacity = 0.9] (1.5,\y,3) circle (0.2cm);
}
\shade[ball color = black!80, opacity = 0.9] (0,1,3) circle (0.2cm);
\foreach \y in {0,1,2}{
	\draw[line width=0.5mm, black!90, opacity=0.9] (1.5,\y,3) -- (1.5,\y,4.5);
}
\draw[line width=0.5mm, black!90, opacity=0.9] (0,1,3) to[out=225,in=45] (0,1,4.5);

% plane 4
\foreach \y in {0,1,2}{
	\draw[line width=0.5mm, black!100, opacity=0.9] (1.5,\y,4.5) -- (2.5,\y,4.5);
}
\draw[line width=0.5mm, black!100, opacity=0.9] (1.5,0,4.5) -- (1.5,2,4.5);
\draw[line width=0.5mm, black!100, opacity=0.9,rounded corners] (1.5,2,4.5) -- (0,2,4.5) -- (0,0,4.5) -- (1.5,0,4.5);
\draw[line width=0.5mm, black!100, opacity=0.9] (0,1,4.5) -- (1.5,1,4.5);
\foreach \y in {0,1,2}{
	\shade[ball color = black!100, opacity = 0.9] (1.5,\y,4.5) circle (0.2cm);
}
\shade[ball color = black!100, opacity = 0.9] (0,1,4.5) circle (0.2cm);

\node[align=center, black!100] at (4,1,2.25) {\large \textbf{$\approx$}};

% plane 1

\draw[line width=0.5mm, black!40, opacity=0.9,rounded corners] (7,2,0) -- (6,2,0) -- (6,0,0) -- (7,0,0);
\draw[line width=0.5mm, black!40, opacity=0.9] (6,1,0) -- (7,1,0);

\shade[ball color = black!40, opacity = 0.9] (6,1,0) circle (0.2cm);
\draw[line width=0.5mm, black!50, opacity=0.9] (6,1,0) to[out=225,in=45] (6,1,1.5);

% plane 2
\draw[line width=0.5mm, black!60, opacity=0.9,rounded corners] (7,2,1.5) -- (6,2,1.5) -- (6,0,1.5) -- (7,0,1.5);
\draw[line width=0.5mm, black!60, opacity=0.9] (6,1,1.5) -- (7,1,1.5);
\shade[ball color = black!60, opacity = 0.9] (6,1,1.5) circle (0.2cm);
\draw[line width=0.5mm, black!70, opacity=0.9] (6,1,1.5) to[out=225,in=45] (6,1,3);

% plane 3
\draw[line width=0.5mm, black!80, opacity=0.9,rounded corners] (7,2,3) -- (6,2,3) -- (6,0,3) -- (7,0,3);
\draw[line width=0.5mm, black!80, opacity=0.9] (6,1,3) -- (7,1,3);
\shade[ball color = black!80, opacity = 0.9] (6,1,3) circle (0.2cm);
\draw[line width=0.5mm, black!90, opacity=0.9] (6,1,3) to[out=225,in=45] (6,1,4.5);

% plane 4
\draw[line width=0.5mm, black!100, opacity=0.9,rounded corners] (7,2,4.5) -- (6,2,4.5) -- (6,0,4.5) -- (7,0,4.5);
\draw[line width=0.5mm, black!100, opacity=0.9] (6,1,4.5) -- (7,1,4.5);
\shade[ball color = black!100, opacity = 0.9] (6,1,4.5) circle (0.2cm);

\end{tikzpicture}
}

\caption{Approximation of environment.}
\label{fig:pepsenv}
\end{figure}

\begin{figure}[!htb]

\tdplotsetmaincoords{0}{0}
\begin{minipage}{0.23\textwidth} 
\centering
\scalebox{0.6}{
\begin{tikzpicture}[font=\tiny,opacity=0.9,tdplot_main_coords]
\tdplotsetrotatedcoords{15}{-60}{-30}
\begin{scope}[tdplot_rotated_coords]

% plane 1
\foreach \y in {0,1,2}{
	\draw[line width=0.5mm, black!40, opacity=0.9] (1.5,\y,0) -- (2.5,\y,0);
}

\draw[line width=0.5mm, black!40, opacity=0.9] (1.5,0,0) -- (1.5,2,0);
\draw[line width=0.5mm, black!40, opacity=0.9,rounded corners] (1.5,2,0) -- (0,2,0) -- (0,0,0) -- (1.5,0,0);
\draw[line width=0.5mm, black!40, opacity=0.9] (0,1,0) -- (1.5,1,0);
\foreach \y in {0,1,2}{
	\shade[ball color = black!40, opacity = 0.9] (1.5,\y,0) circle (0.2cm);
}
\shade[ball color = black!40, opacity = 0.9] (0,1,0) circle (0.2cm);
\foreach \y in {0,1,2}{
	\draw[line width=0.5mm, black!50, opacity=0.9] (1.5,\y,0) -- (1.5,\y,1.5);
}
\draw[line width=0.5mm, black!50, opacity=0.9] (0,1,0) -- (0,1,1.5);

% plane 2
\foreach \y in {0,1,2}{
	\draw[line width=0.5mm, black!60, opacity=0.9] (1.5,\y,1.5) -- (2.5,\y,1.5);
}
\draw[line width=0.5mm, black!60, opacity=0.9] (1.5,0,1.5) -- (1.5,2,1.5);
\draw[line width=0.5mm, black!60, opacity=0.9,rounded corners] (1.5,2,1.5) -- (0,2,1.5) -- (0,0,1.5) -- (1.5,0,1.5);
\draw[line width=0.5mm, black!60, opacity=0.9] (0,1,1.5) -- (1.5,1,1.5);
\foreach \y in {0,1,2}{
	\shade[ball color = black!60, opacity = 0.9] (1.5,\y,1.5) circle (0.2cm);
}
\shade[ball color = black!60, opacity = 0.9] (0,1,1.5) circle (0.2cm);
\foreach \y in {0,1,2}{
	\draw[line width=0.5mm, black!70, opacity=0.9] (1.5,\y,1.5) -- (1.5,\y,3);
}
\draw[line width=0.5mm, black!70, opacity=0.9] (0,1,1.5) -- (0,1,3);

% plane 3
\foreach \y in {0,1,2}{
	\draw[line width=0.5mm, black!80, opacity=0.9] (1.5,\y,3) -- (2.5,\y,3);
}
\draw[line width=0.5mm, black!80, opacity=0.9] (1.5,0,3) -- (1.5,2,3);
\draw[line width=0.5mm, black!80, opacity=0.9,rounded corners] (1.5,2,3) -- (0,2,3) -- (0,0,3) -- (1.5,0,3);
\draw[line width=0.5mm, black!80, opacity=0.9] (0,1,3) -- (1.5,1,3);
\foreach \y in {0,1,2}{
	\shade[ball color = black!80, opacity = 0.9] (1.5,\y,3) circle (0.2cm);
}
\shade[ball color = black!80, opacity = 0.9] (0,1,3) circle (0.2cm);
\foreach \y in {0,1,2}{
	\draw[line width=0.5mm, black!90, opacity=0.9] (1.5,\y,3) -- (1.5,\y,4.5);
}
\draw[line width=0.5mm, black!90, opacity=0.9] (0,1,3) -- (0,1,4.5);

% plane 4
\foreach \y in {0,1,2}{
	\draw[line width=0.5mm, black!100, opacity=0.9] (1.5,\y,4.5) -- (2.5,\y,4.5);
}
\draw[line width=0.5mm, black!100, opacity=0.9] (1.5,0,4.5) -- (1.5,2,4.5);
\draw[line width=0.5mm, black!100, opacity=0.9,rounded corners] (1.5,2,4.5) -- (0,2,4.5) -- (0,0,4.5) -- (1.5,0,4.5);
\draw[line width=0.5mm, black!100, opacity=0.9] (0,1,4.5) -- (1.5,1,4.5);
\foreach \y in {0,1,2}{
	\shade[ball color = blue!100, opacity = 0.9] (1.5,\y,4.5) circle (0.2cm);
}
\shade[ball color = blue!100, opacity = 0.9] (0,1,4.5) circle (0.2cm);

\end{scope}
\end{tikzpicture}
}

\vspace*{0.2cm}
\centerline{\small (a) Initial arrangement.}
\vspace{0.1cm}        
\end{minipage}
\hfill 
\begin{minipage}{0.23\textwidth} 
\centering
\scalebox{0.6}{
\begin{tikzpicture}[font=\tiny,opacity=0.9,tdplot_main_coords]
\tdplotsetrotatedcoords{15}{-60}{-30}
\begin{scope}[tdplot_rotated_coords]

% plane 1
\foreach \y in {0,1,2}{
	\draw[line width=0.5mm, black!40, opacity=0.9] (1.5,\y,0) -- (2.5,\y,0);
}

\draw[line width=0.5mm, black!40, opacity=0.9] (1.5,0,0) -- (1.5,2,0);
\draw[line width=0.5mm, black!40, opacity=0.9,rounded corners] (1.5,2,0) -- (0,2,0) -- (0,0,0) -- (1.5,0,0);
\draw[line width=0.5mm, black!40, opacity=0.9] (0,1,0) -- (1.5,1,0);
\foreach \y in {0,1,2}{
	\shade[ball color = black!40, opacity = 0.9] (1.5,\y,0) circle (0.2cm);
}
\shade[ball color = black!40, opacity = 0.9] (0,1,0) circle (0.2cm);
\foreach \y in {0,1,2}{
	\draw[line width=0.5mm, black!50, opacity=0.9] (1.5,\y,0) -- (1.5,\y,1.5);
}
\draw[line width=0.5mm, black!50, opacity=0.9] (0,1,0) -- (0,1,1.5);

% plane 2
\foreach \y in {0,1,2}{
	\draw[line width=0.5mm, black!60, opacity=0.9] (1.5,\y,1.5) -- (2.5,\y,1.5);
}
\draw[line width=0.5mm, black!60, opacity=0.9] (1.5,0,1.5) -- (1.5,2,1.5);
\draw[line width=0.5mm, black!60, opacity=0.9,rounded corners] (1.5,2,1.5) -- (0,2,1.5) -- (0,0,1.5) -- (1.5,0,1.5);
\draw[line width=0.5mm, black!60, opacity=0.9] (0,1,1.5) -- (1.5,1,1.5);
\foreach \y in {0,1,2}{
	\shade[ball color = black!60, opacity = 0.9] (1.5,\y,1.5) circle (0.2cm);
}
\shade[ball color = black!60, opacity = 0.9] (0,1,1.5) circle (0.2cm);
\foreach \y in {0,1,2}{
	\draw[line width=0.5mm, black!70, opacity=0.9] (1.5,\y,1.5) -- (1.5,\y,3);
}
\draw[line width=0.5mm, black!70, opacity=0.9] (0,1,1.5) -- (0,1,3);

% plane 3
\foreach \y in {0,1,2}{
	\draw[line width=0.5mm, black!80, opacity=0.9] (1.5,\y,3) -- (2.5,\y,3);
}
\draw[line width=0.5mm, black!80, opacity=0.9] (1.5,0,3) -- (1.5,2,3);
\draw[line width=0.5mm, black!80, opacity=0.9,rounded corners] (1.5,2,3) -- (0,2,3) -- (0,0,3) -- (1.5,0,3);
\draw[line width=0.5mm, black!80, opacity=0.9] (0,1,3) -- (1.5,1,3);
\foreach \y in {0,1,2}{
	\shade[ball color = black!80, opacity = 0.9] (1.5,\y,3) circle (0.2cm);
}
\shade[ball color = black!80, opacity = 0.9] (0,1,3) circle (0.2cm);
\foreach \y in {0,1,2}{
	\draw[line width=0.5mm, black!90, opacity=0.9] (1.5,\y,3) -- (1.5,1,4.5);
}
\draw[line width=0.5mm, black!90, opacity=0.9] (0,1,3) -- (1.5,1,4.5);

% plane 4
\draw[line width=0.5mm, black!100, opacity=0.9] (1.5,1,4.5) -- (2.5,1,4.5);
\draw[line width=0.5mm, black!100, opacity=0.9,rounded corners] (2.5,2,4.5) -- (1.5,2,4.5) -- (1.5,0,4.5) -- (2.5,0,4.5);
\shade[ball color = blue!100, opacity = 0.9] (1.5,1,4.5) circle (0.2cm);

\end{scope}
\end{tikzpicture}
}

\vspace*{0.2cm}
\centerline{\small (b) Contraction.}
\vspace*{0.1cm}
\end{minipage}
\\ 
\begin{minipage}{0.23\textwidth}

\centering

\scalebox{0.6}{
\begin{tikzpicture}[font=\tiny,opacity=0.9,tdplot_main_coords]
\tdplotsetrotatedcoords{15}{-60}{-30}
\begin{scope}[tdplot_rotated_coords]

% plane 1
\foreach \y in {0,1,2}{
	\draw[line width=0.5mm, black!40, opacity=0.9] (1.5,\y,0) -- (2.5,\y,0);
}

\draw[line width=0.5mm, black!40, opacity=0.9] (1.5,0,0) -- (1.5,2,0);
\draw[line width=0.5mm, black!40, opacity=0.9,rounded corners] (1.5,2,0) -- (0,2,0) -- (0,0,0) -- (1.5,0,0);
\draw[line width=0.5mm, black!40, opacity=0.9] (0,1,0) -- (1.5,1,0);
\foreach \y in {0,1,2}{
	\shade[ball color = black!40, opacity = 0.9] (1.5,\y,0) circle (0.2cm);
}
\shade[ball color = black!40, opacity = 0.9] (0,1,0) circle (0.2cm);
\foreach \y in {0,1,2}{
	\draw[line width=0.5mm, black!50, opacity=0.9] (1.5,\y,0) -- (1.5,\y,1.5);
}
\draw[line width=0.5mm, black!50, opacity=0.9] (0,1,0) -- (0,1,1.5);

% plane 2
\foreach \y in {0,1,2}{
	\draw[line width=0.5mm, black!60, opacity=0.9] (1.5,\y,1.5) -- (2.5,\y,1.5);
}
\draw[line width=0.5mm, black!60, opacity=0.9] (1.5,0,1.5) -- (1.5,2,1.5);
\draw[line width=0.5mm, black!60, opacity=0.9,rounded corners] (1.5,2,1.5) -- (0,2,1.5) -- (0,0,1.5) -- (1.5,0,1.5);
\draw[line width=0.5mm, black!60, opacity=0.9] (0,1,1.5) -- (1.5,1,1.5);
\foreach \y in {0,1,2}{
	\shade[ball color = black!60, opacity = 0.9] (1.5,\y,1.5) circle (0.2cm);
}
\shade[ball color = black!60, opacity = 0.9] (0,1,1.5) circle (0.2cm);
\foreach \y in {0,1,2}{
	\draw[line width=0.5mm, black!70, opacity=0.9] (1.5,\y,1.5) -- (1.5,\y,3);
}
\draw[line width=0.5mm, black!70, opacity=0.9] (0,1,1.5) -- (0,1,3);

% plane 3
\foreach \y in {0,1,2}{
	\draw[line width=0.5mm, black!80, opacity=0.9] (1.5,\y,3) -- (2.5,\y,3);
}
\draw[line width=0.5mm, black!80, opacity=0.9] (1.5,0,3) -- (1.5,2,3);
\draw[line width=0.5mm, black!80, opacity=0.9,rounded corners] (1.5,2,3) -- (0,2,3) -- (0,0,3) -- (1.5,0,3);
\draw[line width=0.5mm, black!80, opacity=0.9] (0,1,3) -- (1.5,1,3);
\foreach \y in {0,1,2}{
	\shade[ball color = blue!80, opacity = 0.9] (1.5,\y,3) circle (0.2cm);
}
\shade[ball color = blue!80, opacity = 0.9] (0,1,3) circle (0.2cm);
\foreach \y in {0,1,2}{
	\draw[line width=0.5mm, black!90, opacity=0.9] (1.5,\y,3) -- (1.5,1,4.5);
}
\draw[line width=0.5mm, black!90, opacity=0.9] (0,1,3) -- (1.5,1,4.5);

% plane 4
\shade[ball color = blue!100, opacity = 0.9] (1.5,1,4.5) circle (0.2cm);
\draw[line width=0.5mm, black!100, opacity=0.9] (1.5,1,4.5) -- (1.5,1,6);

\draw[line width=0.5mm, black!100, opacity=0.9] (1.5,1,6) -- (2.5,1,6);
\draw[line width=0.5mm, black!100, opacity=0.9,rounded corners] (2.5,2,6) -- (1.5,2,6) -- (1.5,0,6) -- (2.5,0,6);
\shade[ball color = black!100, opacity = 0.9] (1.5,1,6) circle (0.2cm);

\end{scope}
\end{tikzpicture}
}

\vspace*{0.2cm}
\centerline{\small (c) Decomposition.}
\vspace*{0.1cm}

\end{minipage}
\hfill 
\begin{minipage}{0.23\textwidth} 

\centering

\scalebox{0.6}{
\begin{tikzpicture}[font=\tiny,opacity=0.9,tdplot_main_coords]
\tdplotsetrotatedcoords{15}{-60}{-30}
\begin{scope}[tdplot_rotated_coords]

% plane 1
\foreach \y in {0,1,2}{
	\draw[line width=0.5mm, black!40, opacity=0.9] (1.5,\y,0) -- (2.5,\y,0);
}

\draw[line width=0.5mm, black!40, opacity=0.9] (1.5,0,0) -- (1.5,2,0);
\draw[line width=0.5mm, black!40, opacity=0.9,rounded corners] (1.5,2,0) -- (0,2,0) -- (0,0,0) -- (1.5,0,0);
\draw[line width=0.5mm, black!40, opacity=0.9] (0,1,0) -- (1.5,1,0);
\foreach \y in {0,1,2}{
	\shade[ball color = black!40, opacity = 0.9] (1.5,\y,0) circle (0.2cm);
}
\shade[ball color = black!40, opacity = 0.9] (0,1,0) circle (0.2cm);
\foreach \y in {0,1,2}{
	\draw[line width=0.5mm, black!50, opacity=0.9] (1.5,\y,0) -- (1.5,\y,1.5);
}
\draw[line width=0.5mm, black!50, opacity=0.9] (0,1,0) -- (0,1,1.5);

% plane 2
\foreach \y in {0,1,2}{
	\draw[line width=0.5mm, black!60, opacity=0.9] (1.5,\y,1.5) -- (2.5,\y,1.5);
}
\draw[line width=0.5mm, black!60, opacity=0.9] (1.5,0,1.5) -- (1.5,2,1.5);
\draw[line width=0.5mm, black!60, opacity=0.9,rounded corners] (1.5,2,1.5) -- (0,2,1.5) -- (0,0,1.5) -- (1.5,0,1.5);
\draw[line width=0.5mm, black!60, opacity=0.9] (0,1,1.5) -- (1.5,1,1.5);
\foreach \y in {0,1,2}{
	\shade[ball color = black!60, opacity = 0.9] (1.5,\y,1.5) circle (0.2cm);
}
\shade[ball color = black!60, opacity = 0.9] (0,1,1.5) circle (0.2cm);
\foreach \y in {0,1,2}{
	\draw[line width=0.5mm, black!70, opacity=0.9] (1.5,\y,1.5) -- (1.5,1,3);
}
\draw[line width=0.5mm, black!70, opacity=0.9] (0,1,1.5) -- (1.5,1,3);

% plane 3
\draw[line width=0.5mm, black!80, opacity=0.9,rounded corners] (2.5,2,3) -- (1.5,2,3) -- (1.5,0,3) -- (2.5,0,3);
\shade[ball color = blue!80, opacity = 0.9] (1.5,1,3) circle (0.2cm);
\draw[line width=0.5mm, black!80, opacity=0.9] (1.5,1,3) -- (2.5,1,3);

\draw[line width=0.5mm, black!90, opacity=0.9] (1.5,1,3) -- (1.5,1,4.5);

% plane 4
\draw[line width=0.5mm, black!100, opacity=0.9] (1.5,1,4.5) -- (2.5,1,4.5);
\draw[line width=0.5mm, black!100, opacity=0.9,rounded corners] (2.5,2,4.5) -- (1.5,2,4.5) -- (1.5,0,4.5) -- (2.5,0,4.5);
\shade[ball color = black!100, opacity = 0.9] (1.5,1,4.5) circle (0.2cm);

\end{scope}
\end{tikzpicture}
}

\vspace*{0.2cm}
\centerline{\small (d) Contraction.}
\vspace*{0.1cm}
\end{minipage}
\\
\begin{minipage}{0.23\textwidth} 

\centering

\scalebox{0.6}{
\begin{tikzpicture}[font=\tiny,opacity=0.9,tdplot_main_coords]
\tdplotsetrotatedcoords{15}{-60}{-30}
\begin{scope}[tdplot_rotated_coords]

% plane 1
\foreach \y in {0,1,2}{
	\draw[line width=0.5mm, black!40, opacity=0.9] (1.5,\y,0) -- (2.5,\y,0);
}

\draw[line width=0.5mm, black!40, opacity=0.9] (1.5,0,0) -- (1.5,2,0);
\draw[line width=0.5mm, black!40, opacity=0.9,rounded corners] (1.5,2,0) -- (0,2,0) -- (0,0,0) -- (1.5,0,0);
\draw[line width=0.5mm, black!40, opacity=0.9] (0,1,0) -- (1.5,1,0);
\foreach \y in {0,1,2}{
	\shade[ball color = black!40, opacity = 0.9] (1.5,\y,0) circle (0.2cm);
}
\shade[ball color = black!40, opacity = 0.9] (0,1,0) circle (0.2cm);
\foreach \y in {0,1,2}{
	\draw[line width=0.5mm, black!50, opacity=0.9] (1.5,\y,0) -- (1.5,\y,1.5);
}
\draw[line width=0.5mm, black!50, opacity=0.9] (0,1,0) -- (0,1,1.5);

% plane 2
\foreach \y in {0,1,2}{
	\draw[line width=0.5mm, black!60, opacity=0.9] (1.5,\y,1.5) -- (2.5,\y,1.5);
}
\draw[line width=0.5mm, black!60, opacity=0.9] (1.5,0,1.5) -- (1.5,2,1.5);
\draw[line width=0.5mm, black!60, opacity=0.9,rounded corners] (1.5,2,1.5) -- (0,2,1.5) -- (0,0,1.5) -- (1.5,0,1.5);
\draw[line width=0.5mm, black!60, opacity=0.9] (0,1,1.5) -- (1.5,1,1.5);
\foreach \y in {0,1,2}{
	\shade[ball color = blue!60, opacity = 0.9] (1.5,\y,1.5) circle (0.2cm);
}
\shade[ball color = blue!60, opacity = 0.9] (0,1,1.5) circle (0.2cm);
\foreach \y in {0,1,2}{
	\draw[line width=0.5mm, black!70, opacity=0.9] (1.5,\y,1.5) -- (1.5,1,3);
}
\draw[line width=0.5mm, black!70, opacity=0.9] (0,1,1.5) -- (1.5,1,3);

% plane 3
\shade[ball color = blue!80, opacity = 0.9] (1.5,1,3) circle (0.2cm);
\draw[line width=0.5mm, black!80, opacity=0.9] (1.5,1,3) -- (1.5,1,4.5);

\draw[line width=0.5mm, black!80, opacity=0.9,rounded corners] (2.5,2,4.5) -- (1.5,2,4.5) -- (1.5,0,4.5) -- (2.5,0,4.5);
\shade[ball color = black!80, opacity = 0.9] (1.5,1,4.5) circle (0.2cm);
\draw[line width=0.5mm, black!80, opacity=0.9] (1.5,1,4.5) -- (2.5,1,4.5);

\draw[line width=0.5mm, black!90, opacity=0.9] (1.5,1,4.5) -- (1.5,1,6);

% plane 4
\draw[line width=0.5mm, black!100, opacity=0.9] (1.5,1,6) -- (2.5,1,6);
\draw[line width=0.5mm, black!100, opacity=0.9,rounded corners] (2.5,2,6) -- (1.5,2,6) -- (1.5,0,6) -- (2.5,0,6);
\shade[ball color = black!100, opacity = 0.9] (1.5,1,6) circle (0.2cm);

\end{scope}
\end{tikzpicture}
}

\vspace*{0.2cm}
\centerline{\small (e) Decomposition.}
\vspace*{0.1cm}
\end{minipage}
\hfill 
\begin{minipage}{0.23\textwidth}

\centering

\scalebox{0.6}{
\begin{tikzpicture}[font=\tiny,opacity=0.9,tdplot_main_coords]
\tdplotsetrotatedcoords{15}{-60}{-30}
\begin{scope}[tdplot_rotated_coords]

% plane 1
\foreach \y in {0,1,2}{
	\draw[line width=0.5mm, black!40, opacity=0.9] (1.5,\y,0) -- (2.5,\y,0);
}

\draw[line width=0.5mm, black!40, opacity=0.9] (1.5,0,0) -- (1.5,2,0);
\draw[line width=0.5mm, black!40, opacity=0.9,rounded corners] (1.5,2,0) -- (0,2,0) -- (0,0,0) -- (1.5,0,0);
\draw[line width=0.5mm, black!40, opacity=0.9] (0,1,0) -- (1.5,1,0);
\foreach \y in {0,1,2}{
	\shade[ball color = black!40, opacity = 0.9] (1.5,\y,0) circle (0.2cm);
}
\shade[ball color = black!40, opacity = 0.9] (0,1,0) circle (0.2cm);
\foreach \y in {0,1,2}{
	\draw[line width=0.5mm, black!50, opacity=0.9] (1.5,\y,0) -- (1.5,1,1.5);
}
\draw[line width=0.5mm, black!50, opacity=0.9] (0,1,0) -- (1.5,1,1.5);

% plane 2
\draw[line width=0.5mm, black!60, opacity=0.9] (1.5,1,1.5) -- (2.5,1,1.5);
\draw[line width=0.5mm, black!60, opacity=0.9,rounded corners] (2.5,2,1.5) -- (1.5,2,1.5) -- (1.5,0,1.5) -- (2.5,0,1.5);

\shade[ball color = blue!60, opacity = 0.9] (1.5,1,1.5) circle (0.2cm);

\draw[line width=0.5mm, black!70, opacity=0.9] (1.5,1,1.5) -- (1.5,1,3);

% plane 3
\draw[line width=0.5mm, black!80, opacity=0.9,rounded corners] (2.5,2,3) -- (1.5,2,3) -- (1.5,0,3) -- (2.5,0,3);
\shade[ball color = black!80, opacity = 0.9] (1.5,1,3) circle (0.2cm);
\draw[line width=0.5mm, black!80, opacity=0.9] (1.5,1,3) -- (2.5,1,3);

\draw[line width=0.5mm, black!90, opacity=0.9] (1.5,1,3) -- (1.5,1,4.5);

% plane 4
\draw[line width=0.5mm, black!100, opacity=0.9] (1.5,1,4.5) -- (2.5,1,4.5);
\draw[line width=0.5mm, black!100, opacity=0.9,rounded corners] (2.5,2,4.5) -- (1.5,2,4.5) -- (1.5,0,4.5) -- (2.5,0,4.5);
\shade[ball color = black!100, opacity = 0.9] (1.5,1,4.5) circle (0.2cm);

\end{scope}
\end{tikzpicture}
}

\vspace*{0.2cm}
\centerline{\small (f) Contraction.}
\vspace*{0.1cm}
\end{minipage}
\\
\begin{minipage}{0.23\textwidth}

\centering

\scalebox{0.6}{
\begin{tikzpicture}[font=\tiny,opacity=0.9,tdplot_main_coords]
\tdplotsetrotatedcoords{15}{-60}{-30}
\begin{scope}[tdplot_rotated_coords]

% plane 1
\foreach \y in {0,1,2}{
	\draw[line width=0.5mm, black!40, opacity=0.9] (1.5,\y,0) -- (2.5,\y,0);
}

\draw[line width=0.5mm, black!40, opacity=0.9] (1.5,0,0) -- (1.5,2,0);
\draw[line width=0.5mm, black!40, opacity=0.9,rounded corners] (1.5,2,0) -- (0,2,0) -- (0,0,0) -- (1.5,0,0);
\draw[line width=0.5mm, black!40, opacity=0.9] (0,1,0) -- (1.5,1,0);
\foreach \y in {0,1,2}{
	\shade[ball color = blue!40, opacity = 0.9] (1.5,\y,0) circle (0.2cm);
}
\shade[ball color = blue!40, opacity = 0.9] (0,1,0) circle (0.2cm);
\foreach \y in {0,1,2}{
	\draw[line width=0.5mm, black!50, opacity=0.9] (1.5,\y,0) -- (1.5,1,1.5);
}
\draw[line width=0.5mm, black!50, opacity=0.9] (0,1,0) -- (1.5,1,1.5);
\shade[ball color = blue!60, opacity = 0.9] (1.5,1,1.5) circle (0.2cm);
\draw[line width=0.5mm, black!60, opacity=0.9] (1.5,1,1.5) -- (1.5,1,3);

% plane 2
\draw[line width=0.5mm, black!60, opacity=0.9] (1.5,1,3) -- (2.5,1,3);
\draw[line width=0.5mm, black!60, opacity=0.9,rounded corners] (2.5,2,3) -- (1.5,2,3) -- (1.5,0,3) -- (2.5,0,3);

\shade[ball color = black!60, opacity = 0.9] (1.5,1,3) circle (0.2cm);

\draw[line width=0.5mm, black!70, opacity=0.9] (1.5,1,3) -- (1.5,1,4.5);

% plane 3
\draw[line width=0.5mm, black!80, opacity=0.9,rounded corners] (2.5,2,4.5) -- (1.5,2,4.5) -- (1.5,0,4.5) -- (2.5,0,4.5);
\shade[ball color = black!80, opacity = 0.9] (1.5,1,4.5) circle (0.2cm);
\draw[line width=0.5mm, black!80, opacity=0.9] (1.5,1,4.5) -- (2.5,1,4.5);

\draw[line width=0.5mm, black!90, opacity=0.9] (1.5,1,4.5) -- (1.5,1,6);

% plane 4
\draw[line width=0.5mm, black!100, opacity=0.9] (1.5,1,6) -- (2.5,1,6);
\draw[line width=0.5mm, black!100, opacity=0.9,rounded corners] (2.5,2,6) -- (1.5,2,6) -- (1.5,0,6) -- (2.5,0,6);
\shade[ball color = black!100, opacity = 0.9] (1.5,1,6) circle (0.2cm);

\end{scope}
\end{tikzpicture}
}

\vspace*{0.2cm}
\centerline{\small (g) Decomposition.}
\vspace*{0.1cm}
\end{minipage}
\hfill 
\begin{minipage}{0.23\textwidth}

\centering

\scalebox{0.6}{
\begin{tikzpicture}[font=\tiny,opacity=0.9,tdplot_main_coords]
\tdplotsetrotatedcoords{15}{-60}{-30}
\begin{scope}[tdplot_rotated_coords]

% plane 1
\draw[line width=0.5mm, black!40, opacity=0.9] (1.5,1,0) -- (2.5,1,0);
\draw[line width=0.5mm, black!40, opacity=0.9,rounded corners] (2.5,2,0) -- (1.5,2,0) -- (1.5,0,0) -- (2.5,0,0);
\shade[ball color = blue!40, opacity = 0.9] (1.5,1,0) circle (0.2cm);
\draw[line width=0.5mm, black!50, opacity=0.9] (1.5,1,0) -- (1.5,1,1.5);

% plane 2
\draw[line width=0.5mm, black!60, opacity=0.9] (1.5,1,1.5) -- (2.5,1,1.5);
\draw[line width=0.5mm, black!60, opacity=0.9,rounded corners] (2.5,2,1.5) -- (1.5,2,1.5) -- (1.5,0,1.5) -- (2.5,0,1.5);

\shade[ball color = black!60, opacity = 0.9] (1.5,1,1.5) circle (0.2cm);

\draw[line width=0.5mm, black!70, opacity=0.9] (1.5,1,1.5) -- (1.5,1,3);

% plane 3
\draw[line width=0.5mm, black!80, opacity=0.9,rounded corners] (2.5,2,3) -- (1.5,2,3) -- (1.5,0,3) -- (2.5,0,3);
\shade[ball color = black!80, opacity = 0.9] (1.5,1,3) circle (0.2cm);
\draw[line width=0.5mm, black!80, opacity=0.9] (1.5,1,3) -- (2.5,1,3);

\draw[line width=0.5mm, black!90, opacity=0.9] (1.5,1,3) -- (1.5,1,4.5);

% plane 4
\draw[line width=0.5mm, black!100, opacity=0.9] (1.5,1,4.5) -- (2.5,1,4.5);
\draw[line width=0.5mm, black!100, opacity=0.9,rounded corners] (2.5,2,4.5) -- (1.5,2,4.5) -- (1.5,0,4.5) -- (2.5,0,4.5);
\shade[ball color = black!100, opacity = 0.9] (1.5,1,4.5) circle (0.2cm);

\end{scope}
\end{tikzpicture}
}

\vspace*{0.2cm}
\centerline{\small (h) Contraction.}
\vspace*{0.1cm}
\end{minipage}

\caption{Initialization of truncated environment $\tilde{\left| \psi \right>}$.}
\label{fig:pepsenvinit}
\end{figure}
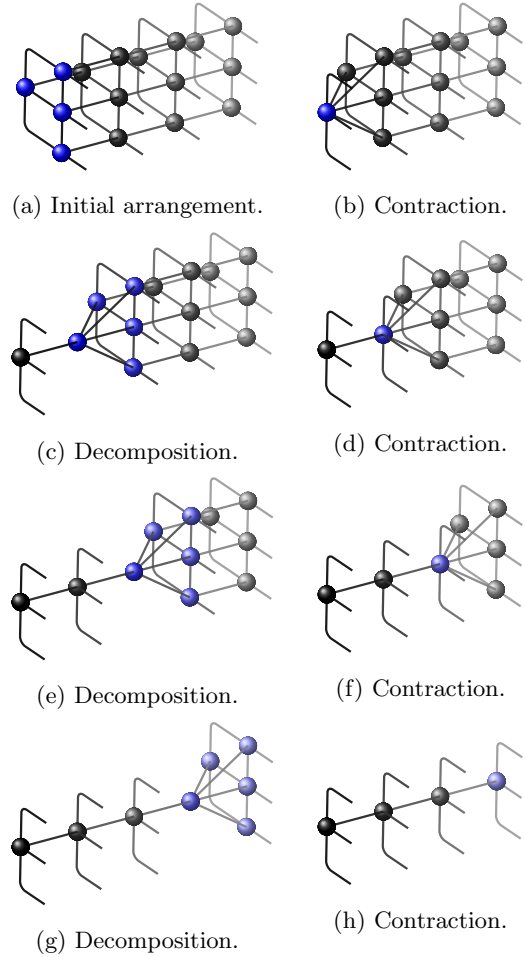

The environment approximation scheme we use here is depicted in
Fig.~\ref{fig:pepsenv};
this scheme is 
is similar to the boundary-MPO construction used in
Ref.~\cite{Lubasch2014Mar}.
Suppose we are given the arrangement of tensors on the left, which consist of 
environment blocks $E_{i-1,j}$, typically retained from a previous
calculation step, fPEPS tensors 
$A_{i,j}$, their adjoints $A^{\dagger}_{i,j}$, and PEPO tensors $W_{i,j}$. 
The goal is to find the tensors $E_{i,j}$ on the right which, as a
whole, approximate the cluster on the left as well as possible
by forming
cumulative indices $\gamma_{i,j}$ 
with a predetermined maximum bond dimension $\chi$
\cite{Verstraete2004Jul}.
If we interpret the former network as a full vector, i.e., 
\begin{align}
\left| \psi \right> & = \left( E_{i-1,1} \cdot A_{i,1} \cdot W_{i,1}
\cdot A^{\dagger}_{i,1} \right) \cdot\nonumber \\
&  \ldots \cdot \left( E_{i-1,N} \cdot A_{i,N} \cdot W_{i,N} \cdot A^{\dagger}_{i,N} \right) \, ,
\end{align}
and the latter as a truncated vector, 
\begin{align}
\tilde{\left| \psi \right>} = E_{i,1} \cdot E_{i,2} \cdot ... \cdot E_{i,N} \, ,
\end{align}
the problem can be stated as finding the maximum of the fidelity
\begin{align}
F = \frac{\langle \psi | \tilde{\psi} \rangle \langle \tilde{\psi} | \psi \rangle}{\langle \tilde{\psi} | \tilde{\psi} \rangle \langle \psi | \psi \rangle} \, ,
\end{align}
whose solution, in general, is given by
\begin{align}
\langle \tilde{\psi} | \tilde{\psi} \rangle = \langle \tilde{\psi} | \psi \rangle. 
\label{eq:optvec}
\end{align}
Using Eq.~\eqref{eq:optvec} to find the optimal
$\tilde{\left| \psi \right>}$ as a whole  
is not possible because $\tilde{\left| \psi \right>}$
consists of
multiple tensors, and contracting all of them is actually what we want
to avoid. 
Instead, we start with 
a trial vector constructed as shown in
Fig.~\ref{fig:pepsenvinit}.
The cluster in Fig.~\ref{fig:pepsenvinit}(a)
is the original 
$\left| \psi \right>$, 
while the subsequent figures illustrate how to systematically forge a truncated 
environment with a bond dimension of one
by alternately carrying out
contractions and truncated SVDs.
The outcome, Fig.~\ref{fig:pepsenvinit}(h),
is then the starting point for an iterative algorithm.
 
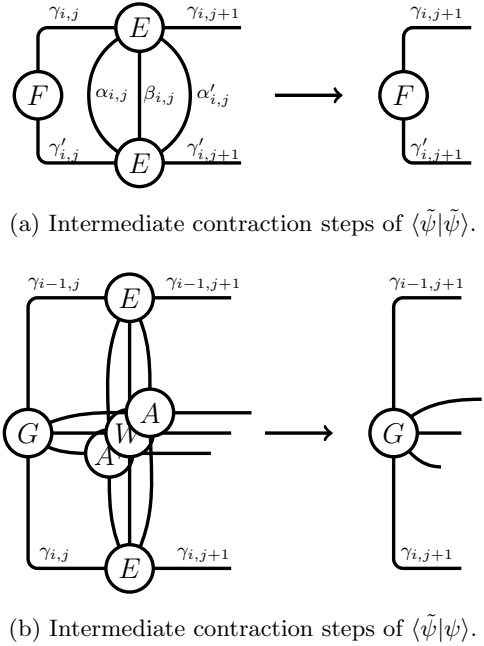
\begin{figure}[!htb]
\begin{minipage}{\widthfac\textwidth}
\centering
\scalebox{0.9}{ 
\begin{tikzpicture}[font=\tiny]

\draw[line width=0.5mm, rounded corners] (3,1) -- (0,1) -- (0,-1) -- (3,-1);
\draw[line width=0.5mm] (1.5,1) to[out=270,in=90] (1.5,-1);
\draw[line width=0.5mm] (1.5,1) to[out=200,in=90] (0.75,0) to[out=270,in=160] (1.5,-1);
\draw[line width=0.5mm] (1.5,1) to[out=340,in=90] (2.25,0) to[out=270,in=20] (1.5,-1);
\draw[fill=white, line width=0.5mm] (0,0) circle [radius=0.35];
\node[align=center] at (0,0) {\large $F$};
\draw[fill=white, line width=0.5mm] (1.5,1) circle [radius=0.35];
\node[align=center] at (1.5,1) {\large $E$};
\draw[fill=white, line width=0.5mm] (1.5,-1) circle [radius=0.35];
\node[align=center] at (1.5,-1) {\large $E$};

\node[align=center] at (0.4,1.2) {\footnotesize $\gamma_{i,j}$};
\node[align=center] at (0.4,-0.8) {\footnotesize $\gamma'_{i,j}$};
\node[align=center] at (2.6,1.2) {\footnotesize $\gamma_{i,j+1}$};
\node[align=center] at (2.6,-0.8) {\footnotesize $\gamma'_{i,j+1}$};
\node[align=center] at (1.1,0) {\footnotesize $\alpha_{i,j}$};
\node[align=center] at (1.8,0) {\footnotesize $\beta_{i,j}$};
\node[align=center] at (2.6,0) {\footnotesize $\alpha'_{i,j}$};

\draw[line width=0.5mm,->] (3.5,0) -- (4.5,0);

\draw[line width=0.5mm, rounded corners] (6.4,1) -- (5.4,1) -- (5.4,-1) -- (6.4,-1);
\draw[fill=white, line width=0.5mm] (5.4,0) circle [radius=0.35];
\node[align=center] at (5.4,0) {\large $F$};
\node[align=center] at (5.9,1.2) {\footnotesize $\gamma_{i,j+1}$};
\node[align=center] at (5.9,-0.8) {\footnotesize $\gamma'_{i,j+1}$};

\end{tikzpicture}
}

\vspace*{0.2cm}
\centerline{\small (a) Intermediate contraction steps of
	$\langle \tilde{\psi} | \tilde{\psi} \rangle$.}
\vspace*{0.1cm}
\end{minipage}

\vspace{3mm}

\begin{minipage}{\widthfac\textwidth}

\centering 
\scalebox{0.9}{
\begin{tikzpicture}[font=\tiny]

\draw[line width=0.5mm, rounded corners] (3,2) -- (0,2) -- (0,-2) -- (3,-2);
\draw[line width=0.5mm] (1.5,2) to[out=240,in=90] (1.2,-0.3);
\draw[line width=0.5mm] (1.5,2) to[out=300,in=90] (1.8,0.3);
\draw[line width=0.5mm] (1.2,-0.3) to[out=270,in=120] (1.5,-2);

\draw[line width=0.5mm] (0,0) to[out=315,in=180] (1.2,-0.3);

\draw[line width=0.5mm] (1.2,-0.3) -- (2.7,-0.3);
\draw[fill=white, line width=0.5mm] (1.2,-0.3) circle [radius=0.35];
\node[align=center] at (1.2,-0.3) {\large $A^{\dagger}$};
\draw[line width=0.5mm] (0,0) -- (1.5,0);

\draw[line width=0.5mm] (1.5,2) -- (1.5,-2);
\draw[line width=0.5mm] (1.5,0) -- (3,0);
\draw[fill=white, line width=0.5mm] (1.5,0) circle [radius=0.35];
\node[align=center] at (1.5,0) {\large $W$};
\draw[line width=0.5mm] (1.8,0.3) -- (3.3,0.3);
\draw[line width=0.5mm] (1.8,0.3) to[out=270,in=60] (1.5,-2);

\draw[fill=white, line width=0.5mm] (1.5,2) circle [radius=0.35];
\node[align=center] at (1.5,2) {\large $E$};
\draw[fill=white, line width=0.5mm] (1.5,-2) circle [radius=0.35];
\node[align=center] at (1.5,-2) {\large $E$};

\node[align=center] at (0.4,2.2) {\footnotesize $\gamma_{i-1,j}$};
\node[align=center] at (0.4,-1.8) {\footnotesize $\gamma_{i,j}$};
\node[align=center] at (2.6,2.2) {\footnotesize $\gamma_{i-1,j+1}$};
\node[align=center] at (2.6,-1.8) {\footnotesize $\gamma_{i,j+1}$};

\draw[line width=0.5mm,->] (3.5,0) -- (4.5,0);

\draw[line width=0.5mm, rounded corners] (6.4,2) -- (5.4,2) -- (5.4,-2) -- (6.4,-2);
\draw[line width=0.5mm] (5.4,0) to[out=50,in=180] (6.7,0.5);
\draw[line width=0.5mm] (5.4,0) -- (6.4,0);
\draw[line width=0.5mm] (5.4,0) to[out=310,in=180] (6.1,-0.5);
\draw[fill=white, line width=0.5mm] (5.4,0) circle [radius=0.35];
\node[align=center] at (5.4,0) {\large $G$};
\node[align=center] at (5.9,2.2) {\footnotesize $\gamma_{i-1,j+1}$};
\node[align=center] at (5.9,-1.8) {\footnotesize $\gamma_{i,j+1}$};

\draw[line width=0.5mm] (0,0) to[out=30,in=180] (1.8,0.3);
\draw[fill=white, line width=0.5mm] (0,0) circle [radius=0.35];
\node[align=center] at (0,0) {\large $G$};
\draw[fill=white, line width=0.5mm] (1.8,0.3) circle [radius=0.35];
\node[align=center] at (1.8,0.3) {\large $A$};

\end{tikzpicture}
}

\vspace*{0.2cm}
\centerline{\small (b) Intermediate contraction steps of $\langle \tilde{\psi} |
  \psi \rangle$.}
\vspace{0.1cm}
\end{minipage}

\caption{Intermediate contraction steps.}
\label{fig:intsteps}
\end{figure}

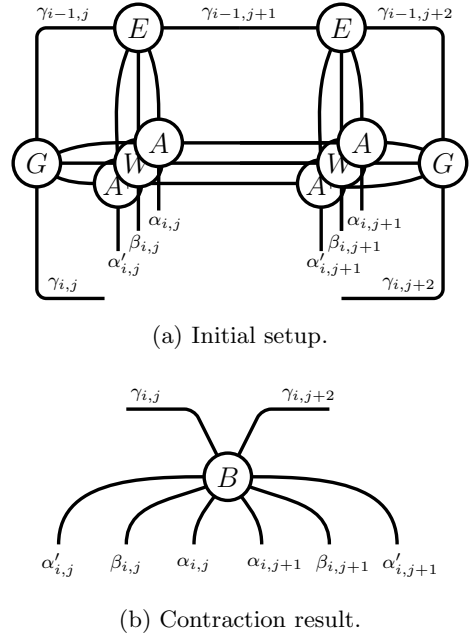
\begin{figure}[!htb]
\begin{minipage}{\widthfac\textwidth}
\centering
\scalebox{0.9}{ 
\begin{tikzpicture}[font=\tiny]

\draw[line width=0.5mm, rounded corners] (3,2) -- (0,2) -- (0,-2) -- (1,-2);
\draw[line width=0.5mm] (1.5,2) to[out=240,in=90] (1.2,-0.3);
\draw[line width=0.5mm] (1.5,2) to[out=300,in=90] (1.8,0.3);
\draw[line width=0.5mm] (4.5,2) to[out=240,in=90] (4.2,-0.3);
\draw[line width=0.5mm] (4.5,2) to[out=300,in=90] (4.8,0.3);

\draw[line width=0.5mm] (1.2,-0.3) -- (1.2,-1.3);

\draw[line width=0.5mm] (0,0) to[out=315,in=180] (1.2,-0.3);

\draw[line width=0.5mm] (1.2,-0.3) -- (4.2,-0.3);
\draw[fill=white, line width=0.5mm] (1.2,-0.3) circle [radius=0.35];
\node[align=center] at (1.2,-0.3) {\large $A^{\dagger}$};
\draw[line width=0.5mm] (0,0) -- (1.5,0);

\draw[line width=0.5mm] (1.5,2) -- (1.5,-1);
\draw[line width=0.5mm] (4.5,2) -- (4.5,-1);
\draw[line width=0.5mm] (1.5,0) -- (4.5,0);
\draw[fill=white, line width=0.5mm] (1.5,0) circle [radius=0.35];
\node[align=center] at (1.5,0) {\large $W$};
\draw[line width=0.5mm] (1.8,0.3) -- (1.8,-0.7);
\draw[line width=0.5mm] (1.8,0.3) -- (4.8,0.3);

\draw[fill=white, line width=0.5mm] (1.5,2) circle [radius=0.35];
\node[align=center] at (1.5,2) {\large $E$};

\node[align=center] at (0.4,2.2) {\footnotesize $\gamma_{i-1,j}$};
\node[align=center] at (0.4,-1.8) {\footnotesize $\gamma_{i,j}$};
\node[align=center] at (3,2.2) {\footnotesize $\gamma_{i-1,j+1}$};
\node[align=center] at (5.6,2.2) {\footnotesize $\gamma_{i-1,j+2}$};
\node[align=center] at (5.5,-1.8) {\footnotesize $\gamma_{i,j+2}$};

\draw[line width=0.5mm] (0,0) to[out=30,in=180] (1.8,0.3);
\draw[fill=white, line width=0.5mm] (0,0) circle [radius=0.35];
\node[align=center] at (0,0) {\large $G$};
\draw[fill=white, line width=0.5mm] (1.8,0.3) circle [radius=0.35];
\node[align=center] at (1.8,0.3) {\large $A$};

\draw[line width=0.5mm] (4.2,-0.3) to[out=180,in=230] (6,0);
\draw[line width=0.5mm] (4.8,0.3) to[out=180,in=120] (6,0);
\draw[line width=0.5mm] (4.5,0) -- (6,0);
\draw[line width=0.5mm, rounded corners] (3,2) -- (6,2) -- (6,-2) -- (4.5,-2);

\draw[line width=0.5mm] (4.2,-0.3) -- (4.2,-1.3);
\draw[fill=white, line width=0.5mm] (4.2,-0.3) circle [radius=0.35];
\node[align=center] at (4.2,-0.3) {\large $A^{\dagger}$};
\draw[line width=0.5mm] (3,0) -- (4.5,0);
\draw[line width=0.5mm] (4.5,0) -- (4.5,-1);
\draw[fill=white, line width=0.5mm] (4.5,0) circle [radius=0.35];
\node[align=center] at (4.5,0) {\large $W$};
\draw[line width=0.5mm] (4.8,0.3) -- (4.8,-0.7);
\draw[line width=0.5mm] (3,0.3) -- (4.8,0.3);
\draw[fill=white, line width=0.5mm] (4.8,0.3) circle [radius=0.35];
\node[align=center] at (4.8,0.3) {\large $A$};
\draw[fill=white, line width=0.5mm] (6,0) circle [radius=0.35];
\node[align=center] at (6,0) {\large $G$};
\draw[fill=white, line width=0.5mm] (4.5,2) circle [radius=0.35];
\node[align=center] at (4.5,2) {\large $E$};

\node[align=center] at (1.3,-1.5) {\footnotesize $\alpha'_{i,j}$};
\node[align=center] at (1.6,-1.2) {\footnotesize $\beta_{i,j}$};
\node[align=center] at (1.9,-0.9) {\footnotesize $\alpha_{i,j}$};
\node[align=center] at (4.4,-1.5) {\footnotesize $\alpha'_{i,j+1}$};
\node[align=center] at (4.7,-1.2) {\footnotesize $\beta_{i,j+1}$};
\node[align=center] at (5,-0.9) {\footnotesize $\alpha_{i,j+1}$};

\end{tikzpicture}
}

\vspace*{0.2cm}
\centerline{\small (a) Initial setup.}
\vspace{0.1cm}        
\end{minipage}

\vspace{3mm}

\begin{minipage}{\widthfac\textwidth}
\centering
\scalebox{0.9}{ 
\begin{tikzpicture}[font=\tiny]

\node[align=center] at (10.3,1.2) {\footnotesize $\gamma_{i,j}$};
\node[align=center] at (12.7,1.2) {\footnotesize $\gamma_{i,j+2}$};

\draw[line width=0.5mm, rounded corners] (10,1) -- (11,1) -- (11.5,0) -- (12,1) -- (13,1);

\draw[line width=0.5mm] (11.5,0) to[out=180,in=90] (9,-1);
\draw[line width=0.5mm] (11.5,0) to[out=210,in=90] (10,-1);
\draw[line width=0.5mm] (11.5,0) to[out=240,in=90] (11,-1);

\draw[line width=0.5mm] (11.5,0) to[out=300,in=90] (12,-1);
\draw[line width=0.5mm] (11.5,0) to[out=330,in=90] (13,-1);
\draw[line width=0.5mm] (11.5,0) to[out=0,in=90] (14,-1);

\draw[fill=white, line width=0.5mm] (11.5,0) circle [radius=0.35];
\node[align=center] at (11.5,0) {\large $B$};

\node[align=center] at (9,-1.3) {\footnotesize $\alpha'_{i,j}$};
\node[align=center] at (10,-1.3) {\footnotesize $\beta_{i,j}$};
\node[align=center] at (11,-1.3) {\footnotesize $\alpha_{i,j}$};

\node[align=center] at (12.2,-1.3) {\footnotesize $\alpha_{i,j+1}$};
\node[align=center] at (13.2,-1.3) {\footnotesize $\beta_{i,j+1}$};
\node[align=center] at (14.2,-1.3) {\footnotesize $\alpha'_{i,j+1}$};

\end{tikzpicture}
}

\vspace*{0.2cm}
\centerline{\small (b) Contraction result.}
\vspace*{0.1cm}
\end{minipage}
\caption{Calculation of the inhomogeneity $B$.}
\label{fig:inhomo}
\end{figure}

\begin{figure}[!htb]
\centering 
\scalebox{0.8}{
\begin{tikzpicture}[font=\tiny]

\draw[line width=0.5mm, rounded corners] (1,1) -- (0,1) -- (0,-1) -- (1,-1);
\draw[fill=white, line width=0.5mm] (0,0) circle [radius=0.35];
\node[align=center] at (0,0) {\large $F$};
\node[align=center] at (0.5,1.2) {\footnotesize $\gamma_{i,j}$};
\node[align=center] at (0.5,-0.8) {\footnotesize $\gamma'_{i,j}$};

\draw[line width=0.5mm, rounded corners] (2,1) -- (3,1) -- (3,-1) -- (2,-1);
\draw[fill=white, line width=0.5mm] (3,0) circle [radius=0.35];
\node[align=center] at (3,0) {\large $F$};
\node[align=center] at (2.5,1.2) {\footnotesize $\gamma_{i,j+2}$};
\node[align=center] at (2.5,-0.8) {\footnotesize $\gamma'_{i,j+2}$};

\draw[line width=0.5mm,->] (4,0) -- (5.3,0);
\node[align=center] at (4.65,0.2) {\footnotesize \textbf{contract}};

\node[align=center] at (6.2,1.2) {\footnotesize $\gamma'_{i,j}$};
\node[align=center] at (6,-0.8) {\footnotesize $\gamma'_{i,j+2}$};
\node[align=center] at (7.8,1.2) {\footnotesize $\gamma_{i,j}$};
\node[align=center] at (7.9,-0.8) {\footnotesize $\gamma_{i,j+2}$};
\draw[line width=0.5mm, rounded corners] (6,1) -- (6.5,1) -- (7,0) -- (6.5,-1) -- (6,-1);
\draw[line width=0.5mm, rounded corners] (8,1) -- (7.5,1) -- (7,0) -- (7.5,-1) -- (8,-1);
\draw[fill=white, line width=0.5mm] (7,0) circle [radius=0.35];
\node[align=center] at (7,0) {\large $M$};

\end{tikzpicture}
}
\caption{Calculation of $M$.}
\label{fig:matrix}
\end{figure}
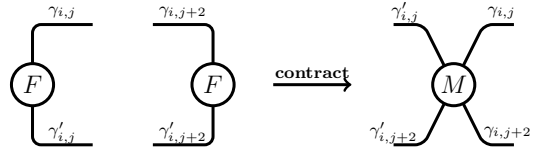

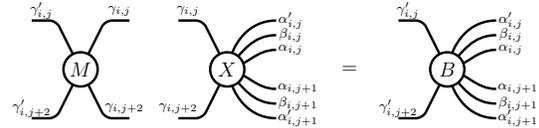
\begin{figure}[!htb]
\scalebox{0.65}{
\centering 
\begin{tikzpicture}[font=\tiny]

\node[align=center] at (9.2,1.2) {\footnotesize $\gamma'_{i,j}$};
\node[align=center] at (9,-0.8) {\footnotesize $\gamma'_{i,j+2}$};
\node[align=center] at (10.8,1.2) {\footnotesize $\gamma_{i,j}$};
\node[align=center] at (10.9,-0.8) {\footnotesize $\gamma_{i,j+2}$};
\draw[line width=0.5mm, rounded corners] (9,1) -- (9.5,1) -- (10,0) -- (9.5,-1) -- (9,-1);
\draw[line width=0.5mm, rounded corners] (11,1) -- (10.5,1) -- (10,0) -- (10.5,-1) -- (11,-1);
\draw[fill=white, line width=0.5mm] (10,0) circle [radius=0.35];
\node[align=center] at (10,0) {\large $M$};

\draw[line width=0.5mm] (13,0) to[out=80,in=180] (14,1);
\draw[line width=0.5mm] (13,0) to[out=50,in=180] (14,0.7);
\draw[line width=0.5mm] (13,0) to[out=10,in=180] (14,0.4);
\draw[line width=0.5mm] (13,0) to[out=350,in=180] (14,-0.4);
\draw[line width=0.5mm] (13,0) to[out=310,in=180] (14,-0.7);
\draw[line width=0.5mm] (13,0) to[out=280,in=180] (14,-1);
\draw[line width=0.5mm, rounded corners] (12,1) -- (12.5,1) -- (13,0) -- (12.5,-1) -- (12,-1);
\draw[fill=white, line width=0.5mm] (13,0) circle [radius=0.35];
\node[align=center] at (13,0) {\large $X$};
\node[align=center] at (12.2,1.2) {\footnotesize $\gamma_{i,j}$};
\node[align=center] at (12,-0.8) {\footnotesize $\gamma_{i,j+2}$};

\node[align=center] at (14.3,1) {\footnotesize $\alpha'_{i,j}$};
\node[align=center] at (14.3,0.7) {\footnotesize $\beta_{i,j}$};
\node[align=center] at (14.3,0.4) {\footnotesize $\alpha_{i,j}$};
\node[align=center] at (14.45,-0.4) {\footnotesize $\alpha_{i,j+1}$};
\node[align=center] at (14.45,-0.7) {\footnotesize $\beta_{i,j+1}$};
\node[align=center] at (14.45,-1) {\footnotesize $\alpha'_{i,j+1}$};

\node[align=center] at (15.5,0) {\large $=$};

\draw[line width=0.5mm] (17.5,0) to[out=80,in=180] (18.5,1);
\draw[line width=0.5mm] (17.5,0) to[out=50,in=180] (18.5,0.7);
\draw[line width=0.5mm] (17.5,0) to[out=10,in=180] (18.5,0.4);
\draw[line width=0.5mm] (17.5,0) to[out=350,in=180] (18.5,-0.4);
\draw[line width=0.5mm] (17.5,0) to[out=310,in=180] (18.5,-0.7);
\draw[line width=0.5mm] (17.5,0) to[out=280,in=180] (18.5,-1);
\draw[line width=0.5mm, rounded corners] (16.5,1) -- (17,1) -- (17.5,0) -- (17,-1) -- (16.5,-1);
\draw[fill=white, line width=0.5mm] (17.5,0) circle [radius=0.35];
\node[align=center] at (17.5,0) {\large $B$};
\node[align=center] at (16.7,1.2) {\footnotesize $\gamma'_{i,j}$};
\node[align=center] at (16.5,-0.8) {\footnotesize $\gamma'_{i,j+2}$};

\node[align=center] at (18.8,1) {\footnotesize $\alpha'_{i,j}$};
\node[align=center] at (18.8,0.7) {\footnotesize $\beta_{i,j}$};
\node[align=center] at (18.8,0.4) {\footnotesize $\alpha_{i,j}$};
\node[align=center] at (18.95,-0.4) {\footnotesize $\alpha_{i,j+1}$};
\node[align=center] at (18.95,-0.7) {\footnotesize $\beta_{i,j+1}$};
\node[align=center] at (18.95,-1) {\footnotesize $\alpha'_{i,j+1}$};

\end{tikzpicture}
}

\caption{Solving for a joint environment tensor.}
\label{fig:solveblock}
\end{figure}

The optimization of $\tilde{\left| \psi \right>}$ proceeds similarly 
to that carried out to find the ground state in the  
density matrix renormalization group (DMRG): Most of its tensors are
fixed in that
$\langle \tilde{\psi} | \tilde{\psi} \rangle$ and 
$\langle \tilde{\psi} | \psi \rangle$
are calculated partially starting from both sides. 
The individual steps are
shown
in Fig.~\ref{fig:intsteps}, which depicts 
an aerial view; here the tensors $F$ and $G$ represent 
intermediate contractions for $\langle \tilde{\psi} | \tilde{\psi} \rangle$ and 
$\langle \tilde{\psi} | \psi \rangle$, respectively. 
Only two sites, labeled
$j$ and $j+1$ here, are not contracted over,
which makes it possible to determine the optimal joint environment block 
$X_{j,j+2}=E_{j,j+1} \cdot E_{j+1,j+2}$ by requiring that
\begin{align}
\langle \tilde{\psi} | \tilde{\psi} \rangle / X^{\dagger}_{j,j+2} 
= \langle \tilde{\psi} | \psi \rangle / X^{\dagger}_{j,j+2} \label{eq:optvec2}
\end{align}
in this reduced vector space.
The calculation necessary to obtain the 
right-hand side is
depicted
in Fig.~\ref{fig:inhomo}, yielding the
inhomogeneity $B$.
The contraction is best performed by calculating the left and right halves of 
the cluster separately,
then multiplying the two results with each other.
Fig.~\ref{fig:matrix} shows how $F_{i,j}$ and $F_{i,j+2}$ are then
contracted, yielding the rank-4 tensor $M$. 
Finding the solution to Eq.~\eqref{eq:optvec2} now amounts to solving
$M \cdot X = B$
for all columns of B, which is depicted in Fig.~\ref{fig:solveblock}.
However, if one performs a QR decomposition on the 
environment tensors $E$, they become isometries, and all $F$ and thus $M$ are 
identity
matrices.
The solution $X$ is then simply the 
inhomogeneity $B$.
Its decomposition into 
$E_{i,j}$ and $E_{i,j+1}$ is depicted in Fig.~\ref{fig:svdblock}.
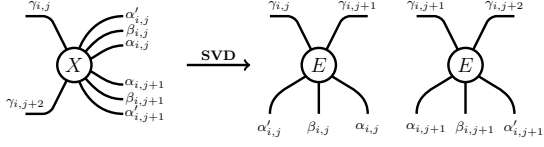
\begin{figure}[!htb]
\centering 

\scalebox{0.65}{
\centering 
\begin{tikzpicture}[font=\tiny]

\draw[line width=0.5mm] (-2,-9) to[out=80,in=180] (-1,-8);
\draw[line width=0.5mm] (-2,-9) to[out=50,in=180] (-1,-8.3);
\draw[line width=0.5mm] (-2,-9) to[out=10,in=180] (-1,-8.6);
\draw[line width=0.5mm] (-2,-9) to[out=350,in=180] (-1,-9.4);
\draw[line width=0.5mm] (-2,-9) to[out=310,in=180] (-1,-9.7);
\draw[line width=0.5mm] (-2,-9) to[out=280,in=180] (-1,-10);
\draw[line width=0.5mm, rounded corners] (-3,-8) -- (-2.5,-8) -- (-2,-9) -- (-2.5,-10) -- (-3,-10);
\draw[fill=white, line width=0.5mm] (-2,-9) circle [radius=0.35];
\node[align=center] at (-2,-9) {\large $X$};
\node[align=center] at (-2.7,-7.8) {\footnotesize $\gamma_{i,j}$};
\node[align=center] at (-3,-9.8) {\footnotesize $\gamma_{i,j+2}$};

\node[align=center] at (-0.7,-8) {\footnotesize $\alpha'_{i,j}$};
\node[align=center] at (-0.7,-8.3) {\footnotesize $\beta_{i,j}$};
\node[align=center] at (-0.7,-8.6) {\footnotesize $\alpha_{i,j}$};
\node[align=center] at (-0.55,-9.4) {\footnotesize $\alpha_{i,j+1}$};
\node[align=center] at (-0.55,-9.7) {\footnotesize $\beta_{i,j+1}$};
\node[align=center] at (-0.55,-10) {\footnotesize $\alpha'_{i,j+1}$};

\draw[line width=0.5mm,->] (0.3,-9) -- (1.6,-9);
\node[align=center] at (0.95,-8.8) {\footnotesize \textbf{SVD}};

\node[align=center] at (2.2,-7.8) {\footnotesize $\gamma_{i,j}$};
\node[align=center] at (3.8,-7.8) {\footnotesize $\gamma_{i,j+1}$};
\draw[line width=0.5mm, rounded corners] (2,-8) -- (2.5,-8) -- (3,-9) -- (3.5,-8) -- (4,-8);
\draw[line width=0.5mm] (3,-9) to[out=230,in=90] (2,-10);
\draw[line width=0.5mm] (3,-9) to[out=270,in=90] (3,-10);
\draw[line width=0.5mm] (3,-9) to[out=310,in=90] (4,-10);
\draw[fill=white, line width=0.5mm] (3,-9) circle [radius=0.35];
\node[align=center] at (3,-9) {\large $E$};

\node[align=center] at (5.2,-7.8) {\footnotesize $\gamma_{i,j+1}$};
\node[align=center] at (6.8,-7.8) {\footnotesize $\gamma_{i,j+2}$};
\draw[line width=0.5mm, rounded corners] (5,-8) -- (5.5,-8) -- (6,-9) -- (6.5,-8) -- (7,-8);
\draw[line width=0.5mm] (6,-9) to[out=230,in=90] (5,-10);
\draw[line width=0.5mm] (6,-9) to[out=270,in=90] (6,-10);
\draw[line width=0.5mm] (6,-9) to[out=310,in=90] (7,-10);
\draw[fill=white, line width=0.5mm] (6,-9) circle [radius=0.35];
\node[align=center] at (6,-9) {\large $E$};

\node[align=center] at (2,-10.3) {\footnotesize $\alpha'_{i,j}$};
\node[align=center] at (3,-10.3) {\footnotesize $\beta_{i,j}$};
\node[align=center] at (4,-10.3) {\footnotesize $\alpha_{i,j}$};

\node[align=center] at (5.2,-10.3) {\footnotesize $\alpha_{i,j+1}$};
\node[align=center] at (6.2,-10.3) {\footnotesize $\beta_{i,j+1}$};
\node[align=center] at (7.2,-10.3) {\footnotesize $\alpha'_{i,j+1}$};

\end{tikzpicture}
}

\caption{Decomposition of joint environment tensor.}
\label{fig:svdblock}
\end{figure}
This generates the composite
index $\gamma_{i,j+1}$, which has
the predetermined maximum bond dimension $\chi$, i.e., the maximum
bond dimension of environment tensors.
It is important not to confuse $\chi$
with the maximum number of virtual states within the PEPS, $D$.
Note that 
algorithms based on MPSs
as well as other
acyclic tensor networks
do not, in general, have a numerical parameter $\chi$,
as the 
corresponding environments are processed exactly.

The scheme presented above still contains one major drawback: If the maximum
bond dimension is $\chi$, the contraction in Fig.~\ref{fig:inhomo} 
generates a temporary number of states $\tilde{\chi} \gg \chi$, most of which are 
dropped in the subsequent truncation in Fig.~\ref{fig:svdblock} and are therefore 
superfluous.
This issue is not restricted to finding the best 
virtual basis within environment tensors in our fPEPS-PEPO-scheme,
but is a general computational bottleneck in tensor network 
algorithms, in particular, in the typical scheme in which 
the 
bond dimension between two adjacent tensors is 
significantly enlarged, then optimized via the maximum overlap or 
minimal energy, and subsequently
truncated again.
In order to reduce the cost of such steps,
Gleis et~al. \cite{Gleis2022Jul} have recently
developed the so-called ``controlled bond expansion'' (CBE) for
calculating ground-state
MPSs.
In this method,
one picks 
$\xi$ orthogonal states with the largest weight, adds them 
to the current $\chi$ states in tangent space, optimizes
the combined set of states, and, 
finally, truncates
it to the desired dimension.
If $\xi$ is chosen such that $\tilde{\chi} \gg \chi + \xi$,
one can thus perform a two-site optimization at one-site cost.
In the following, we will sketch how to use their method for the 
environment approximation described above.

We start by formulating the completeness relations in 
Figs.~\ref{fig:envcomplete}(a) and (b)
for the tensors $E_{i,j}$ and $E_{i,j+1}$, respectively.
\begin{figure}[!htb]

\begin{minipage}{0.5\textwidth}
\scalebox{0.7}{
\begin{tikzpicture}[font=\tiny]

\draw[line width=0.5mm, rounded corners] (-1,0) -- (1,0) -- (1,1.5) -- (-1,1.5);
\node[align=center] at (-0.8,1.2) {\small $\gamma_{i,j}$};
\node[align=center] at (-0.8,0.3) {\small $\gamma'_{i,j}$};
\node[align=center] at (0.5,0.75) {\small $\gamma_{i,j+1}$};

\draw[line width=0.5mm, rounded corners] (0,0) -- (0,-1.5);
\draw[line width=0.5mm] (0,0) to[out=235,in=90] (-1,-1.5);
\draw[line width=0.5mm] (0,0) to[out=315,in=90] (1,-1.5);
\node[align=center] at (-1,-1.8) {\small $\tilde{\alpha}'_{i,j}$};
\node[align=center] at (0,-1.8) {\small $\tilde{\beta}_{i,j}$};
\node[align=center] at (1,-1.8) {\small $\tilde{\alpha}_{i,j}$};

\draw[line width=0.5mm, rounded corners] (0,1.5) -- (0,3);
\draw[line width=0.5mm] (0,1.5) to[out=135,in=270] (-1,3);
\draw[line width=0.5mm] (0,1.5) to[out=45,in=270] (1,3);
\node[align=center] at (-1,3.2) {\small $\alpha'_{i,j}$};
\node[align=center] at (0,3.2) {\small $\beta_{i,j}$};
\node[align=center] at (1,3.2) {\small $\alpha_{i,j}$};

\draw[fill=white, line width=0.5mm] (0,0) circle [radius=0.4];
\node[align=center] at (0,0) {\Large $E$};
\draw[fill=white, line width=0.5mm] (0,1.5) circle [radius=0.4];
\node[align=center] at (0,1.5) {\Large $E$};

\node[align=center] at (1.7,0.75) {\Large $+$};

\draw[line width=0.5mm, rounded corners] (2.2,0) -- (4.2,0) -- (4.2,1.5) -- (2.2,1.5);
\node[align=center] at (2.4,1.2) {\small $\gamma_{i,j}$};
\node[align=center] at (2.4,0.3) {\small $\gamma'_{i,j}$};
\node[align=center] at (3.7,0.75) {\small $\gamma_{i,j+1}$};

\draw[line width=0.5mm, rounded corners] (3.2,0) -- (3.2,-1.5);
\draw[line width=0.5mm] (3.2,0) to[out=235,in=90] (2.2,-1.5);
\draw[line width=0.5mm] (3.2,0) to[out=315,in=90] (4.2,-1.5);

\node[align=center] at (2.2,-1.8) {\small $\tilde{\alpha}'_{i,j}$};
\node[align=center] at (3.2,-1.8) {\small $\tilde{\beta}_{i,j}$};
\node[align=center] at (4.2,-1.8) {\small $\tilde{\alpha}_{i,j}$};

\draw[line width=0.5mm, rounded corners] (3.2,1.5) -- (3.2,3);
\draw[line width=0.5mm] (3.2,1.5) to[out=135,in=270] (2.2,3);
\draw[line width=0.5mm] (3.2,1.5) to[out=45,in=270] (4.2,3);
\node[align=center] at (2.2,3.2) {\small $\alpha'_{i,j}$};
\node[align=center] at (3.2,3.2) {\small $\beta_{i,j}$};
\node[align=center] at (4.2,3.2) {\small $\alpha_{i,j}$};

\draw[fill=white, line width=0.5mm] (3.2,0) circle [radius=0.4];
\node[align=center] at (3.2,0) {\Large $E'$};
\draw[fill=white, line width=0.5mm] (3.2,1.5) circle [radius=0.4];
\node[align=center] at (3.2,1.5) {\Large $E'$};

\node[align=center] at (4.8,0.75) {\Large $=$};

\draw[line width=0.5mm, rounded corners] (5.4,0) -- (6.2,0) -- (6.2,1.5) -- (5.4,1.5);
\node[align=center] at (5.8,1.2) {\small $\gamma_{i,j}$};
\node[align=center] at (5.8,0.3) {\small $\gamma'_{i,j}$};
\node[align=center] at (6.6,0.75) {\Large $\otimes$};
\draw[line width=0.5mm, rounded corners] (7,-1.5) -- (7,3);
\node[align=center] at (7.4,0.75) {\Large $\otimes$};
\draw[line width=0.5mm, rounded corners] (7.8,-1.5) -- (7.8,3);
\node[align=center] at (8.2,0.75) {\Large $\otimes$};
\draw[line width=0.5mm, rounded corners] (8.6,-1.5) -- (8.6,3);
\node[align=center] at (7,3.2) {\small $\alpha'_{i,j}$};
\node[align=center] at (7.8,3.2) {\small $\beta_{i,j}$};
\node[align=center] at (8.6,3.2) {\small $\alpha_{i,j}$};

\node[align=center] at (7,-1.8) {\small $\tilde{\alpha}'_{i,j}$};
\node[align=center] at (7.8,-1.8) {\small $\tilde{\beta}_{i,j}$};
\node[align=center] at (8.6,-1.8) {\small $\tilde{\alpha}_{i,j}$};

\end{tikzpicture}
}

\vspace*{0.2cm}
\centerline{\small (a) Left completeness.}
\vspace{0.1cm}        
\end{minipage}
\\ \mbox{} \\
\begin{minipage}{0.5\textwidth}

\scalebox{0.7}{
\begin{tikzpicture}[font=\tiny]

\draw[line width=0.5mm, rounded corners] (1,0) -- (-1,0) -- (-1,1.5) -- (1,1.5);
\node[align=center] at (0.8,1.2) {\small $\gamma_{i,j+2}$};
\node[align=center] at (0.8,0.3) {\small $\gamma'_{i,j+2}$};
\node[align=center] at (-0.5,0.75) {\small $\gamma_{i,j+1}$};

\draw[line width=0.5mm, rounded corners] (0,0) -- (0,-1.5);
\draw[line width=0.5mm] (0,0) to[out=235,in=90] (-1,-1.5);
\draw[line width=0.5mm] (0,0) to[out=315,in=90] (1,-1.5);
\node[align=center] at (-1,-1.8) {\small $\tilde{\alpha}'_{i,j+1}$};
\node[align=center] at (0,-1.8) {\small $\tilde{\beta}_{i,j+1}$};
\node[align=center] at (1,-1.8) {\small $\tilde{\alpha}_{i,j+1}$};

\draw[line width=0.5mm, rounded corners] (0,1.5) -- (0,3);
\draw[line width=0.5mm] (0,1.5) to[out=135,in=270] (-1,3);
\draw[line width=0.5mm] (0,1.5) to[out=45,in=270] (1,3);
\node[align=center] at (-1,3.2) {\small $\alpha'_{i,j+1}$};
\node[align=center] at (0,3.2) {\small $\beta_{i,j+1}$};
\node[align=center] at (1,3.2) {\small $\alpha_{i,j+1}$};

\draw[fill=white, line width=0.5mm] (0,0) circle [radius=0.4];
\node[align=center] at (0,0) {\Large $E$};
\draw[fill=white, line width=0.5mm] (0,1.5) circle [radius=0.4];
\node[align=center] at (0,1.5) {\Large $E$};

\node[align=center] at (1.7,0.75) {\Large $+$};

\draw[line width=0.5mm, rounded corners] (4.2,0) -- (2.2,0) -- (2.2,1.5) -- (4.2,1.5);
\node[align=center] at (4,1.2) {\small $\gamma_{i,j+2}$};
\node[align=center] at (4,0.3) {\small $\gamma'_{i,j+2}$};
\node[align=center] at (2.7,0.75) {\small $\gamma_{i,j+1}$};

\draw[line width=0.5mm, rounded corners] (3.2,0) -- (3.2,-1.5);
\draw[line width=0.5mm] (3.2,0) to[out=235,in=90] (2.2,-1.5);
\draw[line width=0.5mm] (3.2,0) to[out=315,in=90] (4.2,-1.5);

\node[align=center] at (2.2,-1.8) {\small $\tilde{\alpha}'_{i,j+1}$};
\node[align=center] at (3.2,-1.8) {\small $\tilde{\beta}_{i,j+1}$};
\node[align=center] at (4.2,-1.8) {\small $\tilde{\alpha}_{i,j+1}$};

\draw[line width=0.5mm, rounded corners] (3.2,1.5) -- (3.2,3);
\draw[line width=0.5mm] (3.2,1.5) to[out=135,in=270] (2.2,3);
\draw[line width=0.5mm] (3.2,1.5) to[out=45,in=270] (4.2,3);
\node[align=center] at (2.3,3.2) {\small $\alpha'_{i,j+1}$};
\node[align=center] at (3.3,3.2) {\small $\beta_{i,j+1}$};
\node[align=center] at (4.3,3.2) {\small $\alpha_{i,j+1}$};

\draw[fill=white, line width=0.5mm] (3.2,0) circle [radius=0.4];
\node[align=center] at (3.2,0) {\Large $E'$};
\draw[fill=white, line width=0.5mm] (3.2,1.5) circle [radius=0.4];
\node[align=center] at (3.2,1.5) {\Large $E'$};

\node[align=center] at (4.8,0.75) {\Large $=$};

\draw[line width=0.5mm, rounded corners] (5.4,-1.5) -- (5.4,3);
\node[align=center] at (5.8,0.75) {\Large $\otimes$};
\draw[line width=0.5mm, rounded corners] (6.2,-1.5) -- (6.2,3);
\node[align=center] at (6.6,0.75) {\Large $\otimes$};
\draw[line width=0.5mm, rounded corners] (7,-1.5) -- (7,3);
\node[align=center] at (5.4,3.2) {\small $\alpha'_{i,j+1}$};
\node[align=center] at (6.4,3.2) {\small $\beta_{i,j+1}$};
\node[align=center] at (7.6,2.8) {\small $\alpha_{i,j+1}$};

\node[align=center] at (5.4,-1.8) {\small $\tilde{\alpha}'_{i,j+1}$};
\node[align=center] at (6.4,-1.8) {\small $\tilde{\beta}_{i,j+1}$};
\node[align=center] at (7.6,-1.3) {\small $\tilde{\alpha}_{i,j+1}$};

\node[align=center] at (7.4,0.75) {\Large $\otimes$};
\draw[line width=0.5mm, rounded corners] (8.8,0) -- (7.8,0) -- (7.8,1.5) -- (8.8,1.5);
\node[align=center] at (8.4,1.2) {\small $\gamma_{i,j+2}$};
\node[align=center] at (8.4,0.3) {\small $\gamma'_{i,j+2}$};

\end{tikzpicture}
}

\vspace*{0.2cm}
\centerline{\small (b) Right completeness.}
\vspace{0.1cm}        
\end{minipage}

\caption{Completeness relations of environment tensors.}
\label{fig:envcomplete}
\end{figure}
Due to the 
QR-decomposition (described above), the approximated
environment $\tilde{\left|\psi\right>}$ is in a form
that is essentially 
the mixed canonical form
typically used for 
MPSs, and the rank-8 tensors
$E \cdot E^{\dagger}$ on the left are projectors onto $E$, i.e., onto 
the tangent space.
The right hand sides of both figures consist of
unit tensors, which are formally constructed as the tensor 
product of Kronecker deltas and are simply depicted as
lines.
These configurations
define the orthogonal projectors 
$E' \cdot \left(E'\right)^{\dagger}$, which
are then 
used to enlarge
the bond between two tensors in a controlled manner. 
Fig.~\ref{fig:leftorth}(a) 
shows how the contraction of the left
orthogonal projector with the left half of 
Fig.~\ref{fig:inhomo}(a) 
is carried out.
\begin{figure}[!htb]
\begin{minipage}{\widthfacorth\textwidth}
\centering
\scalebox{\scalefacorth}{ 
\begin{tikzpicture}[font=\tiny]

\draw[line width=0.5mm, rounded corners] (3,2) -- (0,2) -- (0,-2) -- (1,-2);
\draw[line width=0.5mm] (1.5,2) to[out=240,in=90] (1.2,-0.3);
\draw[line width=0.5mm] (1.5,2) to[out=300,in=90] (1.8,0.3);

\draw[line width=0.5mm] (1.2,-0.3) -- (1.2,-1.3);

\draw[line width=0.5mm] (0,0) to[out=315,in=180] (1.2,-0.3);

\draw[line width=0.5mm] (1.2,-0.3) -- (2.5,-0.3);
\draw[fill=white, line width=0.5mm] (1.2,-0.3) circle [radius=0.35];
\node[align=center] at (1.2,-0.3) {\large $A^{\dagger}$};
\draw[line width=0.5mm] (0,0) -- (1.5,0);

\draw[line width=0.5mm] (1.5,2) -- (1.5,-1);
\draw[line width=0.5mm] (1.5,0) -- (2.8,0);
\draw[fill=white, line width=0.5mm] (1.5,0) circle [radius=0.35];
\node[align=center] at (1.5,0) {\large $W$};
\draw[line width=0.5mm] (1.8,0.3) -- (1.8,-0.7);
\draw[line width=0.5mm] (1.8,0.3) -- (3.1,0.3);

\draw[fill=white, line width=0.5mm] (1.5,2) circle [radius=0.35];
\node[align=center] at (1.5,2) {\large $E$};

\node[align=center] at (0.4,2.2) {\normalsize $\gamma_{i-1,j}$};
\node[align=center] at (0.4,-1.8) {\normalsize $\gamma'_{i,j}$};
\node[align=center] at (2.5,2.2) {\normalsize $\gamma_{i-1,j+1}$};

\draw[line width=0.5mm] (0,0) to[out=30,in=180] (1.8,0.3);
\draw[fill=white, line width=0.5mm] (0,0) circle [radius=0.35];
\node[align=center] at (0,0) {\large $G$};
\draw[fill=white, line width=0.5mm] (1.8,0.3) circle [radius=0.35];
\node[align=center] at (1.8,0.3) {\large $A$};

\node[align=center] at (1.3,-1.6) {\normalsize $\tilde{\alpha}'_{i,j}$};
\node[align=center] at (1.6,-1.3) {\normalsize $\tilde{\beta}_{i,j}$};
\node[align=center] at (1.9,-1) {\normalsize $\tilde{\alpha}_{i,j}$};

\node[align=center] at (4,0) {\Large $-$};

\draw[line width=0.5mm, rounded corners] (8,2) -- (5,2) -- (5,-2) -- (8,-2) -- (8,-4) -- (5,-4);
\draw[line width=0.5mm] (6.5,2) to[out=240,in=90] (6.2,-0.3);
\draw[line width=0.5mm] (6.5,2) to[out=300,in=90] (6.8,0.3);

\draw[line width=0.5mm] (6.5,-2) to[out=120,in=270] (6.2,-0.3);
\draw[line width=0.5mm] (6.5,-2) to[out=60,in=270] (6.8,0.3);

\draw[line width=0.5mm] (5,0) to[out=315,in=180] (6.2,-0.3);

\draw[line width=0.5mm] (6.2,-0.3) -- (7.5,-0.3);
\draw[fill=white, line width=0.5mm] (6.2,-0.3) circle [radius=0.35];
\node[align=center] at (6.2,-0.3) {\large $A^{\dagger}$};
\draw[line width=0.5mm] (5,0) -- (6.5,0);

\draw[line width=0.5mm] (6.5,2) -- (6.5,-2);
\draw[line width=0.5mm] (6.5,0) -- (7.8,0);
\draw[fill=white, line width=0.5mm] (6.5,0) circle [radius=0.35];
\node[align=center] at (6.5,0) {\large $W$};
\draw[line width=0.5mm] (6.8,0.3) -- (8.1,0.3);

\draw[fill=white, line width=0.5mm] (6.5,2) circle [radius=0.35];
\node[align=center] at (6.5,2) {\large $E$};

\node[align=center] at (5.1,2.2) {\normalsize $\gamma_{i-1,j}$};
\node[align=center] at (5.4,-1.8) {\normalsize $\gamma_{i,j}$};
\node[align=center] at (7.5,2.2) {\normalsize $\gamma_{i-1,j+1}$};

\draw[line width=0.5mm] (5,0) to[out=30,in=180] (6.8,0.3);
\draw[fill=white, line width=0.5mm] (5,0) circle [radius=0.35];
\node[align=center] at (5,0) {\large $G$};
\draw[fill=white, line width=0.5mm] (6.8,0.3) circle [radius=0.35];
\node[align=center] at (6.8,0.3) {\large $A$};

\draw[line width=0.5mm] (6.5,-4) to[out=240,in=90] (6.2,-5.4);
\draw[line width=0.5mm] (6.5,-4) -- (6.5,-5.1);
\draw[line width=0.5mm] (6.5,-4) to[out=300,in=90] (6.8,-4.8);

\draw[fill=white, line width=0.5mm] (6.5,-2) circle [radius=0.35];
\node[align=center] at (6.5,-2) {\large $E$};

\draw[fill=white, line width=0.5mm] (6.5,-4) circle [radius=0.35];
\node[align=center] at (6.5,-4) {\large $E$};

\node[align=center] at (5.4,-3.8) {\normalsize $\gamma'_{i,j}$};
\node[align=center] at (7.5,-3) {\normalsize $\gamma_{i,j+1}$};
\node[align=center] at (6.3,-5.7) {\normalsize $\tilde{\alpha}'_{i,j}$};
\node[align=center] at (6.6,-5.4) {\normalsize $\tilde{\beta}_{i,j}$};
\node[align=center] at (6.9,-5.1) {\normalsize $\tilde{\alpha}_{i,j}$};

\end{tikzpicture}
}

\vspace*{0.2cm}
\centerline{\small (a) Initial setup.}
\vspace{0.1cm}        
\end{minipage}

\vspace{3mm}

\begin{minipage}{\widthfacorth\textwidth}
\centering
\scalebox{\scalefacorth}{ 
\begin{tikzpicture}[font=\tiny]

\draw[line width=0.5mm, rounded corners] (3,2) -- (0,2) -- (0,-0.8);
\draw[line width=0.5mm, rounded corners] (0,-1.2) -- (0,-3) -- (1,-3);
\draw[line width=0.5mm] (1.5,2) to[out=240,in=90] (1.2,-0.3);
\draw[line width=0.5mm] (1.5,2) to[out=300,in=90] (1.8,0.3);

\draw[line width=0.5mm] (1.2,-0.3) -- (1.2,-1.3);

\draw[line width=0.5mm] (0,0) to[out=315,in=180] (1.2,-0.3);

\draw[line width=0.5mm] (1.2,-0.3) -- (2.5,-0.3);
\draw[fill=white, line width=0.5mm] (1.2,-0.3) circle [radius=0.35];
\node[align=center] at (1.2,-0.3) {\large $A^{\dagger}$};
\draw[line width=0.5mm] (0,0) -- (1.5,0);

\draw[line width=0.5mm] (1.5,2) -- (1.5,-1);
\draw[line width=0.5mm] (1.5,0) -- (2.8,0);
\draw[fill=white, line width=0.5mm] (1.5,0) circle [radius=0.35];
\node[align=center] at (1.5,0) {\large $W$};
\draw[line width=0.5mm] (1.8,0.3) -- (1.8,-0.7);
\draw[line width=0.5mm] (1.8,0.3) -- (3.1,0.3);

\draw[fill=white, line width=0.5mm] (1.5,2) circle [radius=0.35];
\node[align=center] at (1.5,2) {\large $E$};

\node[align=center] at (0.4,2.2) {\normalsize $\gamma_{i-1,j}$};
\node[align=center] at (0.3,-1) {\normalsize $\delta_{i,j}$};
\node[align=center] at (0.4,-2.8) {\normalsize $\gamma'_{i,j}$};
\node[align=center] at (2.5,2.2) {\normalsize $\gamma_{i-1,j+1}$};

\draw[line width=0.5mm] (0,0) to[out=30,in=180] (1.8,0.3);
\draw[fill=white, line width=0.5mm] (0,0) circle [radius=0.5];
\node[align=center] at (0,0) {\large $\Sigma V^T$};
\draw[fill=white, line width=0.5mm] (0,-2) circle [radius=0.35];
\node[align=center] at (0,-2) {\large $U$};
\draw[fill=white, line width=0.5mm] (1.8,0.3) circle [radius=0.35];
\node[align=center] at (1.8,0.3) {\large $A$};

\node[align=center] at (1.3,-1.6) {\normalsize $\tilde{\alpha}'_{i,j}$};
\node[align=center] at (1.6,-1.3) {\normalsize $\tilde{\beta}_{i,j}$};
\node[align=center] at (1.9,-1) {\normalsize $\tilde{\alpha}_{i,j}$};

\node[align=center] at (4,0) {\Large $-$};

\draw[line width=0.5mm, rounded corners] (8,2) -- (5,2) -- (5,-2) -- (8,-2) -- (8,-4) -- (5.7,-4);
\draw[line width=0.5mm, rounded corners] (3.5,-4) -- (5.3,-4);
\draw[line width=0.5mm] (6.5,2) to[out=240,in=90] (6.2,-0.3);
\draw[line width=0.5mm] (6.5,2) to[out=300,in=90] (6.8,0.3);

\draw[line width=0.5mm] (6.5,-2) to[out=120,in=270] (6.2,-0.3);
\draw[line width=0.5mm] (6.5,-2) to[out=60,in=270] (6.8,0.3);

\draw[line width=0.5mm] (5,0) to[out=315,in=180] (6.2,-0.3);

\draw[line width=0.5mm] (6.2,-0.3) -- (7.5,-0.3);
\draw[fill=white, line width=0.5mm] (6.2,-0.3) circle [radius=0.35];
\node[align=center] at (6.2,-0.3) {\large $A^{\dagger}$};
\draw[line width=0.5mm] (5,0) -- (6.5,0);

\draw[line width=0.5mm] (6.5,2) -- (6.5,-2);
\draw[line width=0.5mm] (6.5,0) -- (7.8,0);
\draw[fill=white, line width=0.5mm] (6.5,0) circle [radius=0.35];
\node[align=center] at (6.5,0) {\large $W$};
\draw[line width=0.5mm] (6.8,0.3) -- (8.1,0.3);

\draw[fill=white, line width=0.5mm] (6.5,2) circle [radius=0.35];
\node[align=center] at (6.5,2) {\large $E$};

\node[align=center] at (5.1,2.2) {\normalsize $\gamma_{i-1,j}$};
\node[align=center] at (5.4,-1.8) {\normalsize $\gamma_{i,j}$};
\node[align=center] at (7.5,2.2) {\normalsize $\gamma_{i-1,j+1}$};

\draw[line width=0.5mm] (5,0) to[out=30,in=180] (6.8,0.3);
\draw[fill=white, line width=0.5mm] (5,0) circle [radius=0.35];
\node[align=center] at (5,0) {\large $G$};
\draw[fill=white, line width=0.5mm] (6.8,0.3) circle [radius=0.35];
\node[align=center] at (6.8,0.3) {\large $A$};

\draw[line width=0.5mm] (6.5,-4) to[out=240,in=90] (6.2,-5.4);
\draw[line width=0.5mm] (6.5,-4) -- (6.5,-5.1);
\draw[line width=0.5mm] (6.5,-4) to[out=300,in=90] (6.8,-4.8);

\draw[fill=white, line width=0.5mm] (6.5,-2) circle [radius=0.35];
\node[align=center] at (6.5,-2) {\large $E$};

\draw[fill=white, line width=0.5mm] (6.5,-4) circle [radius=0.5];
\node[align=center] at (6.5,-4) {\large $\Sigma V^T$};

\draw[fill=white, line width=0.5mm] (4.5,-4) circle [radius=0.35];
\node[align=center] at (4.5,-4) {\large $U$};

\node[align=center] at (3.8,-3.8) {\normalsize $\gamma'_{i,j}$};
\node[align=center] at (5.5,-3.8) {\normalsize $\delta_{i,j}$};
\node[align=center] at (7.5,-3) {\normalsize $\gamma_{i,j+1}$};
\node[align=center] at (6.3,-5.7) {\normalsize $\tilde{\alpha}'_{i,j}$};
\node[align=center] at (6.6,-5.4) {\normalsize $\tilde{\beta}_{i,j}$};
\node[align=center] at (6.9,-5.1) {\normalsize $\tilde{\alpha}_{i,j}$};

\end{tikzpicture}
}

\vspace*{0.2cm}
\centerline{\small (b) Truncated decompositions.}
\vspace{0.1cm}        
\end{minipage}

\vspace{3mm}

\begin{minipage}{\widthfacorth\textwidth}
\centering
\scalebox{\scalefacorth}{ 
\begin{tikzpicture}[font=\tiny]

\draw[line width=0.5mm, rounded corners] (2.5,1) -- (1.5,1) -- (1.5,0);
\draw[line width=0.5mm, rounded corners] (0.5,0) -- (1.5,0);

\draw[line width=0.5mm] (1.5,0) to[out=240,in=90] (1.2,-1.3);
\draw[line width=0.5mm] (1.5,0) to[out=270,in=90] (1.5,-1);
\draw[line width=0.5mm] (1.5,0) to[out=300,in=90] (1.8,-0.7);

\draw[line width=0.5mm] (1.5,0) to[out=30,in=180] (2.8,0.3);
\draw[line width=0.5mm] (1.5,0) to[out=0,in=180] (2.5,0);
\draw[line width=0.5mm] (1.5,0) to[out=330,in=180] (2.2,-0.3);

\draw[fill=white, line width=0.5mm] (1.5,0) circle [radius=0.35];
\node[align=center] at (1.5,0) {\large $B_l$};

\node[align=center] at (0.8,0.3) {\normalsize $\delta_{i,j}$};
\node[align=center] at (2,1.2) {\normalsize $\gamma_{i-1,j+1}$};

\node[align=center] at (1.3,-1.6) {\normalsize $\tilde{\alpha}'_{i,j}$};
\node[align=center] at (1.6,-1.3) {\normalsize $\tilde{\beta}_{i,j}$};
\node[align=center] at (1.9,-1) {\normalsize $\tilde{\alpha}_{i,j}$};

\end{tikzpicture}
}

\vspace*{0.2cm}
\centerline{\small (c) Contraction and subtraction.}
\vspace{0.1cm}        
\end{minipage}

\vspace{3mm}

\begin{minipage}{\widthfacorth\textwidth}
\centering
\scalebox{\scalefacorth}{ 
\begin{tikzpicture}[font=\tiny]

\draw[line width=0.5mm, rounded corners] (4.5,1) -- (1.5,1) -- (1.5,0);
\draw[line width=0.5mm, rounded corners] (0.5,0) -- (1.5,0);

\draw[line width=0.5mm] (1.5,0) to[out=240,in=90] (1.2,-1.3);
\draw[line width=0.5mm] (1.5,0) to[out=270,in=90] (1.5,-1);
\draw[line width=0.5mm] (1.5,0) to[out=300,in=90] (1.8,-0.7);

\draw[line width=0.5mm] (1.5,0) to[out=30,in=180] (2.8,0.3);
\draw[line width=0.5mm] (1.5,0) to[out=0,in=180] (2.5,0);
\draw[line width=0.5mm] (1.5,0) to[out=330,in=180] (2.2,-0.3);

\draw[fill=white, line width=0.5mm] (1.5,0) circle [radius=0.35];
\node[align=center] at (1.5,0) {\large $U$};

\node[align=center] at (0.8,0.3) {\normalsize $\delta_{i,j}$};
\node[align=center] at (1.8,1.2) {\normalsize $\epsilon_{i-1,j+1}$};
\node[align=center] at (4.2,1.2) {\normalsize $\gamma_{i-1,j+1}$};

\node[align=center] at (1.3,-1.6) {\normalsize $\tilde{\alpha}'_{i,j}$};
\node[align=center] at (1.6,-1.3) {\normalsize $\tilde{\beta}_{i,j}$};
\node[align=center] at (1.9,-1) {\normalsize $\tilde{\alpha}_{i,j}$};

\draw[fill=white, line width=0.5mm] (3,1) circle [radius=0.5];
\node[align=center] at (3,1) {\large $\Sigma V^T$};

\end{tikzpicture}
}

\vspace*{0.2cm}
\centerline{\small (d) Truncated decomposition.}
\vspace{0.1cm}        
\end{minipage}

\caption{Calculation of left orthogonal block.}
\label{fig:leftorth}
\end{figure}
The first step is to decompose the tensors $G$ and $E$ in the manner
depicted in Fig.~\ref{fig:leftorth}(b), 
while the new index 
$\delta_{i,j}$ is truncated from $\chi$ to $\xi$.
Afterwards, 
both clusters are 
contracted and subtracted, yielding
the left orthogonal block depicted in
Fig.~\ref{fig:leftorth}(c). 
A second truncated decomposition, 
Fig.~\ref{fig:leftorth}(d), 
isolates the weight $\Sigma V^T$ from the left block, which is 
then passed on to the right orthogonal block, 
Fig.~\ref{fig:rightorth}(a). 
\begin{figure}[!htb]
\begin{minipage}{\widthfacorth\textwidth}
\centering
\scalebox{\scalefacorth}{ 
\begin{tikzpicture}[font=\tiny]

\draw[line width=0.5mm, rounded corners] (3,3.5) -- (0,3.5) -- (0,2) -- (3,2) -- (3,-2) -- (2,-2);
\draw[fill=white, line width=0.5mm] (1.5,3.5) circle [radius=0.5];
\node[align=center] at (1.5,3.5) {\large $\Sigma V_T$};

\draw[line width=0.5mm] (1.5,2) to[out=240,in=90] (1.2,-0.3);
\draw[line width=0.5mm] (1.5,2) to[out=300,in=90] (1.8,0.3);
\draw[line width=0.5mm] (1.5,2) -- (1.5,-1);
\draw[line width=0.5mm] (1.8,0.3) -- (1.8,-0.7);
\draw[line width=0.5mm] (1.2,-0.3) -- (1.2,-1.3);

\draw[line width=0.5mm] (3,0) to[out=225,in=180] (1.2,-0.3);
\draw[line width=0.5mm] (3,0) to[out=150,in=0] (1.8,0.3);
\draw[line width=0.5mm] (3,0) to[out=180,in=0] (1.5,0);

\draw[line width=0.5mm] (1.2,-0.3) -- (-0.1,-0.3);
\draw[fill=white, line width=0.5mm] (1.2,-0.3) circle [radius=0.35];
\node[align=center] at (1.2,-0.3) {\large $A^{\dagger}$};

\draw[line width=0.5mm] (1.5,0) -- (0.2,0);
\draw[fill=white, line width=0.5mm] (1.5,0) circle [radius=0.35];
\node[align=center] at (1.5,0) {\large $W$};

\draw[fill=white, line width=0.5mm] (1.5,2) circle [radius=0.35];
\node[align=center] at (1.5,2) {\large $E$};

\node[align=center] at (0.7,2.75) {\normalsize $\gamma_{i-1,j+1}$};
\node[align=center] at (2.7,3.2) {\normalsize $\epsilon_{i-1,j+1}$};
\node[align=center] at (2.5,-1.8) {\normalsize $\gamma'_{i,j+2}$};
\node[align=center] at (2.5,2.2) {\normalsize $\gamma_{i-1,j+2}$};

\draw[fill=white, line width=0.5mm] (3,0) circle [radius=0.35];
\node[align=center] at (3,0) {\large $G$};

\draw[line width=0.5mm] (1.8,0.3) -- (0.5,0.3);
\draw[fill=white, line width=0.5mm] (1.8,0.3) circle [radius=0.35];
\node[align=center] at (1.8,0.3) {\large $A$};

\node[align=center] at (1.4,-1.6) {\normalsize $\tilde{\alpha}'_{i,j+1}$};
\node[align=center] at (1.7,-1.3) {\normalsize $\tilde{\beta}_{i,j+1}$};
\node[align=center] at (2,-1) {\normalsize $\tilde{\alpha}_{i,j+1}$};

\node[align=center] at (4,0) {\Large $-$};

\draw[line width=0.5mm, rounded corners] (8,3.5) -- (5,3.5) -- (5,2) -- (8,2) -- (8,-2) -- (5,-2) -- (5,-4) -- (8,-4);
\draw[fill=white, line width=0.5mm] (6.5,3.5) circle [radius=0.5];
\node[align=center] at (6.5,3.5) {\large $\Sigma V_T$};

\draw[line width=0.5mm] (6.5,2) to[out=240,in=90] (6.2,-0.3);
\draw[line width=0.5mm] (6.5,2) to[out=300,in=90] (6.8,0.3);

\draw[line width=0.5mm] (6.5,-2) to[out=120,in=270] (6.2,-0.3);
\draw[line width=0.5mm] (6.5,-2) to[out=60,in=270] (6.8,0.3);

\draw[line width=0.5mm] (8,0) to[out=225,in=180] (6.2,-0.3);

\draw[line width=0.5mm] (6.2,-0.3) -- (4.9,-0.3);
\draw[fill=white, line width=0.5mm] (6.2,-0.3) circle [radius=0.35];
\node[align=center] at (6.2,-0.3) {\large $A^{\dagger}$};
\draw[line width=0.5mm] (5,0) -- (6.5,0);

\draw[line width=0.5mm] (6.5,2) -- (6.5,-2);
\draw[line width=0.5mm] (6.5,0) -- (7.8,0);
\draw[fill=white, line width=0.5mm] (6.5,0) circle [radius=0.35];
\node[align=center] at (6.5,0) {\large $W$};
\draw[line width=0.5mm] (6.8,0.3) -- (5.5,0.3);

\draw[fill=white, line width=0.5mm] (6.5,2) circle [radius=0.35];
\node[align=center] at (6.5,2) {\large $E$};

\node[align=center] at (5.7,2.75) {\normalsize $\gamma_{i-1,j+1}$};
\node[align=center] at (7.7,3.2) {\normalsize $\epsilon_{i-1,j+1}$};
\node[align=center] at (7.4,-1.8) {\normalsize $\gamma_{i,j+2}$};
\node[align=center] at (7.5,2.2) {\normalsize $\gamma_{i-1,j+2}$};

\draw[line width=0.5mm] (8,0) to[out=150,in=0] (6.8,0.3);
\draw[fill=white, line width=0.5mm] (8,0) circle [radius=0.35];
\node[align=center] at (8,0) {\large $G$};
\draw[fill=white, line width=0.5mm] (6.8,0.3) circle [radius=0.35];
\node[align=center] at (6.8,0.3) {\large $A$};

\draw[line width=0.5mm] (6.5,-4) to[out=240,in=90] (6.2,-5.4);
\draw[line width=0.5mm] (6.5,-4) -- (6.5,-5.1);
\draw[line width=0.5mm] (6.5,-4) to[out=300,in=90] (6.8,-4.8);

\draw[fill=white, line width=0.5mm] (6.5,-2) circle [radius=0.35];
\node[align=center] at (6.5,-2) {\large $E$};

\draw[fill=white, line width=0.5mm] (6.5,-4) circle [radius=0.35];
\node[align=center] at (6.5,-4) {\large $E$};

\node[align=center] at (7.4,-3.8) {\normalsize $\gamma'_{i,j+2}$};
\node[align=center] at (5.5,-3) {\normalsize $\gamma_{i,j+1}$};
\node[align=center] at (6.4,-5.7) {\normalsize $\tilde{\alpha}'_{i,j+1}$};
\node[align=center] at (6.7,-5.4) {\normalsize $\tilde{\beta}_{i,j+1}$};
\node[align=center] at (7,-5.1) {\normalsize $\tilde{\alpha}_{i,j+1}$};

\end{tikzpicture}
}

\vspace*{0.2cm}
\centerline{\small (a) Initial setup.}
\vspace{0.1cm}        
\end{minipage}

\vspace{3mm}

\begin{minipage}{\widthfacorth\textwidth}
\centering
\scalebox{\scalefacorth}{ 
\begin{tikzpicture}[font=\tiny]

\draw[line width=0.5mm, rounded corners] (0,2) -- (3,2) -- (3,-2) -- (2,-2);

\draw[line width=0.5mm] (1.5,2) to[out=240,in=90] (1.2,-0.3);
\draw[line width=0.5mm] (1.5,2) to[out=300,in=90] (1.8,0.3);
\draw[line width=0.5mm] (1.5,2) -- (1.5,-1);
\draw[line width=0.5mm] (1.8,0.3) -- (1.8,-0.7);
\draw[line width=0.5mm] (1.2,-0.3) -- (1.2,-1.3);

\draw[line width=0.5mm] (3,0) to[out=225,in=180] (1.2,-0.3);
\draw[line width=0.5mm] (3,0) to[out=150,in=0] (1.8,0.3);
\draw[line width=0.5mm] (3,0) to[out=180,in=0] (1.5,0);

\draw[line width=0.5mm] (1.2,-0.3) -- (-0.1,-0.3);
\draw[fill=white, line width=0.5mm] (1.2,-0.3) circle [radius=0.35];
\node[align=center] at (1.2,-0.3) {\large $A^{\dagger}$};

\draw[line width=0.5mm] (1.5,0) -- (0.2,0);
\draw[fill=white, line width=0.5mm] (1.5,0) circle [radius=0.35];
\node[align=center] at (1.5,0) {\large $W$};

\draw[fill=white, line width=0.5mm] (1.5,2) circle [radius=0.35];
\node[align=center] at (1.5,2) {\large $E'$};

\node[align=center] at (0.4,2.2) {\normalsize $\epsilon_{i-1,j+1}$};
\node[align=center] at (2.5,-1.8) {\normalsize $\gamma'_{i,j+2}$};
\node[align=center] at (2.5,2.2) {\normalsize $\gamma_{i-1,j+2}$};

\draw[fill=white, line width=0.5mm] (3,0) circle [radius=0.35];
\node[align=center] at (3,0) {\large $G$};

\draw[line width=0.5mm] (1.8,0.3) -- (0.5,0.3);
\draw[fill=white, line width=0.5mm] (1.8,0.3) circle [radius=0.35];
\node[align=center] at (1.8,0.3) {\large $A$};

\node[align=center] at (1.4,-1.6) {\normalsize $\tilde{\alpha}'_{i,j+1}$};
\node[align=center] at (1.7,-1.3) {\normalsize $\tilde{\beta}_{i,j+1}$};
\node[align=center] at (2,-1) {\normalsize $\tilde{\alpha}_{i,j+1}$};

\node[align=center] at (4,0) {\Large $-$};

\draw[line width=0.5mm, rounded corners] (5,2) -- (8,2) -- (8,-2) -- (5,-2) -- (5,-4) -- (8,-4);
\draw[line width=0.5mm] (6.5,2) to[out=240,in=90] (6.2,-0.3);
\draw[line width=0.5mm] (6.5,2) to[out=300,in=90] (6.8,0.3);

\draw[line width=0.5mm] (6.5,-2) to[out=120,in=270] (6.2,-0.3);
\draw[line width=0.5mm] (6.5,-2) to[out=60,in=270] (6.8,0.3);

\draw[line width=0.5mm] (8,0) to[out=225,in=180] (6.2,-0.3);

\draw[line width=0.5mm] (6.2,-0.3) -- (4.9,-0.3);
\draw[fill=white, line width=0.5mm] (6.2,-0.3) circle [radius=0.35];
\node[align=center] at (6.2,-0.3) {\large $A^{\dagger}$};
\draw[line width=0.5mm] (5,0) -- (6.5,0);

\draw[line width=0.5mm] (6.5,2) -- (6.5,-2);
\draw[line width=0.5mm] (6.5,0) -- (7.8,0);
\draw[fill=white, line width=0.5mm] (6.5,0) circle [radius=0.35];
\node[align=center] at (6.5,0) {\large $W$};
\draw[line width=0.5mm] (6.8,0.3) -- (5.5,0.3);

\draw[fill=white, line width=0.5mm] (6.5,2) circle [radius=0.35];
\node[align=center] at (6.5,2) {\large $E'$};

\node[align=center] at (5.1,2.2) {\normalsize $\epsilon_{i-1,j+1}$};
\node[align=center] at (7.4,-1.8) {\normalsize $\gamma_{i,j+2}$};
\node[align=center] at (7.5,2.2) {\normalsize $\gamma_{i-1,j+2}$};

\draw[line width=0.5mm] (8,0) to[out=150,in=0] (6.8,0.3);
\draw[fill=white, line width=0.5mm] (8,0) circle [radius=0.35];
\node[align=center] at (8,0) {\large $G$};
\draw[fill=white, line width=0.5mm] (6.8,0.3) circle [radius=0.35];
\node[align=center] at (6.8,0.3) {\large $A$};

\draw[line width=0.5mm] (6.5,-4) to[out=240,in=90] (6.2,-5.4);
\draw[line width=0.5mm] (6.5,-4) -- (6.5,-5.1);
\draw[line width=0.5mm] (6.5,-4) to[out=300,in=90] (6.8,-4.8);

\draw[fill=white, line width=0.5mm] (6.5,-2) circle [radius=0.35];
\node[align=center] at (6.5,-2) {\large $E$};

\draw[fill=white, line width=0.5mm] (6.5,-4) circle [radius=0.35];
\node[align=center] at (6.5,-4) {\large $E$};

\node[align=center] at (7.4,-3.8) {\normalsize $\gamma'_{i,j+2}$};
\node[align=center] at (5.5,-3) {\normalsize $\gamma_{i,j+1}$};
\node[align=center] at (6.4,-5.7) {\normalsize $\tilde{\alpha}'_{i,j+1}$};
\node[align=center] at (6.7,-5.4) {\normalsize $\tilde{\beta}_{i,j+1}$};
\node[align=center] at (7,-5.1) {\normalsize $\tilde{\alpha}_{i,j+1}$};

\end{tikzpicture}
}

\vspace*{0.2cm}
\centerline{\small (b) Contraction of weight from left block.}
\vspace{0.1cm}        
\end{minipage}

\vspace{3mm}

\begin{minipage}{\widthfacorth\textwidth}
\centering
\scalebox{\scalefacorth}{ 
\begin{tikzpicture}[font=\tiny]

\draw[line width=0.5mm, rounded corners] (0.5,1) -- (1.5,1) -- (1.5,0);
\draw[line width=0.5mm, rounded corners] (2.5,0) -- (1.5,0);

\draw[line width=0.5mm] (1.5,0) to[out=240,in=90] (1.2,-1.3);
\draw[line width=0.5mm] (1.5,0) to[out=270,in=90] (1.5,-1);
\draw[line width=0.5mm] (1.5,0) to[out=300,in=90] (1.8,-0.7);

\draw[line width=0.5mm] (1.5,0) to[out=150,in=0] (0.2,0.3);
\draw[line width=0.5mm] (1.5,0) to[out=180,in=0] (0.5,0);
\draw[line width=0.5mm] (1.5,0) to[out=210,in=0] (0.8,-0.3);

\draw[fill=white, line width=0.5mm] (1.5,0) circle [radius=0.35];
\node[align=center] at (1.5,0) {\large $B_r$};

\node[align=center] at (2.3,0.3) {\normalsize $\gamma'_{i,j+2}$};
\node[align=center] at (1,1.2) {\normalsize $\epsilon_{i-1,j+1}$};

\node[align=center] at (1.4,-1.6) {\normalsize $\tilde{\alpha}'_{i,j+1}$};
\node[align=center] at (1.7,-1.3) {\normalsize $\tilde{\beta}_{i,j+1}$};
\node[align=center] at (2,-1) {\normalsize $\tilde{\alpha}_{i,j+1}$};

\end{tikzpicture}
}

\vspace*{0.2cm}
\centerline{\small (c) Contraction and subtraction.}
\vspace{0.1cm}        
\end{minipage}

\vspace{3mm}

\begin{minipage}{\widthfacorth\textwidth}
\centering
\scalebox{\scalefacorth}{ 
\begin{tikzpicture}[font=\tiny]

\draw[line width=0.5mm, rounded corners] (0.5,1) -- (1.5,1) -- (1.5,0);
\draw[line width=0.5mm, rounded corners] (4.5,0) -- (1.5,0);

\draw[line width=0.5mm] (3.5,0) to[out=240,in=90] (3.2,-1.3);
\draw[line width=0.5mm] (3.5,0) to[out=270,in=90] (3.5,-1);
\draw[line width=0.5mm] (3.5,0) to[out=300,in=90] (3.8,-0.7);

\draw[line width=0.5mm] (1.5,0) to[out=150,in=0] (0.2,0.3);
\draw[line width=0.5mm] (1.5,0) to[out=180,in=0] (0.5,0);
\draw[line width=0.5mm] (1.5,0) to[out=210,in=0] (0.8,-0.3);

\draw[fill=white, line width=0.5mm] (1.5,0) circle [radius=0.45];
\node[align=center] at (1.5,0) {\large $U \Sigma$};

\draw[fill=white, line width=0.5mm] (3.5,0) circle [radius=0.4];
\node[align=center] at (3.5,0) {\large $V^T$};

\node[align=center] at (2.5,0.3) {\normalsize $\gamma'_{i,j+1}$};
\node[align=center] at (4.4,0.3) {\normalsize $\gamma'_{i,j+2}$};
\node[align=center] at (1,1.2) {\normalsize $\epsilon_{i-1,j+1}$};

\node[align=center] at (3.4,-1.6) {\normalsize $\tilde{\alpha}'_{i,j+1}$};
\node[align=center] at (3.7,-1.3) {\normalsize $\tilde{\beta}_{i,j+1}$};
\node[align=center] at (4,-1) {\normalsize $\tilde{\alpha}_{i,j+1}$};

\end{tikzpicture}
}

\vspace*{0.2cm}
\centerline{\small (d) Truncated decomposition.}
\vspace{0.1cm}        
\end{minipage}

\caption{Calculation of right orthogonal block.}
\label{fig:rightorth}
\end{figure}
After the weight is absorbed,
Fig.~\ref{fig:rightorth}(b),
the clusters are contracted and subtracted, as depicted in 
Fig.~\ref{fig:rightorth}(c).
The final decomposition,
Fig.~\ref{fig:rightorth}(d),
yields 
the truncated complement $V^T$, which encodes
those bonds orthogonal to the current basis with the largest
weight \cite{Gleis2022Jul}.
If the current dimension of $\gamma_{i,j+1}$ between
$E_{i,j}$ (Fig.~\ref{fig:envcomplete}(a)) and $E_{i,j+1}$ 
(Fig.~\ref{fig:envcomplete}(a)) is given by
$\text{dim}\left(\gamma_{i,j+1}\right)$, the bond dimension 
of $\gamma'_{i,j+1}$ in Fig.~\ref{fig:rightorth}(d)
is set to
\begin{align}
\text{dim}\left(\gamma'_{i,j+1}\right)
	= \chi + \xi -\text{dim}\left(\gamma_{i,j+1}\right) \, .
	\label{eq:cbedim}
\end{align}
In the final step of the CBE, we
add $V^T$ to $E_{i,j+1}$; here
Eq.~\eqref{eq:cbedim} ensures that the dimension of the bond in the sum does 
not exceed 
$\chi+\xi$.
An additional QR decomposition and the contraction depicted in
Fig.~\ref{fig:intsteps}(b)
allows us to solve for $E_{i,j}$
using 
the cluster depicted in Fig.~\ref{fig:inhomo1}.
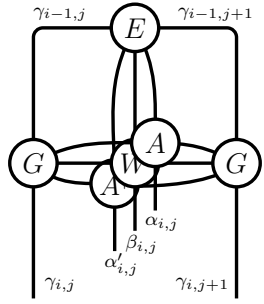
\begin{figure}[!htb]
\centering
\scalebox{0.9}{ 
\begin{tikzpicture}[font=\tiny]

\draw[line width=0.5mm, rounded corners] (3,-2) -- (3,2) -- (0,2) -- (0,-2);
\draw[line width=0.5mm] (1.5,2) to[out=240,in=90] (1.2,-0.3);
\draw[line width=0.5mm] (1.5,2) to[out=300,in=90] (1.8,0.3);

\draw[line width=0.5mm] (1.2,-0.3) -- (1.2,-1.3);

\draw[line width=0.5mm] (0,0) to[out=315,in=180] (1.2,-0.3);
\draw[line width=0.5mm] (3,0) to[out=225,in=180] (1.2,-0.3);

\draw[fill=white, line width=0.5mm] (1.2,-0.3) circle [radius=0.35];
\node[align=center] at (1.2,-0.3) {\large $A^{\dagger}$};
\draw[line width=0.5mm] (0,0) -- (3,0);

\draw[line width=0.5mm] (1.5,2) -- (1.5,-1);
\draw[fill=white, line width=0.5mm] (1.5,0) circle [radius=0.35];
\node[align=center] at (1.5,0) {\large $W$};
\draw[line width=0.5mm] (1.8,0.3) -- (1.8,-0.7);

\draw[fill=white, line width=0.5mm] (1.5,2) circle [radius=0.35];
\node[align=center] at (1.5,2) {\large $E$};

\node[align=center] at (0.4,2.2) {\footnotesize $\gamma_{i-1,j}$};
\node[align=center] at (0.4,-1.8) {\footnotesize $\gamma_{i,j}$};
\node[align=center] at (2.7,2.2) {\footnotesize $\gamma_{i-1,j+1}$};
\node[align=center] at (2.5,-1.8) {\footnotesize $\gamma_{i,j+1}$};

\draw[line width=0.5mm] (0,0) to[out=30,in=180] (1.8,0.3);
\draw[line width=0.5mm] (3,0) to[out=150,in=0] (1.8,0.3);
\draw[fill=white, line width=0.5mm] (0,0) circle [radius=0.35];
\node[align=center] at (0,0) {\large $G$};
\draw[fill=white, line width=0.5mm] (1.8,0.3) circle [radius=0.35];
\node[align=center] at (1.8,0.3) {\large $A$};

\draw[fill=white, line width=0.5mm] (3,0) circle [radius=0.35];
\node[align=center] at (3,0) {\large $G$};

\node[align=center] at (1.3,-1.5) {\footnotesize $\alpha'_{i,j}$};
\node[align=center] at (1.6,-1.2) {\footnotesize $\beta_{i,j}$};
\node[align=center] at (1.9,-0.9) {\footnotesize $\alpha_{i,j}$};

\end{tikzpicture}
}
\caption{Calculation of left environment tensor.}
\label{fig:inhomo1}
\end{figure}
If the dimension of $\gamma_{i,j+1}$ is greater than $\chi$, the
sequence of operations depicted in Fig.~\ref{fig:twoblocksvd} can be
used to truncate the dimension back to $\chi$.
\begin{figure}[!htb]
\begin{minipage}{\widthfac\textwidth}
\centering
\scalebox{0.7}{ 
\begin{tikzpicture}[font=\tiny]

\draw[line width=0.5mm] (1,0) to[out=225,in=90] (0.5,-1.5);
\draw[line width=0.5mm] (1,0) -- (1,-1.5);
\draw[line width=0.5mm] (1,0) to[out=315,in=90] (1.5,-1.5);

\draw[line width=0.5mm] (3,0) to[out=225,in=90] (2.5,-1.5);
\draw[line width=0.5mm] (3,0) -- (3,-1.5);
\draw[line width=0.5mm] (3,0) to[out=315,in=90] (3.5,-1.5);

\draw[line width=0.5mm] (-0.5,0) -- (4.5,0);
\node[align=center] at (0,0.3) {\normalsize $\gamma_j$};
\node[align=center] at (2,0.3) {\normalsize $\gamma_{j+1}$};
\node[align=center] at (4,0.3) {\normalsize $\gamma_{j+2}$};
\draw[fill=white, line width=0.5mm] (1,0) circle [radius=0.5];
\node[align=center] at (1,0) {\normalsize $E$};

\draw[fill=white, line width=0.5mm] (3,0) circle [radius=0.5];
\node[align=center] at (3,0) {\normalsize $E$};

\end{tikzpicture}
}

\vspace*{0.2cm}
\centerline{\small (a) Initial setup.}
\vspace{0.1cm}        
\end{minipage}

\vspace{3mm}

\begin{minipage}{\widthfac\textwidth}
\centering
\scalebox{0.7}{ 
\begin{tikzpicture}[font=\tiny]

\draw[line width=0.5mm] (8,0) to[out=225,in=90] (7.5,-1.5);
\draw[line width=0.5mm] (8,0) -- (8,-1.5);
\draw[line width=0.5mm] (8,0) to[out=315,in=90] (8.5,-1.5);

\draw[line width=0.5mm] (14,0) to[out=225,in=90] (13.5,-1.5);
\draw[line width=0.5mm] (14,0) -- (14,-1.5);
\draw[line width=0.5mm] (14,0) to[out=315,in=90] (14.5,-1.5);

\draw[line width=0.5mm] (6.5,0) -- (15.5,0);
\node[align=center] at (9,0.3) {\normalsize $a$};
\node[align=center] at (13,0.3) {\normalsize $b$};
\node[align=center] at (7,0.3) {\normalsize $\gamma_j$};
\node[align=center] at (11,0.3) {\normalsize $\gamma_{j+1}$};
\node[align=center] at (15,0.3) {\normalsize $\gamma_{j+2}$};
\draw[fill=white, line width=0.5mm] (8,0) circle [radius=0.5];
\node[align=center] at (8,0) {\normalsize $U$};

\draw[fill=white, line width=0.5mm] (10,0) circle [radius=0.5];
\node[align=center] at (10,0) {\normalsize $\Sigma V^T$};

\draw[fill=white, line width=0.5mm] (12,0) circle [radius=0.5];
\node[align=center] at (12,0) {\normalsize $U \Sigma$};

\draw[fill=white, line width=0.5mm] (14,0) circle [radius=0.5];
\node[align=center] at (14,0) {\normalsize $V^T$};

\end{tikzpicture}
}

\vspace*{0.2cm}
\centerline{\small (b) Decompositions.}
\vspace{0.1cm}        
\end{minipage}

\vspace{3mm}

\begin{minipage}{\widthfac\textwidth}
\centering
\scalebox{0.7}{ 
\begin{tikzpicture}[font=\tiny]

\draw[line width=0.5mm] (3.5,-3) to[out=225,in=90] (3,-4.5);
\draw[line width=0.5mm] (3.5,-3) -- (3.5,-4.5);
\draw[line width=0.5mm] (3.5,-3) to[out=315,in=90] (4,-4.5);

\draw[line width=0.5mm] (7.5,-3) to[out=225,in=90] (7,-4.5);
\draw[line width=0.5mm] (7.5,-3) -- (7.5,-4.5);
\draw[line width=0.5mm] (7.5,-3) to[out=315,in=90] (8,-4.5);

\draw[line width=0.5mm] (2,-3) -- (9,-3);
\node[align=center] at (2.5,-2.7) {\normalsize $\gamma_j$};
\node[align=center] at (6.5,-2.7) {\normalsize $b$};
\node[align=center] at (4.5,-2.7) {\normalsize $a$};
\node[align=center] at (8.5,-2.7) {\normalsize $\gamma_{j+2}$};

\draw[fill=white, line width=0.5mm] (3.5,-3) circle [radius=0.5];
\node[align=center] at (3.5,-3) {\normalsize $U$};

\draw[fill=white, line width=0.5mm] (5.5,-3) circle [radius=0.5];
\node[align=center] at (5.5,-3) {\normalsize $T$};

\draw[fill=white, line width=0.5mm] (7.5,-3) circle [radius=0.5];
\node[align=center] at (7.5,-3) {\normalsize $V^T$};

\end{tikzpicture}
}

\vspace*{0.2cm}
\centerline{\small (c) Contraction.}
\vspace{0.1cm}        
\end{minipage}

\vspace{3mm}

\begin{minipage}{\widthfac\textwidth}
\centering
\scalebox{0.7}{ 
\begin{tikzpicture}[font=\tiny]

\draw[line width=0.5mm] (3.5,-6) to[out=225,in=90] (3,-7.5);
\draw[line width=0.5mm] (3.5,-6) -- (3.5,-7.5);
\draw[line width=0.5mm] (3.5,-6) to[out=315,in=90] (4,-7.5);

\draw[line width=0.5mm] (9.5,-6) to[out=225,in=90] (9,-7.5);
\draw[line width=0.5mm] (9.5,-6) -- (9.5,-7.5);
\draw[line width=0.5mm] (9.5,-6) to[out=315,in=90] (10,-7.5);

\draw[line width=0.5mm] (2,-6) -- (11,-6);
\node[align=center] at (2.5,-5.7) {\normalsize $\gamma_j$};
\node[align=center] at (4.5,-5.7) {\normalsize $a$};
\node[align=center] at (8.5,-5.7) {\normalsize $b$};
\node[align=center] at (6.5,-5.7) {\normalsize $\gamma_{j+1}$};
\node[align=center] at (10.5,-5.7) {\normalsize $\gamma_{j+2}$};

\draw[fill=white, line width=0.5mm] (3.5,-6) circle [radius=0.5];
\node[align=center] at (3.5,-6) {\normalsize $U$};

\draw[fill=white, line width=0.5mm] (5.5,-6) circle [radius=0.55];
\node[align=center] at (5.5,-6) {\normalsize $U\sqrt{\Sigma}$};

\draw[fill=white, line width=0.5mm] (7.5,-6) circle [radius=0.55];
\node[align=center] at (7.5,-6) {\normalsize $\sqrt{\Sigma}V^T$};

\draw[fill=white, line width=0.5mm] (9.5,-6) circle [radius=0.5];
\node[align=center] at (9.5,-6) {\normalsize $V^T$};

\end{tikzpicture}
}

\vspace*{0.2cm}
\centerline{\small (d) Truncated decomposition.}
\vspace{0.1cm}        
\end{minipage}

\vspace{3mm}

\begin{minipage}{\widthfac\textwidth}
\centering
\scalebox{0.7}{ 
\begin{tikzpicture}[font=\tiny]

\draw[line width=0.5mm] (3.5,-9) to[out=225,in=90] (3,-10.5);
\draw[line width=0.5mm] (3.5,-9) -- (3.5,-10.5);
\draw[line width=0.5mm] (3.5,-9) to[out=315,in=90] (4,-10.5);

\draw[line width=0.5mm] (5.5,-9) to[out=225,in=90] (5,-10.5);
\draw[line width=0.5mm] (5.5,-9) -- (5.5,-10.5);
\draw[line width=0.5mm] (5.5,-9) to[out=315,in=90] (6,-10.5);

\draw[line width=0.5mm] (2,-9) -- (7,-9);
\node[align=center] at (2.5,-8.7) {\normalsize $\gamma_j$};
\node[align=center] at (4.5,-8.7) {\normalsize $\gamma_{j+1}$};
\node[align=center] at (6.5,-8.7) {\normalsize $\gamma_{j+2}$};

\draw[fill=white, line width=0.5mm] (3.5,-9) circle [radius=0.5];
\node[align=center] at (3.5,-9) {\normalsize $E'$};

\draw[fill=white, line width=0.5mm] (5.5,-9) circle [radius=0.5];
\node[align=center] at (5.5,-9) {\normalsize $E'$};

\end{tikzpicture}
}

\vspace*{0.2cm}
\centerline{\small (e) Contractions.}
\vspace{0.1cm}        
\end{minipage}

\caption{Truncation of two adjacent environment tensors.}
\label{fig:twoblocksvd}
\end{figure}
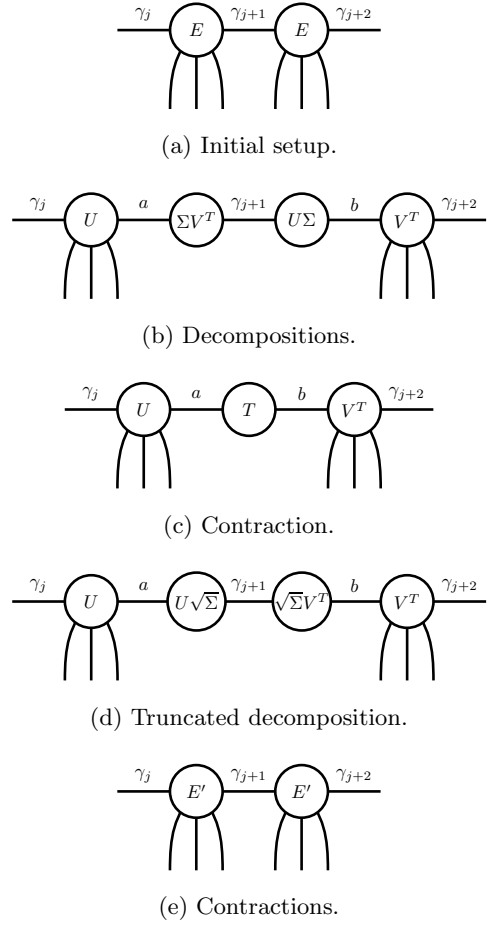

To summarize, 
the CBE allows one to circumvent the operations
depicted in Figs.~\ref{fig:inhomo} and 
\ref{fig:svdblock} by instead using 
orthogonal projectors and a series of contractions and 
truncated SVDs, avoiding operating 
on a rank-8 tensor with
indices of full dimension.

In order to progressively improve the environment,
one can 
sweep repeatedly 
from left to right and from right to left
in a manner similar to the
optimization of an MPS in the DMRG.
The state $\tilde{\left| \psi \right>}$ then converges to the 
best approximation of $\left| \psi \right>$ within the given maximum 
bond dimension $\chi$.
Thus, the scaling of the computational cost of contracting PEPS-based
tensor networks with system size becomes linear rather than
exponential if the contraction is carried out approximately in this
manner.

Note that the approach presented above is not the only possibility for 
approximating environments.
The original algorithm upon which our scheme is based, see
Refs.~\cite{Verstraete2004Nov,Verstraete2004Jul},
optimizes the environment tensors individually in a single-site
manner.
While this 
approach can be used in addition to the CBE optimization
described above, on its own, it
does not change the basis of the environment from that
selected 
by the initialization  
and thus cannot converge to the best solution, as
bonds are not optimized. 
In another approximation introduced in Ref.~\cite{Pizorn2011May},
physical indices rather than 
virtual indexes 
are bundled, which turns out
to be less accurate than the
approach of Ref.~\cite{Lubasch2014Mar}.
The least expensive way to approximate the environment is via a
so-called ``simple update'' \cite{Jiang2008Aug}, which depends solely
on the SVD; however, this 
is also insufficiently accurate. 
Lubasch \textit{et al.}\ \cite{Lubasch2014Mar} have proposed a
more elaborate approximation 
for the calculation of norms by considering larger clusters around the
row or column one wants to process.
In our opinion, 
this algorithm would be too expensive
for our case because one would have 
to sandwich a PEPO
between 
the two fPEPS, leading to a much higher scaling
of the computational cost.

\subsection{Variational optimization} \label{sec:varoptpeps}
Having now covered all 
the necessary preliminaries, 
we discuss in the following two different methods for
variationally optimizing an fPEPS to approximate the ground state.
The first method, which we will term \textit{local updates} involves
contracting the fPEPS-PEPO-fPEPS until a local effective Hamiltonian,
either for one site or for two sites (i.e., a bond),  is formed.
We diagonalize this Hamiltonian, which satisfies a generalized
eigenvalue equation, using an iterative diagonalization scheme and then
optimize the fPEPS locally, keeping the size of the basis fixed.
This local-update optimization method is in the spirit of DMRG
optimization of an MPS using the iterative diagonalization of an
effective Hamiltonian followed by singular value decomposition.
In the second method, termed \textit{gradient updates}, we keep the basis 
fixed and optimize all local tensors in the fPEPS simultaneously
using a gradient-based non-local optimization method.

Both optimization methods have shortcomings.
In particular,
local bond updates 
temporarily expand the Hilbert space
and thus form 
the optimal 
virtual basis in a DMRG-like fashion, but are
limited by the property that 
only two adjacent PEPS tensors are
optimized at each step.
Gradient updates are complementary to local updates in that
all 
PEPS tensors are 
varied simultaneously 
to minimize the energy; however, this is done while keeping the
virtual basis fixed.

We start with a rectangular lattice of a given width and height, the 
corresponding representation of the Hubbard Hamiltonian as a PEPO, and
a particular 
sector of quantum numbers within which we want to find the ground state.
Here we assume that we initialize the algorithm
by converting a product state within that sector into an fPEPS.
(Other initializations, such as 
a suitable randomized state, are also possible.)

\subsubsection{Local updates}
\label{sec:localupdate}
\begin{figure}[!htb]
\centering 
\scalebox{0.8}{
\begin{tikzpicture}[font=\tiny]

\draw[line width=0.5mm, rounded corners] (3,2) -- (-0.5,2) -- (-0.5,-2) -- (3,-2);
\draw[line width=0.5mm] (1.5,2) to[out=240,in=90] (1.2,-0.3);
\draw[line width=0.5mm] (1.5,2) to[out=300,in=90] (1.8,0.3);
\draw[line width=0.5mm] (1.2,-0.3) to[out=270,in=120] (1.5,-2);

\draw[line width=0.5mm] (-0.5,0) to[out=315,in=180] (1.2,-0.3);

\draw[line width=0.5mm] (1.2,-0.3) -- (2.7,-0.3);
\draw[fill=white, line width=0.5mm] (1.2,-0.3) circle [radius=0.5];
\node[align=center] at (1.2,-0.3) {\footnotesize $A^{\dagger}_{i,y}$};
\draw[line width=0.5mm] (-0.5,0) -- (1.5,0);

\draw[line width=0.5mm] (1.5,2) -- (1.5,-2);
\draw[line width=0.5mm] (1.5,0) -- (3,0);
\draw[fill=white, line width=0.5mm] (1.5,0) circle [radius=0.5];
\node[align=center] at (1.5,0) {\footnotesize $W_{i,y}$};
\draw[line width=0.5mm] (1.8,0.3) -- (3.3,0.3);
\draw[line width=0.5mm] (1.8,0.3) to[out=270,in=60] (1.5,-2);

\draw[fill=white, line width=0.5mm] (1.5,2) circle [radius=0.5];
\node[align=center] at (1.5,2) {\footnotesize $E_{i,y+1}$};
\draw[fill=white, line width=0.5mm] (1.5,-2) circle [radius=0.5];
\node[align=center] at (1.5,-2) {\footnotesize $E_{i,y-1}$};

\draw[line width=0.5mm,->] (3.5,0) -- (4.5,0);

\draw[line width=0.5mm, rounded corners] (6.4,2) -- (5.4,2) -- (5.4,-2) -- (6.4,-2);
\draw[line width=0.5mm] (5.4,0) to[out=50,in=180] (6.7,0.5);
\draw[line width=0.5mm] (5.4,0) -- (6.4,0);
\draw[line width=0.5mm] (5.4,0) to[out=310,in=180] (6.1,-0.5);
\draw[fill=white, line width=0.5mm] (5.4,0) circle [radius=0.5];
\node[align=center] at (5.4,0) {\footnotesize $E_{i,y}$};

\draw[line width=0.5mm] (-0.5,0) to[out=30,in=180] (1.8,0.3);
\draw[fill=white, line width=0.5mm] (-0.5,0) circle [radius=0.5];
\node[align=center] at (-0.5,0) {\footnotesize $E_{i-1,y}$};
\draw[fill=white, line width=0.5mm] (1.8,0.3) circle [radius=0.5];
\node[align=center] at (1.8,0.3) {\footnotesize $A_{i,y}$};

\end{tikzpicture}
}
\caption{Zipper contraction of blocks between two approximated environments.}
\label{fig:env3}
\end{figure}
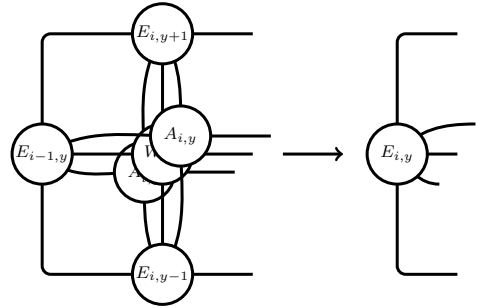

Let us consider first the case of two-site local optimization:
assume that we want to 
optimize two adjacent bulk tensors, say $A_{x,y}$ and $A_{x+1,y}$.
In order to do this, 
we first 
partially contract 
the network of the Hamiltonian,
$\left<\text{fPEPS} \right| \text{PEPO} \left| \text{fPEPS} \right>$,
as depicted in Fig.~\ref{fig:pepstn}(a), 
and also the network of the norm, 
$\left<\text{fPEPS} | \text{fPEPS} \right>$.
We do this by building environments from above and below
using the scheme described in Sec.~\ref{sec:pepsexp}. 
This leaves us 
with four tensor sequences, two for the energy,
$E_{i,y+1}$ and $E_{i,y-1}$, and two for the norm, $N_{i,y+1}$ and $N_{i,y-1}$. 
Subsequently, 
we calculate environment blocks within 
row $y$ 
by starting at both $x=0$ and $x=x_{\text{max}}$ and 
contracting
according to the scheme depicted in 
Fig.~\ref{fig:env3}.
The tensors $A_{i,y}$, $W_{i,y}$, and $A^{\dagger}$ belong to the fPEPS, PEPO, 
and adjoint fPEPS, respectively; note that the latter two are partially
obscured 
in Fig.~\ref{fig:env3}. 
In the left diagram,
$E_{i-1,y}$ 
does not exist 
in the initial step at the edge of 
the lattice but, in subsequent steps, 
contains the outcome of the previous contraction
for $i>0$.
The contractions on the right take place analogously.
One can picture this process as consisting of two zippers
that close 
in on the two sites $\left(x,y\right)$ and $\left(x+1,y\right)$ from
either side.

The configuration at the end of this process 
is depicted in Fig.~\ref{fig:twositeoptpeps}.
The effective Hamiltonian $H_{\text{eff}}$,
Fig.~\ref{fig:twositeoptpeps}(a), 
consists of six environment tensors
$E$ as well as 
the last two tensors of the PEPO that have not been 
contracted, $W_{x,y}$ and $W_{x+1,y}$,
whereas 
the environment for 
the norm $N_{\text{env}}$, Fig.~\ref{fig:twositeoptpeps}(b),
consists of 
six norm tensors $N$ only.
\begin{figure}[!htb]
\tdplotsetmaincoords{65}{0}
\begin{minipage}{0.5\textwidth}
\centering
\scalebox{0.8}{
\begin{tikzpicture}[font=\tiny,tdplot_main_coords]
\tdplotsetrotatedcoords{30}{0}{0}
\begin{scope}[tdplot_rotated_coords]

\draw[line width=0.5mm, rounded corners] (4,2,0) -- (6,2,0) -- (6,0,0);
\draw[line width=0.5mm, rounded corners] (4,2,-0.5) -- (4,2,-1.5) -- (4,1,-1.5);
\shade[ball color = white] (4,2,0) circle (0.5cm);
\node[align=center] at (4,2,0) {\footnotesize $E_{\substack{x+1, \\ y+1}}$};
\draw[line width=0.5mm, rounded corners] (4,2,0.5) -- (4,2,1.5) -- (4,1,1.5);
\draw[line width=0.5mm] (3.5,2,0) -- (2,2,0);
\draw[line width=0.5mm] (4,1.5,0) -- (4,0,0);

\draw[line width=0.5mm, rounded corners] (2,2,-0.5) -- (2,2,-1.5) -- (2,1,-1.5);
\shade[ball color = white] (2,2,0) circle (0.5cm);
\node[align=center] at (2,2,0) {\footnotesize $E_{x,y+1}$};
\draw[line width=0.5mm, rounded corners] (1.5,2,0) -- (0,2,0) -- (0,0,0);
\draw[line width=0.5mm] (2,1.5,0) -- (2,0,0);
\draw[line width=0.5mm, rounded corners] (2,2,0.5) -- (2,2,1.5) -- (2,1,1.5);

\draw[line width=0.5mm, rounded corners] (6,0,-0.5) -- (6,0,-1.5) -- (5,0,-1.5);
\shade[ball color = white] (6,0,0) circle (0.5cm);
\node[align=center] at (6,0,0) {\footnotesize $E_{x+2,y}$};
\draw[line width=0.5mm, rounded corners] (6,0,0.5) -- (6,0,1.5) -- (5,0,1.5);
\draw[line width=0.5mm] (5.5,0,0) -- (4,0,0);

\draw[line width=0.5mm] (4,0,-0.5) -- (4,0,-1);
\shade[ball color = white] (4,0,0) circle (0.5cm);
\node[align=center] at (4,0,0) {\footnotesize $W_{x+1,y}$};
\draw[line width=0.5mm] (4,0,0.5) -- (4,0,1);
\draw[line width=0.5mm] (3.5,0,0) -- (2,0,0);
\draw[line width=0.5mm] (4,-0.5,0) -- (4,-2,0);

\draw[line width=0.5mm] (2,0,-0.5) -- (2,0,-1);
\shade[ball color = white] (2,0,0) circle (0.5cm);
\node[align=center] at (2,0,0) {\footnotesize $W_{x,y}$};
\draw[line width=0.5mm] (1.5,0,0) -- (0,0,0);
\draw[line width=0.5mm] (2,-0.5,0) -- (2,-2,0);
\draw[line width=0.5mm] (2,0,0.5) -- (2,0,1);

\draw[line width=0.5mm, rounded corners] (0,0,-0.5) -- (0,0,-1.5) -- (1,0,-1.5);
\shade[ball color = white] (0,0,0) circle (0.5cm);
\node[align=center] at (0,0,0) {\footnotesize $E_{x-1,y}$};
\draw[line width=0.5mm, rounded corners] (0,0,0.5) -- (0,0,1.5) -- (1,0,1.5);

\draw[line width=0.5mm, rounded corners] (6,-0.5,0) -- (6,-2,0) -- (4,-2,0);
\draw[line width=0.5mm, rounded corners] (4,-2,-0.5) -- (4,-2,-1.5) -- (4,-1,-1.5);
\shade[ball color = white] (4,-2,0) circle (0.5cm);
\node[align=center] at (4,-2,0) {\footnotesize $E_{\substack{x+1, \\ y-1}}$};
\draw[line width=0.5mm, rounded corners] (4,-2,0.5) -- (4,-2,1.5) -- (4,-1,1.5);
\draw[line width=0.5mm] (3.5,-2,0) -- (2,-2,0);
\draw[line width=0.5mm, rounded corners] (2,-2,-0.5) -- (2,-2,-1.5) -- (2,-1,-1.5);
\shade[ball color = white] (2,-2,0) circle (0.5cm);
\node[align=center] at (2,-2,0) {\footnotesize $E_{x,y-1}$};
\draw[line width=0.5mm, rounded corners] (1.5,-2,0) -- (0,-2,0) -- (0,-0.5,0);
\draw[line width=0.5mm, rounded corners] (2,-2,0.5) -- (2,-2,1.5) -- (2,-1,1.5);

\end{scope}
\end{tikzpicture}
}

\vspace*{0.2cm}
\centerline{\small (a) Effective Hamiltonian $H_{\text{eff}}$.}
\vspace{0.1cm}        

\end{minipage}

\vspace{1mm}

\begin{minipage}{0.5\textwidth}
\centering 
\scalebox{0.8}{
\begin{tikzpicture}[font=\tiny,tdplot_main_coords]
\tdplotsetrotatedcoords{30}{0}{0}
\begin{scope}[tdplot_rotated_coords]

\draw[line width=0.5mm, rounded corners] (4,2,0) -- (6,2,0) -- (6,0,0);
\draw[line width=0.5mm, rounded corners] (4,2,-0.5) -- (4,2,-1.5) -- (4,1,-1.5);
\shade[ball color = white] (4,2,0) circle (0.5cm);
\node[align=center] at (4,2,0) {\footnotesize $N_{\substack{x+1, \\ y+1}}$};
\draw[line width=0.5mm, rounded corners] (4,2,0.5) -- (4,2,1.5) -- (4,1,1.5);
\draw[line width=0.5mm] (3.5,2) -- (2,2);

\draw[line width=0.5mm, rounded corners] (2,2,-0.5) -- (2,2,-1.5) -- (2,1,-1.5);
\shade[ball color = white] (2,2,0) circle (0.5cm);
\node[align=center] at (2,2,0) {\footnotesize $N_{x,y+1}$};
\draw[line width=0.5mm, rounded corners] (1.5,2,0) -- (0,2,0) -- (0,0,0);
\draw[line width=0.5mm, rounded corners] (2,2,0.5) -- (2,2,1.5) -- (2,1,1.5);

\draw[line width=0.5mm, rounded corners] (6,0,-0.5) -- (6,0,-1.5) -- (5,0,-1.5);
\shade[ball color = white] (6,0,0) circle (0.5cm);
\node[align=center] at (6,0,0) {\footnotesize $N_{x+2,y}$};
\draw[line width=0.5mm, rounded corners] (6,-0.5,0) -- (6,-2,0) -- (4,-2,0);
\draw[line width=0.5mm, rounded corners] (6,0,0.5) -- (6,0,1.5) -- (5,0,1.5);

\draw[line width=0.5mm, rounded corners] (0,0,-0.5) -- (0,0,-1.5) -- (1,0,-1.5);
\shade[ball color = white] (0,0,0) circle (0.5cm);
\node[align=center] at (0,0,0) {\footnotesize $N_{x-1,y}$};
\draw[line width=0.5mm, rounded corners] (0,0,0.5) -- (0,0,1.5) -- (1,0,1.5);

\draw[line width=0.5mm, rounded corners] (4,-2,-0.5) -- (4,-2,-1.5) -- (4,-1,-1.5);
\shade[ball color = white] (4,-2,0) circle (0.5cm);
\node[align=center] at (4,-2,0) {\footnotesize $N_{\substack{x+1, \\ y-1}}$};
\draw[line width=0.5mm, rounded corners] (4,-2,0.5) -- (4,-2,1.5) -- (4,-1,1.5);
\draw[line width=0.5mm] (3.5,-2) -- (2,-2);

\draw[line width=0.5mm, rounded corners] (2,-2,-0.5) -- (2,-2,-1.5) -- (2,-1,-1.5);
\shade[ball color = white] (2,-2,0) circle (0.5cm);
\node[align=center] at (2,-2,0) {\footnotesize $N_{x,y-1}$};
\draw[line width=0.5mm, rounded corners] (1.5,-2,0) -- (0,-2,0) -- (0,-0.5,0);
\draw[line width=0.5mm, rounded corners] (2,-2,0.5) -- (2,-2,1.5) -- (2,-1,1.5);

\end{scope}

\end{tikzpicture}
}

\vspace*{0.2cm}
\centerline{\small (b) Norm environment $N_{\text{env}}$.}
\vspace{0.1cm}        
\end{minipage}

\caption{Two-site optimization.}
\label{fig:twositeoptpeps}
\end{figure}
Bonds connecting two $E$'s or two $N$'s are cumulative and are created
during the approximation of the environment.
Those connecting both $W$'s or one $W$ and one 
$E$ represent states in an FSM.
The free 
indices protruding from 
$E$'s or $N$'s encode 
quantum-mechanical entanglement within the PEPS, and the dangling links of both 
$W$'s are physical Hilbert spaces.
The clusters $H_{\text{eff}}$ and
$N_{\text{env}}$ satisfy 
the generalized eigenvalue problem
\begin{align}
H_{\text{eff}} \, \left| \psi \right> = E \, N_{\text{env}} \, 
	\left| \psi \right> \, .
\end{align}
Its solution is 
the wave function 
$\left| \psi \right> = A_{x,y} \cdot A_{x+1,y}$ that minimizes the energy $E$.
Note that PEPSs cannot be put into any 
equivalent of the Schmidt form, in which the norm 
of a single tensor is equal to the norm of the overall wave function.
If this were the case, 
we would have been able 
to find a gauge so that 
$\bra{\psi} N_{\text{env}} \ket{\psi} = \bra{\psi} \ket{\psi}$, as in the DMRG.
 
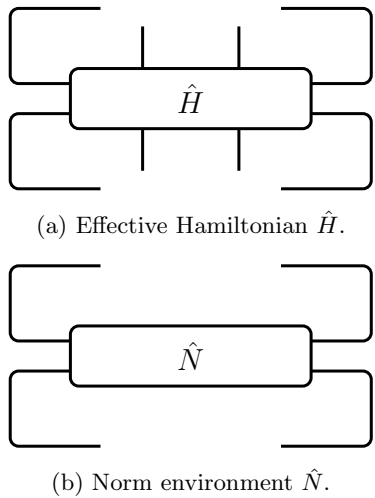
\begin{figure}[!htb]
\begin{minipage}{0.5\textwidth}
\centering 
\scalebox{0.8}{
\begin{tikzpicture}[font=\tiny]
\draw[line width=0.5mm, rounded corners] (0,0.25) -- (-1,0.25) -- (-1,-1)
	-- (0.5,-1);
\draw[line width=0.5mm, rounded corners] (4,0.25) -- (5,0.25) -- (5,-1)
	-- (3.5,-1);
\draw[line width=0.5mm, rounded corners] (0,0.75) -- (-1,0.75) -- (-1,2)
	-- (0.5,2);
\draw[line width=0.5mm, rounded corners] (4,0.75) -- (5,0.75) -- (5,2)
	-- (3.5,2);
\draw[line width=0.5mm, rounded corners] (1.2,1) -- (1.2,1.7);
\draw[line width=0.5mm, rounded corners] (2.8,1) -- (2.8,1.7);
\draw[line width=0.5mm, rounded corners] (1.2,0) -- (1.2,-0.7);
\draw[line width=0.5mm, rounded corners] (2.8,0) -- (2.8,-0.7);

\draw[line width=0.5mm, rounded corners] (1,0) -- (4,0) -- (4,1) -- (0,1)
	-- (0,0) -- (1,0);
\node[align=center] at (2,0.5) {\Large $\hat{H}$};
\end{tikzpicture}
}

\vspace*{0.2cm}
\centerline{\small (a) Effective Hamiltonian $\hat{H}$.}
\vspace{0.1cm}        

\end{minipage}

\vspace{2mm}

\begin{minipage}{0.5\textwidth}
\centering 
\scalebox{0.8}{
\begin{tikzpicture}[font=\tiny]
\draw[line width=0.5mm, rounded corners] (0,0.25) -- (-1,0.25) -- (-1,-1)
	-- (0.5,-1);
\draw[line width=0.5mm, rounded corners] (4,0.25) -- (5,0.25) -- (5,-1)
	-- (3.5,-1);
\draw[line width=0.5mm, rounded corners] (0,0.75) -- (-1,0.75) -- (-1,2)
	-- (0.5,2);
\draw[line width=0.5mm, rounded corners] (4,0.75) -- (5,0.75) -- (5,2)
	-- (3.5,2);

\draw[line width=0.5mm, rounded corners] (1,0) -- (4,0) -- (4,1) -- (0,1)
	-- (0,0) -- (1,0);
\node[align=center] at (2,0.5) {\Large $\hat{N}$};
\end{tikzpicture}
}

\vspace*{0.2cm}
\centerline{\small (b) Norm environment $\hat{N}$.}
\vspace{0.1cm}        
\end{minipage}

\caption{Bond optimization.}
\label{fig:bondoptpeps}
\end{figure}

In principle, one could now proceed with an optimization of 
$\left| \psi \right>$ by carrying out 
iterative diagonalization directly. 
However, it is more efficient to first employ the bond optimization proposed 
by Corboz \cite{Corboz2016Jul}, whose 
initialization is illustrated in Fig. \ref{fig:decomppepsten}.
\begin{figure}[!htb]
\tdplotsetmaincoords{65}{0}
\begin{minipage}{0.5\textwidth}
\centering 
\begin{tikzpicture}[font=\tiny,tdplot_main_coords]
\tdplotsetrotatedcoords{30}{0}{0}
\begin{scope}[tdplot_rotated_coords]

\draw[line width=0.5mm] (4,1.5,0) -- (4,0,0);
\draw[line width=0.5mm] (2,1.5,0) -- (2,0,0);
\draw[line width=0.5mm] (5.5,0,0) -- (4,0,0);

\draw[line width=0.5mm] (4,0,-0.5) -- (4,0,-1);
\shade[ball color = white] (4,0,0) circle (0.5cm);
\node[align=center] at (4,0,0) {\footnotesize $A_{x+1,y}$};
\draw[line width=0.5mm] (3.5,0,0) -- (2,0,0);
\draw[line width=0.5mm] (4,-0.5,0) -- (4,-1.5,0);

\draw[line width=0.5mm] (2,0,-0.5) -- (2,0,-1);
\shade[ball color = white] (2,0,0) circle (0.5cm);
\node[align=center] at (2,0,0) {\footnotesize $A_{x,y}$};
\draw[line width=0.5mm] (1.5,0,0) -- (0.5,0,0);
\draw[line width=0.5mm] (2,-0.5,0) -- (2,-1.5,0);

\end{scope}
\end{tikzpicture}

\vspace*{0.2cm}
\centerline{\small (a) Two adjacent PEPS tensors.}
\vspace{0.1cm}        
\end{minipage}

\vspace{1mm}

\begin{minipage}{0.5\textwidth}
\centering 
\begin{tikzpicture}[font=\tiny,tdplot_main_coords]
\tdplotsetrotatedcoords{30}{0}{0}
\begin{scope}[tdplot_rotated_coords]

\draw[line width=0.5mm] (4.5,1,0) -- (4.5,0,0);
\draw[line width=0.5mm] (5.5,0,0) -- (4.5,0,0);
\shade[ball color = white] (4.5,0,0) circle (0.3cm);
\node[align=center] at (4.5,0,0) {\footnotesize $Y$};
\draw[line width=0.5mm] (4.2,0,0) -- (3.7,0,0);
\draw[line width=0.5mm] (4.5,-1,0) -- (4.5,-0.3,0);

\draw[line width=0.5mm] (3.7,0,0) -- (3.7,0,-1);
\shade[ball color = white] (3.7,0,0) circle (0.3cm);
\node[align=center] at (3.7,0,0) {\footnotesize $q$};
\draw[line width=0.5mm] (3.4,0,0) -- (2.3,0,0);

\draw[line width=0.5mm] (2.3,0,0) -- (2.3,0,-1);
\shade[ball color = white] (2.3,0,0) circle (0.3cm);
\node[align=center] at (2.3,0,0) {\footnotesize $p$};
\draw[line width=0.5mm] (1.5,0,0) -- (2,0,0);

\draw[line width=0.5mm] (1.5,1,0) -- (1.5,0,0);
\shade[ball color = white] (1.5,0,0) circle (0.3cm);
\node[align=center] at (1.5,0,0) {\footnotesize $X$};
\draw[line width=0.5mm] (0.5,0,0) -- (1.2,0,0);
\draw[line width=0.5mm] (1.5,-1,0) -- (1.5,-0.3,0);

\end{scope}
\end{tikzpicture}

\vspace*{0.2cm}
\centerline{\small (b) Decompositions.}
\vspace{0.1cm}        
\end{minipage}

\vspace{1mm}

\begin{minipage}{0.5\textwidth}
\centering 
\begin{tikzpicture}[font=\tiny,tdplot_main_coords]
\tdplotsetrotatedcoords{30}{0}{0}
\begin{scope}[tdplot_rotated_coords]

\draw[line width=0.5mm] (4.5,1,0) -- (4.5,0,0);
\draw[line width=0.5mm] (5.5,0,0) -- (4.5,0,0);
\shade[ball color = white] (4.5,0,0) circle (0.3cm);
\node[align=center] at (4.5,0,0) {\footnotesize $Y$};
\draw[line width=0.5mm] (4.2,0,0) -- (3,0,0);
\draw[line width=0.5mm] (4.5,-1,0) -- (4.5,-0.3,0);

\draw[line width=0.5mm] (3.1,0,0) -- (3.1,0,-1);
\draw[line width=0.5mm] (2.9,0,0) -- (2.9,0,-1);
\shade[ball color = white] (3,0,0) circle (0.3cm);
\node[align=center] at (3,0,0) {\footnotesize $r$};
\draw[line width=0.5mm] (2.7,0,0) -- (1.5,0,0);

\draw[line width=0.5mm] (1.5,1,0) -- (1.5,0,0);
\shade[ball color = white] (1.5,0,0) circle (0.3cm);
\node[align=center] at (1.5,0,0) {\footnotesize $X$};
\draw[line width=0.5mm] (0.5,0,0) -- (1.2,0,0);
\draw[line width=0.5mm] (1.5,-1,0) -- (1.5,-0.3,0);

\end{scope}
\end{tikzpicture}

\vspace*{0.2cm}
\centerline{\small (c) Contraction.}
\vspace{0.1cm}        
\end{minipage}

\caption{Decomposition of PEPS tensors.}
\label{fig:decomppepsten}

\end{figure}
We decompose the PEPS tensors $A_{x,y}$ and $A_{x+1,y}$,
Fig.~\ref{fig:decomppepsten}(a),
into ($X$,$p$) and 
($q$,$Y$), respectively, Fig.~\ref{fig:decomppepsten}(b). 
We then contract the middle tensors, $p$ and $q$,
to form $r$, which carries both of the physical indices,
Fig.~\ref{fig:decomppepsten}(c). 
The tensors $X$ and $Y$ are 
then absorbed into the environment, leading to
the decompositions
$\hat{H} = H_{\text{eff}} \cdot X \cdot X^{\dagger} \cdot Y \cdot Y^{\dagger}$ and
$\hat{N} = N_{\text{env}} \cdot X \cdot X^{\dagger} \cdot Y \cdot Y^{\dagger}$,
which are depicted in Figs.~\ref{fig:bondoptpeps}(a)
and \ref{fig:bondoptpeps}(b), 
respectively. 
After applying the gauge proposed by
Lubasch \textit{et al.}~\cite{Lubasch2014Aug},
the tensor $r$ can be optimized using 
a variation of the Davidson algorithm \cite{davidson} adapted for the
generalized eigenvalue problem.
Subsequently, 
we factorize $r$ back into $p$ and $q$ and then truncate the
bond in between
to a predetermined maximum number of states $D$.
We then carry out a
full update \cite{Corboz2010Apr,Phien2015Jul}, 
which strips the 
wave function of cyclic entanglement \cite{Evenbly2018Aug} and improves
the convergence.

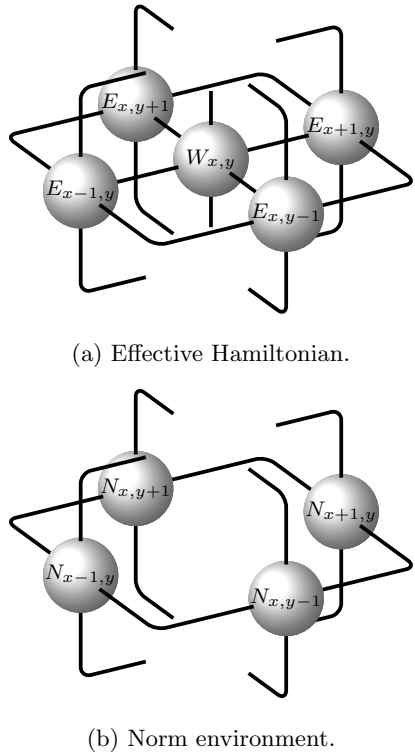
\begin{figure}[!htb]
\tdplotsetmaincoords{65}{0}
\begin{minipage}{0.5\textwidth}
\centering 
\begin{tikzpicture}[font=\tiny,tdplot_main_coords]
\tdplotsetrotatedcoords{30}{0}{0}
\begin{scope}[tdplot_rotated_coords]

\draw[line width=0.5mm, rounded corners] (2,2,0) -- (4,2,0) -- (4,0,0);
\draw[line width=0.5mm, rounded corners] (2,2,-0.5) -- (2,2,-1.5) -- (2,1,-1.5);
\shade[ball color = white] (2,2,0) circle (0.5cm);
\node[align=center] at (2,2,0) {\footnotesize $E_{x,y+1}$};
\draw[line width=0.5mm, rounded corners] (1.5,2,0) -- (0,2,0) -- (0,0,0);
\draw[line width=0.5mm] (2,1.5,0) -- (2,0,0);
\draw[line width=0.5mm, rounded corners] (2,2,0.5) -- (2,2,1.5) -- (2,1,1.5);

\draw[line width=0.5mm, rounded corners] (4,0,-0.5) -- (4,0,-1.5) -- (3,0,-1.5);
\shade[ball color = white] (4,0,0) circle (0.5cm);
\node[align=center] at (4,0,0) {\footnotesize $E_{x+1,y}$};
\draw[line width=0.5mm] (3.5,0,0) -- (2,0,0);
\draw[line width=0.5mm, rounded corners] (4,-0.5,0) -- (4,-2,0) -- (2,-2,0);
\draw[line width=0.5mm, rounded corners] (4,0,0.5) -- (4,0,1.5) -- (3,0,1.5);

\draw[line width=0.5mm] (2,0,-0.5) -- (2,0,-1);
\shade[ball color = white] (2,0,0) circle (0.5cm);
\node[align=center] at (2,0,0) {\footnotesize $W_{x,y}$};
\draw[line width=0.5mm] (1.5,0,0) -- (0,0,0);
\draw[line width=0.5mm] (2,-0.5,0) -- (2,-2,0);
\draw[line width=0.5mm] (2,0,0.5) -- (2,0,1);

\draw[line width=0.5mm, rounded corners] (0,0,-0.5) -- (0,0,-1.5) -- (1,0,-1.5);
\shade[ball color = white] (0,0,0) circle (0.5cm);
\node[align=center] at (0,0,0) {\footnotesize $E_{x-1,y}$};
\draw[line width=0.5mm, rounded corners] (0,0,0.5) -- (0,0,1.5) -- (1,0,1.5);

\draw[line width=0.5mm, rounded corners] (2,-2,-0.5) -- (2,-2,-1.5) -- (2,-1,-1.5);
\shade[ball color = white] (2,-2,0) circle (0.5cm);
\node[align=center] at (2,-2,0) {\footnotesize $E_{x,y-1}$};
\draw[line width=0.5mm, rounded corners] (1.5,-2,0) -- (0,-2,0) -- (0,-0.5,0);
\draw[line width=0.5mm, rounded corners] (2,-2,0.5) -- (2,-2,1.5) -- (2,-1,1.5);

\end{scope}

\end{tikzpicture}

\vspace*{0.2cm}
\centerline{\small (a) Effective Hamiltonian.}
\vspace{0.1cm}        

\end{minipage}

\vspace{1mm}

\begin{minipage}{0.5\textwidth}

\centering 
\begin{tikzpicture}[font=\tiny,tdplot_main_coords]
\tdplotsetrotatedcoords{30}{0}{0}
\begin{scope}[tdplot_rotated_coords]

\draw[line width=0.5mm, rounded corners] (2,2,0) -- (4,2,0) -- (4,0,0);
\draw[line width=0.5mm, rounded corners] (2,2,-0.5) -- (2,2,-1.5) -- (2,1,-1.5);
\shade[ball color = white] (2,2,0) circle (0.5cm);
\node[align=center] at (2,2,0) {\footnotesize $N_{x,y+1}$};
\draw[line width=0.5mm, rounded corners] (1.5,2,0) -- (0,2,0) -- (0,0,0);
\draw[line width=0.5mm, rounded corners] (2,2,0.5) -- (2,2,1.5) -- (2,1,1.5);

\draw[line width=0.5mm, rounded corners] (4,0,-0.5) -- (4,0,-1.5) -- (3,0,-1.5);
\shade[ball color = white] (4,0,0) circle (0.5cm);
\node[align=center] at (4,0,0) {\footnotesize $N_{x+1,y}$};
\draw[line width=0.5mm, rounded corners] (4,-0.5,0) -- (4,-2,0) -- (2,-2,0);
\draw[line width=0.5mm, rounded corners] (4,0,0.5) -- (4,0,1.5) -- (3,0,1.5);

\draw[line width=0.5mm, rounded corners] (0,0,-0.5) -- (0,0,-1.5) -- (1,0,-1.5);
\shade[ball color = white] (0,0,0) circle (0.5cm);
\node[align=center] at (0,0,0) {\footnotesize $N_{x-1,y}$};
\draw[line width=0.5mm, rounded corners] (0,0,0.5) -- (0,0,1.5) -- (1,0,1.5);

\draw[line width=0.5mm, rounded corners] (2,-2,-0.5) -- (2,-2,-1.5) -- (2,-1,-1.5);
\shade[ball color = white] (2,-2,0) circle (0.5cm);
\node[align=center] at (2,-2,0) {\footnotesize $N_{x,y-1}$};
\draw[line width=0.5mm, rounded corners] (1.5,-2,0) -- (0,-2,0) -- (0,-0.5,0);
\draw[line width=0.5mm, rounded corners] (2,-2,0.5) -- (2,-2,1.5) -- (2,-1,1.5);

\end{scope}

\end{tikzpicture}

\vspace*{0.2cm}
\centerline{\small (b) Norm environment.}
\vspace{0.1cm}        

\end{minipage}
\caption{One-site optimization.}
\label{fig:onesiteoptpeps}
\end{figure}

In addition to 
bond 
optimization,
one can also carry out 
one-site optimization, 
in which 
we again start with the configuration of 
Fig.~\ref{fig:twositeoptpeps}, then apply the contraction
of 
Fig.~\ref{fig:env3} to one open site, and, finally, arrive at the setup
depicted 
in Fig.~\ref{fig:onesiteoptpeps} for (a) the effective Hamiltonian and
(b) the norm environment.
Note that 
bond optimizations 
are able to 
explore new subsectors of the Hilbert space,
analogously to 
a two-site DMRG algorithm for MPSs, 
whereas 
one-site optimizations 
can only optimize 
within a given basis.

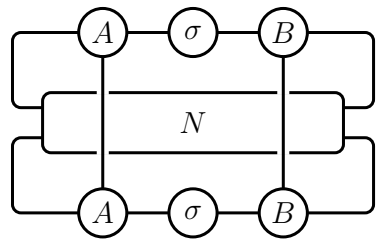
\begin{figure}[!htb]
\centering
\scalebox{0.8}{
\begin{tikzpicture}[font=\tiny]

\draw[line width=0.5mm, rounded corners] (1,0) -- (5,0) -- (5,1) -- (0,1) -- (0,0) -- (1,0);
\node[align=center] at (2.5,0.5) {\Large $N$};

\draw[line width=0.7mm, white] (0.9,1) -- (1.1,1);
\draw[line width=0.7mm, white] (3.9,1) -- (4.1,1);
\draw[line width=0.7mm, white] (0.9,0) -- (1.1,0);
\draw[line width=0.7mm, white] (3.9,0) -- (4.1,0);

\draw[line width=0.5mm, rounded corners] (0,0.75) -- (-0.5,0.75) -- (-0.5,2) -- (5.5,2) -- (5.5,0.75) -- (5,0.75);
\draw[line width=0.5mm, rounded corners] (0,0.25) -- (-0.5,0.25) -- (-0.5,-1) -- (5.5,-1) -- (5.5,0.25) -- (5,0.25);

\draw[line width=0.5mm, rounded corners] (1,2) -- (1,-1);
\draw[line width=0.5mm, rounded corners] (4,2) -- (4,-1);

\draw[fill=white, line width=0.5mm] (1,2) circle [radius=0.4];
\node[align=center] at (1,2) {\Large $A$};
\draw[fill=white, line width=0.5mm] (2.5,2) circle [radius=0.4];
\node[align=center] at (2.5,2) {\Large $\sigma$};
\draw[fill=white, line width=0.5mm] (4,2) circle [radius=0.4];
\node[align=center] at (4,2) {\Large $B$};

\draw[fill=white, line width=0.5mm] (1,-1) circle [radius=0.4];
\node[align=center] at (1,-1) {\Large $A$};
\draw[fill=white, line width=0.5mm] (2.5,-1) circle [radius=0.4];
\node[align=center] at (2.5,-1) {\Large $\sigma$};
\draw[fill=white, line width=0.5mm] (4,-1) circle [radius=0.4];
\node[align=center] at (4,-1) {\Large $B$};

\end{tikzpicture}
}
\caption{Bond setup.}
\label{fig:bondsetup}
\end{figure}

\begin{figure}[!htb]

\begin{minipage}{0.5\textwidth}
\scalebox{0.5}{
\begin{tikzpicture}[font=\tiny]

\draw[line width=0.5mm, rounded corners] (1,0) -- (5,0) -- (5,1) -- (0,1) -- (0,0) -- (1,0);
\node[align=center] at (2.5,0.5) {\Large $N$};

\draw[line width=0.7mm, white] (0.9,1) -- (1.1,1);
\draw[line width=0.7mm, white] (3.9,1) -- (4.1,1);
\draw[line width=0.7mm, white] (0.9,0) -- (1.1,0);
\draw[line width=0.7mm, white] (3.9,0) -- (4.1,0);
\draw[line width=0.7mm, white] (-0.7,0) -- (-0.7,1);

\draw[line width=0.5mm, rounded corners] (0,0.75) -- (-0.5,0.75) -- (-0.5,2) -- (1.6,2);
\draw[line width=0.5mm, rounded corners] (0,0.25) -- (-0.5,0.25) -- (-0.5,-1) -- (1.6,-1);
\draw[line width=0.5mm, rounded corners] (5,0.75) -- (5.5,0.75) -- 
	(5.5,2) -- (1.8,2) -- (1.8,2.7) -- (5.7,2.7) -- (5.7,-1.7) -- 
	(1.8,-1.7) -- (1.8,-1) -- (5.5,-1) -- (5.5,0.25) -- (5,0.25);

\draw[line width=0.5mm, rounded corners] (1,2) -- (1,-1);
\draw[line width=0.5mm, rounded corners] (4,2) -- (4,-1);

\draw[fill=white, line width=0.5mm] (1,2) circle [radius=0.4];
\node[align=center] at (1,2) {\Large $A$};
\draw[fill=white, line width=0.5mm] (2.5,2) circle [radius=0.4];
\node[align=center] at (2.5,2) {\Large $\sigma$};
\draw[fill=white, line width=0.5mm] (4,2) circle [radius=0.4];
\node[align=center] at (4,2) {\Large $B$};

\draw[fill=white, line width=0.5mm] (1,-1) circle [radius=0.4];
\node[align=center] at (1,-1) {\Large $A$};
\draw[fill=white, line width=0.5mm] (2.5,-1) circle [radius=0.4];
\node[align=center] at (2.5,-1) {\Large $\sigma$};
\draw[fill=white, line width=0.5mm] (4,-1) circle [radius=0.4];
\node[align=center] at (4,-1) {\Large $B$};

\node[align=center] at (6.5,0.5) {\Large $=$};

\draw[line width=0.5mm, rounded corners] (8.5,2) -- (8.5,-1);
\draw[line width=0.5mm, rounded corners] (9.5,2) -- (7.5,2) -- (7.5,-1) -- (9.5,-1);
\draw[fill=white, line width=0.5mm] (8.5,2) circle [radius=0.4];
\node[align=center] at (8.5,2) {\Large $A$};
\draw[fill=white, line width=0.5mm] (8.5,-1) circle [radius=0.4];
\node[align=center] at (8.5,-1) {\Large $A$};
\draw[fill=white, line width=0.5mm] (7.5,0.5) circle [radius=0.4];
\node[align=center] at (7.5,0.5) {\Large $N_l$};

\node[align=center] at (11,0.5) {\Large $=\lambda_0 \cdot$};
\draw[line width=0.5mm, rounded corners] (12.5,2) -- (12,2) -- (12,-1) -- (12.5,-1);

\end{tikzpicture}
}

\vspace*{0.2cm}
\centerline{\small (a) Left norm.}
\vspace{0.1cm}        
\end{minipage}

\vspace{1mm}

\begin{minipage}{0.5\textwidth}

\scalebox{0.5}{
\begin{tikzpicture}[font=\tiny]

\draw[line width=0.5mm, rounded corners] (1,0) -- (5,0) -- (5,1) -- (0,1) -- (0,0) -- (1,0);
\node[align=center] at (2.5,0.5) {\Large $N$};

\draw[line width=0.7mm, white] (0.9,1) -- (1.1,1);
\draw[line width=0.7mm, white] (3.9,1) -- (4.1,1);
\draw[line width=0.7mm, white] (0.9,0) -- (1.1,0);
\draw[line width=0.7mm, white] (3.9,0) -- (4.1,0);

\draw[line width=0.5mm, rounded corners] (0,0.75) -- (-0.5,0.75) -- (-0.5,2) -- (3.2,2)
	-- (3.2,2.7) -- (-0.7,2.7) -- (-0.7,-1.7) -- (3.2,-1.7) -- (3.2,-1)
	-- (-0.5,-1) -- (-0.5,0.25) -- (0,0.25);
\draw[line width=0.5mm, rounded corners] (5,0.25) -- (5.5,0.25) -- (5.5,-1) -- (3.4,-1);
\draw[line width=0.5mm, rounded corners] (5,0.75) -- (5.5,0.75) -- (5.5,2) -- (3.4,2);

\draw[line width=0.5mm, rounded corners] (1,2) -- (1,-1);
\draw[line width=0.5mm, rounded corners] (4,2) -- (4,-1);

\draw[fill=white, line width=0.5mm] (1,2) circle [radius=0.4];
\node[align=center] at (1,2) {\Large $A$};
\draw[fill=white, line width=0.5mm] (2.5,2) circle [radius=0.4];
\node[align=center] at (2.5,2) {\Large $\sigma$};
\draw[fill=white, line width=0.5mm] (4,2) circle [radius=0.4];
\node[align=center] at (4,2) {\Large $B$};

\draw[fill=white, line width=0.5mm] (1,-1) circle [radius=0.4];
\node[align=center] at (1,-1) {\Large $A$};
\draw[fill=white, line width=0.5mm] (2.5,-1) circle [radius=0.4];
\node[align=center] at (2.5,-1) {\Large $\sigma$};
\draw[fill=white, line width=0.5mm] (4,-1) circle [radius=0.4];
\node[align=center] at (4,-1) {\Large $B$};

\node[align=center] at (6.5,0.5) {\Large $=$};

\draw[line width=0.5mm, rounded corners] (8.5,2) -- (8.5,-1);
\draw[line width=0.5mm, rounded corners] (7.5,2) -- (9.5,2) -- (9.5,-1) -- (7.5,-1);
\draw[fill=white, line width=0.5mm] (8.5,2) circle [radius=0.4];
\node[align=center] at (8.5,2) {\Large $B$};
\draw[fill=white, line width=0.5mm] (8.5,-1) circle [radius=0.4];
\node[align=center] at (8.5,-1) {\Large $B$};
\draw[fill=white, line width=0.5mm] (9.5,0.5) circle [radius=0.4];
\node[align=center] at (9.5,0.5) {\Large $N_r$};

\node[align=center] at (11,0.5) {\Large $=\lambda_0\cdot$};
\draw[line width=0.5mm, rounded corners] (12,2) -- (12.5,2) -- (12.5,-1) -- (12,-1);

\end{tikzpicture}
}

\vspace*{0.2cm}
\centerline{\small (b) Right norm.}
\vspace{0.1cm}        
\end{minipage}

\caption{Weighted trace gauge.}
\label{fig:wtg}
\end{figure}
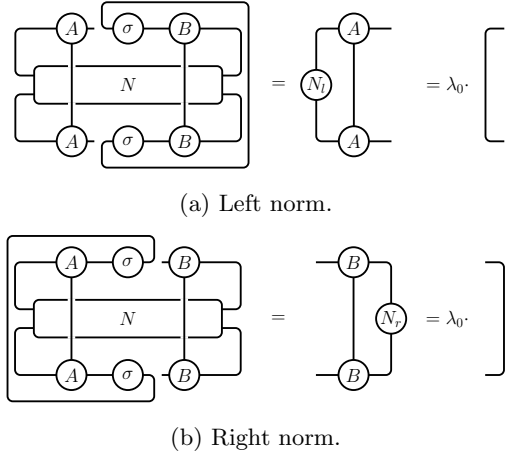

\begin{figure}[!htb]

\begin{minipage}{0.5\textwidth}
\scalebox{0.6}{
\begin{tikzpicture}[font=\tiny]

\node[align=center] at (-1.7,0.75) {\Large $\lambda^{-1}_0\cdot$};
\draw[line width=0.5mm, rounded corners] (-1,0) -- (1,0) -- (1,1.5) -- (-1,1.5);
\draw[line width=0.5mm, rounded corners] (0,0) -- (0,-1);
\draw[line width=0.5mm, rounded corners] (0,1.5) -- (0,2.5);
\draw[fill=white, line width=0.5mm] (0,0) circle [radius=0.4];
\node[align=center] at (0,0) {\Large $A$};
\draw[fill=white, line width=0.5mm] (0,1.5) circle [radius=0.4];
\node[align=center] at (0,1.5) {\Large $A$};

\node[align=center] at (2,0.75) {\Large $+$};

\draw[line width=0.5mm, rounded corners] (3,0) -- (5,0) -- (5,1.5) -- (3,1.5);
\draw[line width=0.5mm, rounded corners] (4,0) -- (4,-1);
\draw[line width=0.5mm, rounded corners] (4,1.5) -- (4,2.5);
\draw[fill=white, line width=0.5mm] (4,0) circle [radius=0.4];
\node[align=center] at (4,0) {\Large $A'$};
\draw[fill=white, line width=0.5mm] (4,1.5) circle [radius=0.4];
\node[align=center] at (4,1.5) {\Large $A'$};

\node[align=center] at (6,0.75) {\Large $=$};

\draw[line width=0.5mm, rounded corners] (6.5,0) -- (7.5,0) -- (7.5,1.5) -- (6.5,1.5);
\draw[fill=white, line width=0.5mm] (7.5,0.75) circle [radius=0.65];
\node[align=center] at (7.5,0.75) {\Large $N^{-1}_l$};
\node[align=center] at (8.6,0.75) {\Large $\otimes$};
\draw[line width=0.5mm, rounded corners] (9,-1) -- (9,2.5);

\end{tikzpicture}
}

\vspace*{0.2cm}
\centerline{\small (a) Left completeness.}
\vspace{0.1cm}        
\end{minipage}

\vspace{1mm}

\begin{minipage}{0.5\textwidth}

\scalebox{0.6}{
\begin{tikzpicture}[font=\tiny]

\node[align=center] at (-1.7,0.75) {\Large $\lambda^{-1}_0\cdot$};
\draw[line width=0.5mm, rounded corners] (1,0) -- (-1,0) -- (-1,1.5) -- (1,1.5);
\draw[line width=0.5mm, rounded corners] (0,0) -- (0,-1);
\draw[line width=0.5mm, rounded corners] (0,1.5) -- (0,2.5);
\draw[fill=white, line width=0.5mm] (0,0) circle [radius=0.4];
\node[align=center] at (0,0) {\Large $B$};
\draw[fill=white, line width=0.5mm] (0,1.5) circle [radius=0.4];
\node[align=center] at (0,1.5) {\Large $B$};

\node[align=center] at (2,0.75) {\Large $+$};

\draw[line width=0.5mm, rounded corners] (5,0) -- (3,0) -- (3,1.5) -- (5,1.5);
\draw[line width=0.5mm, rounded corners] (4,0) -- (4,-1);
\draw[line width=0.5mm, rounded corners] (4,1.5) -- (4,2.5);
\draw[fill=white, line width=0.5mm] (4,0) circle [radius=0.4];
\node[align=center] at (4,0) {\Large $B'$};
\draw[fill=white, line width=0.5mm] (4,1.5) circle [radius=0.4];
\node[align=center] at (4,1.5) {\Large $B'$};

\node[align=center] at (6,0.75) {\Large $=$};

\draw[line width=0.5mm, rounded corners] (6.5,-1) -- (6.5,2.5);
\node[align=center] at (6.9,0.75) {\Large $\otimes$};
\draw[line width=0.5mm, rounded corners] (9,0) -- (8,0) -- (8,1.5) -- (9,1.5);
\draw[fill=white, line width=0.5mm] (8,0.75) circle [radius=0.65];
\node[align=center] at (8,0.75) {\Large $N^{-1}_r$};

\end{tikzpicture}
}

\vspace*{0.2cm}
\centerline{\small (b) Right completeness.}
\vspace{0.1cm}        
\end{minipage}

\caption{Completeness relations.}
\label{fig:complete}
\end{figure}
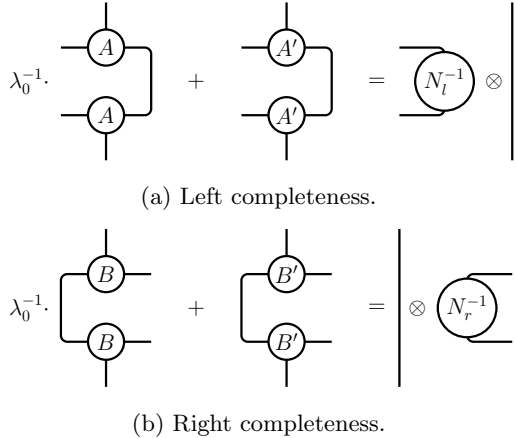

Bond optimization is beset by 
the same problem as the environment 
approximation described in Sec.~\ref{sec:pepsexp}:
If the maximum
bond dimension is $D$, the effective Hamiltonian of 
Fig.~\ref{fig:bondoptpeps}(a) 
will
generate 
$\tilde{D} \gg D$ states, most of which will be 
dropped in the subsequent truncation.
To avoid this, 
we again apply the CBE to circumvent contracting
large tensors so that we can 
perform the bond optimization at one-site cost.
Here we will not reprise 
the details of the CBE, but will instead
only describe 
the adaptations necessary to apply it to 
PEPS tensors.
First, we have to reexamine 
the bond between two tensors, as shown in
Fig.~\ref{fig:bondsetup}.
The product $r = A \cdot \sigma \cdot B$ is what we would normally optimize as 
a whole within the bond optimization.
Note that, for MPSs, 
the norm environment $N$ factorizes
into two terms. 
We now choose the bond matrix $\sigma$ so 
that it satisfies the weighted trace gauge 
\cite{Evenbly2018Aug}, which 
can be seen as the generalization of left- and right-normalization for
cyclic tensor networks.
The resulting orthogonality conditions 
are depicted in Fig. \ref{fig:wtg} for (a) the left norm and (b) the
right norm.
Subsequently, 
the completeness relations, depicted in Fig.~\ref{fig:complete},
can be used to flesh out the orthogonal projectors
$A' \cdot \left(A'\right)^{\dagger}$ and  
$B' \cdot \left(B'\right)^{\dagger}$.
Aside from 
these two customizations, the CBE for PEPS tensors is structurally identical 
to the original version developed for MPSs \cite{Gleis2022Jul} and the adaptation
for environment tensors described in 
Sec.~\ref{sec:pepsexp}. 

\newcommand{\widthfacb}{0.08}
\newcommand{\scalefacb}{0.55}

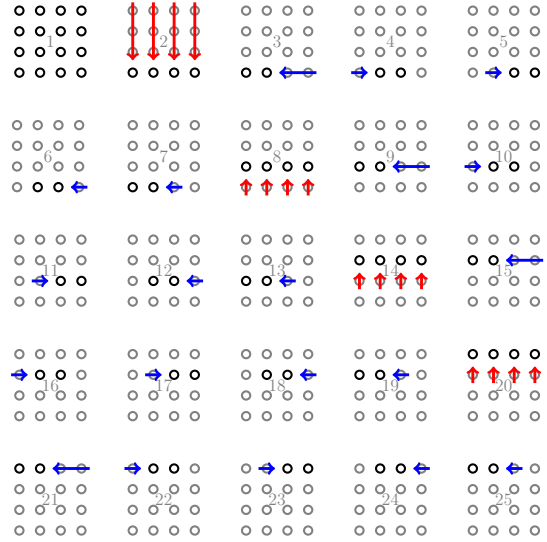
\begin{figure}[!htb]

% FIRST ROW
\begin{minipage}{\widthfacb\textwidth}
\centering
\scalebox{\scalefacb}{
\begin{tikzpicture}[font=\tiny]
\node[align=center,color=black!50] at (0.75,0.75) {\large $1$};
\draw[line width=0.7mm, white] (-0.2,-0.2) -- (-0.1,-0.1);
\draw[line width=0.7mm, white] (1.7,1.7) -- (1.6,1.6);
\foreach \x in {0,0.5,1,1.5}{
	\foreach \y in {0,0.5,1,1.5}{
		\draw[black!100,fill=white, line width=0.5mm] (\x,\y) circle [radius=0.1];
	}
}
\end{tikzpicture}
}
\end{minipage}
\hfill 
\begin{minipage}{\widthfacb\textwidth}
\centering
\scalebox{\scalefacb}{
\begin{tikzpicture}[font=\tiny]
\node[align=center,color=black!50] at (0.75,0.75) {\large $2$};
\draw[line width=0.7mm, white] (-0.2,-0.2) -- (-0.1,-0.1);
\draw[line width=0.7mm, white] (1.7,1.7) -- (1.6,1.6);
\foreach \x in {0,0.5,1,1.5}{
	\foreach \y in {0.5,1,1.5}{
		\draw[black!50,fill=white, line width=0.5mm] (\x,\y) circle [radius=0.1];
	}
}
\foreach \x in {0,0.5,1,1.5}{
	\draw[black!100,fill=white, line width=0.5mm] (\x,0) circle [radius=0.1];
}
\draw[line width=0.7mm, red,->] (0,1.7) -- (0,0.3);
\draw[line width=0.7mm, red,->] (0.5,1.7) -- (0.5,0.3);
\draw[line width=0.7mm, red,->] (1,1.7) -- (1,0.3);
\draw[line width=0.7mm, red,->] (1.5,1.7) -- (1.5,0.3);
\end{tikzpicture}
}
\end{minipage}
\hfill 
\begin{minipage}{\widthfacb\textwidth}
\centering
\scalebox{\scalefacb}{
\begin{tikzpicture}[font=\tiny]
\node[align=center,color=black!50] at (0.75,0.75) {\large $3$};
\draw[line width=0.7mm, white] (-0.2,-0.2) -- (-0.1,-0.1);
\draw[line width=0.7mm, white] (1.7,1.7) -- (1.6,1.6);
\foreach \x in {0,0.5,1,1.5}{
	\foreach \y in {0.5,1,1.5}{
		\draw[black!50,fill=white, line width=0.5mm] (\x,\y) circle [radius=0.1];
	}
}
\draw[black!100,fill=white, line width=0.5mm] (0,0) circle [radius=0.1];
\draw[black!100,fill=white, line width=0.5mm] (0.5,0) circle [radius=0.1];
\draw[black!50,fill=white, line width=0.5mm] (1,0) circle [radius=0.1];
\draw[black!50,fill=white, line width=0.5mm] (1.5,0) circle [radius=0.1];
\draw[line width=0.7mm, blue,->] (1.7,0) -- (0.8,0);
\end{tikzpicture}
}
\end{minipage}
\hfill 
\begin{minipage}{\widthfacb\textwidth}
\centering
\scalebox{\scalefacb}{
\begin{tikzpicture}[font=\tiny]
\node[align=center,color=black!50] at (0.75,0.75) {\large $4$};
\draw[line width=0.7mm, white] (-0.2,-0.2) -- (-0.1,-0.1);
\draw[line width=0.7mm, white] (1.7,1.7) -- (1.6,1.6);
\foreach \x in {0,0.5,1,1.5}{
	\foreach \y in {0.5,1,1.5}{
		\draw[black!50,fill=white, line width=0.5mm] (\x,\y) circle [radius=0.1];
	}
}
\draw[black!50,fill=white, line width=0.5mm] (0,0) circle [radius=0.1];
\draw[black!100,fill=white, line width=0.5mm] (0.5,0) circle [radius=0.1];
\draw[black!100,fill=white, line width=0.5mm] (1,0) circle [radius=0.1];
\draw[black!50,fill=white, line width=0.5mm] (1.5,0) circle [radius=0.1];
\draw[line width=0.7mm, blue,->] (-0.2,0) -- (0.2,0);
\end{tikzpicture}
}
\end{minipage}
\hfill 
\begin{minipage}{\widthfacb\textwidth}
\centering
\scalebox{\scalefacb}{
\begin{tikzpicture}[font=\tiny]
\node[align=center,color=black!50] at (0.75,0.75) {\large $5$};
\draw[line width=0.7mm, white] (-0.2,-0.2) -- (-0.1,-0.1);
\draw[line width=0.7mm, white] (1.7,1.7) -- (1.6,1.6);
\foreach \x in {0,0.5,1,1.5}{
	\foreach \y in {0.5,1,1.5}{
		\draw[black!50,fill=white, line width=0.5mm] (\x,\y) circle [radius=0.1];
	}
}
\draw[black!50,fill=white, line width=0.5mm] (0,0) circle [radius=0.1];
\draw[black!50,fill=white, line width=0.5mm] (0.5,0) circle [radius=0.1];
\draw[black!100,fill=white, line width=0.5mm] (1,0) circle [radius=0.1];
\draw[black!100,fill=white, line width=0.5mm] (1.5,0) circle [radius=0.1];
\draw[line width=0.7mm, blue,->] (0.3,0) -- (0.7,0);
\end{tikzpicture}
}
\end{minipage}
\\
\mbox{} \\
\mbox{} \\
%%%%%%%%%%%%%%%%%%%%%%%%%%%%%%%%%%%%%%%%%%%%%%%%%%%%%%%%%%%%%%%%
\begin{minipage}{\widthfacb\textwidth}
\centering
\scalebox{\scalefacb}{
\begin{tikzpicture}[font=\tiny]
\node[align=center,color=black!50] at (0.75,0.75) {\large $6$};
\foreach \x in {0,0.5,1,1.5}{
	\foreach \y in {0.5,1,1.5}{
		\draw[black!50,fill=white, line width=0.5mm] (\x,\y) circle [radius=0.1];
	}
}
\draw[black!50,fill=white, line width=0.5mm] (0,0) circle [radius=0.1];
\draw[black!100,fill=white, line width=0.5mm] (0.5,0) circle [radius=0.1];
\draw[black!100,fill=white, line width=0.5mm] (1,0) circle [radius=0.1];
\draw[black!50,fill=white, line width=0.5mm] (1.5,0) circle [radius=0.1];
\draw[line width=0.7mm, blue,->] (1.7,0) -- (1.3,0);
\end{tikzpicture}
}
\end{minipage}
\hfill
\begin{minipage}{\widthfacb\textwidth}
\centering
\scalebox{\scalefacb}{
\begin{tikzpicture}[font=\tiny]
\node[align=center,color=black!50] at (0.75,0.75) {\large $7$};
\draw[line width=0.7mm, white] (-0.2,-0.2) -- (-0.1,-0.1);
\draw[line width=0.7mm, white] (1.7,1.7) -- (1.6,1.6);
\foreach \x in {0,0.5,1,1.5}{
	\foreach \y in {0.5,1,1.5}{
		\draw[black!50,fill=white, line width=0.5mm] (\x,\y) circle [radius=0.1];
	}
}
\draw[black!100,fill=white, line width=0.5mm] (0,0) circle [radius=0.1];
\draw[black!100,fill=white, line width=0.5mm] (0.5,0) circle [radius=0.1];
\draw[black!50,fill=white, line width=0.5mm] (1,0) circle [radius=0.1];
\draw[black!50,fill=white, line width=0.5mm] (1.5,0) circle [radius=0.1];
\draw[line width=0.7mm, blue,->] (1.2,0) -- (0.8,0);
\end{tikzpicture}
}
\end{minipage}
\hfill
\begin{minipage}{\widthfacb\textwidth}
\centering
\scalebox{\scalefacb}{
\begin{tikzpicture}[font=\tiny]
\node[align=center,color=black!50] at (0.75,0.75) {\large $8$};
\draw[line width=0.7mm, white] (-0.2,-0.2) -- (-0.1,-0.1);
\draw[line width=0.7mm, white] (1.7,1.7) -- (1.6,1.6);
\foreach \x in {0,0.5,1,1.5}{
	\foreach \y in {1,1.5}{
		\draw[black!50,fill=white, line width=0.5mm] (\x,\y) circle [radius=0.1];
	}
}
\draw[black!100,fill=white, line width=0.5mm] (0,0.5) circle [radius=0.1];
\draw[black!100,fill=white, line width=0.5mm] (0.5,0.5) circle [radius=0.1];
\draw[black!100,fill=white, line width=0.5mm] (1,0.5) circle [radius=0.1];
\draw[black!100,fill=white, line width=0.5mm] (1.5,0.5) circle [radius=0.1];
\foreach \x in {0,0.5,1,1.5}{
	\draw[black!50,fill=white, line width=0.5mm] (\x,0) circle [radius=0.1];
}
\draw[line width=0.7mm, red,->] (0,-0.2) -- (0,0.2);
\draw[line width=0.7mm, red,->] (0.5,-0.2) -- (0.5,0.2);
\draw[line width=0.7mm, red,->] (1,-0.2) -- (1,0.2);
\draw[line width=0.7mm, red,->] (1.5,-0.2) -- (1.5,0.2);
\end{tikzpicture}
}
\end{minipage}
\hfill
\begin{minipage}{\widthfacb\textwidth}
\centering
\scalebox{\scalefacb}{
\begin{tikzpicture}[font=\tiny]
\node[align=center,color=black!50] at (0.75,0.75) {\large $9$};
\draw[line width=0.7mm, white] (-0.2,-0.2) -- (-0.1,-0.1);
\draw[line width=0.7mm, white] (1.7,1.7) -- (1.6,1.6);
\foreach \x in {0,0.5,1,1.5}{
	\foreach \y in {1,1.5}{
		\draw[black!50,fill=white, line width=0.5mm] (\x,\y) circle [radius=0.1];
	}
}
\draw[black!100,fill=white, line width=0.5mm] (0,0.5) circle [radius=0.1];
\draw[black!100,fill=white, line width=0.5mm] (0.5,0.5) circle [radius=0.1];
\draw[black!50,fill=white, line width=0.5mm] (1,0.5) circle [radius=0.1];
\draw[black!50,fill=white, line width=0.5mm] (1.5,0.5) circle [radius=0.1];
\foreach \x in {0,0.5,1,1.5}{
	\draw[black!50,fill=white, line width=0.5mm] (\x,0) circle [radius=0.1];
}
\draw[line width=0.7mm, blue,->] (1.7,0.5) -- (0.8,0.5);
\end{tikzpicture}
}
\end{minipage}
\hfill 
\begin{minipage}{\widthfacb\textwidth}
\centering
\scalebox{\scalefacb}{
\begin{tikzpicture}[font=\tiny]
\node[align=center,color=black!50] at (0.75,0.75) {\large $10$};
\draw[line width=0.7mm, white] (-0.2,-0.2) -- (-0.1,-0.1);
\draw[line width=0.7mm, white] (1.7,1.7) -- (1.6,1.6);
\foreach \x in {0,0.5,1,1.5}{
	\foreach \y in {1,1.5}{
		\draw[black!50,fill=white, line width=0.5mm] (\x,\y) circle [radius=0.1];
	}
}
\draw[black!50,fill=white, line width=0.5mm] (0,0.5) circle [radius=0.1];
\draw[black!100,fill=white, line width=0.5mm] (0.5,0.5) circle [radius=0.1];
\draw[black!100,fill=white, line width=0.5mm] (1,0.5) circle [radius=0.1];
\draw[black!50,fill=white, line width=0.5mm] (1.5,0.5) circle [radius=0.1];
\foreach \x in {0,0.5,1,1.5}{
	\draw[black!50,fill=white, line width=0.5mm] (\x,0) circle [radius=0.1];
}
\draw[line width=0.7mm, blue,->] (-0.2,0.5) -- (0.2,0.5);
\end{tikzpicture}
}
\end{minipage}
\\
\mbox{} \\
\mbox{} \\
%%%%%%%%%%%%%%%%%%%%%%%%%%%%%%%%%%%%%%%%%%%%%%%%%%%%%%%%%%%%%%%%
\begin{minipage}{\widthfacb\textwidth}
\centering
\scalebox{\scalefacb}{
\begin{tikzpicture}[font=\tiny]
\node[align=center,color=black!50] at (0.75,0.75) {\large $11$};
\draw[line width=0.7mm, white] (-0.2,-0.2) -- (-0.1,-0.1);
\draw[line width=0.7mm, white] (1.7,1.7) -- (1.6,1.6);
\foreach \x in {0,0.5,1,1.5}{
	\foreach \y in {1,1.5}{
		\draw[black!50,fill=white, line width=0.5mm] (\x,\y) circle [radius=0.1];
	}
}
\draw[black!50,fill=white, line width=0.5mm] (0,0.5) circle [radius=0.1];
\draw[black!50,fill=white, line width=0.5mm] (0.5,0.5) circle [radius=0.1];
\draw[black!100,fill=white, line width=0.5mm] (1,0.5) circle [radius=0.1];
\draw[black!100,fill=white, line width=0.5mm] (1.5,0.5) circle [radius=0.1];
\foreach \x in {0,0.5,1,1.5}{
	\draw[black!50,fill=white, line width=0.5mm] (\x,0) circle [radius=0.1];
}
\draw[line width=0.7mm, blue,->] (0.3,0.5) -- (0.7,0.5);

\end{tikzpicture}
}
\end{minipage}
\hfill 
\begin{minipage}{\widthfacb\textwidth}
\centering
\scalebox{\scalefacb}{
\begin{tikzpicture}[font=\tiny]
\node[align=center,color=black!50] at (0.75,0.75) {\large $12$};
\draw[line width=0.7mm, white] (-0.2,-0.2) -- (-0.1,-0.1);
\draw[line width=0.7mm, white] (1.7,1.7) -- (1.6,1.6);
\foreach \x in {0,0.5,1,1.5}{
	\foreach \y in {1,1.5}{
		\draw[black!50,fill=white, line width=0.5mm] (\x,\y) circle [radius=0.1];
	}
}
\draw[black!50,fill=white, line width=0.5mm] (0,0.5) circle [radius=0.1];
\draw[black!100,fill=white, line width=0.5mm] (0.5,0.5) circle [radius=0.1];
\draw[black!100,fill=white, line width=0.5mm] (1,0.5) circle [radius=0.1];
\draw[black!50,fill=white, line width=0.5mm] (1.5,0.5) circle [radius=0.1];
\foreach \x in {0,0.5,1,1.5}{
	\draw[black!50,fill=white, line width=0.5mm] (\x,0) circle [radius=0.1];
}
\draw[line width=0.7mm, blue,->] (1.7,0.5) -- (1.3,0.5);
\end{tikzpicture}
}
\end{minipage}
\hfill 
\begin{minipage}{\widthfacb\textwidth}
\centering
\scalebox{\scalefacb}{
\begin{tikzpicture}[font=\tiny]
\node[align=center,color=black!50] at (0.75,0.75) {\large $13$};
\draw[line width=0.7mm, white] (-0.2,-0.2) -- (-0.1,-0.1);
\draw[line width=0.7mm, white] (1.7,1.7) -- (1.6,1.6);
\foreach \x in {0,0.5,1,1.5}{
	\foreach \y in {1,1.5}{
		\draw[black!50,fill=white, line width=0.5mm] (\x,\y) circle [radius=0.1];
	}
}
\draw[black!100,fill=white, line width=0.5mm] (0,0.5) circle [radius=0.1];
\draw[black!100,fill=white, line width=0.5mm] (0.5,0.5) circle [radius=0.1];
\draw[black!50,fill=white, line width=0.5mm] (1,0.5) circle [radius=0.1];
\draw[black!50,fill=white, line width=0.5mm] (1.5,0.5) circle [radius=0.1];
\foreach \x in {0,0.5,1,1.5}{
	\draw[black!50,fill=white, line width=0.5mm] (\x,0) circle [radius=0.1];
}
\draw[line width=0.7mm, blue,->] (1.2,0.5) -- (0.8,0.5);
\end{tikzpicture}
}
\end{minipage}
\hfill 
\begin{minipage}{\widthfacb\textwidth}
\centering
\scalebox{\scalefacb}{
\begin{tikzpicture}[font=\tiny]
\node[align=center,color=black!50] at (0.75,0.75) {\large $14$};
\draw[line width=0.7mm, white] (-0.2,-0.2) -- (-0.1,-0.1);
\draw[line width=0.7mm, white] (1.7,1.7) -- (1.6,1.6);
\foreach \x in {0,0.5,1,1.5}{
	\draw[black!50,fill=white, line width=0.5mm] (\x,1.5) circle [radius=0.1];
}
\draw[black!100,fill=white, line width=0.5mm] (0,1) circle [radius=0.1];
\draw[black!100,fill=white, line width=0.5mm] (0.5,1) circle [radius=0.1];
\draw[black!100,fill=white, line width=0.5mm] (1,1) circle [radius=0.1];
\draw[black!100,fill=white, line width=0.5mm] (1.5,1) circle [radius=0.1];
\foreach \x in {0,0.5,1,1.5}{
	\foreach \y in {0,0.5}{
		\draw[black!50,fill=white, line width=0.5mm] (\x,\y) circle [radius=0.1];
	}
}
\draw[line width=0.7mm, red,->] (0,0.3) -- (0,0.7);
\draw[line width=0.7mm, red,->] (0.5,0.3) -- (0.5,0.7);
\draw[line width=0.7mm, red,->] (1,0.3) -- (1,0.7);
\draw[line width=0.7mm, red,->] (1.5,0.3) -- (1.5,0.7);
\end{tikzpicture}
}
\end{minipage}
\hfill 
\begin{minipage}{\widthfacb\textwidth}
\centering
\scalebox{\scalefacb}{
\begin{tikzpicture}[font=\tiny]
\node[align=center,color=black!50] at (0.75,0.75) {\large $15$};
\draw[line width=0.7mm, white] (-0.2,-0.2) -- (-0.1,-0.1);
\draw[line width=0.7mm, white] (1.7,1.7) -- (1.6,1.6);
\foreach \x in {0,0.5,1,1.5}{
	\draw[black!50,fill=white, line width=0.5mm] (\x,1.5) circle [radius=0.1];
}
\draw[black!100,fill=white, line width=0.5mm] (0,1) circle [radius=0.1];
\draw[black!100,fill=white, line width=0.5mm] (0.5,1) circle [radius=0.1];
\draw[black!50,fill=white, line width=0.5mm] (1,1) circle [radius=0.1];
\draw[black!50,fill=white, line width=0.5mm] (1.5,1) circle [radius=0.1];
\foreach \x in {0,0.5,1,1.5}{
	\foreach \y in {0,0.5}{
		\draw[black!50,fill=white, line width=0.5mm] (\x,\y) circle [radius=0.1];
	}
}
\draw[line width=0.7mm, blue,->] (1.7,1) -- (0.8,1);
\end{tikzpicture}
}
\end{minipage}
\\ 
\mbox{} \\ 
\mbox{} \\ 
\begin{minipage}{\widthfacb\textwidth}
\centering
\scalebox{\scalefacb}{
\begin{tikzpicture}[font=\tiny]
\node[align=center,color=black!50] at (0.75,0.75) {\large $16$};
\draw[line width=0.7mm, white] (-0.2,-0.2) -- (-0.1,-0.1);
\draw[line width=0.7mm, white] (1.7,1.7) -- (1.6,1.6);
\foreach \x in {0,0.5,1,1.5}{
	\draw[black!50,fill=white, line width=0.5mm] (\x,1.5) circle [radius=0.1];
}
\draw[black!50,fill=white, line width=0.5mm] (0,1) circle [radius=0.1];
\draw[black!100,fill=white, line width=0.5mm] (0.5,1) circle [radius=0.1];
\draw[black!100,fill=white, line width=0.5mm] (1,1) circle [radius=0.1];
\draw[black!50,fill=white, line width=0.5mm] (1.5,1) circle [radius=0.1];
\foreach \x in {0,0.5,1,1.5}{
	\foreach \y in {0,0.5}{
		\draw[black!50,fill=white, line width=0.5mm] (\x,\y) circle [radius=0.1];
	}
}
\draw[line width=0.7mm, blue,->] (-0.2,1) -- (0.2,1);
\end{tikzpicture}
}
\end{minipage}
\hfill
\begin{minipage}{\widthfacb\textwidth}
\centering
\scalebox{\scalefacb}{
\begin{tikzpicture}[font=\tiny]
\node[align=center,color=black!50] at (0.75,0.75) {\large $17$};
\draw[line width=0.7mm, white] (-0.2,-0.2) -- (-0.1,-0.1);
\draw[line width=0.7mm, white] (1.7,1.7) -- (1.6,1.6);
\foreach \x in {0,0.5,1,1.5}{
	\draw[black!50,fill=white, line width=0.5mm] (\x,1.5) circle [radius=0.1];
}
\draw[black!50,fill=white, line width=0.5mm] (0,1) circle [radius=0.1];
\draw[black!50,fill=white, line width=0.5mm] (0.5,1) circle [radius=0.1];
\draw[black!100,fill=white, line width=0.5mm] (1,1) circle [radius=0.1];
\draw[black!100,fill=white, line width=0.5mm] (1.5,1) circle [radius=0.1];
\foreach \x in {0,0.5,1,1.5}{
	\foreach \y in {0,0.5}{
		\draw[black!50,fill=white, line width=0.5mm] (\x,\y) circle [radius=0.1];
	}
}
\draw[line width=0.7mm, blue,->] (0.3,1) -- (0.7,1);
\end{tikzpicture}
}
\end{minipage}
\hfill
\begin{minipage}{\widthfacb\textwidth}
\centering
\scalebox{\scalefacb}{
\begin{tikzpicture}[font=\tiny]
\node[align=center,color=black!50] at (0.75,0.75) {\large $18$};
\draw[line width=0.7mm, white] (-0.2,-0.2) -- (-0.1,-0.1);
\draw[line width=0.7mm, white] (1.7,1.7) -- (1.6,1.6);
\foreach \x in {0,0.5,1,1.5}{
	\draw[black!50,fill=white, line width=0.5mm] (\x,1.5) circle [radius=0.1];
}
\draw[black!50,fill=white, line width=0.5mm] (0,1) circle [radius=0.1];
\draw[black!100,fill=white, line width=0.5mm] (0.5,1) circle [radius=0.1];
\draw[black!100,fill=white, line width=0.5mm] (1,1) circle [radius=0.1];
\draw[black!50,fill=white, line width=0.5mm] (1.5,1) circle [radius=0.1];
\foreach \x in {0,0.5,1,1.5}{
	\foreach \y in {0,0.5}{
		\draw[black!50,fill=white, line width=0.5mm] (\x,\y) circle [radius=0.1];
	}
}
\draw[line width=0.7mm, blue,->] (1.7,1) -- (1.3,1);
\end{tikzpicture}
}
\end{minipage}
\hfill 
\begin{minipage}{\widthfacb\textwidth}
\centering
\scalebox{\scalefacb}{
\begin{tikzpicture}[font=\tiny]
\node[align=center,color=black!50] at (0.75,0.75) {\large $19$};
\draw[line width=0.7mm, white] (-0.2,-0.2) -- (-0.1,-0.1);
\draw[line width=0.7mm, white] (1.7,1.7) -- (1.6,1.6);
\foreach \x in {0,0.5,1,1.5}{
	\draw[black!50,fill=white, line width=0.5mm] (\x,1.5) circle [radius=0.1];
}
\draw[black!100,fill=white, line width=0.5mm] (0,1) circle [radius=0.1];
\draw[black!100,fill=white, line width=0.5mm] (0.5,1) circle [radius=0.1];
\draw[black!50,fill=white, line width=0.5mm] (1,1) circle [radius=0.1];
\draw[black!50,fill=white, line width=0.5mm] (1.5,1) circle [radius=0.1];
\foreach \x in {0,0.5,1,1.5}{
	\foreach \y in {0,0.5}{
		\draw[black!50,fill=white, line width=0.5mm] (\x,\y) circle [radius=0.1];
	}
}
\draw[line width=0.7mm, blue,->] (1.2,1) -- (0.8,1);
\end{tikzpicture}
}
\end{minipage}
\hfill
\begin{minipage}{\widthfacb\textwidth}
\centering
\scalebox{\scalefacb}{
\begin{tikzpicture}[font=\tiny]
\node[align=center,color=black!50] at (0.75,0.75) {\large $20$};
\draw[line width=0.7mm, white] (-0.2,-0.2) -- (-0.1,-0.1);
\draw[line width=0.7mm, white] (1.7,1.7) -- (1.6,1.6);
\draw[black!100,fill=white, line width=0.5mm] (0,1.5) circle [radius=0.1];
\draw[black!100,fill=white, line width=0.5mm] (0.5,1.5) circle [radius=0.1];
\draw[black!100,fill=white, line width=0.5mm] (1,1.5) circle [radius=0.1];
\draw[black!100,fill=white, line width=0.5mm] (1.5,1.5) circle [radius=0.1];
\foreach \x in {0,0.5,1,1.5}{
	\foreach \y in {0,0.5,1}{
		\draw[black!50,fill=white, line width=0.5mm] (\x,\y) circle [radius=0.1];
	}
}
\draw[line width=0.7mm, red,->] (0,0.8) -- (0,1.2);
\draw[line width=0.7mm, red,->] (0.5,0.8) -- (0.5,1.2);
\draw[line width=0.7mm, red,->] (1,0.8) -- (1,1.2);
\draw[line width=0.7mm, red,->] (1.5,0.8) -- (1.5,1.2);
\end{tikzpicture}
}
\end{minipage}
\\
\mbox{} \\
\mbox{} \\
\begin{minipage}{\widthfacb\textwidth}
\centering
\scalebox{\scalefacb}{
\begin{tikzpicture}[font=\tiny]
\node[align=center,color=black!50] at (0.75,0.75) {\large $21$};
\draw[line width=0.7mm, white] (-0.2,-0.2) -- (-0.1,-0.1);
\draw[line width=0.7mm, white] (1.7,1.7) -- (1.6,1.6);
\draw[black!100,fill=white, line width=0.5mm] (0,1.5) circle [radius=0.1];
\draw[black!100,fill=white, line width=0.5mm] (0.5,1.5) circle [radius=0.1];
\draw[black!50,fill=white, line width=0.5mm] (1,1.5) circle [radius=0.1];
\draw[black!50,fill=white, line width=0.5mm] (1.5,1.5) circle [radius=0.1];
\foreach \x in {0,0.5,1,1.5}{
	\foreach \y in {0,0.5,1}{
		\draw[black!50,fill=white, line width=0.5mm] (\x,\y) circle [radius=0.1];
	}
}
\draw[line width=0.7mm, blue,->] (1.7,1.5) -- (0.8,1.5);
\end{tikzpicture}
}
\end{minipage}
\hfill 
\begin{minipage}{\widthfacb\textwidth}
\centering
\scalebox{\scalefacb}{
\begin{tikzpicture}[font=\tiny]
\node[align=center,color=black!50] at (0.75,0.75) {\large $22$};
\draw[line width=0.7mm, white] (-0.2,-0.2) -- (-0.1,-0.1);
\draw[line width=0.7mm, white] (1.7,1.7) -- (1.6,1.6);
\draw[black!50,fill=white, line width=0.5mm] (0,1.5) circle [radius=0.1];
\draw[black!100,fill=white, line width=0.5mm] (0.5,1.5) circle [radius=0.1];
\draw[black!100,fill=white, line width=0.5mm] (1,1.5) circle [radius=0.1];
\draw[black!50,fill=white, line width=0.5mm] (1.5,1.5) circle [radius=0.1];
\foreach \x in {0,0.5,1,1.5}{
	\foreach \y in {0,0.5,1}{
		\draw[black!50,fill=white, line width=0.5mm] (\x,\y) circle [radius=0.1];
	}
}
\draw[line width=0.7mm, blue,->] (-0.2,1.5) -- (0.2,1.5);
\end{tikzpicture}
}
\end{minipage}
\hfill
\begin{minipage}{\widthfacb\textwidth}
\centering
\scalebox{\scalefacb}{
\begin{tikzpicture}[font=\tiny]
\node[align=center,color=black!50] at (0.75,0.75) {\large $23$};
\draw[line width=0.7mm, white] (-0.2,-0.2) -- (-0.1,-0.1);
\draw[line width=0.7mm, white] (1.7,1.7) -- (1.6,1.6);
\draw[black!50,fill=white, line width=0.5mm] (0,1.5) circle [radius=0.1];
\draw[black!50,fill=white, line width=0.5mm] (0.5,1.5) circle [radius=0.1];
\draw[black!100,fill=white, line width=0.5mm] (1,1.5) circle [radius=0.1];
\draw[black!100,fill=white, line width=0.5mm] (1.5,1.5) circle [radius=0.1];
\foreach \x in {0,0.5,1,1.5}{
	\foreach \y in {0,0.5,1}{
		\draw[black!50,fill=white, line width=0.5mm] (\x,\y) circle [radius=0.1];
	}
}
\draw[line width=0.7mm, blue,->] (0.3,1.5) -- (0.7,1.5);
\end{tikzpicture}
}
\end{minipage}
\hfill 
\begin{minipage}{\widthfacb\textwidth}
\centering
\scalebox{\scalefacb}{
\begin{tikzpicture}[font=\tiny]
\node[align=center,color=black!50] at (0.75,0.75) {\large $24$};
\draw[line width=0.7mm, white] (-0.2,-0.2) -- (-0.1,-0.1);
\draw[line width=0.7mm, white] (1.7,1.7) -- (1.6,1.6);
\draw[black!50,fill=white, line width=0.5mm] (0,1.5) circle [radius=0.1];
\draw[black!100,fill=white, line width=0.5mm] (0.5,1.5) circle [radius=0.1];
\draw[black!100,fill=white, line width=0.5mm] (1,1.5) circle [radius=0.1];
\draw[black!50,fill=white, line width=0.5mm] (1.5,1.5) circle [radius=0.1];
\foreach \x in {0,0.5,1,1.5}{
	\foreach \y in {0,0.5,1}{
		\draw[black!50,fill=white, line width=0.5mm] (\x,\y) circle [radius=0.1];
	}
}
\draw[line width=0.7mm, blue,->] (1.7,1.5) -- (1.3,1.5);
\end{tikzpicture}
}
\end{minipage}
\hfill
\begin{minipage}{\widthfacb\textwidth}
\centering
\scalebox{\scalefacb}{
\begin{tikzpicture}[font=\tiny]
\node[align=center,color=black!50] at (0.75,0.75) {\large $25$};
\draw[line width=0.7mm, white] (-0.2,-0.2) -- (-0.1,-0.1);
\draw[line width=0.7mm, white] (1.7,1.7) -- (1.6,1.6);
\draw[black!100,fill=white, line width=0.5mm] (0,1.5) circle [radius=0.1];
\draw[black!100,fill=white, line width=0.5mm] (0.5,1.5) circle [radius=0.1];
\draw[black!50,fill=white, line width=0.5mm] (1,1.5) circle [radius=0.1];
\draw[black!50,fill=white, line width=0.5mm] (1.5,1.5) circle [radius=0.1];
\foreach \x in {0,0.5,1,1.5}{
	\foreach \y in {0,0.5,1}{
		\draw[black!50,fill=white, line width=0.5mm] (\x,\y) circle [radius=0.1];
	}
}
\draw[line width=0.7mm, blue,->] (1.2,1.5) -- (0.8,1.5);
\end{tikzpicture}
}
\end{minipage}

\caption{Horizontal sweep in variational optimization of fPEPS on a
  $4\times 4$ 
  lattice.
}
\label{fig:pepssweep}
\end{figure}

\quad The overall procedure for variational optimization
via local updates for a $4\times 4$ lattice is illustrated in 
Fig. \ref{fig:pepssweep}. 
The sketches are to be read row-wise
and from left to right, as indicated by the numbering. 
Sketch $1$ in the upper left corner is a reduced 
aerial projection 
of Fig.~\ref{fig:pepstn}(a) 
and depicts 
the starting point. 
The red arrows in sketch $2$ denote three successive constructions of 
approximate environments for both the energy and the norm, which isolates 
the lowest row.
The blue arrow in sketch $3$ represents 
the zipper contraction of the two sites on the right, which
prepares the system for 
the two-site optimization of the lower-left corner
using the scheme 
described above.
Subsequently, 
we make 
two steps to the right, optimize both times via
bond optimization or CBE, reverse direction, and again optimize
twice.  
In sketch $8$, the treatment of the next
row is prepared by building the environment of the recently processed first row. 
Sketches $9$ to $13$ illustrate the optimization of row $2$ in the same order. 
The remaining sketches 
follow the same pattern, repeating until the uppermost row is 
reached.
As a whole, the $25$ sketches illustrate one horizontal sweep, which 
contains 
four 
DMRG-like optimizations of each row and, in the case
of the Hubbard model, moves charge and spin across the
lattice to minimize the energy. 
A vertical sweep can be implemented simply by rotating every sketch in Fig. 
\ref{fig:pepssweep} by $90$ degrees.
We denote 
a horizontal sweep followed by a vertical sweep a ``full sweep''.

As a final remark, we note that, due to the approximate construction of 
environments, the
Rayleigh-Ritz 
variational principle is, unfortunately, violated.
Therefore, 
the energy 
of the PEPS is not necessarily bounded from below by the ground-state
energy of the Hamiltonian. 

\subsubsection{Gradient updates}
\label{sec:gradientupdate}
Assuming that we have 
an fPEPS within a particular, fixed, virtual basis, we now consider the
functional of the energy, varying 
one particular site
$i$: 
\begin{align}
f\left(A_i\right) = \frac{A_i \, H_i \, A_i}{A_i \, N_i \, A_i} = \frac{e}{n} \, .
\end{align}
Here the $A_i$ are elements of a PEPS tensor reshaped to a vector,
while 
$H_i$ and $N_i$ are the effective Hamiltonian
(Fig.~\ref{fig:onesiteoptpeps}(a)) 
and norm environments (Fig.~\ref{fig:onesiteoptpeps}(b)),
respectively,
reshaped to matrices.
Note that $H_i$ and $N_i$ depend on all $A_{j \neq i}$.
The gradient with respect to one site reads
\begin{align}
\nabla f\left(A_i\right) = \frac{2}{n^2} \left( n \, H_i - e \, N_i \right) A_i \, .
\end{align}
The complete 
gradient can be constructed from one-site gradients
by concatenation:
\begin{align}
\nabla f = \left( \nabla f\left(A_1\right), \nabla f\left(A_2\right), ... , \nabla f\left(A_N\right) \right) \, .
\end{align}
Once 
$f$ and $\nabla f$ are available, one can employ
any suitable 
gradient-based optimization algorithms
to optimize the set of local tensors $A_i$.
In the context of tensor networks, the conjugate gradient method 
was 
used by Vanderstraeten \textit{et al.}~\cite{Vanderstraeten2016Oct}
to carry out such an optimization within 
iPEPS.
Here we instead
utilize 
the L-BFGS algorithm \cite{Buckley1985Jun,Liu1989Aug}, which is 
a memory-efficient, quasi-Newton method.
We find the L-BFGS method to be substantially more effective, as might
be expected since it 
is a second-order method, 
while conjugate gradient is first-order.

\section{results}
\label{sec:results}
In this section, we apply the fPEPS methods described above to the
two-dimensional Hubbard model with Hamiltonian
\eqref{eq:usualhubbardham} and open boundary conditions on lattices
ranging from $3\times 3$ to $8\times 8$.
Our goal is to test the efficacy and performance of the method and to
investigate its convergence as a function of the algorithmic
parameters: the bond dimension $D$, the environment dimension $\chi$,
and the nature and number of update steps, where both local updates
(Sec.~\ref{sec:localupdate}) and a combination of local and gradient updates
(Sec.~\ref{sec:gradientupdate}) will be carried out.
As a measure of convergence, we use primarily the ground-state energy,
which we compare to (numerically) exact or, for
larger lattice sizes, accurate variational results.
As an accurate variational estimate of the 
ground-state energy $E_0$, we use
a highly efficient 
DMRG program developed by
G.~Ehlers~\footnote{This code was developed for momentum-space and
    hybrid momentum-real space calculations
    \cite{Ehlers2015Dec,Ehlers2017Mar,Ehlers2018Jan}, but is also
    quite efficient for pure real-space calculations.}
keeping 
$4000$ states for all simulations.
This estimate is numerically exact for $3\times 3$ and
$4\times 4$ lattices, should be quite accurate for $6\times 6$
lattices, and will give a fairly accurate upper bound to the
ground-state energy for $8\times 8$ lattices.
In addition to ground-state energies, we will also investigate the
behavior of local observables, in particular, local hole density
$1- \langle n_i \rangle$, with  $n_i = n_{i,\uparrow} + n_{i,\downarrow}$
and local spin density
\[
\langle \mathbf{S}_i^2  \rangle = \frac{3}{4} \left< n_i - 2
n_{i,\uparrow} n_{i,\downarrow} \right> \, .
\]
For the case of U(1) symmetry, we will also calculate
$\langle S_i^z\rangle = \langle \frac{1}{2} ( n_{i,\uparrow}
- n_{i,\downarrow} )\rangle$, which is identically zero for an
SU(2)-symmetric state, but can be nonzero for a U(1)-symmetric state,
as an approximate numerical algorithm such as fPEPS can numerically
break spin-inversion symmetry.

Here we treat the ground state in zero magnetic field only, so that
both the total spin $S$ and its $z$-component $S_z$
are taken to be 
zero for all calculations. 
This setup leads to two possible choices of symmetry groups:
The first
is 
$\text{U(1)}_{\text{spin}} \otimes \text{U(1)}_{\text{charge}}$, describing
the local conservation of both the deviation from half-filling, $c_z$, 
and the $z$-component of the spin, $s_z$.
(We use small letters to denote conserved quantum numbers on sites,
virtual indices, and state kets and capital letters to denote the
corresponding conserved quantum numbers for states of the entire lattice.)
The corresponding bonds of tensors are then
parameterized by $\left| s_z, c_z, t \right\rangle$, with $t$
iterating over any 
additional degeneracy .
within a symmetry sector of given $c_z$ and $s_z$.

The second 
choice is to take advantage of the 
full spin-rotation symmetry and classify all states using 
$\text{SU(2)}_{\text{spin}} \otimes \text{U(1)}_{\text{charge}}$.
This yields the far richer states $\left| s, c_z, t \right>$
that, for a given total 
spin $s$, encompass 
the entire spin 
multiplet, that is, the set of states with $s_z=-s, \ldots ,s_z=s$.
In the following, we designate the use of the
$\text{U(1)}_{\text{spin}} \otimes \text{U(1)}_{\text{charge}}$
symmetry as ``U(1)'' and the use of the $\text{SU(2)}_{\text{spin}}
\otimes \text{U(1)}_{\text{charge}}$ symmetry as ``SU(2)''.

The central 
parameter for controlling the variational accuracy in 
fPEPS simulations (and in tensor network
algorithms in general)
is the number of virtual states $D$, usually 
referred to as 
the ``bond dimension''.
Since we have to approximate the contraction of the tensor
network for any optimization scheme, 
we must 
also choose the dimension $\chi$ of the environment tensors, as defined in
Sec.~\ref{sec:pepsexp} for a given set of model parameters, a given
symmetry group, and a given bond dimension $D$.
We do this empirically by choosing $\chi$ to be sufficiently large so
that the stability of local updates and the accuracy are satisfactory, i.e., so
that they are not significantly improved by increasing $\chi$.
For local updates (Sec.~\ref{sec:localupdate}),
we optimize  the 
environment tensors by carrying out sweeps
consisting 
of two CBE sweeps followed by one one-site sweep.
For gradient updates
(Sec.~\ref{sec:gradientupdate}),
the virtual basis of the $D$ states is fixed;
we therefore choose to keep 
the basis of the $\chi$-states fixed as well and perform
only one one-site sweep to optimize the environment.

\begin{figure}[!htb]
\begin{minipage}{0.23\textwidth}
\centering
\begin{tikzpicture}[font=\tiny]
\draw[very thick] (1,0) circle [radius=0.30];
\draw[very thick] (1,1) circle [radius=0.30];
\draw[very thick] (1,2) circle [radius=0.30];
\draw[very thick] (2,1) circle [radius=0.30];
\draw[very thick] (2,2) circle [radius=0.30];
\draw[->,very thick,blue] (-0.1,-0.25)--(-0.1,0.25);
\draw[->,very thick,red] (0.1,0.25)--(0.1,-0.25);
\draw[->,very thick,blue] (-0.1,0.75)--(-0.1,1.25);
\draw[->,very thick,red] (0.1,1.25)--(0.1,0.75);
\draw[->,very thick,blue] (-0.1,1.75)--(-0.1,2.25);
\draw[->,very thick,red] (0.1,2.25)--(0.1,1.75);
\draw[->,very thick,blue] (1.9,-0.25)--(1.9,0.25);
\draw[->,very thick,red] (2.1,0.25)--(2.1,-0.25);
\draw[thick,rounded corners] (0, -0.5)--(2.5, -0.5)--(2.5, 2.5)--(-0.5, 2.5)--(-0.5, -0.5)--(0, -0.5);
\end{tikzpicture}

\vspace*{0.1cm}
\centerline{\small (a) Columnwise SU(2).}
\vspace{0.3cm}
\end{minipage}
\hfill
\begin{minipage}{0.23\textwidth}
\centering
\begin{tikzpicture}[font=\tiny]

\draw[very thick] (0,0) circle [radius=0.30];
\draw[very thick] (2,0) circle [radius=0.30];
\draw[very thick] (1,1) circle [radius=0.30];
\draw[very thick] (0,2) circle [radius=0.30];
\draw[very thick] (2,2) circle [radius=0.30];

\draw[->,very thick,blue] (-0.1,0.75)--(-0.1,1.25);
\draw[->,very thick,red] (0.1,1.25)--(0.1,0.75);

\draw[->,very thick,blue] (0.9,-0.25)--(0.9,0.25);
\draw[->,very thick,red] (1.1,0.25)--(1.1,-0.25);

\draw[->,very thick,blue] (1.9,0.75)--(1.9,1.25);
\draw[->,very thick,red] (2.1,1.25)--(2.1,0.75);

\draw[->,very thick,blue] (0.9,1.75)--(0.9,2.25);
\draw[->,very thick,red] (1.1,2.25)--(1.1,1.75);

\draw[thick,rounded corners] (0, -0.5)--(2.5, -0.5)--(2.5, 2.5)--(-0.5, 2.5)--(-0.5, -0.5)--(0, -0.5);
\end{tikzpicture}

\vspace*{0.1cm}
\centerline{\small (b) Checkerboard SU(2).}
\vspace{0.3cm}
\end{minipage}
\\
\begin{minipage}{0.23\textwidth}
\centering
\begin{tikzpicture}[font=\tiny]
\draw[very thick] (1,0) circle [radius=0.30];
\draw[very thick] (0,1) circle [radius=0.30];
\draw[very thick] (1,2) circle [radius=0.30];
\draw[very thick] (2,1) circle [radius=0.30];
\draw[very thick] (2,2) circle [radius=0.30];

\draw[->,very thick,blue] (-0.1,-0.25)--(-0.1,0.25);
\draw[->,very thick,red] (0.1,0.25)--(0.1,-0.25);

\draw[->,very thick,blue] (1.9,-0.25)--(1.9,0.25);
\draw[->,very thick,red] (2.1,0.25)--(2.1,-0.25);

\draw[->,very thick,blue] (0.9,0.75)--(0.9,1.25);
\draw[->,very thick,red] (1.1,1.25)--(1.1,0.75);

\draw[->,very thick,blue] (-0.1,1.75)--(-0.1,2.25);
\draw[->,very thick,red] (0.1,2.25)--(0.1,1.75);

\draw[thick,rounded corners] (0, -0.5)--(2.5, -0.5)--(2.5, 2.5)--(-0.5, 2.5)--(-0.5, -0.5)--(0, -0.5);
\end{tikzpicture}

\vspace*{0.1cm}
\centerline{\small (c) Triangle SU(2).}
\vspace{0.3cm}
\end{minipage}
\hfill
\begin{minipage}{0.23\textwidth}
\centering
\begin{tikzpicture}[font=\tiny]
\draw[very thick] (1,1) circle [radius=0.30];
\draw[-,ultra thick,violet] (0,2)--(1,2);
\draw[-,ultra thick,violet] (0,0)--(0,1);
\draw[-,ultra thick,violet] (2,0)--(1,0);
\draw[-,ultra thick,violet] (2,2)--(2,1);
\draw[thick,rounded corners] (0, -0.5)--(2.5, -0.5)--(2.5, 2.5)--(-0.5, 2.5)--(-0.5, -0.5)--(0, -0.5);
\end{tikzpicture}

\vspace*{0.1cm}
\centerline{\small (d) Pair SU(2).}
\vspace{0.3cm}
\end{minipage}
\\
\begin{minipage}{0.23\textwidth}
\centering
\begin{tikzpicture}[font=\tiny]
\draw[very thick] (2,2) circle [radius=0.30];
\draw[->,very thick,blue] (0,-0.25)--(0,0.25);
\draw[->,very thick,red] (1,0.25)--(1,-0.25);
\draw[->,very thick,blue] (2,-0.25)--(2,0.25);
\draw[->,very thick,red] (0,1.25)--(0,0.75);
\draw[->,very thick,blue] (1,0.75)--(1,1.25);
\draw[->,very thick,red] (2,1.25)--(2,0.75);
\draw[->,very thick,blue] (0,1.75)--(0,2.25);
\draw[->,very thick,red] (1,2.25)--(1,1.75);
\draw[thick,rounded corners] (0, -0.5)--(2.5, -0.5)--(2.5, 2.5)--(-0.5, 2.5)--(-0.5, -0.5)--(0, -0.5);
\end{tikzpicture}

\vspace*{0.1cm}
\centerline{\small (e) Columnwise U(1).}
\vspace{0.3cm}
\end{minipage}
\caption{
  Depiction of initial product states for the $3\times 3$ lattice,
  where states (a), (b), (c), and (d) are used with SU(2) symmetry and
  state (e) with U(1) symmetry.
  Here a black circle depicts the empty local state $\left|0\right>$, an
  isolated blue uparrow the local state $\left|\uparrow\right>$, an
  isolated  red downarrow the local state $\left|\downarrow\right>$, a pair
  of arrows the doubly occupied local state
  $\left|\uparrow\downarrow\right>$, and a thick purple line the state
  $\left|\text{BS}_{\langle i,j \rangle }\right>$, a
  bond singlet between nearest-neighbor sites $i$ and $j$.
}
\label{fig:3x3_product_states}
\end{figure}
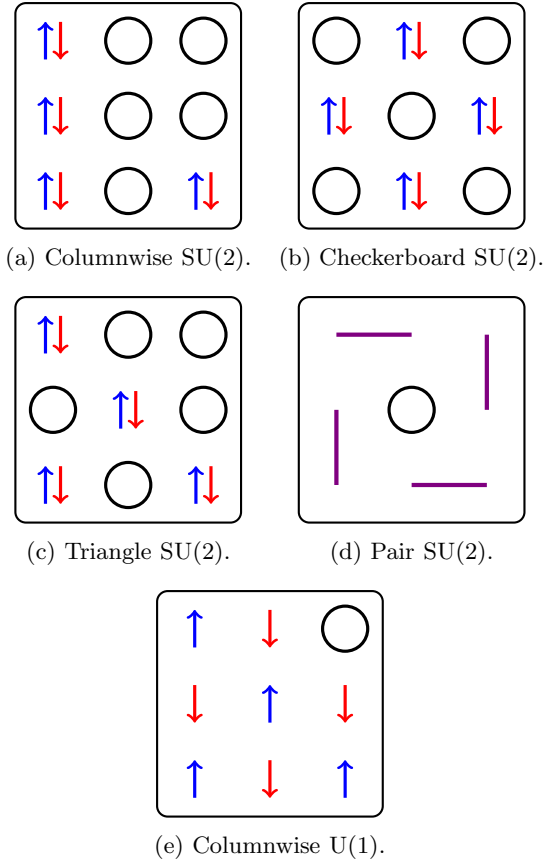

To initiate the simulations, 
we use 
product states made up of patterns of different local states defined
either on a site or on a bond.
For U(1)-spin-symmetric simulations, we make up the initial product states
out of the states
$\left|\uparrow\right> \, = \,\left|s_z=\frac{1}{2},c_z=0\right>$, 
$\left|\downarrow\right>\, = \,\left|s_z=-\frac{1}{2},c_z=0\right>$
and $\left|0\right> \, = \,\left|s_z=0,c_z=-\frac{1}{2}\right>$ (the
empty local state).
For SU(2)-spin-symmetric simulations, we make up the initial product states
out of the local states
$\left|0\right> \, = \,\left|s=0,c_z=-\frac{1}{2}\right>$ and 
$\left|\uparrow\downarrow\right>\, = \,\left|s=0,c_z=\frac{1}{2}\right>$
and also out of the bond-singlet state
\begin{align}
  \left|\text{BS}_{\langle i,j \rangle }\right>\, = \,
  \ \ \ \ \ \ \ \ \ \ \ \  & \nonumber \\
  -  \left|s={\textstyle\frac{1}{2}},c_z=0\right>_i \, & \otimes \,
  \left|s={\textstyle\frac{1}{2}},c_z=0\right>_j \, ,
\end{align}
where $\langle i,j \rangle$ denotes the pair of nearest-neighbor sites
$i$ and $j$.
As we will discuss in the following, we find that the convergence
behavior of fPEPS does depend on the choice of initial state.
We will therefore describe the particular initial product state used
for each simulation as well as the effect of the choice of initial
state on the convergence.
As a reference, we depict product states used to initialize 
the $3\times 3$ lattice with $\langle n \rangle= 8/9$ for SU(2)
as well as for U(1) symmetry in Fig.~\ref{fig:3x3_product_states}.

In order to investigate the fundamental convergence of the fPEPS
algorithm, we start with the minimal case for which all aspects of the
algorithm come into play; this is the $3\times 3$ lattice.
Furthermore, we first take $U=0$ as this does not engender trivial
convergence of fPEPS so that we can test the convergence behavior and compare
the computational 
cost for a case that is very easy to calculate 
exactly 
with almost any numerical method.
Note that $U=0$ is also a nontrivial case for a real-space DMRG
calculation.

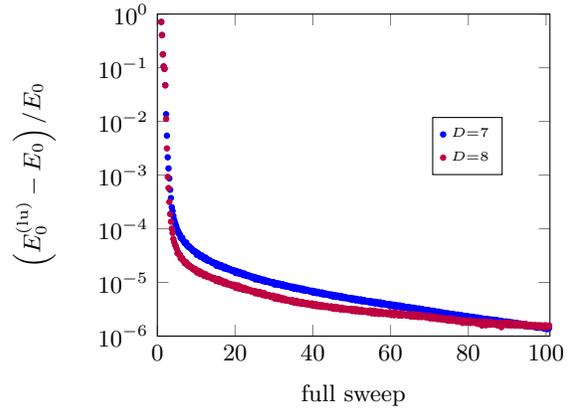
\begin{figure}[!htb]
\centering
\begin{tikzpicture}
\begin{axis}[width=0.45\textwidth,
                     x label style={at={(axis description cs:0.5,-0.12)},anchor=south},
                     y label style={at={(axis description cs:-0.001,.5)},anchor=south},
                     xlabel={full sweep},
                     ylabel={$\left(E_0^{\text{(lu)}}-E_0\right)/E_0$},
                     legend cell align={left},
                     legend style={font=\tiny,at={(0.70,.50)},anchor=south west}, 
                     ymode=log,
                     ytick={1,1e-1,1e-2,1e-3,1e-4,1e-5,1e-6},
                     xmin=0, xmax=101,
                     ymin=1e-6, ymax=1]

\addplot[only marks,blue,mark options={scale=0.5}]
   table[x=x,y=y]{data/3x3_U0/3x3_su2_7_states_U0_raw_3.asc};
   \addlegendentry{
     $D$$=$$7$} 

\addplot[only marks,purple,
  mark options={scale=0.5}]
   table[x=x,y=y]{data/3x3_U0/3x3_su2_8_states_U0_raw_3.asc};
   \addlegendentry{
     $D$$=$$8$} 

\end{axis}
\end{tikzpicture}

\caption{
  Relative error in the ground-state energy of the
  Hubbard model  on a $3\times 3$ lattice with open boundary
  conditions for $U=0$,
  $S=0$, 
  and $N=8$ (i.e., $\langle n\rangle=8/9$), calculated with fPEPS with
  SU(2) symmetry 
  using local optimizations only, $E_0^{\text{(lu)}}$, with respect to the exact
  ground-state energy $E_0$ plotted as a function of the sequence of
  full local update sweeps.
  For both indicated bond dimensions $D$, the
  environment dimension $\chi=500$.
}
\label{fig:3x3U0}
\end{figure}

We first carry out a simulation that only uses local
optimization (i.e., no gradient optimization step) in order to explore
the efficacy of the local optimization scheme.
In 
Fig.~\ref{fig:3x3U0}, we display the convergence of the ground-state
energy for 
$100$ full sweeps of local optimization two different bond dimensions,
$D=7$ and $D=8$, utilizing SU(2) symmetry.
Here $N=8$ electrons, so that $\langle n \rangle=  8/9 \approx 0.89$,
a value as close to half filling as is possible for a $3\times 3$ lattice.
As the initial state, we take the columnwise SU(2) product state,
Fig.~\ref{fig:3x3_product_states}(a);
we will discuss the choice of the initial state in more
detail in the context of the combined local and gradient (supersweep)
algorithm below.
As can be seen, for both values of $D$, there is systematic
convergence to the ground-state energy to a relative accuracy of
$10^{-6}$ as the number of local update full sweeps is increased for
both values of $D$.
Furthermore, at least up to approximately 60 update sweeps, there is a
systematic improvement of the energy with $D$.
(A possible cause for the poorer convergence of the
$D=8$ calculation relative to the $D=7$ calculation above 60 sweeps
could be that the 
environment dimension, $\chi=500$, is too small for the higher bond
dimension, $D=8$.)
Particularly notable, however, is how slow the convergence is,
especially with the number of sweeps, but also with $D$.
As described in Sec.~\ref{sec:localupdate}, each full sweep
encompasses 18 bond optimizations, where each includes an iterative
diagonalization.
Thus, carrying out 100 sweeps is computationally quite expensive---here
ca.\ 8 days of wall time for the $D=7$ and 58 days for the $D=8$ runs
exclusively using all 16 cores in parallel on a 2.5 GHz Intel Xeon 4215
processor-based compute node.
(We will use this compute-node configuration, utilized exclusively and
in parallel, as a measure for computational cost for all calculations described
in this paper; we will term this measure our ``reference compute node''.) 
In contrast, the ground-state energy for this system can be obtained
numerically to essentially arbitrary accuracy with exact
diagonalization or DMRG in under a second of computer time.

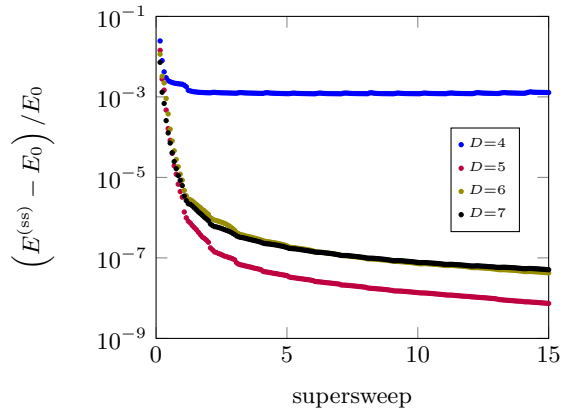
\begin{figure}[!htb]
\begin{tikzpicture}
\begin{axis}[width=0.45\textwidth,
                     x label style={at={(axis description cs:0.5,-0.12)},anchor=south},
                     y label style={at={(axis description cs:-0.001,.5)},anchor=south},
                     xlabel={supersweep},
                     ylabel={$\left(E^{\text{(ss)}}-E_0\right)/E_0$},
                     legend style={font=\tiny,at={(0.75,0.33)},anchor=south west},
                     ymode=log,
                     ytick={1e-1,1e-3,1e-5,1e-7,1e-9},
                     xmin=0, xmax=15,
                     ymin=1e-9, ymax=0.1]
                     ]

\addplot[only marks,blue, 
  mark options={scale=0.4}]
   table[x=x,y=y]{data/3x3_U0_checker2/3x3_D4_chi500_U0_checker2_SU2_proc_RMN_2.asc};
\addlegendentry{$D$$=$$4$}

\addplot[only marks,
  purple, 
  mark options={scale=0.4}]
   table[x=x,y=y]{data/3x3_U0_checker2/3x3_D5_chi500_U0_checker2_SU2_proc_RMN_2.asc};
\addlegendentry{$D$$=$$5$}

\addplot[only marks,olive, 
  mark options={scale=0.4}]
   table[x=x,y=y]{data/3x3_U0_checker2/3x3_D6_chi500_U0_checker2_SU2_proc_RMN_2.asc};
\addlegendentry{$D$$=$$6$}

\addplot[only marks, black, 
  mark options={scale=0.4}]
   table[x=x,y=y]{data/3x3_U0_checker2/3x3_D7_chi500_U0_checker2_SU2_proc_RMN_2.asc};
\addlegendentry{$D$$=$$7$}

\end{axis}
\end{tikzpicture}
\caption{
  Relative error in the ground-state energy calculated with fPEPS with
  SU(2) symmetry
  using supersweeps consisting of 3 full local update sweeps
  followed by 100 gradient updates, $E_0^{\text{(ss)}}$, with respect to the exact
  ground-state energy $E_0$ on the $3\times 3$ Hubbard model with open
  boundary conditions, $U=0$,
  $S=0$, 
  and $N=8$ for the indicated values of bond dimension $D$ and $\chi$$=$$500$.
  The lowest energy for every local update sweep and
  the energy of every tenth gradient update is plotted.
}
\label{fig:3x3U0conv}
\end{figure}

Nevertheless, we emphasize that Fig.~\ref{fig:3x3U0}
demonstrates that systematic 
variational convergence does occur in fPEPS with local optimization, a
result that we have found to be quite hard to achieve.
In particular, we have found that if the approximate contraction is
not carried out sufficiently carefully (e.g., with a value of $\chi$
that is too small), the variationality of the
optimization procedure is disturbed, and the variational determination
of the energy through iterative diagonalization of the effective
Hamiltonian (with a given norm environment), see
Fig.~\ref{fig:twositeoptpeps}, becomes unstable.

We now address 
the question of whether the speed and/or the degree
of convergence can be improved by the addition of gradient
optimization.
Note that the gradient updates described in
Sec.~\ref{sec:gradientupdate} optimize all local PEPS tensors $A_i$
simultaneously rather than only two at once for local bond updates and one
at once for single-site update (see Sec.~\ref{sec:localupdate}), but,
like single-site updates, cannot change the structure of the basis.
Thus, we have found that it is favorable to carry out a mixture of
local and gradient optimization, 
i.e., combinations of the two types of sweeps, which we denote ``supersweeps''.
In particular, we perform successive supersweeps, each consisting of  
$3$ full local update sweeps followed by $100$ gradient updates, repeating
until either a saturation of the convergence or a reasonable limit of
computer time has been reached.
We have empirically found this supersweep configuration to be optimal
and use it for all subsequent calculations described below.

The results for the relative error in the ground-state energy are
plotted in Fig.~\ref{fig:3x3U0conv} for values of the bond dimension
ranging from $D=4$ to $D=7$.
As can be seen, there is a regular, systematic convergence with
supersweep steps for all values of $D$.
The $D=4$ curve saturates after
approximately two supersweeps (26 update steps), while the curves for
higher values of $D$ continue to go down up to the maximum of
15 supersweeps 
displayed.
Going from $D=4$ to $D=5$, there is a very large jump in convergence;
$D=4$ saturates at a relative accuracy slightly above $10^{-3}$,
while the highest relative accuracy of $D=5$ is below $10^{-8}$.
Further increasing $D$, as can be seen from the $D=6$ and $7$ curves,
actually reduces the accuracy, with both curves lying approximately
on top of one another.
Thus, for a $U=0$ on a $3\times 3$ lattice, there seems to be a
breakdown of systematic convergence with $D$ above $D=5$.

\begin{figure}[!htb]
\centering
\begin{tikzpicture}[font=\tiny]
        \draw[fill=green] (0,0) circle [radius=0.24999736903480596];
\draw[fill=green] (1,0) circle [radius=1.8663769261184626e-06];
\draw[fill=green] (2,0) circle [radius=0.250000838378713];
\draw[fill=green] (0,1) circle [radius=2.5388245971091905e-06];
\draw[fill=green] (1,1) circle [radius=-4.346497571861008e-06];
\draw[fill=green] (2,1) circle [radius=2.2172452927016195e-06];
\draw[fill=green] (0,2) circle [radius=0.2499977214667537];
\draw[fill=green] (1,2) circle [radius=3.762931203632469e-06];
\draw[fill=green] (2,2) circle [radius=0.24999803223926964];
\draw[->,thick,blue] (0,-0.11718873665343503)--(0,0.11718873665343503);
\draw[->,thick,blue] (1,-0.12500047626956826)--(1,0.12500047626956826);
\draw[->,thick,blue] (2,-0.1171881284851239)--(2,0.1171881284851239);
\draw[->,thick,blue] (0,0.8749994501513323)--(0,1.1250005498486677);
\draw[->,thick,blue] (1,0.8750002617848601)--(1,1.1249997382151398);
\draw[->,thick,blue] (2,0.8749999247174376)--(2,1.1250000752825624);
\draw[->,thick,blue] (0,1.8828120910000719)--(0,2.117187908999928);
\draw[->,thick,blue] (1,1.8750002755715025)--(1,2.1249997244284975);
\draw[->,thick,blue] (2,1.8828116995096353)--(2,2.1171883004903647);
\draw[rounded corners] ((0, -0.5)--(2.5, -0.5)--(2.5, 2.5)--(-0.5, 2.5)--(-0.5, -0.5)--(0, -0.5);
\node[align=center,text width=6mm] at (0, 2.7) { \footnotesize  0.36} %%% 0.24
;\node[align=center,text width=6mm] at (1, 2.7) { \footnotesize 0.38} %%% 0.25
;\node[align=center,text width=6mm] at (2, 2.7) { \footnotesize 0.36} %%% 0.24
;\node[align=center,text width=6mm,color=black!30!green] at (0, -0.7) { \footnotesize 0.17}
;\node[align=center,text width=6mm,color=black!30!green] at (1, -0.7) { \footnotesize 0.0}
;\node[align=center,text width=6mm,color=black!30!green] at (2, -0.7) { \footnotesize 0.17}
;
        \end{tikzpicture}
\caption{
  Local spin density $\langle\mathbf{S}^2_i\rangle$ (size of
  blue arrows)
  and local hole density $1 - \langle n_i \rangle$ (diameter of
  green-shaded circles) on a $3\times 3$
  lattice for $U=0$, $S=0$,
  and $N=8$, calculated using combined local and gradient
  updates and SU(2) symmetry with $D=5$ and
  $\chi=500$.
  The black numbers on the top edge are the averages of the spin densities
  on the column of sites below, and the green numbers on the bottom
  edge are the averages of the hole densities on the column of sites above.
}
\label{fig:3x3_U0_D5_spin_charge}
\end{figure}
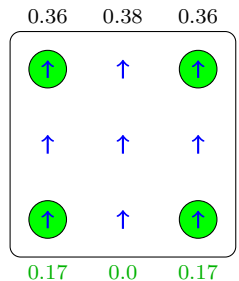

In order to understand the origin of this breakdown, we have examined
the state structure of the bond truncation at which the divergence
occurs in detail for $D=5$ and $D=6$.
The $D$ states in the retained set are distributed over various
$s,c_z$ quantum numbers, with one or, at most, two states per quantum
number.
These sets of quantum numbers differ, with a set being retained for
$D=5$ that is not retained for $D=6$.
In view of this, it is our hypothesis that the granularity of the
state selection is very sensitive to which states are selected at
critical steps; this selection seems to be particularly fortuitous for
$D=5$ in the $U=0$ case treated here.
An examination of the local spin and hole densities, displayed in
Fig.~\ref{fig:3x3_U0_D5_spin_charge}, shows that the 
single hole is almost exclusively distributed over the four corner
sites of the $3\times 3$ lattice.
(The spin and charge densities displayed in
Fig.~\ref{fig:3x3_U0_D5_spin_charge} are calculated using fPEPS so
that the SU(2) symmetry is explicitly preserved; for
$D=5$, they are essentially numerically exact on the scale of the
plot.)
This unevenness of this hole distribution is an artifact of the lack
of interaction, i.e., that $U=0$; the distribution of the exact
values of the hole densities between the different corner sites is
also quite unstable; calculations of lesser accuracy or without SU(2)
symmetry result in uneven distributions.
Such an uneven hole distribution that is so sensitive to fine details of
the variational state is consistent with the sensitivity to granular
state selection described above.

We also note that the convergence behavior is significantly dependent
on the initial product state taken.
Here we use the checkerboard SU(2) state,
Fig.~\ref{fig:3x3_product_states}(b), which yields the most accurate
energies for the $U=0$, $3\times 3$ system for all $D$ values except $D=5$.
For $D=5$, using a columnwise SU(2) initial state,
Fig.~\ref{fig:3x3_product_states}(a), yields an energy curve similar
to the $D=5$ curve in Fig.~\ref{fig:3x3U0conv}, but with a ground-state energy
that is approximately a factor of two more accurate (not shown).
For all other $D$ values, however, the energies obtained with a
columnwise initial state are significantly less
accurate than those in Fig.~\ref{fig:3x3U0conv}.
We therefore also ascribe the anomalously accurate energy for the
$D=5$ case with a columnwise initial state to a fortuitous
bond truncation similar to that of the checkerboard
initial state, but one that happens to be slightly more fortuitous.

We now turn on the interaction to $U=8$, 
treating the same system size,
$3\times 3$, and band filling, $\langle n \rangle = 8/9$.
For this system, we have found that using local optimization alone
does not lead to systematic convergence.
We therefore only use the supersweep scheme here.
However, we do carry out and compare calculations using both the U(1)-symmetric
and the SU(2)-symmetric fPEPS states.
For the U(1) calculations, we initialize all runs with the
columnwise U(1) product state, Fig.~\ref{fig:3x3_product_states}(e),
while for the SU(2) calculations, we use the triangle SU(2) product state,
Fig.~\ref{fig:3x3_product_states}(c).
Empirically, we have found these initial states to give the best
overall convergence for each symmetry.

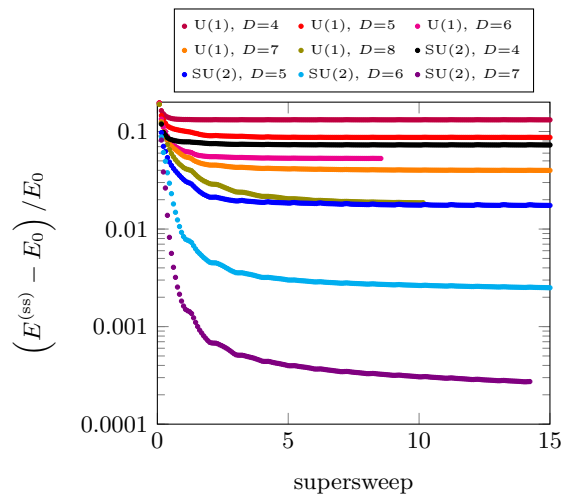
\begin{figure}[!htb]
\centering
\begin{tikzpicture}
  \begin{axis}[width=0.45\textwidth,
                     x label style={at={(axis description cs:0.5,-0.12)},anchor=south},
                     y label style={at={(axis description cs:-0.001,.5)},anchor=south},
                     xlabel={supersweep},
                     ylabel={$\left(E^{\text{(ss)}}-E_0\right)/E_0$},
                     legend columns=3,
                     legend style={font=\tiny,at={(0.042,1.03)},anchor=south west},
                     ymode=log,
                     log ticks with fixed point,
                     xmin=0, xmax=15,
                     ymin=0.0001, ymax=0.2]

\addplot[only marks,purple,mark options={scale=0.4}]
   table[x=x,y=y]{data/3x3_U8_SU2_pair_U1_upperright/3x3_D4_chi500_U8_column_U1_proc_RMN_2.asc};
\addlegendentry{U(1), $D$$=$$4$~~~}

\addplot[only marks,red,mark options={scale=0.4}]
   table[x=x,y=y]{data/3x3_U8_SU2_pair_U1_upperright/3x3_D5_chi500_U8_column_U1_proc_RMN_2.asc};
\addlegendentry{U(1), $D$$=$$5$~~~}

\addplot[only marks,magenta,mark options={scale=0.4}]
   table[x=x,y=y]{data/3x3_U8_SU2_pair_U1_upperright/3x3_D6_chi500_U8_column_U1_proc_RMN_2.asc};
\addlegendentry{U(1), $D$$=$$6$~~}

\addplot[only marks,orange,mark options={scale=0.4}]
   table[x=x,y=y]{data/3x3_U8_SU2_pair_U1_upperright/3x3_D7_chi500_U8_column_U1_proc_RMN_2.asc};
\addlegendentry{U(1), $D$$=$$7$~~~}

\addplot[only marks,olive,mark options={scale=0.4}]
   table[x=x,y=y]{data/3x3_U8_SU2_pair_U1_upperright/3x3_D8_chi500_U8_column_U1_proc_RMN_2.asc};
\addlegendentry{U(1), $D$$=$$8$~~~}

\addplot[only marks,black,
  mark options={scale=0.4}]
   table[x=x,y=y]{data/3x3_U8/3x3_D4_chi500_U8_SU2_proc_RMN_2.asc};
\addlegendentry{SU(2), $D$$=$$4$}

\addplot[only marks,blue,
  mark options={scale=0.4}]
table[x=x,y=y]{data/3x3_U8/3x3_D5_chi500_U8_SU2_proc_RMN_2.asc};
\addlegendentry{SU(2), $D$$=$$5$~~}

\addplot[only marks,cyan,
  mark options={scale=0.4}]
   table[x=x,y=y]{data/3x3_U8/3x3_D6_chi500_U8_SU2_proc_RMN_2.asc};
\addlegendentry{SU(2), $D$$=$$6$~~}

\addplot[only marks,violet,
  mark options={scale=0.4}]
   table[x=x,y=y]{data/3x3_U8/3x3_D7_chi500_U8_SU2_proc_RMN_2.asc};
\addlegendentry{SU(2), $D$$=$$7$}

\end{axis}
\end{tikzpicture}
\caption{
  Relative error in the ground-state energy of the Hubbard model
  on a $3\times 3$ lattice with open
  boundary conditions, $U=8$,
  $S=0$, and $N=8$ ($\langle n\rangle=8/9$),
  calculated with fPEPS with U(1) and SU(2) symmetry, as indicated,
  using supersweeps consisting of 3 full local optimization sweeps
  followed by 100 gradient optimizations, $E_0^{\text{(ss)}}$, with respect to the exact
  ground-state energy $E_0$ for $\chi$$=$$500$ and various $D$.
  The lowest energy for every local update sweep and
  the energy of every tenth gradient update is plotted.
}
\label{fig:3x3U8conv}
\end{figure}

We display the convergence of the relative error in the ground-state energy as
the algorithm progresses in Fig.~\ref{fig:3x3U8conv}.
For both U(1) and SU(2) symmetry, there is initial rapid convergence
for the first few supersweeps, with runs with larger values of $D$
converging to a higher accuracy in that the rapid convergence
continues for a larger number of supersweeps.
For both symmetries, small wave-like fluctuations are visible at the
end of each supersweep, especially for larger values of $D$.
These are the points at which the optimization switches from gradient
updates to local updates.
At these points, the slope of the convergence of the energy becomes
flatter or, in some cases, particularly where a plateau in the
convergence has been reached, actually becomes larger.
However, when gradient sweeps begin again, after 3 full local update
sweeps, corresponding to 3 points, the slope of the convergence
becomes steeper again.
The behavior demonstrates that neither type of optimization alone
displays systematic convergence.
The local updates do lead to rapid convergence at the beginning as the
basis is adapted from a product state to a more suitable basis for the
many-body ground state, but the convergence then slows down
significantly.
Once a new basis is formed, gradient updates initially lead to a rapid
convergence within that basis, but the convergence then saturates, as
the basis cannot be changed within the gradient updates.
Subsequent local updates do change the local basis, but do not
necessarily improve the energy; however, following gradient updates can
then continue to improve the energy until saturation is reached once
again.
Note that this behavior can also be seen for the $U=0$ case in
Fig.~\ref{fig:3x3U0conv}, albeit at a slightly smaller scale so that
it is not as visible.

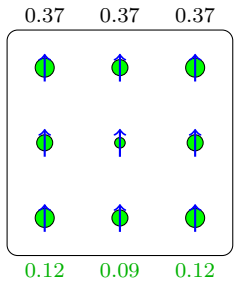
\begin{figure}[!htb]
\centering
\begin{tikzpicture}[font=\tiny]
\draw[fill=green] (0,0) circle [radius=0.12621570866124998];
\draw[fill=green] (1,0) circle [radius=0.10631325714494999];
\draw[fill=green] (2,0) circle [radius=0.12619987642145003];
\draw[fill=green] (0,1) circle [radius=0.10631319812393003];
\draw[fill=green] (1,1) circle [radius=0.06991516172078005];
\draw[fill=green] (2,1) circle [radius=0.10631344573252999];
\draw[fill=green] (0,2) circle [radius=0.12619970376080003];
\draw[fill=green] (1,2) circle [radius=0.10631347388127];
\draw[fill=green] (2,2) circle [radius=0.12621617455305];

% S = (3/4) * (n - 2*n_up*n_down)
\draw[->,thick,blue] (0,-0.18451304904134175)--(0,0.18451304904134175); % S=0.3690260980826835
\draw[->,thick,blue] (1,-0.18538077963274213)--(1,0.18538077963274213); % S=0.37076155926548426
\draw[->,thick,blue] (2,-0.18451379831896936)--(2,0.18451379831896936); % S=0.3690275966379387
\draw[->,thick,blue] (0,0.8146192180592011)--(0,1.1853807819407989); % S=0.37076156388159787
\draw[->,thick,blue] (1,0.8134165243299407)--(1,1.1865834756700593); % S=0.3731669513401187
\draw[->,thick,blue] (2,0.8146192278635418)--(2,1.1853807721364582); % S=0.3707615442729164
\draw[->,thick,blue] (0,1.8154861934404478)--(0,2.1845138065595522);  % S=0.3690276131191043
\draw[->,thick,blue] (1,1.8146192290222791)--(1,2.185380770977721); % S=0.37076154195544186
\draw[->,thick,blue] (2,1.8154869730300687)--(2,2.184513026969931); % S=0.36902605393986265

\draw[rounded corners] ((0, -0.5)--(2.5, -0.5)--(2.5, 2.5)--(-0.5, 2.5)--(-0.5, -0.5)--(0, -0.5);
\node[align=center,text width=6mm] at (0, 2.7) { \footnotesize 0.37}
;\node[align=center,text width=6mm] at (1, 2.7) { \footnotesize 0.37}
;\node[align=center,text width=6mm] at (2, 2.7) { \footnotesize 0.37}
;
;\node[align=center,text width=6mm,color=black!30!green] at (0, -0.7) { \footnotesize 0.12} % 0.11957620351532668
;\node[align=center,text width=6mm,color=black!30!green] at (1, -0.7) { \footnotesize 0.09} % 0.09418063091566668
;\node[align=center,text width=6mm,color=black!30!green] at (2, -0.7) { \footnotesize 0.12} % 0.11957649890234334
;
        \end{tikzpicture}
\caption{
  Local spin density $\langle\mathbf{S}^2_i\rangle$ and local hole
  density $1-\langle n_i \rangle$
  on a $3\times 3$
  lattice for $U=8$, $S=0$,
  and $N=8$, 
  calculated using
  the DMRG with $D=4000$ states kept
  and plotted using the same scheme as in Fig.~\ref{fig:3x3_U0_D5_spin_charge}.
}
\label{fig:3x3_U8_DMRG_spin_charge}
\end{figure}

For the U(1) calculations, the relative error reaches successively
lower plateaus for all values of $D$ with the largest value, $D=8$,
reaching a value of approximately $0.02$ after 5 supersweeps.
The SU(2) calculations also have an initial rapid convergence for all
$D$ values that
crosses over to either a plateau for $D=4$ and $D=5$ or a more
gradual decline for $D=6$ and $D=7$ on a scale of approximately 5
supersweeps.
The accuracy improves systematically and rapidly with $D$.
The irregular convergence with $D$
seen for the $U=0$ calculations with SU(2) symmetry in
Fig.~\ref{fig:3x3U0conv}, where accuracy in the energy is much higher
for $D=5$ than for $D=6$ and $D=7$, is not present here.
We observe that the distribution of charge and spin for the $U=8$
case, displayed in Fig.~\ref{fig:3x3_U8_DMRG_spin_charge} (here
calculated numerically exactly using the DMRG), is much
more uniform than that for the $U=0$ case,
Fig.~\ref{fig:3x3_U0_D5_spin_charge}.
Evidently, turning on the interaction leads to a much more uniform
distribution and, presumably, reduces the problems due to granularity
in state selection within the bond truncation.
Note, however, that the relative error in the ground-state energy is,
in fact, much smaller in best case for $U=0$, $D=5$, than in the most accurate
$U=8$ calculation, which has the highest bond dimension, $D=7$.

In all cases, a particular
SU(2) calculation is significantly more accurate than the U(1)
calculation with the same $D$.
This significant improvement in going from U(1) to SU(2) symmetry is
expected, as one SU(2) state comprises an entire spin 
multiplet, as discussed in Sec.~\ref{sec:su2}.
Thus, a given SU(2) 
bond dimension $D$ effectively encompasses a region of
the total Hilbert space that could only be encompassed with a
U(1)-symmetric fPEPS with a significantly higher bond dimension.
We also note that SU(2)-symmetric calculations explicitly preserve the
fundamental spin-rotation symmetry present in the system, which is not
the case for U(1) calculations.

The minimum relative error reached, for the $D=7$, SU(2) calculation,
is just under $3\times 10^{-4}$.
This calculation took approximately 9 days of wall time
on our reference compute node.
Thus, the fPEPS method is certainly not competitive either in accuracy
or in computational efficiency with other
numerical methods that are numerically exact on this small system, such as
exact diagonalization and the DMRG.

We now turn to larger system sizes.
Based on the results for the $3\times 3$ system discussed above, we
expect the computational effort needed to obtain reasonable convergence,
i.e., at least to the level of Fig.~\ref{fig:3x3U8conv}, to be unreachably
large for larger lattice sizes.
Nevertheless, we have carried out runs that are as long as practicably
possible in order to gain knowledge about the convergence behavior.

We first consider the $4\times 4$ lattice, whose
ground state is still accessible using exact diagonalization.
Our primary purpose here (and for the $6\times 6$ lattice treated
below) is to test the convergence behavior.
Thus, we try to minimize
detailed physical effects due to charge inhomogeneity and treat the
half-filled system, i.e., take $\langle n\rangle=1$, that is, $N=16$
electrons on the $4 \times 4$ lattice.
This choice also simplifies the choice of the initial state: for U(1)
symmetry, we start the simulation with a N\'eel state, i.e.,
a product state with antiferromagnetically alternating local spins, a
pattern corresponding to the spin pattern in
Fig.~\ref{fig:3x3_product_states}(e) (i.e., with no hole).
For the SU(2) calculations, we take a columnwise SU(2) state, i.e.,
the half-filled analog of Fig.~\ref{fig:3x3_product_states}(a) on a
$4\times 4$ lattice, that is, two columns of doubly occupied local
states interspersed with two columns of unoccupied states.
We have ascertained that choosing a physically reasonable alternate
initial product state, in particular, one in which the lattice is
covered by vertically oriented bond-singlet pairs, 
cf.\  Fig.~\ref{fig:3x3_product_states}(d), does not
significantly effect the convergence behavior.
In this sense, fPEPS
for this larger, half-filled system seems to be significantly less
sensitive to choice of initial states than the $3\times 3$ systems
with one hole discussed above.

\begin{figure}[!htb]
\centering
\begin{tikzpicture}
  \begin{axis}[width=0.45\textwidth,
                     x label style={at={(axis description cs:0.5,-0.12)},anchor=south},
                     y label style={at={(axis description cs:-0.001,.5)},anchor=south},
                     xlabel={supersweep},
                     ylabel={$\left(E^{\text{(ss)}}-E_0\right)/E_0$},
                     legend columns=2,
                     legend style={font=\tiny,at={(0.02,1.03)},anchor=south west},
                     ymode=log,
                     log ticks with fixed point,
                     xmin=0, xmax=11,
                     ymin=0.01, ymax=0.2]
                     ]

\addplot[only marks,purple,mark options={scale=0.4}]
   table[x=x,y=y]{data/4x4/4x4_D4_chi250_U8_U1_proc_RMN_2.asc};
\addlegendentry{U(1), $D$$=$$4$, $\chi$$=$$250$~~}

\addplot[only marks,red,mark options={scale=0.4}]
   table[x=x,y=y]{data/4x4/4x4_D5_chi250_U8_U1_proc_RMN_2.asc};
\addlegendentry{U(1), $D$$=$$5$, $\chi$$=$$250$~~}

\addplot[only marks,magenta,mark options={scale=0.4}]
   table[x=x,y=y]{data/4x4/4x4_D6_chi350_U8_U1_proc_RMN_2.asc};
\addlegendentry{U(1), $D$$=$$6$, $\chi$$=$$350$~~}

\addplot[only marks,orange,mark options={scale=0.4}]
   table[x=x,y=y]{data/4x4/4x4_D7_chi350_U8_U1_proc_RMN_2.asc};
\addlegendentry{U(1), $D$$=$$7$, $\chi$$=$$350$~~}

\addplot[only marks,olive,mark options={scale=0.4}]
   table[x=x,y=y]{data/4x4/4x4_D8_chi350_U8_U1_proc_RMN_2.asc};
\addlegendentry{U(1), $D$$=$$8$, $\chi$$=$$350$~~}

\addplot[only marks,
  black,mark options={scale=0.4}]
   table[x=x,y=y]{data/4x4/4x4_D4_chi250_U8_SU2_proc_RMN_2.asc};
\addlegendentry{SU(2), $D$$=$$4$, $\chi$$=$$250$~~}

\addplot[only marks,blue,
  mark options={scale=0.4}]
   table[x=x,y=y]{data/4x4/4x4_D5_chi300_U8_SU2_proc_RMN_2.asc};
\addlegendentry{SU(2), $D$$=$$5$, $\chi$$=$$300$~~}

\addplot[only marks,
  cyan,mark options={scale=0.4}]
   table[x=x,y=y]{data/4x4/4x4_D6_chi300_U8_SU2_proc_RMN_2.asc};
\addlegendentry{SU(2), $D$$=$$6$, $\chi$$=$$300$~~}

\end{axis}
\end{tikzpicture}
\caption{
  Relative error in the ground-state energy of the Hubbard model
  on a $4\times 4$ lattice with open
  boundary conditions, $U=8$,
  $S=0$, and $N=16$ (half filling),
  calculated with fPEPS with U(1) and SU(2) symmetry, as indicated,
  using supersweeps consisting of 3 full local optimization sweeps
  followed by 100 gradient optimizations, $E_0^{\text{(ss)}}$, with respect to the exact
  ground-state energy $E_0$ for various $D$ with the indicated $\chi$.
  The lowest energy for every local update sweep and
  the energy of every tenth gradient update is plotted.
}
\label{fig:4x4conv}
\end{figure}
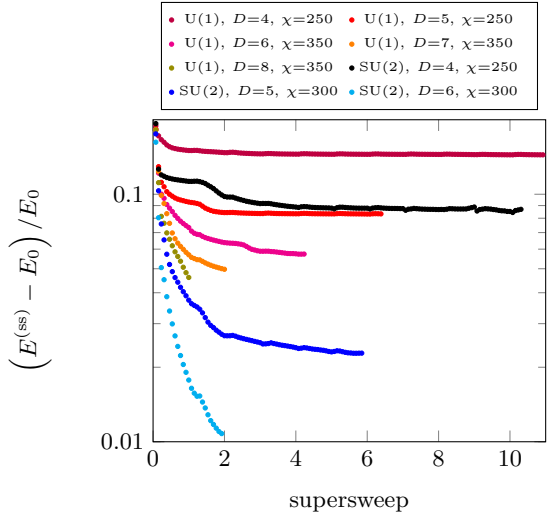

Fig.~\ref{fig:4x4conv} depicts the 
the relative error in the ground-state energy on a half-filled $4\times 4$
lattice at $U=8$ as a
function of the optimization step for runs  with 
both U(1) and SU(2) symmetry and various bond dimensions ranging from
$D=4$ to $D=8$.
Consistent with the behavior for the $3\times 3$, $U=8$ system, there is systematic
convergence both with increasing supersweep step and with increasing $D$.
As before, utilizing SU(2) rather than U(1) symmetry for a given
fixed $D$ also leads to a significant 
improvement in the relative accuracy for the $4\times 4$ system.
The best relative accuracy in energy of about $0.01$ is obtained from
the SU(2) simulation with $D=6$ states, which took approximately $4$
days of wall time on our reference compute node; this is the largest
bond dimension that we could practicably retain for this system.
Taking the $3\times 3$, $U=8$ calculations as a guide, at least 15
supersweeps would likely be required to achieve
reasonable convergence for $D=6$.
This would correspond to a wall time of approximately one month on our
reference compute node.

\begin{figure}[!htb]
\centering
\begin{tikzpicture}
\begin{axis}[width=0.45\textwidth,
                     x label style={at={(axis description cs:0.5,-0.12)},anchor=south},
                     y label style={at={(axis description cs:-0.001,.5)},anchor=south},
                     xlabel={supersweep},
                     ylabel={$\left(E^{\text{(ss)}}-E_0\right)/E_0$},
                     legend cell align={left},
                     legend columns=2,
                     legend style={font=\tiny,at={(0.02,1.03)},anchor=south west},
                     ymode=log,
                     log ticks with fixed point,
                     ytick={10,1,0.1,0.05},
                     xmin=0, xmax=3,
                     ymin=0.05, ymax=0.2]
                     ]

\addplot[only marks,purple,mark options={scale=0.6}]
   table[x=x,y=y]{data/6x6/6x6_D4_chi300_U8_U1_proc_RMN_2.asc};
\addlegendentry{U(1), $D$$=$$4$, $\chi$$=$$300$~~}

\addplot[only marks,red,mark options={scale=0.6}]
   table[x=x,y=y]{data/6x6/6x6_D5_chi300_U8_U1_proc_RMN_2.asc};
\addlegendentry{U(1), $D$$=$$5$, $\chi$$=$$300$~~}

\addplot[only marks,magenta,mark options={scale=0.6}]
   table[x=x,y=y]{data/6x6/6x6_D6_chi300_U8_U1_proc_RMN_2.asc};
\addlegendentry{U(1), $D$$=$$6$, $\chi$$=$$300$~~}

\addplot[only marks,orange,mark options={scale=0.6}]
   table[x=x,y=y]{data/6x6/6x6_D7_chi300_U8_U1_proc_RMN_2.asc};
\addlegendentry{U(1), $D$$=$$7$, $\chi$$=$$300$~~}

\addplot[only marks,
  black,mark options={scale=0.6}]
   table[x=x,y=y]{data/6x6/6x6_D4_chi300_U8_SU2_proc_RMN_2.asc};
\addlegendentry{SU(2), $D$$=$$4$, $\chi$$=$$300$~~}

\addplot[only marks, blue,
  mark options={scale=0.6}]
   table[x=x,y=y]{data/6x6/6x6_D5_chi350_U8_SU2_proc_RMN_2.asc};
\addlegendentry{SU(2), $D$$=$$5$, $\chi$$=$$350$~~}

\addplot[only marks,
  cyan,mark options={scale=0.6}]
   table[x=x,y=y]{data/6x6/6x6_D6_chi400_U8_SU2_proc_RMN_2.asc};
\addlegendentry{SU(2), $D$$=$$6$, $\chi$$=$$400$~~}

\end{axis}
\end{tikzpicture}
\caption{
  Relative error in the ground-state energy of the Hubbard model
  on a $6\times 6$ lattice with open
  boundary conditions, $U=8$,
  $S=0$, and $N=36$ (half filling),
  calculated with fPEPS with U(1) and SU(2) symmetry, as indicated,
  using supersweeps consisting of 3 full local optimization sweeps
  followed by 100 gradient optimizations, $E_0^{\text{(ss)}}$, with respect to the
  ground-state energy calculated using the DMRG, $E_0$, for various
  $D$ with the indicated $\chi$.
  The lowest energy for every local update sweep and
  the energy of every tenth gradient update is plotted.
}
\label{fig:6x6conv}
\end{figure}
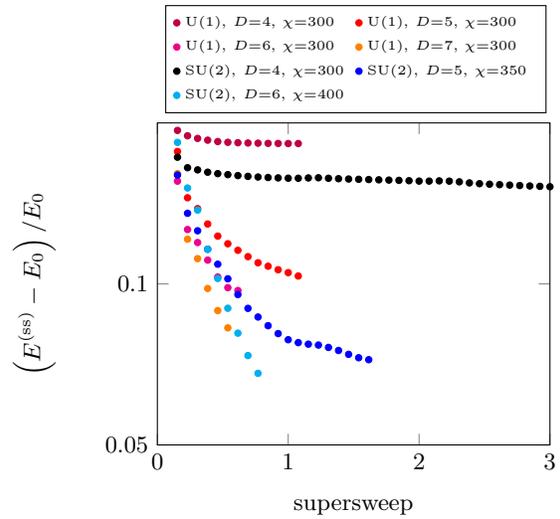

Fig.~\ref{fig:6x6conv} displays corresponding results for the 
accuracy of the ground-state energy on a $6\times 6$ lattice relative
to DMRG calculations.
(Note that the DMRG energy is no longer numerically exact on a
$6\times 6$ lattice, but should be
accurate to at least 2 or 3 significant digits more than the best
fPEPS calculation.)
The initial states are scaled-up versions of those used for the
half-filled $4\times 4$ calculations: columnwise states with
alternating columns of double- and zero-occupied sites for SU(2)
symmetry and a N\'eel state for U(1) symmetry.
The overall features of the energy convergence found in the
calculations for $U=8$ on the $3\times 3$ and $4\times 4$ lattices are
retained for this larger system size: there is systematic convergence
in supersweep step and in the energy convergence curves as $D$ is increased.
For a given bond dimension $D$, the SU(2) calculations are
substantially more accurate than the U(1) calculations, albeit at a
higher computational cost.
Due to the cost of the calculations, none of the runs could
practicably be carried out for more than two supersweeps except for the
computationally least expensive $D=4$, U(1) calculation.
The best accuracy of the energy relative to DMRG achieved in this
scope, for the $D=6$, SU(2) calculations, is approximately 7\%, almost
an order of magnitude less accurate 
than the best accuracy obtained for the $4 \times 4$ lattice.
Note that, as for the $4\times 4$ system, we vary $\chi$ from run to
run in order to balance stability and accuracy with computational
cost.
In particular, the $D=5$ and $D=6$, SU(2)-symmetric runs require higher
values of $\chi$ to ensure stability than runs with the corresponding
$D$ for the $4\times 4$ lattice.
As on the $4\times 4$ lattice, the SU(2), $D = 6$ case is far from
converged in the number of supersweeps.
The wall time needed for the calculation shown, which comprises less
than one full supersweep, is approximately 14 days, so that reasonable
convergence in the number of supersweeps would take at least several
months if not years of wall time on our reference compute node.
While it would be useful to be able to extract systematics in the
scaling of accuracy and computational cost with system size, $D$, and
$\chi$ for half-filled systems, it is not practical to do this here,
as we have not been able to achieve sufficient convergence with
reasonable computational resources for the two smallest system sizes;
treating larger systems is clearly outside of reasonable scope
for the computational resources available to us.

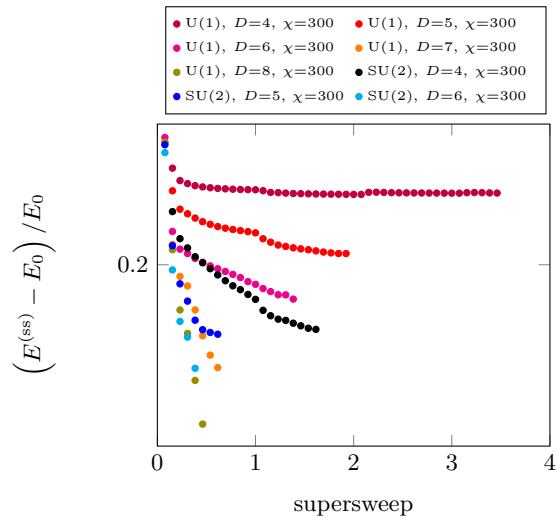
\begin{figure}[!htb]
\centering
\begin{tikzpicture}
\begin{axis}[width=0.45\textwidth,
                     x label style={at={(axis description cs:0.5,-0.12)},anchor=south},
                     y label style={at={(axis description cs:-0.001,.5)},anchor=south},
                     xlabel={supersweep},
                     ylabel={$\left(E^{\text{(ss)}}-E_0\right)/E_0$},
                     legend cell align={left},
                     legend columns=2,
                     legend style={font=\tiny,at={(0.02,1.03)},anchor=south west}, 
                     ymode=log,
                     log ticks with fixed point,
                     ytick={10,1,0.2,0.1},
                     xmin=0, xmax=4,
                     ymin=0.15, ymax=0.25]
                     ]

\addplot[only marks,purple,mark options={scale=0.6}]
   table[x=x,y=y]{data/8x8/8x8_D4_chi300_U8_U1_proc_RMN_2.asc};
\addlegendentry{U(1), $D$$=$$4$, $\chi$$=$$300$~~}

\addplot[only marks,red,mark options={scale=0.6}]
   table[x=x,y=y]{data/8x8/8x8_D5_chi300_U8_U1_proc_RMN_2.asc};
\addlegendentry{U(1), $D$$=$$5$, $\chi$$=$$300$~~}

\addplot[only marks,magenta,mark options={scale=0.6}]
   table[x=x,y=y]{data/8x8/8x8_D6_chi300_U8_U1_proc_RMN_2.asc};
\addlegendentry{U(1), $D$$=$$6$, $\chi$$=$$300$~~}

\addplot[only marks,orange,mark options={scale=0.6}]
   table[x=x,y=y]{data/8x8/8x8_D7_chi300_U8_U1_proc_RMN_2.asc};
\addlegendentry{U(1), $D$$=$$7$, $\chi$$=$$300$~~}

\addplot[only marks,olive,mark options={scale=0.6}]
   table[x=x,y=y]{data/8x8/8x8_D8_chi300_U8_U1_proc_RMN_2.asc};
\addlegendentry{U(1), $D$$=$$8$, $\chi$$=$$300$~~}

\addplot[only marks,
  black,mark options={scale=0.6}]
   table[x=x,y=y]{data/8x8_su2_pair_init/8x8_D4_chi300_U8_SU2_proc_RMN_2.asc};
\addlegendentry{SU(2), $D$$=$$4$, $\chi$$=$$300$~~}

\addplot[only marks,
  blue,mark options={scale=0.6}]
   table[x=x,y=y]{data/8x8_su2_pair_init/8x8_D5_chi300_U8_SU2_proc_RMN_2.asc};
\addlegendentry{SU(2), $D$$=$$5$, $\chi$$=$$300$~~}

\addplot[only marks,
  cyan,mark options={scale=0.6}]
   table[x=x,y=y]{data/8x8_su2_pair_init/8x8_D6_chi300_U8_SU2_proc_RMN_2.asc};
\addlegendentry{SU(2), $D$$=$$6$, $\chi$$=$$300$~~}

\end{axis}
\end{tikzpicture}
\caption{
  Relative error in the ground-state energy for the Hubbard model
  on an $8\times 8$ lattice for $U=8$,
  $S=0$, and $N=56$ ($\langle n \rangle = 0.875$),
  calculated with fPEPS with U(1) and SU(2) symmetry, as indicated,
  using supersweeps consisting of 3 full local optimization sweeps
  followed by 100 gradient optimizations, $E_0^{\text{(ss)}}$, with respect to the
  ground-state energy from a DMRG calculation (see text), $E_0$.
  For every local update sweep, the lowest energy is plotted, and
  the energy of every tenth gradient update is plotted.
}
\label{fig:8x8conv}
\end{figure}

However, in order to explore to what extent the fPEPS method can
capture the essential physics of the doped two-dimensional Hubbard
model qualitatively, we nevertheless present here simulation results for
the $8\times 8$ lattice at $1/8$-doping, i.e., with $N=56$
electrons so that $\langle n\rangle = 0.875$.
The simulations were initialized with scaled-up versions of the pair
SU(2) state, Fig.~\ref{fig:3x3_product_states}(d), for the SU(2)
simulations and with scaled-up versions of columnwise U(1) state for
the U(1) simulations.
We have found that the choice of initial state does have a substantial
effect on the initial convergence of the SU(2) simulations especially;
for example, the accuracy is substantially lower if a columnwise SU(2)
state, i.e., a scaled-up version of
Fig.~\ref{fig:3x3_product_states}(a), is used for the SU(2)
simulations.
Fig.~\ref{fig:8x8conv} displays the error in the ground-state energy
relative to DMRG calculations.
Here we expect the DMRG calculations to have substantially lower
relative accuracy than then $6\times 6$ calculations, so that we
estimate that the 
relative error in the ground-state energy obtained using the DMRG
could be up to the order of 1\%.
However, as can be seen in Fig.~\ref{fig:8x8conv}, the deviation of
the ground-state energy calculated with fPEPS from that calculated
using the DMRG does not go below 10\%.
In this sense, it is not necessary to have more accurate variational
estimates of the ground-state energy for comparison.
Surprisingly, we find that a slightly lower $\chi$, as compared to the 
$6\times 6$, half-filled system, is sufficient for
stable SU(2) simulations, despite the fact  
that the system size is larger.
Thus, doping seems to lead to greater stability in the
variational optimization for given $D$ and $\chi$.
The origins of this behavior are unclear: while the Hilbert space of
the doped system is smaller than the half-filled system, the 1/8-doped
$8\times 8$ system should nevertheless have a larger Hilbert space than
the half-filled $6\times 6$ system.
It could be that the breaking of translational invariance in both the
charge and spin sectors due to stripe structures (see below) helps
increase the stability of the calculations.

The smaller environment dimensions $\chi$ required allow us to carry
out calculations with comparable values of $D$
(including even a higher value, $D = 8$, in the U(1) calculations) and
a comparable number of supersweep steps as for the half-filled
$6\times 6$ system.
The convergence behavior is qualitatively similar to that
of the $U=8$ calculations on all other system sizes in that the convergence
systematically improves as a function of supersweep step, of $D$, and
in going from U(1) to SU(2) symmetry.
The wavelike variations at the boundaries between supersweeps is also
present.
Note however, that the lowest ground-state energy obtained (actually
for the $D=8$ U(1) calculation) is about 10\% above the DMRG
energy.
The wall time required for the $D=8$, U(1) calculation, which consists
of approximately half of a supersweep, was a little more than 21 days
on our reference compute node. 
For the $D=6$, SU(2) calculation, the wall time was almost 29 days.

In order to gain insight into the convergence behavior and understand
to what extent the fPEPS calculations can capture the physical
behavior of the doped two-dimensional Hubbard model qualitatively, we
now examine the local spin and hole densities for both U(1) and SU(2)
calculations.
We note that doing this on the $8\times 8$ rather than on smaller
lattices is appropriate because the ground state is expected to have a
stripe structure with wavelength $\lambda = 8$
stripes~\cite{Zheng2017Dec}.
Such a stripe structure would be frustrated on smaller lattices,
leading to complicated patterns in the local spin and hole densities
and, potentially, to instabilities in the energetically most favorable
configurations.

We depict the local spin density
$\langle S^z_i \rangle = \frac{1}{2}\left (n_{i,\uparrow} - n_{i,\downarrow}  \right )$
and local hole density $1-\langle n_i \rangle$ graphically on the
$8\times 8$ lattice for the most accurate U(1)-symmetric calculation,
the one with $D=8$ and $\chi=300$, in Fig.~\ref{fig:8x8u1}.
(Since the U(1) calculation breaks spin-rotation symmetry, the
magnetic structure can best be seen using maps of $\langle S^z_i \rangle$.)
The stripe structures found in Ref.~\cite{Zheng2017Dec} using a
variety of independent, state-of-the-art numerical methods is well
reproduced:
There is a vertical site-centered hole stripe separating antiferromagnetic
regions with an antiphase boundary between the regions that has a total
wavelength of $\lambda=8$.
Note that this state is significantly evolved from the initial state,
which is a product state with a N\'eel configuration in the first
seven columns and a vertical line of holes in the eighth column.
If one examines the corresponding spin and hole densities as the
convergence progresses, i.e., as more supersweep steps are carried out,
one can see that the initial configuration evolves step-by-step to the
configuration depicted in Fig.~\ref{fig:8x8u1} in that the line of
holes on the eighth column are moved to the left and spread out and
that the antiferromagnetic spin structure with opposite phase forms to
the right of the stripe.
Stripe structures intermediate between the initial product state and
those of Fig.~\ref{fig:8x8u1} are also found in U(1) calculations
with smaller bond dimension $D$.

\begin{figure}[!htb]
\centering
\scalebox{0.7}{
	\begin{tikzpicture}[font=\tiny]
	\draw[fill=green] (0,0) circle [radius=0.005668341665729049];
\draw[fill=green] (1,0) circle [radius=0.024409086111898848];
\draw[fill=green] (2,0) circle [radius=0.0867757593246911];
\draw[fill=green] (3,0) circle [radius=0.19460983189472703];
\draw[fill=green] (4,0) circle [radius=0.23098213729162348];
\draw[fill=green] (5,0) circle [radius=0.2395969479141349];
\draw[fill=green] (6,0) circle [radius=0.1328507157975486];
\draw[fill=green] (7,0) circle [radius=0.06878327378341503];
\draw[fill=green] (0,1) circle [radius=-0.0025561452871475376];
\draw[fill=green] (1,1) circle [radius=0.0167432492824149];
\draw[fill=green] (2,1) circle [radius=0.07868140242553057];
\draw[fill=green] (3,1) circle [radius=0.18392152317036503];
\draw[fill=green] (4,1) circle [radius=0.27712463770399953];
\draw[fill=green] (5,1) circle [radius=0.24910952191192892];
\draw[fill=green] (6,1) circle [radius=0.1325412039123339];
\draw[fill=green] (7,1) circle [radius=0.05616878401821279];
\draw[fill=green] (0,2) circle [radius=0.0035971373344185348];
\draw[fill=green] (1,2) circle [radius=0.014390623244099787];
\draw[fill=green] (2,2) circle [radius=0.06677330870013536];
\draw[fill=green] (3,2) circle [radius=0.23747288366681485];
\draw[fill=green] (4,2) circle [radius=0.30568438766263034];
\draw[fill=green] (5,2) circle [radius=0.2471216069515361];
\draw[fill=green] (6,2) circle [radius=0.09814496359696412];
\draw[fill=green] (7,2) circle [radius=0.04404340211932972];
\draw[fill=green] (0,3) circle [radius=0.0015860244565806259];
\draw[fill=green] (1,3) circle [radius=0.023793967705829963];
\draw[fill=green] (2,3) circle [radius=0.07745837161040359];
\draw[fill=green] (3,3) circle [radius=0.2359779365348079];
\draw[fill=green] (4,3) circle [radius=0.3274610801720208];
\draw[fill=green] (5,3) circle [radius=0.19241662159546769];
\draw[fill=green] (6,3) circle [radius=0.07707005765116964];
\draw[fill=green] (7,3) circle [radius=0.030702828219246525];
\draw[fill=green] (0,4) circle [radius=0.004140853193836697];
\draw[fill=green] (1,4) circle [radius=0.023124996437977385];
\draw[fill=green] (2,4) circle [radius=0.10095180710510447];
\draw[fill=green] (3,4) circle [radius=0.27055956491201727];
\draw[fill=green] (4,4) circle [radius=0.292402587070052];
\draw[fill=green] (5,4) circle [radius=0.20632921217410163];
\draw[fill=green] (6,4) circle [radius=0.0927902407561757];
\draw[fill=green] (7,4) circle [radius=0.03770858941596926];
\draw[fill=green] (0,5) circle [radius=0.005253255216019936];
\draw[fill=green] (1,5) circle [radius=0.024857435621919027];
\draw[fill=green] (2,5) circle [radius=0.08565821659893336];
\draw[fill=green] (3,5) circle [radius=0.19201010385611836];
\draw[fill=green] (4,5) circle [radius=0.2626148114752568];
\draw[fill=green] (5,5) circle [radius=0.24891258050765053];
\draw[fill=green] (6,5) circle [radius=0.1221506216022927];
\draw[fill=green] (7,5) circle [radius=0.04473466742226728];
\draw[fill=green] (0,6) circle [radius=0.009154461498779881];
\draw[fill=green] (1,6) circle [radius=0.03398558677997965];
\draw[fill=green] (2,6) circle [radius=0.1008756061312725];
\draw[fill=green] (3,6) circle [radius=0.2562293712009589];
\draw[fill=green] (4,6) circle [radius=0.27843863232818705];
\draw[fill=green] (5,6) circle [radius=0.20782157669401521];
\draw[fill=green] (6,6) circle [radius=0.09848821330696711];
\draw[fill=green] (7,6) circle [radius=0.041190491136071525];
\draw[fill=green] (0,7) circle [radius=0.010845680552274417];
\draw[fill=green] (1,7) circle [radius=0.03317823743905013];
\draw[fill=green] (2,7) circle [radius=0.10043032522122064];
\draw[fill=green] (3,7) circle [radius=0.1838059210143978];
\draw[fill=green] (4,7) circle [radius=0.23956290303547217];
\draw[fill=green] (5,7) circle [radius=0.24267040726751488];
\draw[fill=green] (6,7) circle [radius=0.1321618976719045];
\draw[fill=green] (7,7) circle [radius=0.05784930928029819];
\draw[->,thick,blue] (0,-0.17674315682525116)--(0,0.17674315682525116);
\draw[->,thick,red] (1,0.16494132232599507)--(1,-0.16494132232599507);
\draw[->,thick,blue] (2,-0.16874176117489756)--(2,0.16874176117489756);
\draw[->,thick,red] (3,0.1000987273290548)--(3,-0.1000987273290548);
\draw[->,thick,blue] (4,-0.0240523714169453)--(4,0.0240523714169453);
\draw[->,thick,blue] (5,-0.06685886261260626)--(5,0.06685886261260626);
\draw[->,thick,red] (6,0.08320724443966869)--(6,-0.08320724443966869);
\draw[->,thick,blue] (7,-0.13672142692938571)--(7,0.13672142692938571);
\draw[->,thick,red] (0,1.1918796275183958)--(0,0.8081203724816042);
\draw[->,thick,blue] (1,0.8076786321055529)--(1,1.192321367894447);
\draw[->,thick,red] (2,1.185847060670905)--(2,0.8141529393290948);
\draw[->,thick,blue] (3,0.8680713111564968)--(3,1.1319286888435032);
\draw[->,thick,red] (4,1.03356592644736)--(4,0.96643407355264);
\draw[->,thick,red] (5,1.0834325256759392)--(5,0.9165674743240607);
\draw[->,thick,blue] (6,0.8659137259741665)--(6,1.1340862740258335);
\draw[->,thick,red] (7,1.1759953497281863)--(7,0.8240046502718137);
\draw[->,thick,blue] (0,1.799674525332934)--(0,2.200325474667066);
\draw[->,thick,red] (1,2.1813328246919816)--(1,1.8186671753080184);
\draw[->,thick,blue] (2,1.8107007035839635)--(2,2.1892992964160363);
\draw[->,thick,red] (3,2.1283517853963447)--(3,1.8716482146036555);
\draw[->,thick,blue] (4,1.9909715148148557)--(4,2.0090284851851443);
\draw[->,thick,blue] (5,1.8892964740112577)--(5,2.1107035259887423);
\draw[->,thick,red] (6,2.1616861822704307)--(6,1.8383138177295693);
\draw[->,thick,blue] (7,1.808075305579448)--(7,2.191924694420552);
\draw[->,thick,red] (0,3.1952644456741295)--(0,2.8047355543258705);
\draw[->,thick,blue] (1,2.8148291295055814)--(1,3.1851708704944186);
\draw[->,thick,red] (2,3.1863626599171164)--(2,2.8136373400828836);
\draw[->,thick,blue] (3,2.8874191839087113)--(3,3.1125808160912887);
\draw[->,thick,blue] (4,2.9825619275248667)--(4,3.0174380724751333);
\draw[->,thick,red] (5,3.1309059391083305)--(5,2.8690940608916695);
\draw[->,thick,blue] (6,2.836816696421577)--(6,3.163183303578423);
\draw[->,thick,red] (7,3.1930881196480123)--(7,2.8069118803519877);
\draw[->,thick,blue] (0,3.800716433887223)--(0,4.1992835661127765);
\draw[->,thick,red] (1,4.187920846824399)--(1,3.812079153175601);
\draw[->,thick,blue] (2,3.8230956759151944)--(2,4.176904324084806);
\draw[->,thick,red] (3,4.09029300080464)--(3,3.9097069991953606);
\draw[->,thick,red] (4,4.021948964534853)--(4,3.978051035465147);
\draw[->,thick,blue] (5,3.871897166242899)--(5,4.128102833757101);
\draw[->,thick,red] (6,4.165389582674855)--(6,3.834610417325145);
\draw[->,thick,blue] (7,3.8123033794066608)--(7,4.187696620593339);
\draw[->,thick,red] (0,5.193486763969204)--(0,4.806513236030796);
\draw[->,thick,blue] (1,4.827871223152402)--(1,5.172128776847598);
\draw[->,thick,red] (2,5.175268083516201)--(2,4.824731916483799);
\draw[->,thick,blue] (3,4.885952384563741)--(3,5.114047615436259);
\draw[->,thick,red] (4,5.028472578292085)--(4,4.971527421707915);
\draw[->,thick,red] (5,5.082718585113141)--(5,4.917281414886859);
\draw[->,thick,blue] (6,4.846762602591797)--(6,5.153237397408203);
\draw[->,thick,red] (7,5.187564548883955)--(7,4.812435451116045);
\draw[->,thick,blue] (0,5.812404885290361)--(0,6.187595114709639);
\draw[->,thick,red] (1,6.159946063576622)--(1,5.840053936423378);
\draw[->,thick,blue] (2,5.8358564781868045)--(2,6.1641435218131955);
\draw[->,thick,red] (3,6.087265836532785)--(3,5.912734163467215);
\draw[->,thick,red] (4,6.038771422954876)--(4,5.961228577045124);
\draw[->,thick,blue] (5,5.873111181847606)--(5,6.126888818152394);
\draw[->,thick,red] (6,6.17302805965759)--(6,5.82697194034241);
\draw[->,thick,blue] (7,5.812939510633344)--(7,6.187060489366656);
\draw[->,thick,red] (0,7.1688335033772494)--(0,6.8311664966227506);
\draw[->,thick,blue] (1,6.863796495268076)--(1,7.136203504731924);
\draw[->,thick,red] (2,7.1559289189281525)--(2,6.8440710810718475);
\draw[->,thick,blue] (3,6.904553090401734)--(3,7.095446909598266);
\draw[->,thick,red] (4,7.022714747910186)--(4,6.977285252089814);
\draw[->,thick,red] (5,7.066716054536247)--(5,6.933283945463753);
\draw[->,thick,blue] (6,6.865060693643791)--(6,7.134939306356209);
\draw[->,thick,red] (7,7.172559741257425)--(7,6.827440258742575);
\draw[rounded corners] ((0, -0.5)--(7.5, -0.5)--(7.5, 7.5)--(-0.5, 7.5)--(-0.5, -0.5)--(0, -0.5);\node[align=center,text width=6mm] at (0, 7.7) { \footnotesize 0.38}
;\node[align=center,text width=6mm] at (1, 7.7) { \footnotesize 0.34}
;\node[align=center,text width=6mm] at (2, 7.7) { \footnotesize 0.35}
;\node[align=center,text width=6mm] at (3, 7.7) { \footnotesize 0.22}
;\node[align=center,text width=6mm] at (4, 7.7) { \footnotesize 0.05}
;\node[align=center,text width=6mm] at (5, 7.7) { \footnotesize 0.2}
;\node[align=center,text width=6mm] at (6, 7.7) { \footnotesize 0.29}
;\node[align=center,text width=6mm] at (7, 7.7) { \footnotesize 0.36}
;\node[align=center,text width=6mm,color=black!30!green] at (0, -0.7) { \footnotesize 0.01}
;\node[align=center,text width=6mm,color=black!30!green] at (1, -0.7) { \footnotesize 0.02}
;\node[align=center,text width=6mm,color=black!30!green] at (2, -0.7) { \footnotesize 0.09}
;\node[align=center,text width=6mm,color=black!30!green] at (3, -0.7) { \footnotesize 0.22}
;\node[align=center,text width=6mm,color=black!30!green] at (4, -0.7) { \footnotesize 0.28}
;\node[align=center,text width=6mm,color=black!30!green] at (5, -0.7) { \footnotesize 0.23}
;\node[align=center,text width=6mm,color=black!30!green] at (6, -0.7) { \footnotesize 0.11}
;\node[align=center,text width=6mm,color=black!30!green] at (7, -0.7) { \footnotesize 0.05}
;
	\end{tikzpicture}
}
\caption{
  Local $z$-component of the spin $\langle S^z_i \rangle =
  \frac{1}{2}\left (n_{i,\uparrow} - n_{i,\downarrow}  \right )$
  (size, color, and direction of arrows) and local hole density
  $1-\langle n_i \rangle$ (diameter of
  green-shaded circles) on an
  $8\times 8$ lattice with open boundary conditions calculated with
  U(1) symmetry, bond dimension $D=8$, and $\chi=300$.
  Here $U=8$, $S_z=0$, and $N=56$ so that $\langle n\rangle = 0.875$.
  The black numbers are the average $\langle S^z_i \rangle$ for the column of
  sites below, 
  and the green numbers on the bottom
  edge are the average hole densities for the column of sites above.
}
\label{fig:8x8u1}
\end{figure}
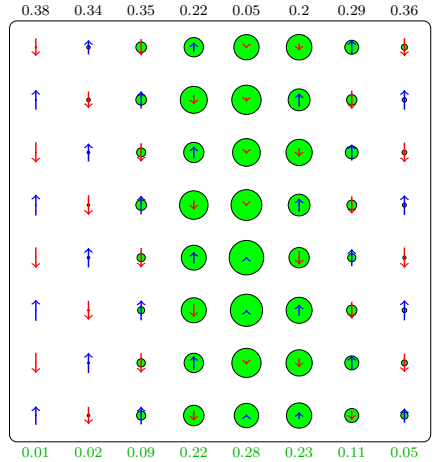

\begin{figure}[!htb]
\centering
\scalebox{0.7}{
	\begin{tikzpicture}[font=\tiny]
	\draw[fill=green] (0,0) circle [radius=0.02994078920422616];
\draw[fill=green] (1,0) circle [radius=0.06029988299308231];
\draw[fill=green] (2,0) circle [radius=0.16682398119101183];
\draw[fill=green] (3,0) circle [radius=0.1731468521911136];
\draw[fill=green] (4,0) circle [radius=0.17039646763922434];
\draw[fill=green] (5,0) circle [radius=0.17070323456833536];
\draw[fill=green] (6,0) circle [radius=0.06134622681132462];
\draw[fill=green] (7,0) circle [radius=0.034359435560335494];
\draw[fill=green] (0,1) circle [radius=0.03411723486375051];
\draw[fill=green] (1,1) circle [radius=0.07093809544358765];
\draw[fill=green] (2,1) circle [radius=0.20265188892012442];
\draw[fill=green] (3,1) circle [radius=0.18980022355398374];
\draw[fill=green] (4,1) circle [radius=0.21647702779690525];
\draw[fill=green] (5,1) circle [radius=0.25870741913802253];
\draw[fill=green] (6,1) circle [radius=0.10673866195045412];
\draw[fill=green] (7,1) circle [radius=0.04793111756395685];
\draw[fill=green] (0,2) circle [radius=0.010938486361200872];
\draw[fill=green] (1,2) circle [radius=0.02728876866763741];
\draw[fill=green] (2,2) circle [radius=0.10004776904943813];
\draw[fill=green] (3,2) circle [radius=0.1417966668744418];
\draw[fill=green] (4,2) circle [radius=0.27953087250556086];
\draw[fill=green] (5,2) circle [radius=0.27153956187776973];
\draw[fill=green] (6,2) circle [radius=0.11907840948503878];
\draw[fill=green] (7,2) circle [radius=0.06500503792883638];
\draw[fill=green] (0,3) circle [radius=0.002744343150714923];
\draw[fill=green] (1,3) circle [radius=0.013268358218714638];
\draw[fill=green] (2,3) circle [radius=0.07401887098327653];
\draw[fill=green] (3,3) circle [radius=0.11990621304491089];
\draw[fill=green] (4,3) circle [radius=0.2719864279085047];
\draw[fill=green] (5,3) circle [radius=0.3089233987077401];
\draw[fill=green] (6,3) circle [radius=0.13023614206019551];
\draw[fill=green] (7,3) circle [radius=0.06487516018029393];
\draw[fill=green] (0,4) circle [radius=0.02227908041838733];
\draw[fill=green] (1,4) circle [radius=0.04722084817675287];
\draw[fill=green] (2,4) circle [radius=0.15679517582685476];
\draw[fill=green] (3,4) circle [radius=0.16371368651215323];
\draw[fill=green] (4,4) circle [radius=0.2356670107219141];
\draw[fill=green] (5,4) circle [radius=0.26245293438438233];
\draw[fill=green] (6,4) circle [radius=0.08044200877163221];
\draw[fill=green] (7,4) circle [radius=0.03690576699643];
\draw[fill=green] (0,5) circle [radius=0.01386204265508273];
\draw[fill=green] (1,5) circle [radius=0.0333654748445521];
\draw[fill=green] (2,5) circle [radius=0.12272805399074371];
\draw[fill=green] (3,5) circle [radius=0.14578158891404258];
\draw[fill=green] (4,5) circle [radius=0.24156554614505743];
\draw[fill=green] (5,5) circle [radius=0.27529394253518524];
\draw[fill=green] (6,5) circle [radius=0.10896824859921028];
\draw[fill=green] (7,5) circle [radius=0.05783867993164615];
\draw[fill=green] (0,6) circle [radius=0.007871748761180797];
\draw[fill=green] (1,6) circle [radius=0.020962078006003915];
\draw[fill=green] (2,6) circle [radius=0.08224815351766979];
\draw[fill=green] (3,6) circle [radius=0.11143520780592453];
\draw[fill=green] (4,6) circle [radius=0.2856193536378021];
\draw[fill=green] (5,6) circle [radius=0.2555327816553564];
\draw[fill=green] (6,6) circle [radius=0.10191441431555304];
\draw[fill=green] (7,6) circle [radius=0.06891374934283201];
\draw[fill=green] (0,7) circle [radius=0.004043369084150705];
\draw[fill=green] (1,7) circle [radius=0.01705838534969306];
\draw[fill=green] (2,7) circle [radius=0.08847190887468748];
\draw[fill=green] (3,7) circle [radius=0.12398662312958275];
\draw[fill=green] (4,7) circle [radius=0.27634889909657256];
\draw[fill=green] (5,7) circle [radius=0.31454508417775484];
\draw[fill=green] (6,7) circle [radius=0.15383057252483867];
\draw[fill=green] (7,7) circle [radius=0.08678829099851348];
\draw[->,thick,blue] (0,-0.22891267637327292)--(0,0.22891267637327292);
\draw[->,thick,blue] (1,-0.2176084096388781)--(1,0.2176084096388781);
\draw[->,thick,blue] (2,-0.19206771386608779)--(2,0.19206771386608779);
\draw[->,thick,blue] (3,-0.19241113198512216)--(3,0.19241113198512216);
\draw[->,thick,blue] (4,-0.19262290374986607)--(4,0.19262290374986607);
\draw[->,thick,blue] (5,-0.19383279823041746)--(5,0.19383279823041746);
\draw[->,thick,blue] (6,-0.2178001250276402)--(6,0.2178001250276402);
\draw[->,thick,blue] (7,-0.22931277597892638)--(7,0.22931277597892638);
\draw[->,thick,blue] (0,0.7723396258167896)--(0,1.2276603741832104);
\draw[->,thick,blue] (1,0.7851706320656122)--(1,1.2148293679343878);
\draw[->,thick,blue] (2,0.8147746201402528)--(2,1.1852253798597472);
\draw[->,thick,blue] (3,0.8115495860358194)--(3,1.1884504139641805);
\draw[->,thick,blue] (4,0.8182484738745299)--(4,1.18175152612547);
\draw[->,thick,blue] (5,0.8256441726774323)--(5,1.1743558273225676);
\draw[->,thick,blue] (6,0.7927706077920522)--(6,1.2072293922079478);
\draw[->,thick,blue] (7,0.7731033824199971)--(7,1.226896617580003);
\draw[->,thick,blue] (0,1.7652008879848857)--(0,2.2347991120151143);
\draw[->,thick,blue] (1,1.774732919004713)--(1,2.225267080995287);
\draw[->,thick,blue] (2,1.7914738639219685)--(2,2.2085261360780315);
\draw[->,thick,blue] (3,1.8004438523156607)--(3,2.1995561476843393);
\draw[->,thick,blue] (4,1.8314427689400645)--(4,2.1685572310599355);
\draw[->,thick,blue] (5,1.8298003726826226)--(5,2.1701996273173774);
\draw[->,thick,blue] (6,1.7961457162343009)--(6,2.203854283765699);
\draw[->,thick,blue] (7,1.7786164399478779)--(7,2.221383560052122);
\draw[->,thick,blue] (0,2.763571243074329)--(0,3.236428756925671);
\draw[->,thick,blue] (1,2.7711726550425437)--(1,3.2288273449574563);
\draw[->,thick,blue] (2,2.7847990399223326)--(2,3.2152009600776674);
\draw[->,thick,blue] (3,2.794346125626837)--(3,3.205653874373163);
\draw[->,thick,blue] (4,2.8293365029639106)--(4,3.1706634970360894);
\draw[->,thick,blue] (5,2.837663716562589)--(5,3.162336283437411);
\draw[->,thick,blue] (6,2.797598651401329)--(6,3.202401348598671);
\draw[->,thick,blue] (7,2.77899347308165)--(7,3.22100652691835);
\draw[->,thick,blue] (0,3.767746531798384)--(0,4.232253468201616);
\draw[->,thick,blue] (1,3.779088953191896)--(1,4.220911046808104);
\draw[->,thick,blue] (2,3.803622204098904)--(2,4.196377795901096);
\draw[->,thick,blue] (3,3.807418402444543)--(3,4.192581597555456);
\draw[->,thick,blue] (4,3.8216271028231685)--(4,4.178372897176831);
\draw[->,thick,blue] (5,3.82734892802099)--(5,4.17265107197901);
\draw[->,thick,blue] (6,3.7865477830451715)--(6,4.2134522169548285);
\draw[->,thick,blue] (7,3.7735961256210744)--(7,4.226403874378925);
\draw[->,thick,blue] (0,4.765771860068591)--(0,5.234228139931409);
\draw[->,thick,blue] (1,4.775394369463556)--(1,5.224605630536444);
\draw[->,thick,blue] (2,4.796179098313632)--(2,5.203820901686368);
\draw[->,thick,blue] (3,4.802198951152544)--(3,5.197801048847456);
\draw[->,thick,blue] (4,4.822260346156675)--(4,5.177739653843325);
\draw[->,thick,blue] (5,4.830892300991397)--(5,5.169107699008603);
\draw[->,thick,blue] (6,4.7930101239718885)--(6,5.2069898760281115);
\draw[->,thick,blue] (7,4.777243314154315)--(7,5.222756685845685);
\draw[->,thick,blue] (0,5.7656878684492305)--(0,6.2343121315507695);
\draw[->,thick,blue] (1,5.772888933631676)--(1,6.227111066368324);
\draw[->,thick,blue] (2,5.787438065005878)--(2,6.212561934994122);
\draw[->,thick,blue] (3,5.793838072268816)--(3,6.206161927731184);
\draw[->,thick,blue] (4,5.832553092082895)--(4,6.167446907917105);
\draw[->,thick,blue] (5,5.825712358501642)--(5,6.174287641498358);
\draw[->,thick,blue] (6,5.793441754798883)--(6,6.206558245201117);
\draw[->,thick,blue] (7,5.780779511698652)--(7,6.219220488301348);
\draw[->,thick,blue] (0,6.766375962919804)--(0,7.233624037080196);
\draw[->,thick,blue] (1,6.773610436183203)--(1,7.226389563816797);
\draw[->,thick,blue] (2,6.7888255260033965)--(2,7.2111744739966035);
\draw[->,thick,blue] (3,6.799078708985974)--(3,7.200921291014026);
\draw[->,thick,blue] (4,6.829501299083483)--(4,7.170498700916517);
\draw[->,thick,blue] (5,6.838779668613752)--(5,7.161220331386248);
\draw[->,thick,blue] (6,6.8018288743696305)--(6,7.1981711256303695);
\draw[->,thick,blue] (7,6.782833066110139)--(7,7.217166933889861);
\draw[rounded corners] ((0, -0.5)--(7.5, -0.5)--(7.5, 7.5)--(-0.5, 7.5)--(-0.5, -0.5)--(0, -0.5);
\node[align=center,text width=6mm] at (0, 7.7) { \footnotesize 0.71}    %%% 0.47
;\node[align=center,text width=6mm] at (1, 7.7) { \footnotesize 0.68}   %%% 0.45
;\node[align=center,text width=6mm] at (2, 7.7) { \footnotesize 0.62}   %%% 0.41
;\node[align=center,text width=6mm] at (3, 7.7) { \footnotesize 0.6}    %%% 0.4
;\node[align=center,text width=6mm] at (4, 7.7) { \footnotesize 0.53}   %%% 0.35
;\node[align=center,text width=6mm] at (5, 7.7) { \footnotesize 0.51}   %%% 0.34
;\node[align=center,text width=6mm] at (6, 7.7) { \footnotesize 0.62}   %%% 0.41
;\node[align=center,text width=6mm] at (7, 7.7) { \footnotesize 0.68}   %%% 0.45
;\node[align=center,text width=6mm,color=black!30!green] at (0, -0.7) { \footnotesize 0.02}
;\node[align=center,text width=6mm,color=black!30!green] at (1, -0.7) { \footnotesize 0.04}
;\node[align=center,text width=6mm,color=black!30!green] at (2, -0.7) { \footnotesize 0.12}
;\node[align=center,text width=6mm,color=black!30!green] at (3, -0.7) { \footnotesize 0.15}
;\node[align=center,text width=6mm,color=black!30!green] at (4, -0.7) { \footnotesize 0.25}
;\node[align=center,text width=6mm,color=black!30!green] at (5, -0.7) { \footnotesize 0.26}
;\node[align=center,text width=6mm,color=black!30!green] at (6, -0.7) { \footnotesize 0.11}
;\node[align=center,text width=6mm,color=black!30!green] at (7, -0.7) { \footnotesize 0.06}
;
	\end{tikzpicture}
}
\caption{
  Local spin density $\langle\mathbf{S}^2_i\rangle$ (size of
  blue arrows)
  and local hole density $1 - \langle n_i \rangle$ (diameter of
  green-shaded circles)  for the Hubbard model on an $8\times 8$
  lattice with open boundary conditions and
  $U=8$, $S=0$, and $N=56$ so that $\langle n\rangle = 0.875$,
  calculated with SU(2) symmetry, bond dimension $D=6$, and $\chi=300$.
  The black numbers are the average spin density for the column of
  sites below, 
  and the green numbers on the bottom
  edge are the average hole densities for the column of sites above.
}
\label{fig:8x8su2}
\end{figure}
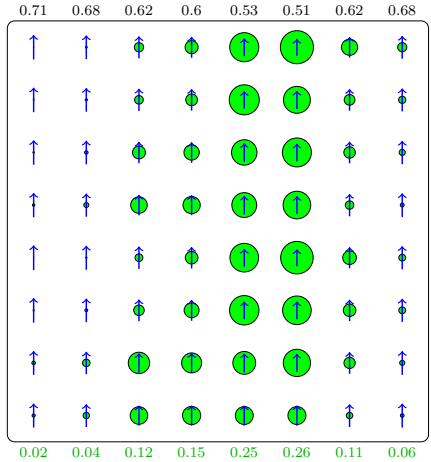

We also examine the spin and hole densities for the most accurate
SU(2)-symmetric calculations, for $D=6$ and $\chi=300$, in
Fig.~\ref{fig:8x8su2}.
Since $\langle S^z_i \rangle = 0$ identically on all sites $i$ for
SU(2)-symmetric calculations, we instead depict the local spin moment
$\langle\mathbf{S}^2_i\rangle$, which is large in magnetically ordered
regions, but cannot depict the antiferromagnetic structure explicitly.
(This could be done by calculating spin-spin correlations
between pairs of sites.)
The initial product state for the SU(2) calculation is a pair-bond
state of the type depicted in Fig.~\ref{fig:3x3_product_states}(d) for
the $3\times 3$ lattice, but with vertical bond-singlet states
covering the first seven columns and a vertical line of zero-occupied
local states in the eighth column.
As can be seen in Fig.~\ref{fig:8x8su2}, this state evolves towards a
stripe configuration roughly consistent with that of the U(1)
calculation, Fig.~\ref{fig:8x8u1}.
We note that the distribution of holes is, however, not quite as
symmetric in both the horizontal and in the vertical directions as
that of the U(1) calculation and that the hole stripe is not as
clearly site-centered on the fifth column of sites.
It is important to realize that this calculation is fairly far from
convergence in that only three local update sweeps and 30 gradient updates have been
carried out within the first supersweep and that the energy is
substantially higher than the lowest energy for the U(1),
$D=8$ calculation in Fig.~\ref{fig:8x8conv}.
Thus, we ascribe these discrepancies to the fact that the SU(2)
calculation is, relatively speaking, more poorly converged than the
U(1) calculation.

\section{Summary and Discussion}
\label{sec:discussion}
In this paper, we have developed the groundwork to carry out
variational optimization of finite projected entangled pair
states.
We treat 
the two-dimensional Hubbard model
in this work because we are interested in developing methods for
short-range itinerant fermionic models in which a high degree of symmetry is
present.
The single-band model with nearest-neighbor hopping displays
interesting and complicated physical behavior, which is partially, but
not fully, understood, and is accessible to a set of other numerical
methods, albeit at a level that pushes the limits of these methods.
In this sense, it is suited to be a stringent and demanding test bed for
our methods.
However, we emphasize that the fPEPS method is
applicable to wide set of short-ranged, two-dimensional quantum
lattice models: in particular, to quantum-spin-based models such as
Heisenberg and $t$-$J$ models, to spinless-fermion models, as well as
to extended versions of the Hubbard model that include features such as
longer-range hopping or interaction and multiple bands.

Our method is based on the following
essential building blocks:
\textit{(i)} a general framework to create and handle projected
entangled pair operators for arbitrary local Hamiltonians (Sec.~\ref{sec:pepo}),
\textit{(ii)} a generic scheme to incorporate
$\text{SU(2)}_{\text{spin}}$-symmetry (Sec.~\ref{sec:su2}) into PEPS
states at the level of tensor representations,
\textit{(iii)} a new procedure which optimizes the environment within
PEPS-based contractions using a two-block configuration (Sec.~\ref{sec:pepsexp}),
and \textit{(iv)} the utilization of a generalized version of the controlled bond
expansion \citep{Gleis2022Jul} to circumvent contracting large tensors.
We have integrated these technical contributions with
methods and representations from the extensive
existing body of knowledge about tensor networks
in order to formulate a comprehensive,
PEPO-based framework for calculating ground-state properties of
two-dimensional quantum lattice systems.

Given this framework, which enables us to represent variational states
as finite PEPS and operators on these states as PEPOs, the crucial
component to build an algorithm to approximate the ground state is a
method to optimize the fPEPS tensor network for a given Hamiltonian,
i.e., given a PEPO.
Here we have formulated two optimization methods:
a local-update optimization procedure that optimized two adjacent local tensors
within the fPEPS via iterative diagonalization
and a gradient-update optimization scheme that
optimizes the elements of all local tensors simultaneously.
We find that the methods are complementary in that local optimization
can change the structure of the basis (i.e., what states with which
quantum numbers are included) on a particular bond between
local tensors in addition to optimizing the elements of these two
tensors, whereas gradient optimization carries out its simultaneous
optimization of all local tensor elements within a fixed basis of all
tensor indices.

Test simulations, presented in Sec.~\ref{sec:results},
show that our implementation of the fPEPS algorithm does, in fact,
achieve systematic variational improvement in the number of
optimization steps if combined local and gradient optimizations are
carried out (what we call the ``supersweep'' procedure), as long as
care is taken that the environment dimension $\chi$ is chosen to be high
enough for all systems and that a suitable initial state is chosen.
For small values of the bond dimension $D$, the convergence saturates
after a certain number of optimization steps; for the larger values of
$D$, such a
saturation does not occur on the scale of computer time accessible to
our calculations.
For all systems with nonzero Hubbard interaction $U$, increasing the bond
dimension $D$ for calculations with either U(1) or SU(2) symmetry
systematically lowers the variational approximation to the
ground-state energy.
Note that this is not so for the $3\times 3$, $U=0$ test case, for which the
best numerical accuracy is achieved for bond dimension $D=5$, and
further increasing $D$ actually increases the approximate ground-state
energy.
A detailed analysis suggests, however, that this behavior can likely
be ascribed to particularities of the $U=0$ state with one hole, namely,
a very inhomogeneous distribution of the single hole over the corner
sites and a near-degeneracy to variations of the hole density on
these sites.
For all systems, calculations with SU(2) symmetry with a given bond
dimension $D$ are substantially more accurate than U(1)-symmetric
calculations with the same $D$.
This is as expected because a single SU(2) state, in general corresponds to a
multiplet of states in the U(1) representation.
Note, however, that an SU(2) calculation is generally substantially
more expensive computationally, both in computer time and in memory,
than a U(1) calculation with the same $D$.
Nevertheless, for all systems except for $8\times 8$, $U=8$, we have
achieved the most accurate results using SU(2) symmetry.

Despite these generally satisfying aspects of the convergence
behavior, the results show that fPEPS is not competitive with 
other numerical methods, in particular, with MPS-based
method such as the DMRG, in calculating the ground state of
the two-dimensional Hubbard model and other two-dimensional quantum
lattice models.
The main obstacle seems to be the poor convergence of the optimization process.
Except for $U=0$, local updates alone do not seem to be able to reach the lowest 
possible energy for a given bond dimension $D$.
Adding gradient updates within the supersweep procedure overcomes this
barrier, but the variational ground-state energy for a given $D$ only
approaches the exact ground state energy $E_0$ after 
an unreasonably large number of optimization steps, as is exemplified
for the $3\times 3$, $U=8$ test system in Fig.~\ref{fig:3x3U8conv};
the runs with larger values of $D$ do not converge in optimization
steps at all for reachable computer time, even for this small test system.
For larger systems, we do observe that increasing $D$ leads to lower
energies, but, due to the computational cost of carrying out these
calculations, it is hard to estimate at what point and at what level of
accuracy of the energy would converge in these runs.
The best relative error in the ground-state energy reached is, in absolute
terms, quite bad for a variational calculation: ca.~1\% for the
$4\times 4$ and ca.~7\% for the $6\times 6$, $U=8$, half-filled
systems, and ca.~10\% for the $8\times 8$, $U=8$, 1/8-doped systems.
Nevertheless, the results do display some aspects that are
encouraging: for the lowest values of $D$, there does seem to be
convergence in optimization steps, even for the $8\times 8$ system (at
least with U(1) symmetry).
The behavior of the spin and hole densities of the doped Hubbard
model on the $8\times 8$ lattice, Figs.~\ref{fig:8x8u1} and
\ref{fig:8x8su2}, is in qualitative agreement with the 
expected stripe configuration known from calculations with other
methods, both for U(1) symmetry, and, to a lesser extent because of
the rather incomplete convergence, for the SU(2) symmetry, despite
the relatively inaccurate ground-state energies.

Thus, we are lead to the question of whether it is possible to
substantially improve the convergence behavior of fPEPS.
This question can be divided into two aspects: First, how well does an fPEPS
approximate a particular many-body ground state on a fundamental
level?
Second, can existing optimization procedures be improved or alternative
optimization procedures be developed so that the optimal fPEPS
for a given bond dimension $D$ can efficiently be found?
As to the first question, our results have both encouraging and
cautionary aspects.
Encouraging is that increasing $D$ or going from U(1) to SU(2) symmetry
does seem to systematically improve the fPEPS approximation to the
ground state.
How this improvement takes place and what $D$ is required to obtain a
given accuracy for given Hamiltonian parameters, particle number,
and system size could only be explored to a limited extent due to the
poor convergence for all but the smallest systems.
Cautionary is the rather sensitive dependence of not only the rate of
convergence, but also the level of apparent convergence after many
optimization steps, on the initial state.

As to the second question, 
the fact that both local optimization and gradient optimization
have features that work in a complementary fashion
indicates that neither method alone is sufficient to carry out an
efficient optimization.
Successive local optimizations of adjacent sites do not seem to lead 
towards a minimum of the entire fPEPS,
whereas gradient-based optimization is, by construction, unable to modify
the bases on the local tensor bonds to form an fPEPS that efficiently
approximates the many-body ground state.
Combining the methods partially overcomes these problems, but the
synergy between the two methods is limited, so that convergence is
still very slow.
A more effective optimization algorithm would presumably have to
combine the basis adaptation of the local updates and the non-local
optimization of PEPS tensors of the gradient updates in a coherent
way.

Note that a large amount of previous work on both fPEPS and iPEPS
algorithms uses imaginary time evolution to optimize the PEPSs.
Usually, this is done by carrying out a Trotter decomposition and applying
terms of the Hamiltonian locally on sites and bonds.
We therefore expect imaginary time evolution to have, at best, a
convergence along the lines of, but poorer than that of our local
optimization scheme.
In addition, imaginary time evolution introduces an additional
systematic Trotter error.

While the DMRG may still be faster and more accurate than fPEPS for all system
sizes treated here, it is limited in scaling two-dimensional systems
to the thermodynamic limit due to the exponential
increase of states needed to maintain accuracy as the lattice width is
increased, which
is due to the entropy area law \cite{Hastings2007Aug}.
PEPSs, on the other hand, are, in principle, capable of efficiently
describing states of arbitrarily
large two-dimensional lattices as long as they satisfy the entropy
area law \cite{Verstraete2004Jul}.
Thus, should it be possible to overcome the poor convergence of fPEPS using a
better optimization scheme, fPEPS could potentially be competitive
with MPS-based methods as two-dimensional lattices are scaled to
larger size.

An additional possible direction for further development is in
applying the methods for representing, manipulating, and optimizing
PEPS-like states developed in this work to the iPEPS algorithm.
The iPEPS method carries out the approximate contraction of the PEPS
tensor network using a corner-transfer matrix rather than the row-wise
contraction into an effective MPS used for fPEPS, so that a number of
aspects of the methods developed here would have to be adapted to the
iPEPS contraction scheme.
iPEPS usually works with either a single translationally invariant
local PEPS tensor or with a unit cell of such tensors of limited size,
used when the physics of the system is expected to break translational
invariance.
The convergence problems in fPEPS occur only for system sizes of
$3\times 3$ or larger because in smaller fPEPS, the environment
reduces to what is essentially a pure MPS.
It would be interesting to investigate the convergence behavior
of local and gradient optimization methods when applied to iPEPS when
a unit cell of $3\times 3$ independent local tensors or more is used.

\begin{acknowledgments}
The authors thank Jan von Delft, Laurens Vanderstraeten and Maarten Van Damme
for fruitful discussions.

\end{acknowledgments}

%\appendix

%\section{Appendixes}

\bibliography{references}

\end{document}